\documentclass[12pt]{article}
\pdfoutput=1

\usepackage[usenames,dvipsnames,svgnames,table]{xcolor} 
\usepackage[obeyspaces,hyphens,spaces]{url}
\usepackage{jcapmod}
\usepackage{verbatim}
\usepackage{graphics}
\usepackage{epsfig}
\usepackage{booktabs}
\usepackage{booktabs}
\usepackage[english]{babel}
\usepackage{amsmath, amssymb, amsbsy, amstext, amsthm, simplewick}
\usepackage{hyperref}
\usepackage{graphicx}
\usepackage{amsfonts}
\usepackage{upgreek}
\usepackage{framed}
\usepackage{tensor}
\usepackage{pifont}
\usepackage{latexsym, mathrsfs}
\usepackage{array}
\usepackage{hyperref}
\usepackage{xspace}
\usepackage{longtable}
\usepackage{multirow}
\usepackage{cases}
\usepackage{empheq}

\usepackage{bm}
\usepackage{bbm}
\usepackage{float}
\usepackage{afterpage}
\usepackage{setspace}
\usepackage{caption}
\usepackage[load-configurations=astronomy, range-units=brackets, range-phrase=-, per-mode=reciprocal, mode=math]{siunitx}
\usepackage{threeparttable}
\usepackage{subfigure}
\usepackage{tcolorbox}
\allowdisplaybreaks 

\usepackage{braket}

\usepackage{ragged2e}

\usepackage{hhline}
\usepackage{array}
\newcolumntype{P}[1]{>{\centering\arraybackslash}p{#1}}
\usepackage{booktabs}
\usepackage{tikz}
\usetikzlibrary{calc,shadings,patterns,tikzmark,fadings}
\usepackage{cleveref} 
\Crefname{equation}{Eq.}{Eqs.}
\Crefname{section}{Sec.}{Secs.}
\Crefname{figure}{Fig.}{Figs.}
\Crefname{table}{Table}{Tables}

\definecolor{Blue}{rgb}{0.25, 0.41, 0.88}
\definecolor{Red}{rgb}{0.92,0.,0.}
\definecolor{darkorange}{rgb}{1.0,0.549,0.}
\definecolor{cobalt}{RGB}{44, 98, 120}
\definecolor{Mathematica1}{rgb}{0.368417, 0.506779, 0.709798}
\definecolor{Mathematica2}{rgb}{0.880722, 0.611041, 0.142051}
\definecolor{Mathematica3}{rgb}{0.560181, 0.691569, 0.194885}
\definecolor{Mathematica4}{rgb}{0.922526, 0.385626, 0.209179}
\definecolor{Mathematica5}{rgb}{0.528488, 0.470624, 0.701351}
\definecolor{Mathematica6}{rgb}{0.772079, 0.431554, 0.102387}
\definecolor{Mathematica7}{rgb}{0.363898, 0.618501, 0.782349}
\definecolor{Mathematica8}{rgb}{1, 0.75, 0}
\definecolor{Mathematica9}{rgb}{0.647624, 0.37816, 0.614037}
\definecolor{plotBlue}{RGB}{94, 130, 181}
\definecolor{plotRed}{RGB}{233, 85, 54}
\definecolor{plotGreen}{RGB}{142, 176, 50}
\definecolor{plotPurple}{RGB}{135, 120, 178}

\newcolumntype{C}[1]{>{\centering\let\newline\\\arraybackslash\hspace{0pt}}m{#1}}

\def\H{{\cal H}}

\def\k{{\bf k}}

\makeatletter
\newlength{\apb@width}
\newcommand{\autoparbox}[2][c]{\settowidth{\apb@width}{#2}\parbox[#1]{\apb@width}{#2}}

\makeatother

\makeatletter
\newsavebox\myboxA
\newsavebox\myboxB
\newlength\mylenA

\newcommand*\xoverline[2][0.75]{
	\sbox{\myboxA}{$\m@th#2$}%
	\setbox\myboxB\null
	\ht\myboxB=\ht\myboxA%
	\dp\myboxB=\dp\myboxA%
	\wd\myboxB=#1\wd\myboxA
	\sbox\myboxB{$\m@th\overline{\copy\myboxB}$}
	\setlength\mylenA{\the\wd\myboxA}
	\addtolength\mylenA{-\the\wd\myboxB}%
	ifdim\wd\myboxB<\wd\myboxA%
	\rlap{\hskip 0.5\mylenA\usebox\myboxB}{\usebox\myboxA}%
	\else
	\hskip -0.5\mylenA\rlap{\usebox\myboxA}{\hskip 0.5\mylenA\usebox\myboxB}%
	\fi}
\makeatother

\numberwithin{equation}{section}
\numberwithin{figure}{section}
\numberwithin{table}{section}

\def\beq{\begin{equation}}
\def\eeq{\end{equation}}

\def\bea{\begin{eqnarray}}
\def\eea{\end{eqnarray}}

\def\beq{\begin{equation}}
\def\eeq{\end{equation}}
\def\bea{\begin{eqnarray}}
\def\eea{\end{eqnarray}}

\def\C{{\cal C}}
\def\N{{\cal N}}

\numberwithin{equation}{section}
\def\beq{\begin{equation}}
\def\eeq{\end{equation}}

\def\bea{\begin{eqnarray}}
\def\eea{\end{eqnarray}}

\def\H{{\cal H}}

\DeclareRobustCommand{\SkipTocEntry}[4]{}

\usepackage{colortbl}
\definecolor{blue2}{cmyk}{1, 0.1, 0.1, 0.1}

\newcommand{\DB}[1]{\textcolor{red}{#1}}

\newcommand{\HS}[1]{\textcolor{blue}{#1}}

\definecolor{pyBlue}{RGB}{31, 119, 180}
\definecolor{pyRed}{RGB}{214, 39, 40}
\definecolor{pyGreen}{RGB}{44, 160, 44}
\definecolor{pyBlue2}{RGB}{0, 111, 237}
\definecolor{pyRed2}{RGB}{224, 52, 36}

\newcolumntype{P}[1]{>{\centering\arraybackslash}p{#1}}
\newcolumntype{M}[1]{>{\centering\arraybackslash}m{#1}}

\begin{document}
	
	\tcbset{colframe=black,arc=0mm,box align=center,halign=left,valign=center,top=-10pt}
	
	\renewcommand{\thefootnote}{\fnsymbol{footnote}}
	
	\pagenumbering{roman}
	\begin{titlepage}
		\baselineskip=5.5pt \thispagestyle{empty}
		
		\bigskip\
		
		\vspace{0.2cm}
		\begin{center}
			{\Huge \textcolor{Sepia}{\bf \sffamily Q}}{\large \textcolor{Sepia}{\bf\sffamily UANTUM}} ~{\Huge \textcolor{Sepia}{\bf \sffamily A}}{\large \textcolor{Sepia}{\bf\sffamily SPECTS~OF}} ~{\Huge \textcolor{Sepia}{\bf \sffamily C}}{\large \textcolor{Sepia}{\bf\sffamily HAOS}} ~{\large \textcolor{Sepia}{\bf\sffamily AND}}~ {\Huge \textcolor{Sepia}{\bf\sffamily C}}{\large \textcolor{Sepia}{\bf\sffamily OMPLEXITY}}\vspace{0.45cm} \\{\large \textcolor{Sepia}{\bf\sffamily F}}{\large \textcolor{Sepia}{\bf\sffamily ROM}} \\ \vspace{0.25cm} {\Huge \textcolor{Sepia}{\bf\sffamily B}}{\large \textcolor{Sepia}{\bf\sffamily OUNCING }}~
			{\Huge \textcolor{Sepia}{\bf\sffamily C}}{\large \textcolor{Sepia}{\bf\sffamily OSMOLOGY}}\\  \vspace{0.25cm}
			{\fontsize{13}{16}\selectfont  \bfseries \textcolor{Sepia}{A study with two-mode single field squeezed state formalism}}
		\end{center}
		
		\vspace{0.05cm}

			\begin{center}
				{\fontsize{12}{18}\selectfont Parth Bhargava${}^{\textcolor{Sepia}{1}}$},
				{\fontsize{12}{18}\selectfont Sayantan Choudhury${}^{\textcolor{Sepia}{2},{3},{4}}$\footnote{{\sffamily \textit{ Corresponding author, E-mail}} : {\ttfamily sayantan.choudhury@aei.mpg.de, sayanphysicsisi@gmail.com}}}${{}^{,}}$
				\footnote{{\sffamily \textit{ NOTE: This project is the part of the non-profit virtual international research consortium ``Quantum Aspects of Space-Time \& Matter" (QASTM)} }. }, 
				{\fontsize{12}{18}\selectfont Satyaki Chowdhury${}^{\textcolor{Sepia}{3,4}}$},
				{\fontsize{12}{18}\selectfont Anurag Mishara${}^{\textcolor{Sepia}{5}}$},
				{\fontsize{12}{18}\selectfont Sachin Panneer Selvam${}^{\textcolor{Sepia}{6}}$},
				{\fontsize{12}{18}\selectfont Sudhakar Panda ${}^{\textcolor{Sepia}{3,4}}$}
				{\fontsize{12}{18}\selectfont Gabriel D. Pasquino${}^{\textcolor{Sepia}{7}}$},
				
			\end{center}

		
		\begin{center}
			\vskip4pt
			\scriptsize{ \sffamily
				\textit{${}^{1}$Institute for Theoretical Particle Physics and Cosmology(TTK), RWTH Aachen University, D-52056, Aachen, Germany}\\
				\textit{${}^{2}$Quantum Gravity and Unified Theory and Theoretical Cosmology Group, \\Max Planck Institute for Gravitational Physics (Albert Einstein Institute),\\
					Am M$\ddot{u}$hlenberg 1,
					14476 Potsdam-Golm, Germany.}
				\\
				\textit{${}^{3}$National Institute of Science Education and Research, Bhubaneswar, Odisha - 752050, India}\\
				\textit{${}^{4}$Homi Bhabha National Institute, Training School Complex, Anushakti Nagar, Mumbai - 400085, India}\\
				\textit{${}^{5}$ Department of Physics and Astronomy, National Institute of Technology, Rourkela, Odisha, 769001}\\
				\textit{${}^{6}$ Department of Physics, Birla Institute of Technology and Science, Pilani, Hyderabad Campus, Hyderabad - 500078, India}\\
				\textit{${}^{7}$University  of Waterloo, 200 University Ave W, Waterloo, ON, Canada, N2L 3G1}\\
			}
		\end{center}
		
		\vspace{0.2cm}
		\hrule 
		\begin{center}
			\textbf{Abstract}
		\end{center}
Circuit Complexity, a well known computational technique has recently become the backbone of the physics community to probe the chaotic behaviour and random quantum fluctuations of quantum fields. This paper is devoted to the study of out-of-equilibrium aspects and quantum chaos appearing in the universe from the paradigm of two well known bouncing cosmological solutions viz. \textit{Cosine hyperbolic} and \textit{Exponential} models of scale factors. Besides circuit complexity, we use the Out-of-Time Ordered correlation (OTOC) functions for probing the random behaviour of the universe both at early and the late times. In particular, we use the techniques of well known two-mode squeezed state formalism in cosmological perturbation theory as a key ingredient for the purpose of our computation. To give an appropriate theoretical interpretation that is consistent with the observational perspective we use the scale factor and the number of e-foldings as a dynamical variable instead of conformal time for this computation. From this study, we found that the period of post bounce is the most interesting one. Though it may not be immediately visible but an exponential rise can be seen in the complexity once the post bounce feature is extrapolated to the present time scales. We also find within the very small acceptable error range a universal connecting relation between Complexity computed from two different kinds of cost functionals-linearly weighted and geodesic weighted with the OTOC. Furthermore, from the complexity computation obtained from both the cosmological models under consideration and also using the well known Maldacena (M) Shenker (S) Stanford (S) bound on quantum Lyapunov exponent, $\lambda\leq 2\pi/\beta$ for the saturation of chaos, we estimate the lower bound on the equilibrium temperature of our universe at the late time scale. Finally, we provide a rough estimation of the scrambling time scale in terms of the conformal time. 
		\noindent

		\vskip10pt
		\hrule
		\vskip10pt
		
		\noindent
		\text{Keywords: Complexity, Bouncing Cosmology, Cosmology beyond the standard model.} 
		
	\end{titlepage}

\newpage
\setcounter{tocdepth}{2}

  \makebox[0pt][l]{%
  \raisebox{-0.934\totalheight}[0pt][0pt]{%
\hskip -20pt
    \includegraphics[width=20cm,height=27.5cm]{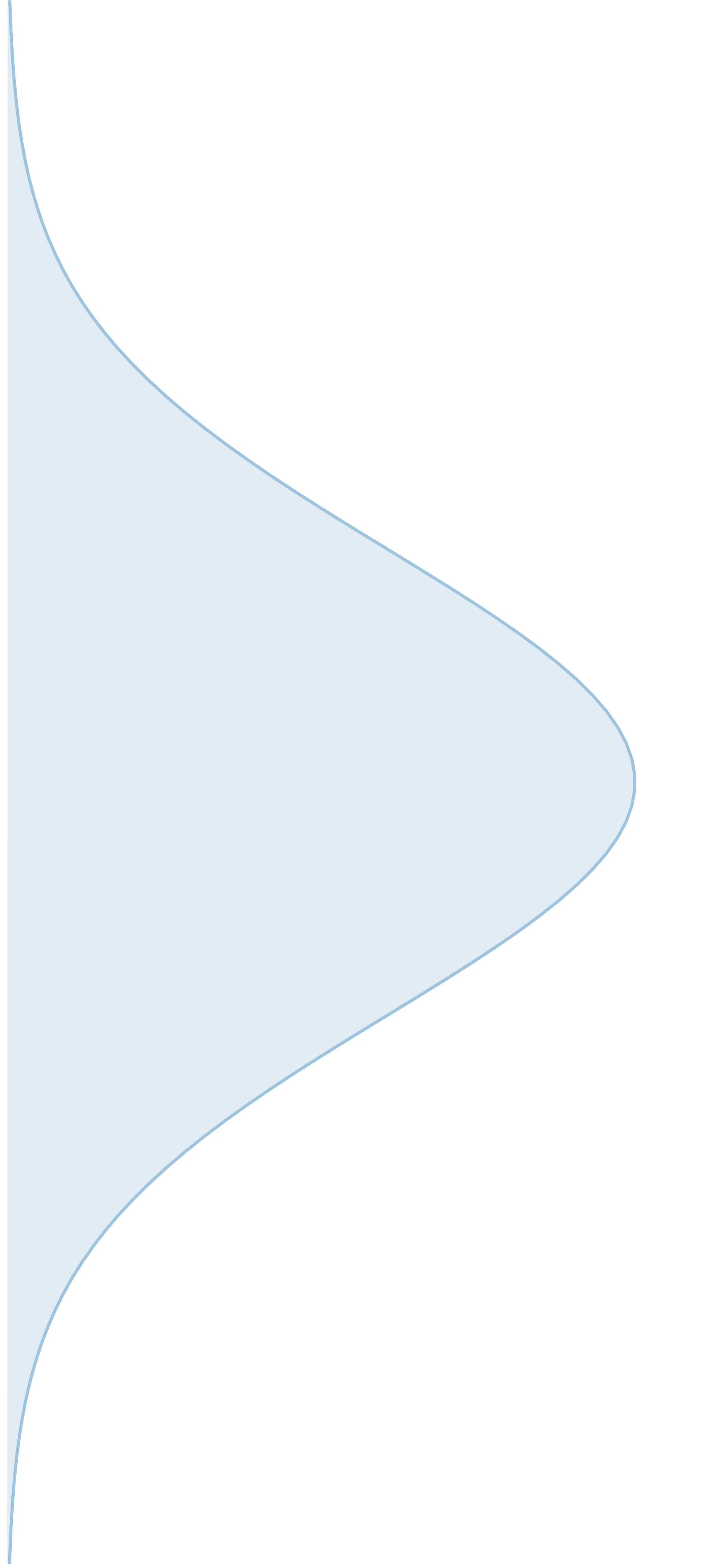}}}%
\vskip 0cm
\tableofcontents

\newpage
	
	\clearpage
	\pagenumbering{arabic}
	\setcounter{page}{1}
	
	\renewcommand{\thefootnote}{\arabic{footnote}}
	
\textcolor{Sepia}{\section{\sffamily Introduction}\label{sec:introduction}}
The idea of \textit{circuit complexity} \cite{Susskind:2018pmk,Jefferson:2017sdb,Guo:2018kzl,Chapman:2016hwi,Chapman:2018dem,Chapman:2018lsv,Caceres:2019pgf,Carmi:2017jqz,Bueno:2019ajd,Bhattacharyya:2019kvj,Bhattacharyya:2020kgu,Bhattacharyya:2020rpy,Guo:2020dsi,Jiang:2018nzg,Adhikari:2021pvv,Choudhury:2020hil,Choudhury:2020lja,Choudhury:2021qod} has recently gained huge attraction of the theoretical physics community and is recently used as a diagnostic for Quantum chaos \cite{Maldacena:2015waa,Shenker:2013pqa,Stanford:2014jda,Choudhury:2017tax,Maldacena:2016hyu,Xu:2018dfp,Gharibyan:2018fax,Zhuang:2019jyq,Yoshida:2019qqw,Couch:2019zni,Sahu:2020xdn}. The absence of a proper tool to develop a wholesome understanding about the AdS/CFT correspondence\cite{Maldacena:1997re} in certain black hole settings is what motivated the high energy theoretical physics community to apply this computational concept in the context of Quantum Field Theory (QFT). The information about the bulk geometry that can be extracted from the boundary Conformal Field Theory (CFT) remains very much incomplete and is one of the toughest challenges that one faces when probing black hole physics beyond the horizon. One of the main difficulties in boundary field theories is that it reaches thermal equilibrium very quickly while the Einstein-Rosen bridge continues to grow. These challenges motivated Leonard Susskind and collaborators to propose the \textit{Complexity=Volume} and \textit{Compexity=Action} conjectures to probe gravity beyond the horizon of black holes and have led to the development of enormous new ideas about the application of complexity and other information theoretic measures in the gravity sector \cite{Susskind:2018pmk,Susskind:2018vql,Susskind:2019ddc,Susskind:2020gnl,Brown:2016wib,Brown:2017jil}. The CV conjecture suggests that the holographic complexity of the boundary field theory is equal to the volume of an extremal codimension one surface extending the boundary time slice into the bulk whereas the CA conjecture suggests that the complexity of a boundary state is dual to the gravitational action evaluated on the Wheeler-DeWitt patch. However the traditional way of computing complexity has certain shortcomings when applied to holography and QFT states. Generally in these contexts, one considers a continuum of states and a proper way to define complexity in this continuum of states faces several questions that need to be addressed. To name some of them, selecting the initial reference state, a set of infinitesimal unitary generators or quantum gates, a proper measure for understanding the role of these gates in minimizing the distance function and the procedure it follows. One of the proposals for facing these issues is to compute quantum complexity using the path length obtained by integrating the {\it Fubini study line element} joining the reference and the target state. The reference state is mainly chosen to be Gaussian because the ground states of free field theories are in general Gaussian. For Gaussian quantum states, a geometric way of computing the complexity was given in \cite{nielsen2006quantum,dowling2006geometry,nielsen2005geometric}. It includes two different methods commonly known as the wave-function approach\cite{Jefferson:2017sdb} or the covariance matrix approach \cite{Khan:2018rzm,Hackl:2018ptj}. The wave function approach has been found to be the most insightful one to probe the underlying physics, especially in the context of time evolution.

Sharing an intimate relation with the \textit{Out-Of-Time-Ordered-Correlation} functions \cite{larkin1969quasiclassical,Bhagat:2020pcd,Choudhury:2020yaa}, abbreviated as OTOC, these two measures has been the recent tools to probe quantum randomness and chaos in various quantum mechanical systems. OTOC's which first appeared in literature in the context of superconductivity \cite{larkin1969quasiclassical} soon became popular as a theoretical probe to explore the out of equilibrium phenomenon in finite temperature field theories, bulk gravitational theories and many-body quantum systems. A lot of investigation has followed since then to conclude that whether OTOC's can be considered as a good measure to study stochastic randomness and chaos of quantum systems at out of equilibrium phase. Together with OTOCs, complexity is now considered to be an integral part of the machinery used in the diagnosis of quantum randomness and chaos. Both of these measures have been found to provide information like {\it Lyapunov exponent}, {\it scrambling time} etc., which are by far the most essential quantities required to comment on the chaoticity of any quantum mechanical system.  

In this work, our attempt will be to apply this quantum information theoretic measure to the framework of bouncing cosmological paradigm. \textit{Bouncing cosmology} is gaining traction to resolve the problem of Big Bang Singularity in recent years \cite{Lehners:2009qu,Lehners:2011ig,Lehners:2015mra,Anabalon:2019equ,Fertig:2015dva,Fertig:2016czu,Cai:2007zv,Cai:2008qw,Cai:2011zx,Cai:2012va,Cai:2013vm,Cook:2020oaj,Bars:2011aa,Bars:2012mt,Battarra:2014tga,Bramberger:2019zez,Brandenberger:2009rs,Brandenberger:2010dk,Brandenberger:2012zb,Brandenberger:2013zea,Koehn:2013upa,Koehn:2015vvy,Karouby:2010wt,Ijjas:2015hcc,Xue:2011nw,Alexander:2007zm,Gao:2009wn}. A solid model in bouncing cosmology can resolve the \textit{Horizon problem}, \textit{Flatness problem}, the \textit{CMB Inhomgeneity} and other problems that are prevalent in the current model of Big Bang and Inflationary cosmology\cite{Panda:2006mw,Panda:2007ie,Baumann:2010nu,Baumann:2010ys,Baumann:2014cja,Baumann:2014nda,Baumann:2015nta,Baumann:2015xxa,Baumann:2017jvh,Choudhury:2003vr,Choudhury:2011jt,Choudhury:2011rz,Choudhury:2011sq,Choudhury:2012ib,Choudhury:2012yh,Choudhury:2013iaa,Choudhury:2014kma,Choudhury:2015pqa,Mazumdar:2001mm,Assassi:2012et,Assassi:2012zq,Choudhury:2014sxa,Choudhury:2014uxa,Choudhury:2013jya,Choudhury:2013zna,Choudhury:2015yna,Mazumdar:2010sa,Allahverdi:2010xz,Biswas:2012bp,Biswas:2013dry,Salvio:2015kka,Silverstein:2013wua,Liddle:2000cg,Abazajian:2013vfg,Baumann:2009ds,Baumann:2018muz,Senatore:2016aui,Choudhury:2014sua,Choudhury:2015hvr,Choudhury:2013woa}. One way of getting a non-singular ghost free bouncing models is through non-local infinite derivative gravity theories with an addition of appropriate non-local function in the Einstein-Hilbert action in the ultraviolet regime that captures all the derivative terms \cite{Brandenberger_2017,Lin:2010pf,Kumar:2020xsl,Battefeld:2004cd,Battefeld:2014uga}. Moreover non singular bouncing solutions of a positive cosmological constant can make inflation geodesically complete \cite{Biswas:2010zk}. The primary motivation to apply the formalism of cosmological complexity in bouncing background is that the study of complexity can give great insight about a given model in bouncing cosmology and the explicit calculation of the Lyapunov exponent and the corresponding lower bound on equilibrium temperature \cite{Maldacena:2015waa} during the bouncing period can be very useful in our understanding of primordial cosmology. 
In this paper, we intend to apply this concept of cosmological complexity under a squeezed state formalism with scalar cosmological perturbations to two well known bouncing solutions - the cosine hyperbolic bounce \cite{Biswas:2010zk,Biswas_2006} and the exponential bounce \cite{Kumar:2020xsl}, which we have derived from usual Einstein gravity with two different models of dynamical scalar matter field embedded in spatially flat ($k=0$) Friedmann-Lemaitre-Robertson-Walker (FLRW) cosmological background in $3+1$ dimensions. However, the exact same solutions can also be derived from higher derivative non-local gravity theory admitting isotropic and homogeneous bouncing universes in the absence of matter \cite{Biswas:2010zk,Biswas_2006}.

\begin{figure}[!htb]
	\centering
	\includegraphics[width=18cm,height=20.5cm]{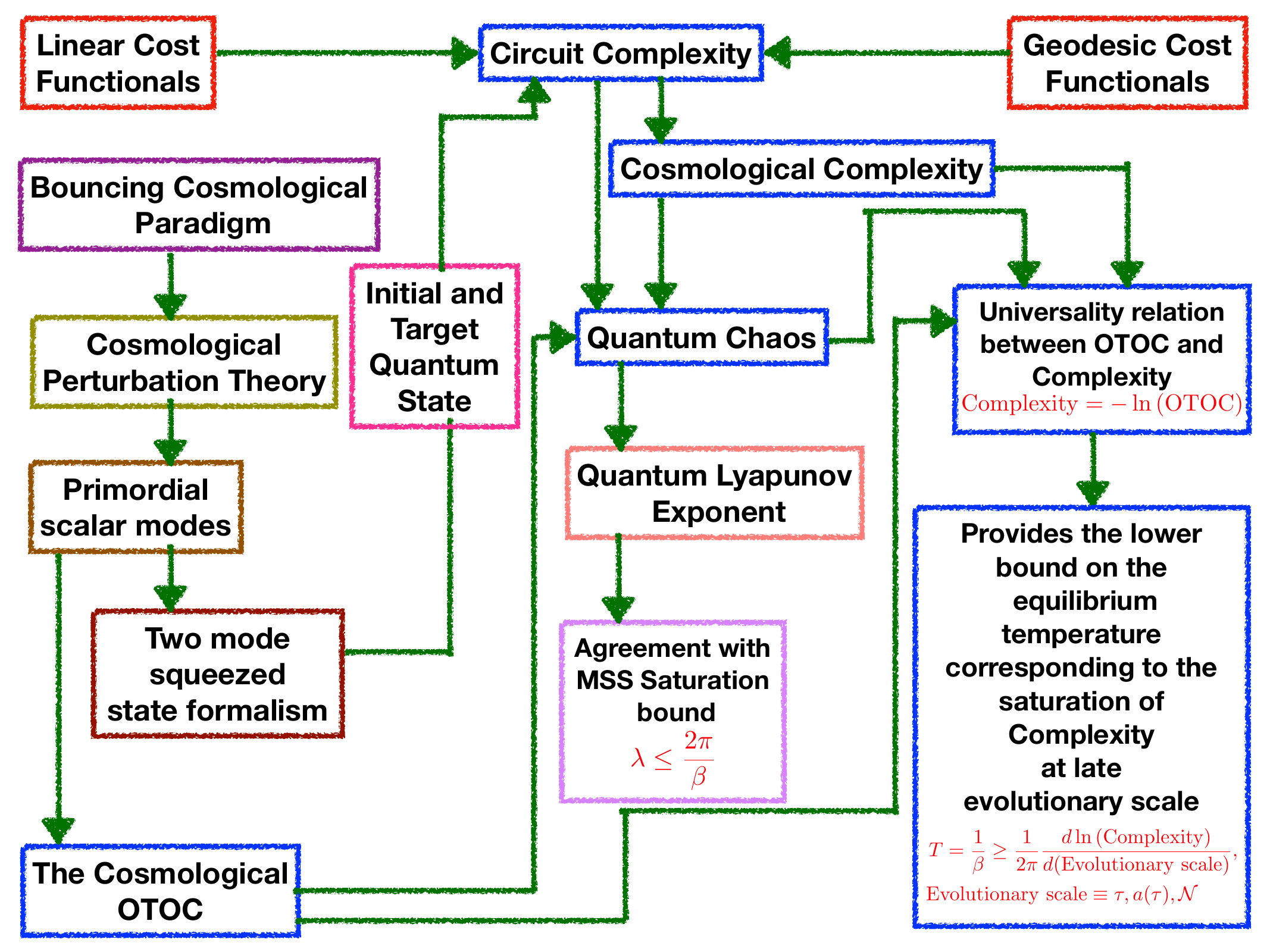}
	\caption{Flowchart showing the plan of the entire work}
	\label{bounce}
\end{figure}

We have developed a framework for bouncing cosmology from potentials derived from String theory descriptions at very high energy scale, that can be treated with the squeezed state formalism \cite{Albrecht:1992kf,Grishchuk:1990bj,Martin:2003bt,Martin:2016nrr,Martin:2016tbd,Ando:2020kdz,Grain:2019vnq}, and using that result the cosmological complexity can be further analyzed. We write a generalized scalar perturbation in the framework of bouncing cosmology and expressed the action, and its parameters including the dispersion relation without truncating higher order terms initially and then give the limiting solutions in the sub-Hubble, Horizon crossing and the super-Hubble regions. The Hamiltonian is also written in its most general form, as compared to \cite{Albrecht:1992kf} before fixing the initial conditions at the horizon crossing scale at $k\tau = -1$ and formulating the squeezed states with a next-to-leading order time dependent slowly varying term in the dispersion relation that we found after the quantization of the Hamiltonian to be more relevant in the context of bouncing cosmology~\footnote{\underline{\textcolor{red}{\bf Note:}}~~In refs.~\cite{Albrecht:1992kf,Bhattacharyya:2020rpy,Bhattacharyya:2020kgu} the authors have not considered the slowly varying contribution in the evolution in the sub-Hubble region ($-k\tau \gg 1$) in their computation. During describing the inflationary paradigm all of them have considered the exact de Sitter solution, which is in realistic cosmological analysis is not very useful and also appropriate. The prime reason is using exact de Sitter solution one cannot able to stop the inflation at all in the evolutionary time scale or equivalently in the field space. To stop inflation in an appropriate field space one needs to include slow-roll parameters, which basically considering the small but significant deviation from exact de Sitter solution. When the slow roll parameter reaches the unity the end of inflation is ensured. }. Other works in cosmological complexity \cite{Bhattacharyya:2020rpy,Bhattacharyya:2020kgu}, have only considered the leading constants in the dispersion relation and squeezed state formalism under the assumptions of stationary background space time. We have then focused our further analysis with bounce in the sub-Hubble region ($-k\tau \gg 1$) to get a better analysis of the quantum fluctuations as compared to the super-Hubble region which falls under the classical domain. This is where the necessary approximations to the dispersion relation is made and the complexity cost functions based on an early general description of family of cost functions is derived. A universality relation between the OTOC and the complexity has also been given under certain conditions. We make certain key observations from our numerical analysis including:\\
\begin{itemize}
	\item \underline{\textcolor{red}{\bf Observation~I:}}\\
	Behaviour of squeeze parameters in and around the bounce and at late times.
	\item \underline{\textcolor{red}{\bf Observation~II:}}\\
	Initially fluctuating complexity that grows at later times and achieves a saturation at very large time.
	\item \underline{\textcolor{red}{\bf Observation~III:}}\\
	There exists a smooth transition between the non-equilibrium growing phase and the equilibrium saturating phase.
	\item \underline{\textcolor{red}{\bf Observation~IV:}}\\
	The saturation at late times indicates a bound on chaos, which makes it possible to describe the Lyapunov exponent and the lower bound of the equilibrium temperature using the well known, {\it Maldacena(M) Shenker(S) Stanford(S) saturation bound} on 
quantum Lyapunov exponent \cite{Maldacena:2015waa}~\footnote{It is important to note that, some other extension of this bound have been studied in the refs.~\cite{Murthy:2019fgs}. },	
	 \bea &&\textcolor{blue}{\textbf{MSS Bound}:} ~~~\lambda\leq \frac{2\pi}{\beta}, ~~~~{\rm where}~~~ \beta=\frac{1}{T}~~~~{\rm with }~~k_{\rm B}=1, \hbar=1,\eea 
	 where $T$ is the equilibrium temperature corresponding to saturation of quantum chaos at the late time scale.
	\item \underline{\textcolor{red}{\bf Observation~V:}}\\
	The two different measures used for complexity point to Lyapunov exponent whose fractional deviation is under ten percent, and hence it is safe to assume that our universality relation holds perfectly, in the context of our study.	
	\item \underline{\textcolor{red}{\bf Observation~VI:}}\\
	We have also very roughly estimated the scrambling time \cite{Sekino:2008he,Yoshida:2017non} for both models and found them to be decent indicator of the time that the OTOC may take to attain equilibrium in both cases. 
\end{itemize}
We expect the bound on quantum chaos and hence the resulting Lyapunov exponent from the two measures of complexity to be much more closer in value by doing the analysis with a full dispersion relation given in the paper. 
We had initially done the numerical analysis against scale factor for simplicity, but to connect with the observational constraints we have extended the analysis of the complexity in the bouncing background with respect to the co-moving Hubble radius as well, which can be further expressed in terms of the number of e-foldings. It is expected from the present study that this theoretical formulation and the corresponding analysis of cosmological complexity and, its connection with quantum chaos through OTOC could act as a very strong theoretical indicator for future observational probes for studying non-equilibrium physics within the framework of bouncing cosmology.\\

 \textcolor{blue}{\textbf{\underline{Organization of the Paper:}}}
\begin{itemize}
\item In \Cref{{sec:circuitcomplexity}} a brief review of the concept of circuit complexity has been given and how it can be used to probe new areas of physics in the context of Cosmology.

\item  \Cref{{sec:bouncingcosmology}} introduces the reader to the framework of Bouncing cosmology and the models that we have considered for the computation of complexity. 

\item In \Cref{{sec:cosperwsqueezedQS}} a detailed computation of the cosmological scalar perturbations in the bouncing cosmology framework has been provided along with the origin of the squeezed quantum states and its various solutions. 

\item In \Cref{{sec:coscompwsqueezedQS}} a discussion on the complexity for the squeezed quantum states has been given.

\item Finally in \Cref{{sec:tybouncesol}} the computational details of the considered models has been provided with all the relevant discussions. We conclude with all our major observations and future prospects in this direction.

\end{itemize}

\textcolor{Sepia}{\section{\sffamily Circuit Complexity for dummies }\label{sec:circuitcomplexity}}
The concept of circuit complexity was primarily used in the field of Computer Science to know the depth of different circuits. It is basically defined as the effort required to carry out a given task or the difficulty in implementing a given task. The task at hand is essentially to prepare a desired quantum field theoretic target state from a reference state.
It is generally an optimization technique. Technically, it refers to the minimum number of unitary operations required to implement a given task.  The process of carrying out the task involves constructing a unitary transformation that takes a given reference state to the desired final state. The unitary operator being referred to here usually represents the sequences of quantum gates $\{g_{i_1},g_{i_2},....g_{i_n}\}$ required to achieve the desired the target state.
\begin{align}
\ket{\psi_T}=U\ket{\psi_R}= g_{i_1},g_{i_2},....g_{i_n}\ket{\psi_R}
\end{align}
Of course, there exist infinite such sequences that produce the desired target state from the given reference state, but the complexity of a quantum circuit provides the sequence which requires the minimum number of gates to do so. This optimal number will depend on the choice of the reference state, $\ket{\psi_R}$ and the gate set $\{g_{1},g_{2},....g_{n}\}$. The construction of the unitary operator involves finding a time-dependent Hamiltonian that produces the desired $U$. The unitary operator is then constructed from a continuous sequence of parametrized path-ordered exponential of the chosen Hamiltonian,
\begin{align}
U(s)=\overleftarrow{\mathcal{P}}\exp\left(\displaystyle -i\int_{0}^{s}ds' ~H(s')\right).
\end{align}  
The variable $s$ parametrizes a path in the space of unitaries. The Hamiltonian $H(s)$ can be expanded in terms of generalized Pauli matrices i.e.
\begin{align}
H(s)= \sum_{I}Y^{I}(s)M_{I},
\end{align}
where $M_{I}$ are the basis in which the Hamiltonian is expanded and the coefficients $Y^{I}(s)$ are the control functions that decide the gate acting at certain values of the parameters. The control function basically represents a tangent in the space of unitaries and acts as the Hamiltonian in the Schr\"{o}dinger equation satisfied by the unitarity operator $U$,
\begin{align}
\frac{dU(s)}{ds}=-iY(s)^I M_I U(s).
\end{align}
The idea then is to define a cost for the various possible paths, minimizing which leads to the identification of the optimal circuit. The cost functional is defined as follows:
\begin{equation}
\label{costfunc}
\mathcal{D}(U(s))=\int_{0}^{1} dt~ F(U(s),\dot{U}(s)),
\end{equation} 
where $F$ is a local cost function depending on the position $U(s)$ and the tangent vector $Y^I(s)$. Once the concept of cost function is introduced, the problem is identical to finding the trajectory of a particle by minimizing the action from the Lagrangian $F(U(s), Y^I(s))$.
There are certain desirable features for $F$ to be a cost functional \cite{Jefferson:2017sdb} viz. \textit{smoothness, positivity, triangle inequality} and \textit{positive homogenity}. Some of the simplest cost functionals which satisfy the above properties and the ones which we have considered in this paper are the linear and the quadratic cost functionals defined as \cite{Jefferson:2017sdb}:
\bea
&&\textcolor{red}{\bf Linear~cost~functional:}~~~~~~~~~~~~~~~~~F_1:=\sum_I |Y^I(s)| ,\\
&&\textcolor{red}{\bf Quadratic~cost~functional:}~~~~~~~~~~~ F_2:= \sqrt{\sum_I (Y^I(s))^2},
\eea 
where the degree of homogeneity is $1$ for both of them. 

To be precise the cost function $F_1$ comes closest to counting the number of gates required to make the optimal circuit. The measure $F_2$ however brings in a notion of proper distance in Riemannian geometry and converts the problem of constructing the optimal circuit to finding the shortest curve connecting the initial and the final states in that geometry. Some other types of cost functionals are also discussed in \cite{Bueno:2019ajd,Jefferson:2017sdb}. 

On the other hand, a general class of inhomogeneous and homogeneous family of functionals are represented by the following expression \cite{Bueno:2019ajd}:
\bea &&\textcolor{red}{\bf F_{\kappa}~family~cost~functional:}~~~~~~~~~~~~~~~~F_{\kappa}:=\sum_{I} |Y^I(s)|^{\kappa},\\
&&\textcolor{red}{\bf \left[F_{\kappa}\right]^{\frac{1}{\kappa}}~family~cost~functional:}~~~~~~~~~~~\left[F_{\kappa}\right]^{\frac{1}{\kappa}}=\left[\sum_{I} |Y^I(s)|^{\kappa}\right]^{\frac{1}{\kappa}},\eea
where for all family members, the degree of homogeneity is represented by the superscript, $\kappa>1$. Here the inhomogeneous family of functionals, $F_{\kappa}$ was introduced to match the results obtained from both the leading order UV divergences appearing from the well known, 
“complexity= action” \cite{Brown:2016wib,Brown:2017jil} and “complexity=volume” \cite{Brown:2016wib,Brown:2017jil} conjectures proposed within the framework of holography.
Though the results agreed with with the holographic complexity results, these cost functions do not satisfy the homogeneity property i.e \Cref{costfunc} is not invariant under the reparametrizations of s for these $\kappa$ family cost functions.
Apart from these previously mentioned measures, one can further introduce the following sets of basis independent and state independent cost functionals, which are given by \cite{Bueno:2019ajd}:
\bea &&\textcolor{red}{\bf Trace~norm~cost~functional:}~~~~~~~~~~~~~~~~F_{|{\rm Tr}~H(s)|}:=|{\rm Tr}~H(s)|,~~~~~\\
&&\textcolor{red}{\bf Schatten~ norm~cost~functional:}~~~~~~~~~~~F_{\rm Sch}:=\left[{\rm Tr}\left((H^2(s))^{\frac{p}{2}}\right)\right]^{\frac{1}{p}}.\eea

The Schatten norm cost functional helps to express the circuit complexity in a basis independent way, a problem which occurs with the general $\kappa$ family of cost functions including the linear and the quadratic ones.
Further, one can construct few more state dependent cost functionals which are given by the following expressions \cite{Bueno:2019ajd}:
\bea &&F_{\langle H^2\rangle}=\sqrt{\langle \psi(s)|H^2(s)|\psi(s)\rangle},\\
&&F_{|\langle H\rangle|}=\left|\langle \psi(s)|H(s)|\psi(s)\rangle\right|,\\ 
&&F_{\rm FS}=\sqrt{F_{\langle H^2\rangle}-F^2_{|\langle H\rangle|}}\equiv F_{\sigma^2}\eea

In the context of cosmology, using the quantum squeezed state formalism in the perturbation picture enables one to compute the expression for the cosmological complexity.  Scalar perturbations on an expanding background can naturally be described with the formalism of squeezed quantum
	states. The ground state is chosen as the reference state while the mode is inside the horizon,
	and a target state consisting of the time-evolved cosmological perturbation on the expanding
	background. Thus squeezed state formalism gives an elegant way of defining the reference and the target state between which the circuit complexity can be computed. The squeezed state formalism also enables to translate the entire problem in terms of just two quantities known as the squeezed state parameter and the squeezed angle. The whole idea of squeezed state formalism can be easily understood using the well known model of inverted harmonic oscillator. In this formalism, the wave function is squeezed with a large uncertainty in one direction and with a small uncertainty in another direction. Similar observations can be
	found if one looks into the phase space trajectories of an inverted harmonic oscillator. The presence of one growing and one decaying solution produces a squeezing effect even in the classical level. The main idea behind the squeezed states is to re-parametrize the unitary operator as the product of a squeezed and a rotation operator. The squeezed and the rotation operator can further be expressed entirely in terms of the creation and annihilation operators. The significance of the rotation operator is less as it mainly
	produces a phase factor. However, the squeezing operator is of prime significance
	as the entire problem and all the important observables can eventually be expressed
	in terms of two quantities, the squeezing parameter and the squeezing angle. 
	Thus, the squeezed state formalism not only gives an elegant way of finding the
	target and the reference state but also helps to express all the important observables
	in terms of only two quantities as will be seen in the upcoming sections.

Here, the complexity can be defined in terms of all the previously mentioned different types of cost functionals and one can test as to which ones do give out the best features in terms of the study of quantum chaos for a given cosmological model of our universe. However, in this paper we have restricted our computation by considering only the cost functionals, $F_1$ and $F_2$ from which we compute the expression for cosmological complexity. Using the universality relation, we have further computed the expression for OTOC, {\it Quantum Lyapunov exponent} and the lower bound on the equilibrium temperature of a system within the framework of bouncing cosmological paradigm.\\
\textcolor{Sepia}{\section{\sffamily A simple framework for Bouncing Cosmology}\label{sec:bouncingcosmology}}
In this section, our prime objective is to construct a bouncing cosmological framework that can further participate in the computation of cosmological complexity. In the present context, we start with the following representative action, given by:
\bea S=\frac{1}{2}\int d^4x\sqrt{-g}\left[R-\left(\partial\phi\right)^2-2V(\phi)\right],\eea
where we have fixed the reduced Planck mass $M_p=1$ for the simplification of the computation. We have introduced a single scalar field with a kinetic term that is minimally coupled with the classical gravitational background. Here $V(\phi)$ is the effective potential for the scalar field $\phi$ in $3+1$ dimensions from which we will describe pre-bounce, bounce, and post bounce scenario. We consider here two models which can serve our purpose:
\bea
&&\underline{\textcolor{blue}{\bf Model~I: Cosine~Hyperbolic~model}}\nonumber\\
\displaystyle\displaystyle &&V(\phi)= \left\{ 
     \begin{array}{lr}
\displaystyle   \frac{48V_0}{(1+r_1)^2}~\exp\left(-\sqrt{3(1+r_1)}~\phi\right)&~ \text{\textcolor{red}{\bf Pre-Bounce~($t<-t_{\rm B}$)}}\\ 
    \displaystyle\frac{3r_1V_0}{2}\left[1-\cosh^2\left(\frac{2}{3\sqrt{r_1}}\phi\right)\right]~~~~~~~~~&~~~~~ \displaystyle\text{\textcolor{red}{\bf Bounce~~~($-t_{\rm B}<t<t_{\rm B}$)}}\\ 
  \displaystyle  \frac{48V_0}{(1+r_1)^2}~\exp\left(\sqrt{3(1+r_1)}~\phi\right)~~~~&\text{\textcolor{red}{\bf Post-Bounce~($t>t_{\rm B}$)}} \end{array}
   \right.~~~~~~\eea

\bea
&&\underline{\textcolor{blue}{\bf Model~II: Exponential~model}}\nonumber\\
\displaystyle\displaystyle && V(\phi)=  \left\{ 
     \begin{array}{lr}
\displaystyle    \frac{4V_0}{3(1+r_1)^2}~\exp\left(-\sqrt{3(1+r_1)}~\phi\right)   &~ \text{\textcolor{red}{\bf Pre-Bounce~($t<-t_{\rm B}$)}}\\ 
    \displaystyle\frac{1}{2}m^2_{\phi}\phi^2~~~{\rm with}~~m^2_{\phi}=9\sqrt{3}r_1V_0~~~~&\displaystyle\text{\textcolor{red}{\bf Bounce~~~($-t_{\rm B}<t<t_{\rm B}$)}}\\ 
  \displaystyle  \frac{4V_0}{3(1+r_1)^2}~\exp\left(\sqrt{3(1+r_1)}~\phi\right)  ~~~~~~~&\text{\textcolor{red}{\bf Post-Bounce~($t>t_{\rm B}$)}} \end{array}
   \right.~~~~~~
\eea
where $r_1$ is the dimensionless parameter in the Planckian units for both of the bouncing models and $V_0$ represents the overall energy scale of the potential which mimics the role of Cosmological Constant at a very high energy scale.
\begin{figure}[!htb]
	\centering
	\subfigure[][Model I]{\includegraphics[height=9.2cm,width=7.7cm]{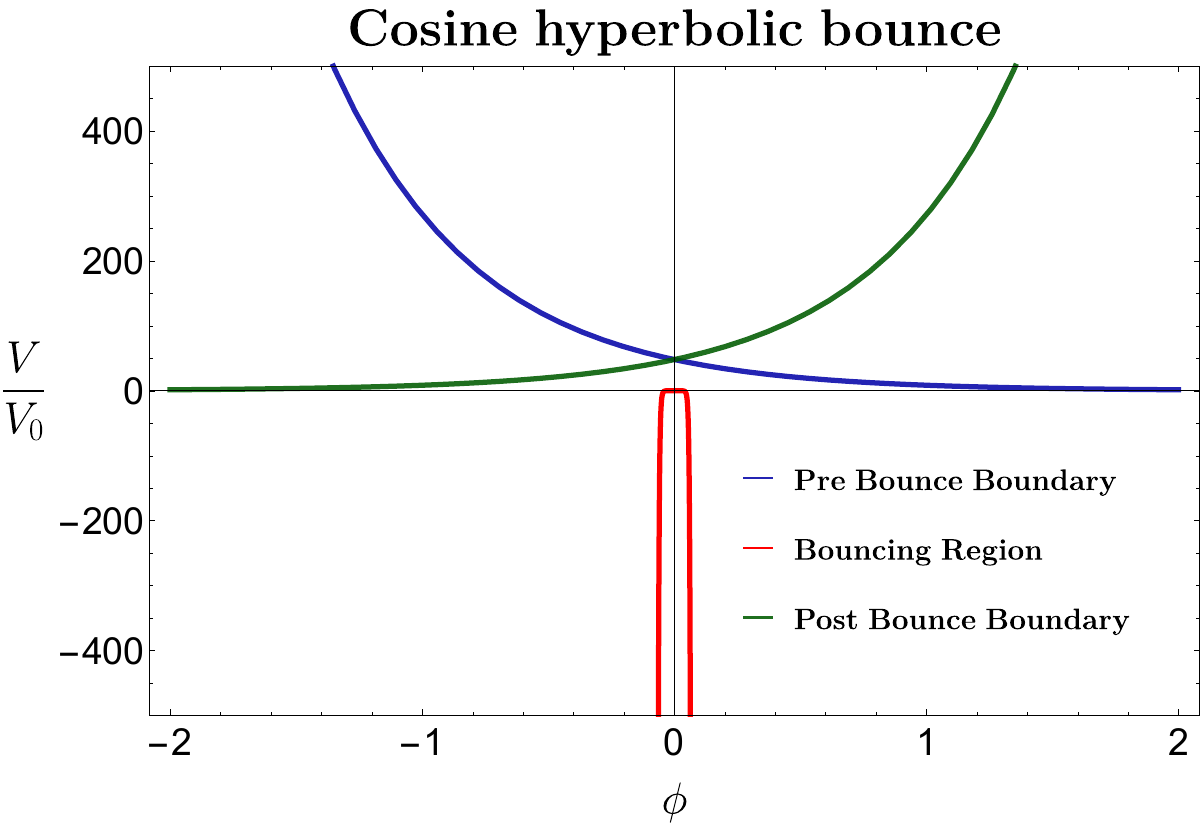}\label{fig:coshpot}}\hfill
	\subfigure[][Model II]{\includegraphics[height=9.2cm,width=7.7cm]{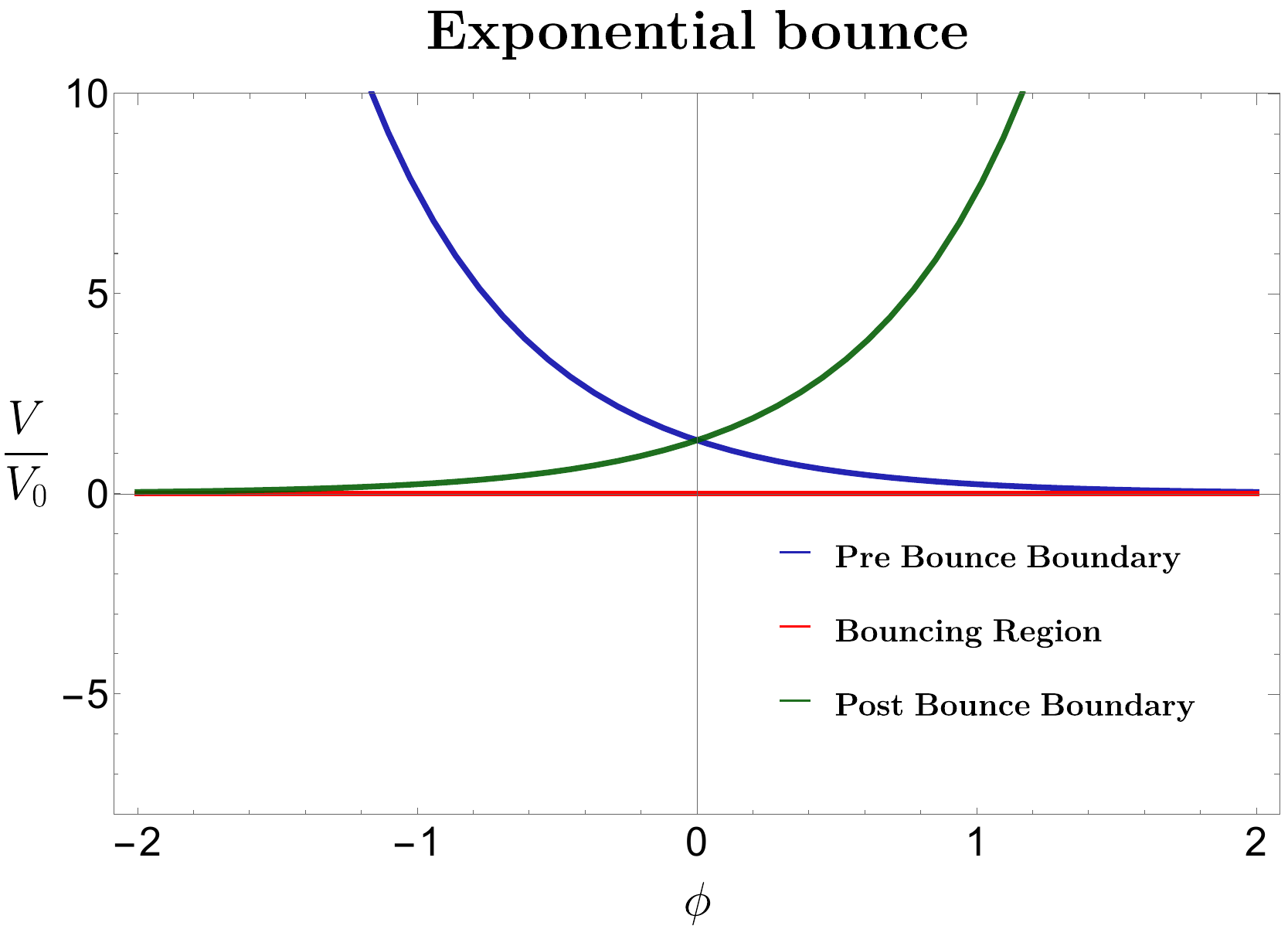}\label{fig:exppot}}
	\caption{Behaviour of the potentials of the considered models with respect to the field variable $\phi$}
	\label{fig:potentialwrtfield}
\end{figure}

 In \Cref{fig:potentialwrtfield}, the potentials of the two bouncing cosmology models considered in this paper have been studied with respect to the field variable $\phi$. For both the models, the potential for the pre-bounce region decreases exponentially to negligible values as the value of the field variable increases. An exponential increase in the potential is seen in both the cases as $\phi$ is increased. However, for the bouncing region, the behaviour of the potentials is widely different. For the Cosine hyperbolic model, the potential of the bouncing region is negative and goes to large negative values for a slight change in the field variable $\phi$. For the Exponential model, the potential of the bouncing region does also vary even though it is not apparent in \Cref{fig:potentialwrtfield}.
 
The aforementioned potentials used to describe the pre-bounce region, bouncing region and post-bounce region can be derived from String Theory descriptions at a very high scale. On the other hand, one can think of another equivalent situation where without introducing a scalar field in the classical gravitational background, one can also study the cosmological bouncing framework. Originally, the concept of cosmological bounce was proposed to resolve the coordinate intrinsic singularity of space-time at the time scale of the Big Bang, which is $t=0$. This is because the inflationary paradigm cannot resolve this issue. Not only that, the well known {\it Swampland Criteria} and {\it Trans-Planckian Censorship Criteria} \cite{Vafa:2005ui,Agrawal:2018own,Achucarro:2018vey,Heisenberg:2018yae,Kehagias:2018uem,Matsui:2018bsy,Lin:2018kjm,Dasgupta:2018rtp,Lin:2018rnx,Kawasaki:2018daf,Christodoulidis:2019jsx,Rudelius:2019cfh,Taylor:2019ots,Elizalde:2018dvw,Cai:2018ebs,Dasgupta:2019gcd,Tosone:2018qei,Heisenberg:2019qxz,Brandenberger:2019jbs,Wu:2019xtv,Bedroya:2020rac,Ooguri:2006in,Ooguri:2018wrx,Garg:2018reu,Palti:2019pca,Ooguri:2016pdq,Brown:2015iha,Kinney:2018nny,Kats:2006xp,Denef:2018etk,Palti:2017elp,Cicoli:2018kdo,Bedroya:2019snp,Blumenhagen:2017cxt,Murayama:2018lie,Heidenreich:2018kpg,Lee:2018urn,Roupec:2018mbn,Green:2007zzb,Blumenhagen:2018nts,Lee:2018spm,Dvali:2018jhn,Conlon:2018eyr,Ashoorioon:2018sqb,Hebecker:2018vxz,Kumar:2009us,Grimm:2018cpv,Heckman:2018jxk,Andriot:2018mav,Choi:2018rze,Han:2018yrk,Bena:2018fqc,Lust:2019zwm,Danielsson:2018qpa,Dimopoulos:2018upl,Ben-Dayan:2018mhe,Blumenhagen:2018hsh,Corvilain:2018lgw,Kinney:2018kew,Fukuda:2018haz,Olguin-Tejo:2018pfq,Cvetic:2017epq,Hamaguchi:2018vtv,Raveri:2018ddi,Reece:2018zvv,Kallosh:2019axr,Huang:2007qz,Agrawal:2019dlm,Junghans:2018gdb,Font:2019cxq,Blumenhagen:2019qcg,Hebecker:2018ofv,Blaback:2018hdo,Ibanez:2017kvh,Klaewer:2018yxi,Gonzalo:2019gjp,Buratti:2018xjt,Gonzalo:2018guu,Scalisi:2018eaz,Joshi:2019nzi,Chiang:2018lqx,Agrawal:2018rcg,Berera:2019zdd,Hamada:2017yji,Park:2018fuj,Ibanez:2017oqr} which are very useful to construct a physically consistent Effective Field Theory framework at a relatively lower scale than the very high UV cut-off scale of quantum gravity, commonly fixed at the Planck scale, can be described by bouncing paradigm more consistently than the inflation. Additionally, the bouncing cosmological paradigm can be done in presence of higher derivative quantum gravity corrections to the Einstein-Hilbert action. If such corrections are only a function of Ricci scalar then it is known as, $f(R)$ gravity, and within this class $R+\alpha R^2$, which is known as the {\it Starobinsky model} is the most famous one~\footnote{In the {\it Jordan frame} one can actually compute the corresponding mathematical form of the $f(R)$ gravity by making use of the following equations in $M_p=1$ unit:
\bea f(R)=\exp\left(\frac{2\sqrt{2}}{\sqrt{3}}\phi\right)\left[\sqrt{6}\frac{dV(\phi)}{d\phi}+2V(\phi)\right]~~~{\rm with}~~~R=\exp\left(\frac{2\sqrt{2}}{\sqrt{3}}\phi\right)\left[\sqrt{6}\frac{dV(\phi)}{d\phi}+4V(\phi)\right].\eea
For an example, for the potential $V(\phi)=\frac{1}{2}m^2_{\phi}\phi^2$, with $\phi\gg 1$ we get, $f(R)=R^2$ and with $\phi\ll 1$ we get, $f(R)=R$. So by considering both the limiting contribution one can construct a $f(R)$ function which is basically made up of both $R$ and $R^2$ contributions and they are appearing with appropriate coefficients i.e., $f(R)=\alpha R+ \beta R^2$. For $\phi\ll 1$, we have $\alpha \gg \beta$ and for $\phi \gg 1$ we have $\alpha \ll \beta$.}. One can show that using this model, along with infinite derivative non-local correction to the gravity sector of the form, $R+R{\cal F}(\Box)R$ ,\cite{Biswas:2010zk,Koshelev_2019,Koshelev:2018rau,Koshelev:2017bxd} and a Cosmological Constant term $\Lambda$, can produce the same type of bouncing solution in the spatially flat Friedmann-Lemaitre-Robertson-Walker (FLRW) metric in $3+1$ dimensional space-time, which is described by the following line element:
\begin{equation}
\label{eq:FLRWmetric}
ds^{2} =-dt^2+a^2(t)d{\bf x}^2= a^2(\tau)(-d\tau^{2} + {\bf dx}^{2}),
\end{equation}
where $\tau$ is the conformal time coordinate which is related to the physical time coordinate $t$ through the following replacement relation in the line element:
\bea d\tau= \frac{dt}{a(t)}.\eea

The prime objective to include such non-local correction was to produce a ghost-free renormalizable theory of gravity whose classical limit will be consistent with the local Einstein-Hilbert gravity contribution. Apart from this, the bouncing framework is very important in the context of primordial cosmology because the Big Bang singularity can be removed from the theory by imposing the bouncing condition on the related scale factors in the spatially flat FLRW background, which can be explicitly computed by making use of the {\it Friedmann equation} and the {\it Klein-Gordon equation} for the scalar field $\phi$. At the cosmological bounce scale $t=t_{\rm B}$ one has to satisfy the following constraint conditions to find out the appropriate dynamical solutions of the field equations:
\bea &&\underline{\textcolor{red}{\bf Bouncing~condition~I:}}~~~\nonumber\\
&&~~~~~~~~~~~~~~~~~~~~ \dot{a}_{\rm B}=\dot{a}(t_{\rm B})=\left(\frac{da(t)}{dt}\right)_{t=t_{\rm B}}=0~~\Longrightarrow~~H_B=H(t_{\rm B})=0,~~~~~~~~~\\
&&\underline{\textcolor{red}{\bf Bouncing~condition~II:}}~~~\nonumber\\
&&~~~~~~~~~~~~~~~~~~~~ \ddot{a}_{\rm B}=\ddot{a}(t_{\rm B})=\left(\frac{d^2a(t)}{dt^2}\right)_{t=t_{\rm B}}>0~~\Longrightarrow~~\dot{H}_{\rm B}=\dot{H}(t_{\rm B})>0. \eea
This same condition for the bounce at the conformal time scale $\tau=\tau_{\rm B}$ can be further translated in the following simplified form:
\bea &&\underline{\textcolor{red}{\bf Bouncing~condition~I:}}~~~\nonumber\\
&&~~~~~~~~~~~~~~~~~~~~  {a}'_{\rm B}={a}'(\tau_{\rm B})=\left(\frac{da(\tau)}{d\tau}\right)_{\tau=\tau_{\rm B}}=0~~\Longrightarrow~~{\cal H}_{\rm B}={\cal H}(\tau_{\rm B})=0,~~~~~~\\
&&\underline{\textcolor{red}{\bf Bouncing~condition~II:}}~~~\nonumber\\
&&~~~~~~~~~~~~~~~~~~~~ {a}''_{\rm B}={a}''(\tau_{\rm B})=\left(\frac{d^2a(\tau)}{d\tau^2}\right)_{\tau=\tau_{\rm B}}>0~~\Longrightarrow~~{\cal H}'_{\rm B}={\cal H}'(\tau_{\rm B})>0. \eea
This implies that the mathematical structure of the bouncing conditions remains the same in physical time and the conformal time coordinates, though they are not exactly the same as we have pointed earlier. One can also write constraint conditions on the potential function at the point of bounce, which is given by the following expressions:
\bea V_{\rm B}=V(\phi_{\rm B})=0,~~~~V_{,\phi,{\rm B}}=\left(\frac{dV(\phi)}{d\phi}\right)_{\phi=\phi_{\rm B}}=0,~~~~V_{,\phi\phi,{\rm B}}=\left(\frac{d^2V(\phi)}{d\phi^2}\right)_{\phi=\phi_{\rm B}}<0.~~~~~~\eea
Consequently, around the point of bounce if we expand the potential function in Taylor series in the field space, we get:
\bea V(\phi)=\frac{V_{,\phi\phi,{\rm B}}}{2}(\phi-\phi_{\rm B})^2+\frac{V_{,\phi\phi\phi,{\rm B}}}{6}(\phi-\phi_{\rm B})^3+\frac{V_{,\phi\phi\phi\phi,{\rm B}}}{24}(\phi-\phi_{\rm B})^4+\cdots,\eea
where the first three terms are the renormalizable contributions and other $\cdots$ represent non-renormalizable terms. 

From the previously mentioned models the scale factors can be computed in terms of the physical time coordinate as:
\bea
\underline{\textcolor{blue}{\bf Model~I:}}&&\nonumber\\
\displaystyle\displaystyle a(t)&=&  \left\{ 
     \begin{array}{lr}
\displaystyle  a_{\rm Pre} \cosh\left(\sqrt{\frac{r_1}{2}}~t_{\rm B}\right)~\left(-\frac{t}{t_{\rm B}}\right)^{\frac{2}{3(1+r_1)}}&~ \text{\textcolor{red}{\bf Pre-Bounce~($t<-t_{\rm B}$)}}\\  
    \displaystyle~a_{\rm B} \cosh\left(\sqrt{\frac{r_1}{2}}~t\right)~~~~~~~~~&~~~~~ \displaystyle\text{\textcolor{red}{\bf Bounce~~~($-t_{\rm B}<t<t_{\rm B}$)}}\\ 
  \displaystyle  a_{\rm Post} \cosh\left(\sqrt{\frac{r_1}{2}}~t_{\rm B}\right)~\left(\frac{t}{t_{\rm B}}\right)^{\frac{2}{3(1+r_1)}}~~~~&\text{\textcolor{red}{\bf Post-Bounce~($t>t_{\rm B}$)}} \end{array}
   \right.~~~~~~
\\ \\
\underline{\textcolor{blue}{\bf Model~II:}}&&\nonumber\\
\displaystyle\displaystyle a(t)&=& \left\{ 
     \begin{array}{lr}
\displaystyle    a_{\rm Pre}~\exp\left(\frac{9}{2}r_1~t_{\rm B}^2\right)~\left(-\frac{t}{t_{\rm B}}\right)^{\frac{2}{3(1+r_1)}}   &~ \text{\textcolor{red}{\bf Pre-Bounce~($t<-t_{\rm B}$)}}\\ 
    \displaystyle a_{\rm B}~\exp\left(\frac{9}{2}r_1~t^2\right)~~~~&\displaystyle\text{\textcolor{red}{\bf Bounce~~~($-t_{\rm B}<t<t_{\rm B}$)}}\\ 
  \displaystyle  a_{\rm Post}~\exp\left(\frac{9}{2}r_1~t_{\rm B}^2\right)~\left(\frac{t}{t_{\rm B}}\right)^{\frac{2}{3(1+r_1)}} ~~~~~~~&\text{\textcolor{red}{\bf Post-Bounce~($t>t_{\rm B}$)}} \end{array}
   \right.~~~~~~
\eea

\begin{figure}[!htb]
	\centering
	\subfigure[][Model I]{\includegraphics[height=9cm,width=7.7cm]{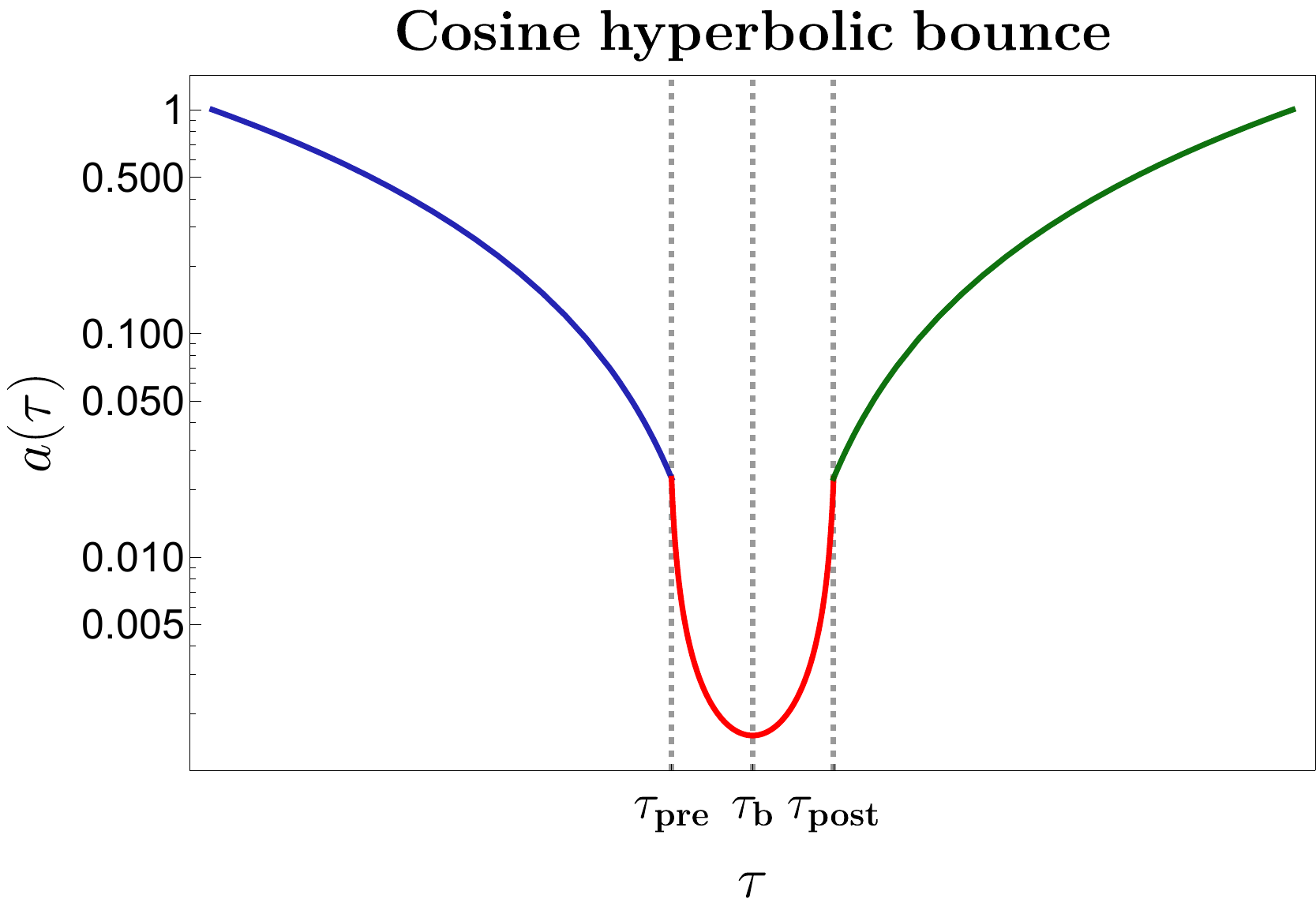}\label{fig:avstaucosh}}\hfill
	\subfigure[][Model II]{\includegraphics[height=9cm,width=7.7cm]{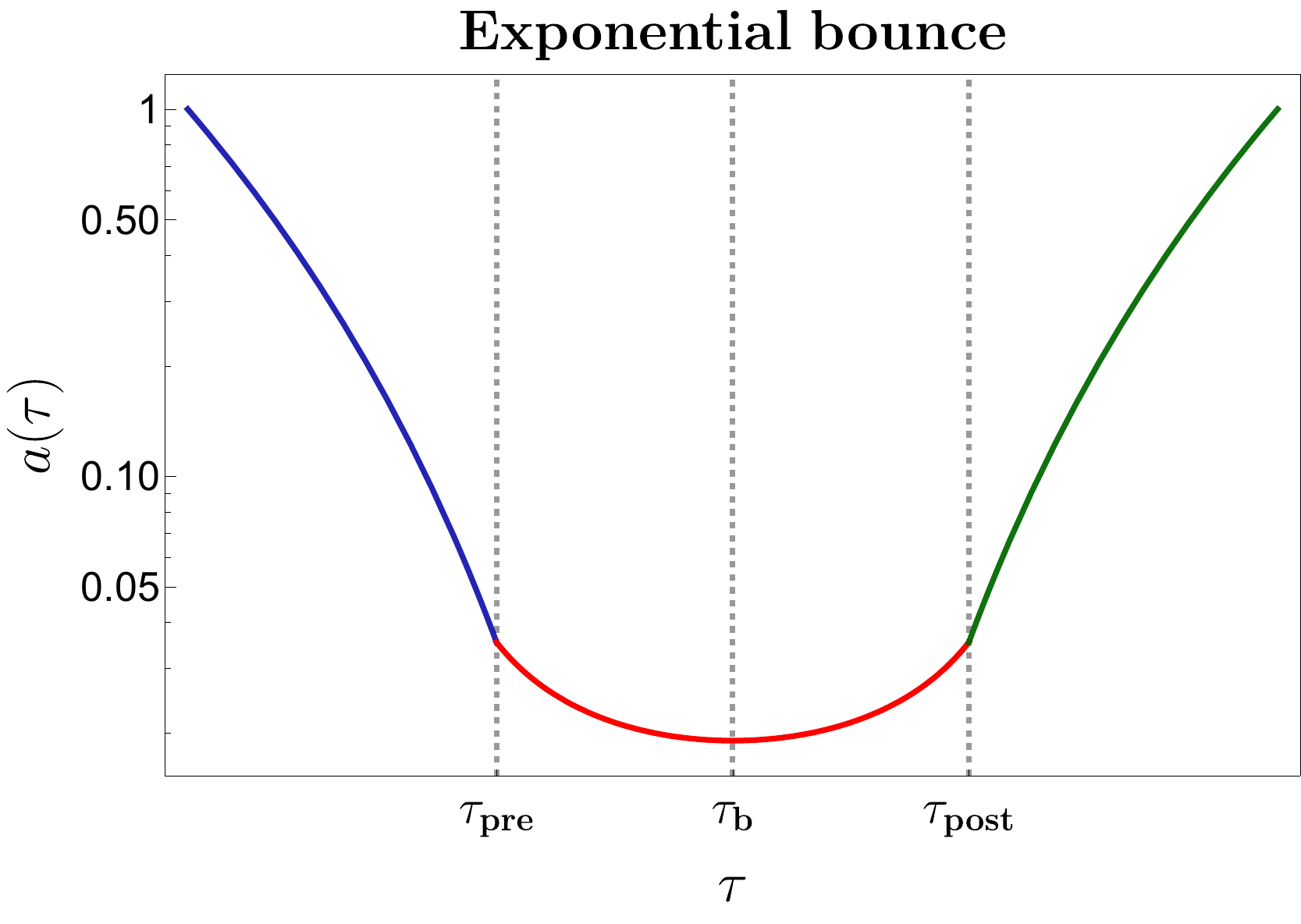}\label{fig:avstauexp}}
	\caption{Variation of derivative of scale factor wrt the conformal time showing three different regions viz. Pre-bounce, Bouncing and the Post-bounce regions}
	\label{fig:avstau}
\end{figure}

\begin{figure}[!htb]
	\centering
	\subfigure[][Model I]{\includegraphics[height=8cm,width=8.0cm]{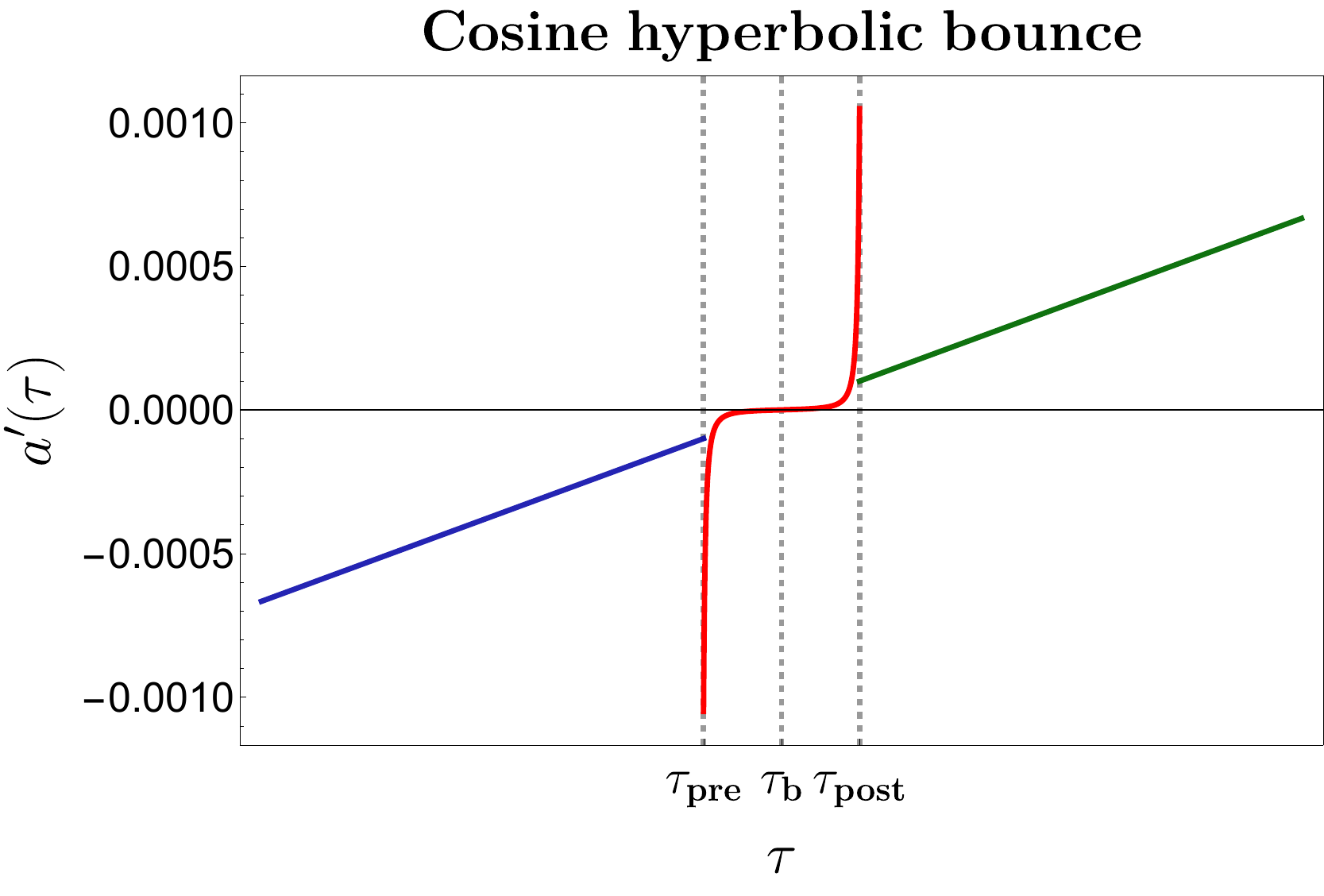}\label{fig:aprimevstaucosh}}\hfill
	\subfigure[][Model II]{\includegraphics[height=8cm,width=8.0cm]{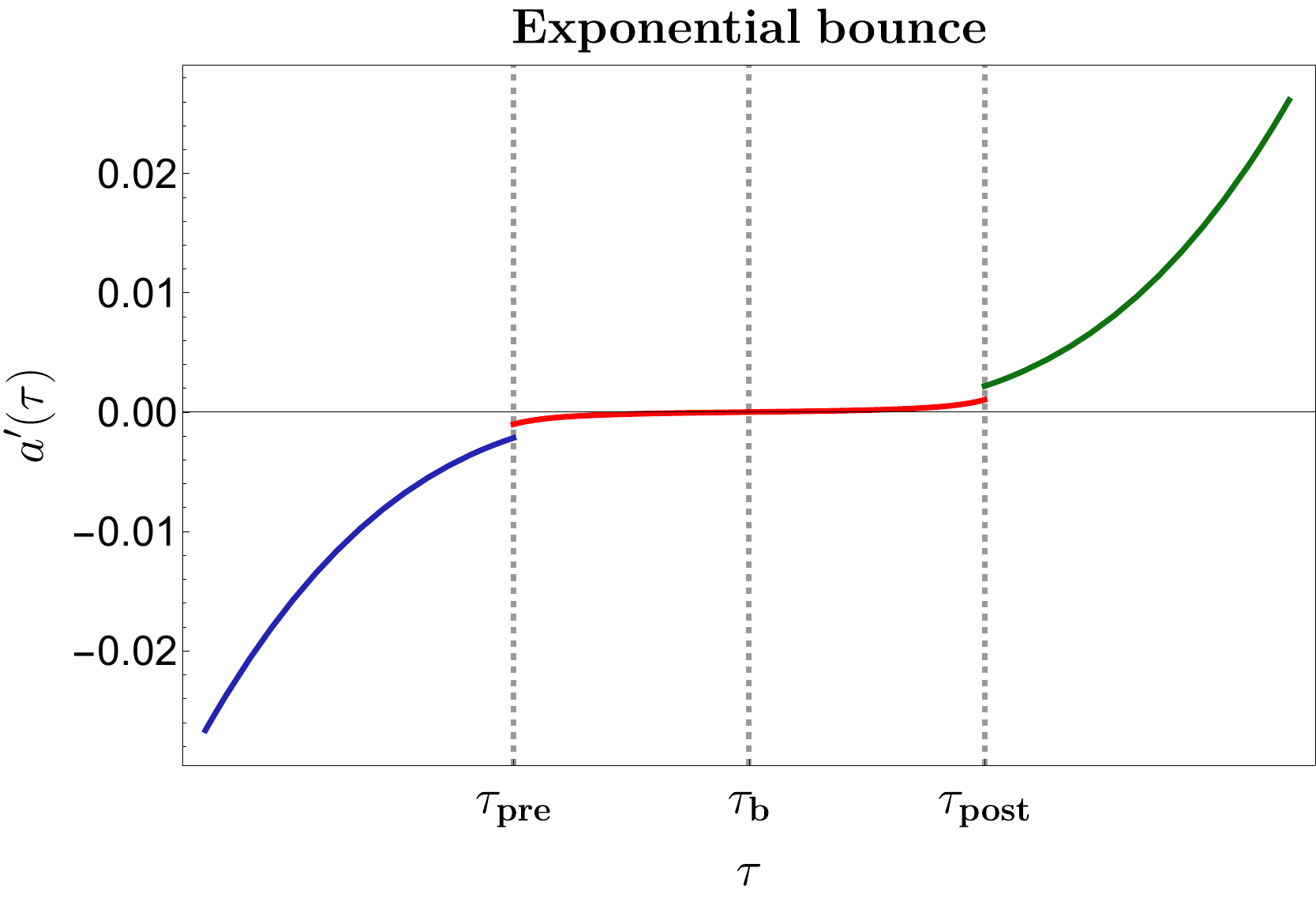}\label{fig:aprimevstauexp}}
\caption{Variation of derivative of scale factor wrt the conformal time showing three different regions viz. Pre-bounce, Bouncing and the Post-bounce regions}
	\label{fig:aprimevstau}
\end{figure}

\begin{figure}[!htb]
	\centering
	\subfigure[][Model I]{\includegraphics[height=7.9cm,width=8cm]{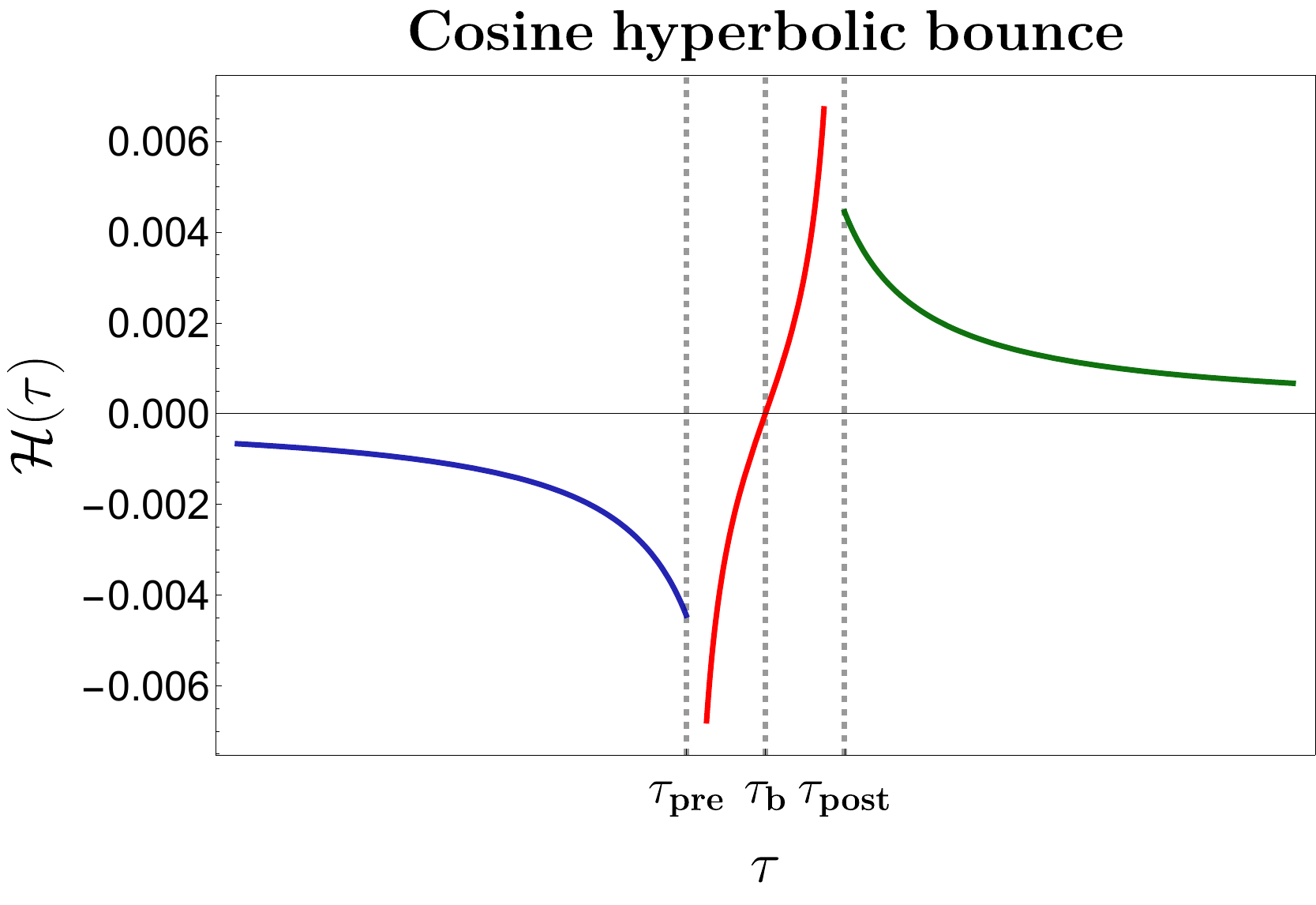}\label{fig:Hvstaucosh}}
	\hfill
	\subfigure[][Model II]{\includegraphics[height=7.9cm,width=8cm]{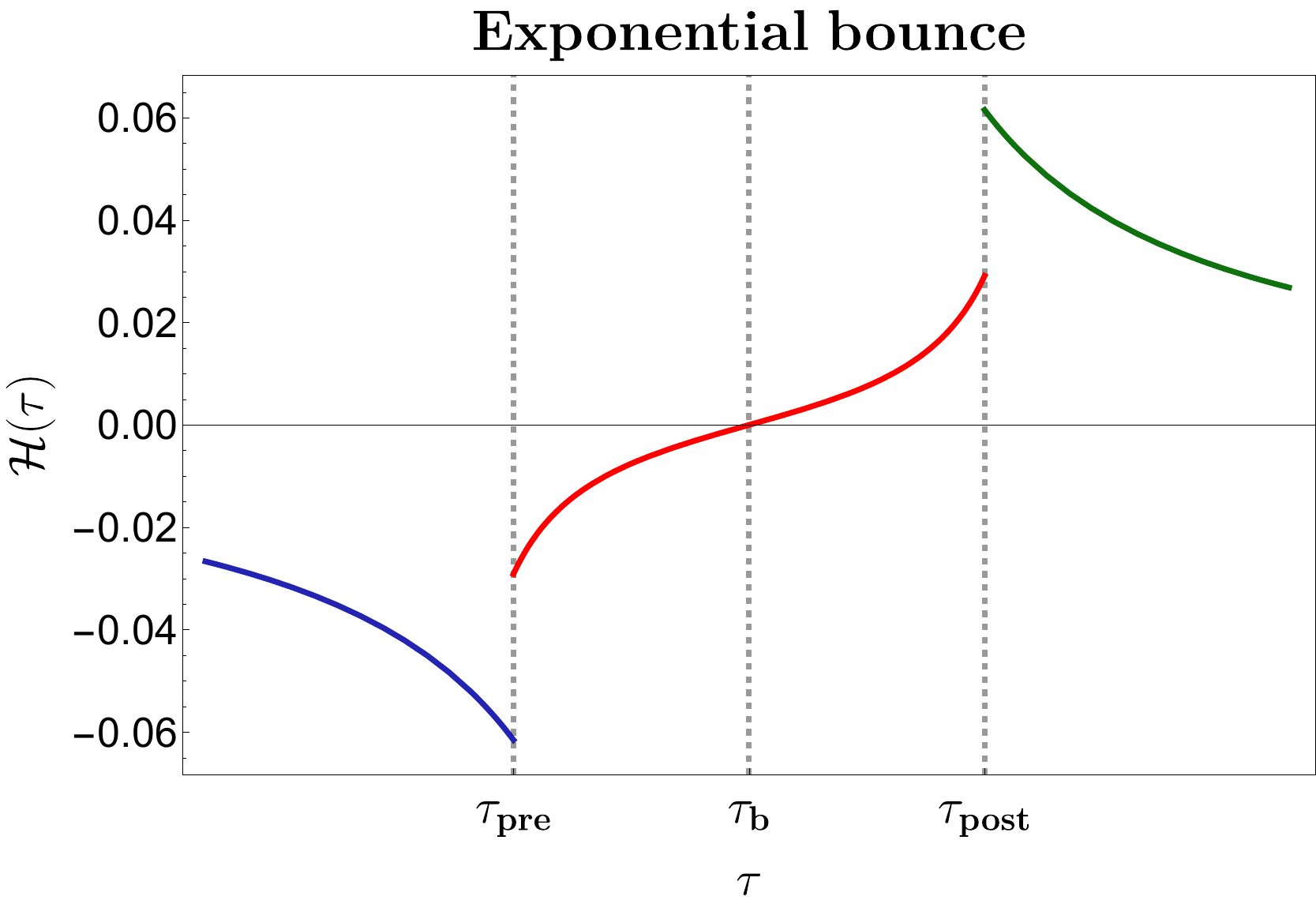}\label{fig:Hvstauexp}}
	\caption{Variation of Hubble paramter wrt the conformal time showing three different regions viz. Pre-bounce, Bouncing and the Post-bounce regions}
	\label{fig:Hvstau}
\end{figure}

\begin{figure}[!htb]
	\centering
	\subfigure[][Model I]{\includegraphics[height=7.9cm,width=8cm]{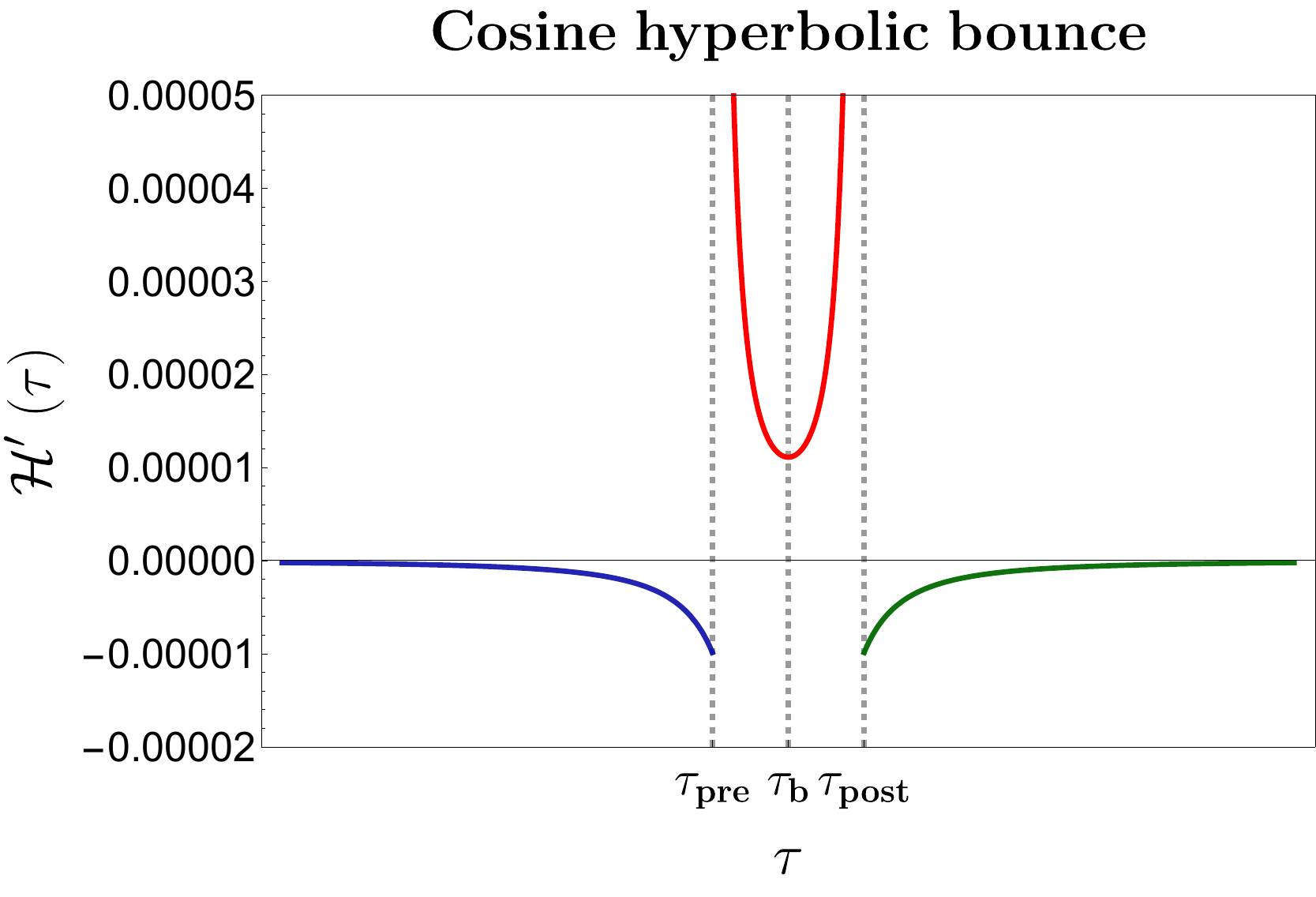}\label{fig:Hprimevstaucosh}}
	\hfill
	\subfigure[][Model II]{\includegraphics[height=7.9cm,width=8cm]{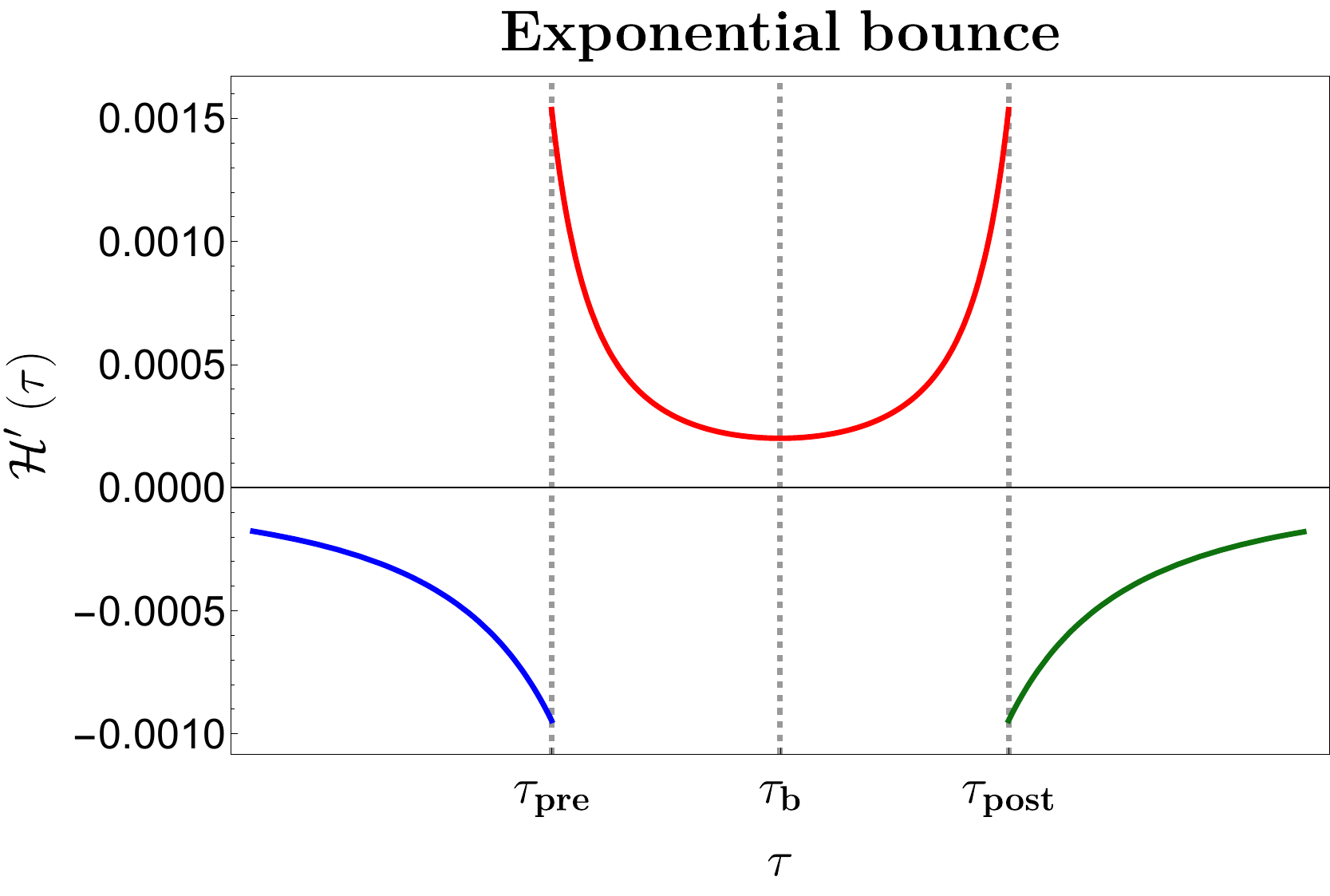}\label{fig:Hvstauexp}}
	\caption{Variation of the derivative of the Hubble parameter wrt the conformal time showing three different regions viz. Pre-bounce, Bouncing and the Post-bounce regions}
	\label{fig:Hprimevstau}
\end{figure}
In terms of conformal time coordinate one can further compute the expression for the scale factors, which are given by:
\bea
\underline{\textcolor{blue}{\bf Model~I:}}&&\nonumber\\
\footnotesize a(\tau)&=&  \left\{ 
     \begin{array}{lr}
\displaystyle    a_{\rm Pre}\cosh\left(\sqrt{\frac{r_1}{2}} t_{\rm B}\right)\left(- \frac{(1+3r_1)}{3t_{\rm B} (1+r_1)}\cosh\left(\sqrt{\frac{r_1}{2}} t_{\rm B}\right)(\tau - \tau_b) \right)^{\frac{2}{(1+r_1)}} &~ \text{\textcolor{red}{\bf Pre-Bounce}}\\ 
    \displaystyle a_{\rm B}\sec\left(\sqrt{\frac{r_1}{2}} (\tau - \tau_b)\right)&\displaystyle\text{\textcolor{red}{\bf Bounce}}~~~\\ 
  \displaystyle  a_{\rm Post}\cosh\left(\sqrt{\frac{r_1}{2}} t_{\rm B}\right)\left( \frac{(1+3r_1)}{3t_{\rm B} (1+r_1)}\cosh\left(\sqrt{\frac{r_1}{2}} t_{\rm B}\right)(\tau - \tau_b) \right)^{\frac{2}{(1+r_1)}}~&\text{\textcolor{red}{\bf Post-Bounce}} \end{array}
   \right.~~~~~~
\eea
\bea
\underline{\textcolor{blue}{\bf Model~II:}}&&\nonumber\\
\footnotesize a(\tau)&=&  \left\{ 
     \begin{array}{lr}
\displaystyle    a_{\rm Pre}~\exp\biggl({\frac{9}{2}r_1t^2_B}\biggr) \biggl(\frac{-(\tau-\tau_b)~\exp(\frac{9}{2}r_1t^2_B)}{t_B} \frac{1+3r_1}{3(1+r_1)}\biggr)^{\frac{4}{1+3r_1}} 
   &~ \text{\textcolor{red}{\bf Pre-Bounce}}\\ 
    \displaystyle a_{\rm B}~\exp\biggl(\text{InverseErf}\biggl(\frac{2(\tau-\tau_b)3\sqrt{r_1}}{\sqrt{2\pi}}\biggr)^2\biggr)&~\displaystyle\text{\textcolor{red}{\bf Bounce}}\\ 
  \displaystyle  a_{\rm Post}~\exp\biggl({\frac{9}{2}r_1t^2_B}\biggr) \biggl(\frac{(\tau-\tau_b)~\exp(\frac{9}{2}r_1t^2_B)}{t_B} \frac{1+3r_1}{3(1+r_1)}\biggr)^{\frac{4}{1+3r_1}}~&\text{\textcolor{red}{\bf Post-Bounce}} \end{array}
   \right.~~~~~~~~
\eea

The scale factors have been plotted in the Logarithmic scale to show the rising values near the boundary of the bouncing region as can be seen in \Cref{fig:avstau}. We also expect the models to satisfy the bouncing conditions given before, in \Cref{fig:Hvstau} and \Cref{fig:Hprimevstau}, and for both the models $\mathcal{H}_B = 0$ and $\mathcal{H}'_B > 0$ can be verified. As we will see later, the behaviour of $\mathcal{H}$ and the difference in signs on either side of the point of bounce will require two different squeezing parameters that describe the bouncing region, one before the point of bounce and one after it, which will result in differing behaviour of the complexity.

In the next section using these solutions our prime objective is to perform the cosmological perturbation and find the explicit role of these class of solutions to construct the squeezed vacuum sates.
\textcolor{Sepia}{\section{\sffamily Perturbation with squeezed quantum states in Bouncing Cosmology}
\label{sec:cosperwsqueezedQS}}
\textcolor{Sepia}{\subsection{\sffamily Scalar perturbation in Bouncing Cosmology}}

 In this section we will study squeezed state formalism within the framework of cosmological perturbation theory \cite{Lyth:2009zz,Mukhanov:2007zz,Langlois:2010vx,Langlois:2009jp,Langlois:2010xc,Langlois:2011jt} for FLRW spatially flat background specifically for post-bounce, bounce and pre bounce region. 
 In this context one needs to consider the following perturbation in the scalar field in the De Sitter background:
\bea \phi({\bf x},\tau)=\phi(\tau)+\delta\phi({\bf x},\tau)~~~~~~~~~\eea
and to express the whole dynamics in terms of a gauge invariant description through a variable:
\bea\zeta({\bf x},\tau)=-\frac{{\cal H}(\tau)}{\displaystyle\left(\frac{d\phi(\tau)}{d\tau}\right)}\delta\phi({\bf x},t).~~~~~~~~\eea 
At the level of first order perturbation theory in a spatially flat FLRW background metric, we fix the following gauge constraints:
\bea \delta\phi({\bf x},\tau)=0,~~
 g_{ij}({\bf x},\tau)= a^2(\tau)\left[\left(1+2\zeta({\bf x},\tau)\right)\delta_{ij}+h_{ij}({\bf x},\tau)\right],~~
\partial_{i}h_{ij}({\bf x},\tau)=0=h^{i}_{i}({\bf x},\tau),~~~~~~~~\eea
which fix the space-time re-parametrization. In this gauge, the spatial curvature of constant hyper-surface vanishes, which implies curvature perturbation variable is conserved outside the horizon. 

Applying the ADM formalism one can further compute the second-order perturbed action for scalar modes. The action, after gauge fixing, can then be expressed by the following:
\bea \delta^{(2)}S=\frac{1}{2}\int d\tau~d^3{\bf x}~\frac{a^2(\tau)}{{\cal H}^2}\left(\frac{d{\phi}(\tau)}{d\tau}\right)^2\left[\left(\partial_{\tau}\zeta({\bf x},\tau)\right)^2-\left(\partial_{i}\zeta({\bf x},\tau)\right)^2\right].\eea
Now, to re-parametrize the above mentioned second-order perturbed action expressed for primordial scalar perturbation, we introduce the following space-time dependent variable:
\bea v({\bf x},\tau)=z(\tau)~\zeta({\bf x},\tau),~~~~{\rm where}~~z(\tau)=a(\tau)\sqrt{\epsilon(\tau)},\eea which helps transform the perturbed action to that of the familiar mathematical form of canonical scalar field. In the cosmology literature, this is known as the {\it Mukhanov variable}, in terms of which we will perform the rest of the computation. Additionally, it is important to note that the newly defined quantity, $\epsilon(\tau)$ is the conformal time dependent slowly varying parameter, which is defined as:
 \bea \epsilon(\tau):=-\frac{\dot{H}}{H^2} =-\frac{a(\tau)}{{\cal H}^2}\frac{d}{d\tau}\left(\frac{{\cal H}}{a(\tau)}\right)= 1 - \frac{\mathcal{H'}}{\mathcal{H}^2}. \eea
 Consequently, the new version of the second order perturbed action for the scalar perturbation after re-parametrization in terms of the {\it Mukhanov variable} can be written as:
\begin{equation}
\label{eq4:action}
\delta^{(2)}S = \frac{1}{2} \int d\tau~d^3{\bf x} \biggl[ v'^{2}({\bf x},\tau)-(\partial_{i}v({\bf x},\tau))^{2} +\biggl(\frac{z'(\tau)}{z(\tau)}\biggr)^{2}v^{2}({\bf x},\tau) - 2\biggl(\frac{z'(\tau)}{z(\tau)}\biggr)v'({\bf x},\tau)v({\bf x},\tau) \biggr].
\end{equation}
Now, we explicitly compute the following crucial conformal time dependent contribution, which plays a  significant role to explore various unknown physical facts of the primordial universe:
\bea\frac{z'(\tau)}{z(\tau)}&=&\frac{a'(\tau)}{a(\tau)}+\frac{1}{2}\frac{\epsilon'(\tau)}{\epsilon(\tau)}\nonumber\\
&=&{\cal H}+\frac{1}{2}\frac{1}{\displaystyle\left(1 - \frac{\mathcal{H'}}{\mathcal{H}^2}\right)}\left( - \frac{\mathcal{H''}}{\mathcal{H}^2}+2 \frac{\mathcal{H'}^2}{\mathcal{H}^3}\right)\nonumber\\
&=&{\cal H}\left[1+\frac{1}{2}\frac{1}{\epsilon(\tau)}\left(2(1-\epsilon(\tau))^2- \frac{\mathcal{H''}}{\mathcal{H}^3}\right)\right]\nonumber\\
&=&{\cal H}\left[\frac{1}{\epsilon(\tau)}-1+\epsilon(\tau)- \frac{1}{2}\frac{1}{\epsilon(\tau)}\frac{\mathcal{H''}}{\mathcal{H}^3}\right]\eea

Our job is now to further convert the second-order perturbed action for the scalar degrees of freedom in terms of the Fourier modes, by implementing the following $ansatz$ for the Fourier transformation:
\begin{equation}
\label{eq:fouriermodes}
v({\bf x},\tau):=\int \frac{d^{3}{\bf k}}{(2\pi)^{3}}v_{{\bf k}}(\tau)~\exp(-i{\bf k}.{\bf x}),
\end{equation}
using which one can compute the following contributions from the time and space derivative of the perturbed field variable appearing in the second-order action :
\bea 
&& v'({\bf x},\tau):=\int \frac{d^{3}{\bf k}}{(2\pi)^{3}}v'_{{\bf k}}(\tau)~\exp(-i{\bf k}.{\bf x}),\\
&&\partial_jv({\bf x},\tau):=i\int \frac{d^{3}{\bf k}}{(2\pi)^{3}}v_{{\bf k}}(\tau)~k_j\exp(-i{\bf k}.{\bf x}).
\eea
After the substitution of all the aforementioned expressions, the simplified version of the second-order perturbation for the scalar modes in Fourier space can be further recast as:
\bea \delta^{(2)}S = \frac{1}{2} \int d\tau~d^3{\bf k}\underbrace{ \biggl[ |v'_{\bf k}(\tau)|^2+\Biggl(k^2+\biggl(\frac{z'(\tau)}{z(\tau)}\biggr)^{2}\Biggr)|v_{\bf k}(\tau)|^{2} - 2\biggl(\frac{z'(\tau)}{z(\tau)}\biggr)v'_{\bf k}(\tau)v_{-{\bf k}}(\tau) \biggr]}_{\textcolor{red}{\bf Lagrangian~density~{\cal L}^{(2)}(v_{\bf k}(\tau),v'_{\bf k}(\tau),\tau)}},~~~\eea
where it is important to note that:
\bea  |v'_{\bf k}(\tau)|^2=v^{'*}_{-{\bf k}}(\tau)v^{'}_{\bf k}(\tau),~~~~|v_{\bf k}(\tau)|^{2}=v^{*}_{-{\bf k}}(\tau)v_{\bf k}(\tau).\eea
Now after varying the second-order perturbed action with respect to the perturbed field variable expressed in the Fourier space, we get the following equation of motion:
\bea v''_{\bf k}(\tau)+\omega^2(k,\tau)v_{\bf k}(\tau)=0.\eea
This is commonly known as the {\it Mukhanov-Sasaki equation} and actually represents the classical equation of motion of a parametric oscillator where the frequency of the oscillator is conformal time dependent and in the present context of discussion, be explicitly given by :
\bea \omega^2(k,\tau):=k^2+m^2_{\rm eff}(\tau),\eea
where we have introduced a conformal time dependent effective mass in the present computation, which is quantified by the following expression:
\bea m^2_{\rm eff}(\tau)&=&-\frac{z''(\tau)}{z(\tau)}\nonumber\\
&=&{\cal H}^2\Biggl(-\frac{2}{\epsilon^2(\tau)}+\frac{5}{\epsilon(\tau)}-2(1-\epsilon(\tau))+\epsilon^2(\tau)-\left(1-\frac{1}{\epsilon^2(\tau)}\right)\frac{{\cal H}''}{{\cal H}^3}\nonumber\\
&&~~~~~~~~~~~~~~~~~~~~~~~~~~~~~~~~~~~~~+\left[\frac{1}{\epsilon(\tau)}-1+\epsilon(\tau)- \frac{1}{2}\frac{1}{\epsilon(\tau)}\frac{\mathcal{H''}}{\mathcal{H}^3}\right]^2\Biggr)\nonumber\\
&&~~~~-\frac{1}{2\epsilon(\tau){\cal H}^2}\Biggl({\cal H}''' -2\frac{{\cal H}'{\cal H}''}{{\cal H}}-\frac{{\cal H}''}{\epsilon(\tau)}\left(2(1-\epsilon(\tau))^2- \frac{\mathcal{H''}}{\mathcal{H}^3}\right)\Biggr)\nonumber\\
&=&\frac{1}{\tau^2}\left(\nu^2_{\rm B}(\tau)-\frac{1}{4}\right)\eea
where for the purpose of simplification of computation we have introduced a conformal time dependent mass parameter, $\nu_{\rm B}(\tau)$, which is defined as:
\bea \nu_{\rm B}(\tau):&=&\Biggl\{\tau^2{\cal H}^2\Biggl(-\frac{2}{\epsilon^2(\tau)}+\frac{5}{\epsilon(\tau)}-2(1-\epsilon(\tau))+\epsilon^2(\tau)-\left(1-\frac{1}{\epsilon^2(\tau)}\right)\frac{{\cal H}''}{{\cal H}^3}\nonumber\\
&&~~~~~~~~~~~~~~~~~~~~~~~~~~~~~~~~~~~~~+\left[\frac{1}{\epsilon(\tau)}-1+\epsilon(\tau)- \frac{1}{2}\frac{1}{\epsilon(\tau)}\frac{\mathcal{H''}}{\mathcal{H}^3}\right]^2\Biggr)\nonumber\\
&&~~~~-\frac{1}{2\epsilon(\tau){\cal H}^2}\Biggl({\cal H}''' -2\frac{{\cal H}'{\cal H}''}{{\cal H}}-\frac{{\cal H}''}{\epsilon(\tau)}\left(2(1-\epsilon(\tau))^2- \frac{\mathcal{H''}}{\mathcal{H}^3}\right)\Biggr)+\frac{1}{4}\Biggr\}^{\frac{1}{2}}\nonumber\\
&=&\frac{1}{2}+\left(1-\frac{1}{\epsilon(\tau)}\right)\frac{{\cal H}''}{{\cal H}^2}+\cdots,\eea 
where $\cdots$ is the contribution which is varying very slowly in the context of our present discussion. 

\textcolor{Sepia}{\subsection{\sffamily Scalar mode function}}

As a result, the {\it Mukhanov-Sasaki equation} can be translated into the following simplified form:
\bea v''_{\bf k}(\tau)+\Biggl(k^2-\frac{1}{\tau^2}\left(\nu^2_{\rm B}(\tau)-\frac{1}{4}\right)\Biggr)v_{\bf k}(\tau)=0.\eea
The most general analytical solution of the above equation can be expressed as:
\bea v_{\bf k}(\tau):=\sqrt{-\tau}\left[{\cal C}_1~{\cal H}^{(1)}_{\nu_{\rm B}}(-k\tau)+{\cal C}_2~{\cal H}^{(2)}_{\nu_{\rm B}}(-k\tau)\right]\eea
where ${\cal H}^{(1)}_{\nu_{\rm B}}(-k\tau)$ and ${\cal H}^{(2)}_{\nu_{\rm B}}(-k\tau)$ are Hankel functions of the first and second kind,respectively, with argument $-k\tau$ and order $\nu_{\rm B}$. During this computation, we have also used the fact that the conformal time-dependent quantity $\nu_{\rm B}$ is varying very slowly with respect to the evolutionary time scale of our universe. Additionally it is important to note that, the two integration constants, ${\cal C}_1$ and ${\cal C}_2$ can be fixed by the choice of the initial quantum vacuum state in the present context. In this work, we choose the most popular and the simplest initial quantum vacuum state, which is known as {\it Bunch Davies vacuum} or {\it Hartle Hawking vacuum} or {\it Chernkov vacuum}, and can be fixed by choosing ${\cal C}_1=1$ and ${\cal C}_2=0$. 

Consequently, we get the following solution:
\bea v_{\bf k}(\tau)=\sqrt{-\tau}~{\cal H}^{(1)}_{\nu_{\rm B}}(-k\tau).\eea
Upon further considering $-k\tau\rightarrow 0$ and $-k\tau\rightarrow \infty$ asymptotic limits, one can write the following simplified form of the Hankel functions of the first kind:
\begin{equation}
	{\lim_{-k\tau\rightarrow \infty}{\cal H}^{(1)}_{\nu_{\rm B}}(-k\tau)=\sqrt{\frac{2}{\pi}}\frac{1}{\sqrt{-k\tau}}\exp\left(-i\left\{k\tau+\frac{\pi}{2}\left(\nu_{\rm B}+\frac{1}{2}\right)\right\}\right)}.
\end{equation}
 Using these asymptotic results of the Hankel functions of the first kind the most general solution for the perturbed field can be expressed as:
\begin{eqnarray} 
&&~~~~v_{{\bf k}}(\tau)=\frac{2^{\nu_{\rm B}-\frac{3}{2}}(-k\tau)^{\frac{3}{2}-\nu_{\rm B}}}{\sqrt{2k}}\left|\frac{\Gamma(\nu_{\rm B})}{\Gamma\left(\frac{3}{2}\right)}\right|~\left(1-\frac{i}{k\tau}\right)~\exp\left(-i\left\{k\tau+\frac{\pi}{2}\left(\nu_{\rm B}-\frac{3}{2}\right)\right\}\right).~~~~~~~~~~~ 
\end{eqnarray}
In the present solution, the slowly varying time-dependent mass parameter $\nu_{\rm B}(\tau)$ is a completely model-dependent one. For this reason, to fix the value and the behaviour of the slowly-varying function with respect to the underlying conformal time scale we need to explicitly compute this expression for different models which are describing the pre-bounce, bounce, post-bounce, and the away from the bounce region~\footnote{\textcolor{red}{\bf Note:}~~Here it is important to note that during inflation the mass parameter $\nu_{\rm B}=\frac{3}{2}$, if we exactly follow the De Sitter expansion in the spatially flat FLRW background. But in order to stop inflation, one needs to consider a slight deviation from exact De Sitter expansion during inflation, and technically this slight amount of deviation has been taken by considering the slowly varying time dependent slow-roll parameters. So it is expected that for exact De Sitter expansion, the factor $\nu_{\rm B}-\frac{3}{2}$ will exactly vanish, and for the quasi-De Sitter expansion, this difference will be proportional to the amount of deviation from the exact De Sitter expansion. \textcolor{blue}{\bf But} in the present context we are interested in the pre-bounce, bounce, post-bounce, and away from bounce, where it appears to us that the analytical solution of the scalar mode function appearing from the cosmological perturbation in the spatially flat FLRW background is identical to the structure that one may compute by solving the equation of motion of the scalar mode fluctuation, which is the {\it Mukhanov-Sasaki equation} in the context of inflation. The significant difference can be observed clearly if we look into the mathematical structure and the leading , sub-leading order contribution appearing in the expression of the mass parameter in both of the cases separately. For inflation, this value is slightly larger than $\frac{3}{2}$, which as we told can demonstrate the quasi-De Sitter expansion. On the other hand, for the alternative to the inflationary paradigm - which is described by pre-bounce, bounce, post-bounce, etc., it is expected that the value of the mass parameter will be completely different from $\frac{3}{2}$ and the amount of deviation from the exact De Sitter is very large. This is because the slowly varying parameter $\epsilon$ and its derivatives are significantly large compared to the value obtained for this parameter, which is smaller than unity during inflation and approximately unity at the end of inflation. Apart from this underlying significant difference, for the sake of consistency with the previous works and their findings, we have expressed the solution of the scalar mode function for the pre-bounce, bounce, post-bounce, and away from the bounce phases like the result obtained from inflation.}.  

One can further consider two asymptotic cases, super-Hubble and the sub-Hubble which might be extremely useful to study the physical impact of the mode function obtained for the scalar fluctuations in the two different physical regions as mentioned before. In terms of the representative dynamical scale, the super-Hubble and the sub-Hubble limit is described by $-k\tau\ll 1$ and $-k\tau\gg1$, respectively. Additionally, it is important to note that in this context of the discussion, the cosmological horizon crossing is described by $-k\tau=1$. Now we shall implement all the discussed limits to get simplified results from the scalar mode function obtained previously within the framework of bouncing cosmological paradigm. These limiting results are appended below:
\begin{eqnarray} 
&&\underline{\textcolor{red}{\bf Sub-Hubble~limiting~solution:}}\nonumber\\
&&~~~~v_{{\bf k}}(-k\tau\gg 1)=\frac{2^{\nu_{\rm B}-\frac{3}{2}}(-k\tau)^{\frac{3}{2}-\nu_{\rm B}}}{\sqrt{2k}}\left|\frac{\Gamma(\nu_{\rm B})}{\Gamma\left(\frac{3}{2}\right)}\right|~\exp\left(-i\left\{k\tau+\frac{\pi}{2}\left(\nu_{\rm B}-\frac{3}{2}\right)\right\}\right).~~~~~~~~~~~ 
\\
&&\underline{\textcolor{red}{\bf Horizon~crossing~solution:}}\nonumber\\
&&~~~~v_{{\bf k}}(-k\tau=1)=\frac{2^{\nu_{\rm B}-1}}{\sqrt{2k}}\left|\frac{\Gamma(\nu_{\rm B})}{\Gamma\left(\frac{3}{2}\right)}\right|~\exp\left(-i\left\{\frac{\pi}{2}\left(\nu_{\rm B}-2\right)-1\right\}\right).~~~~~~~~~~~ 
\\
&&\underline{\textcolor{red}{\bf Super-Hubble~limiting~solution:}}\nonumber\\
&&~~~~v_{{\bf k}}(-k\tau \ll 1)=\frac{2^{\nu_{\rm B}-\frac{3}{2}}(-k\tau)^{\frac{1}{2}-\nu_{\rm B}}}{\sqrt{2k}}\left|\frac{\Gamma(\nu_{\rm B})}{\Gamma\left(\frac{3}{2}\right)}\right|~\exp\left(-i\left\{\frac{\pi}{2}\left(\nu_{\rm B}-\frac{5}{2}\right)\right\}\right).~~~~~~~~~~~ 
\end{eqnarray}

\textcolor{Sepia}{\subsection{\sffamily Quantization of Hamiltonian for scalar modes}}

Using these solutions, one can further compute the expression for the derivatives of these field variables with respect to the conformal time scale, which will be helpful for the further computation in the present context:
\bea
 &&\nonumber v'_{{\bf k}}(\tau)=i\sqrt{\frac{k}{2}}~2^{\nu_{\rm B}-\frac{3}{2}}(-k\tau)^{\frac{3}{2}-\nu_{\rm B}}\left|\frac{\Gamma(\nu_{\rm B})}{\Gamma\left(\frac{3}{2}\right)}\right|~\left\{1-\left(\nu_{\rm B}-\frac{1}{2}\right)\frac{i}{k\tau}\left(1-\frac{i}{k\tau}\right)\right\}\nonumber\\
 &&~~~~~~~~~~~~~~~~~~~~~~~~~~~~~~~~~~~~~~~~~~~\exp\left(-i\left\{k\tau+\frac{\pi}{2}\left(\nu_{\rm B}-\frac{1}{2}\right)\right\}\right).~~~~~~~~~~~\eea
 
As mentioned in the previous subsection, one needs to further consider two asymptotic cases, the super-Hubble and the sub-Hubble limiting situation which might be extremely useful to study the physical impact of the obtained mode function for the scalar fluctuations in the present context. In terms of the representative dynamical scale, the super-Hubble and the sub-Hubble limit is described by $-k\tau\ll 1$ and $-k\tau\gg1$, respectively. Additionally, it is important to note that in this context of the discussion, the cosmological horizon crossing is described by $-k\tau=1$. By following the same logical reasoning one can write down the following expressions for the conformal time derivative of the mode functions from scalar fluctuations which will explicitly contribute further in the expression for the canonically conjugate momenta associated with these scalar modes:
\bea
&&\underline{\textcolor{red}{\bf Sub-Hubble~limiting~solution:}}\nonumber\\
 && v'_{{\bf k}}(-k\tau\gg 1)=i\sqrt{\frac{k}{2}}~2^{\nu_{\rm B}-\frac{3}{2}}(-k\tau)^{\frac{3}{2}-\nu_{\rm B}}\left|\frac{\Gamma(\nu_{\rm B})}{\Gamma\left(\frac{3}{2}\right)}\right|~\exp\left(-i\left\{k\tau+\frac{\pi}{2}\left(\nu_{\rm B}-\frac{1}{2}\right)\right\}\right).~~~~~~~~~~~\\
 &&\underline{\textcolor{red}{\bf Horizon~crossing~solution:}}\nonumber\\
 &&\nonumber v'_{{\bf k}}(-k\tau=1)=i\sqrt{\frac{k}{2}}~2^{\nu_{\rm B}-\frac{3}{2}}\left|\frac{\Gamma(\nu_{\rm B})}{\Gamma\left(\frac{3}{2}\right)}\right|~\left\{1-\sqrt{2}\left(\nu_{\rm B}-\frac{1}{2}\right)\exp\left(-\frac{i\pi}{4}\right)\right\}\nonumber\\
 &&~~~~~~~~~~~~~~~~~~~~~~~~~~~~~~~~~~~~~~~~~~~\exp\left(-i\left\{\frac{\pi}{2}\left(\nu_{\rm B}-\frac{1}{2}\right)-1\right\}\right).~~~~~~~~~~~\\
 &&\underline{\textcolor{red}{\bf Super-Hubble~limiting~solution:}}\nonumber\\
 && v'_{{\bf k}}(-k\tau\ll 1)=i\sqrt{\frac{k}{2}}~2^{\nu_{\rm B}-\frac{3}{2}}(-k\tau)^{-\left(\nu_{\rm B}+\frac{1}{2}\right)}\left|\frac{\Gamma(\nu_{\rm B})}{\Gamma\left(\frac{3}{2}\right)}\right|~\left(\nu_{\rm B}-\frac{1}{2}\right)\exp\left(-i\left\{\frac{\pi}{2}\left(\nu_{\rm B}-\frac{5}{2}\right)\right\}\right).~~~~~~~~~~~\eea
\begin{figure}[!htb]
	\centering
	\subfigure[][Model I]{\includegraphics[height=7cm,width=15cm]{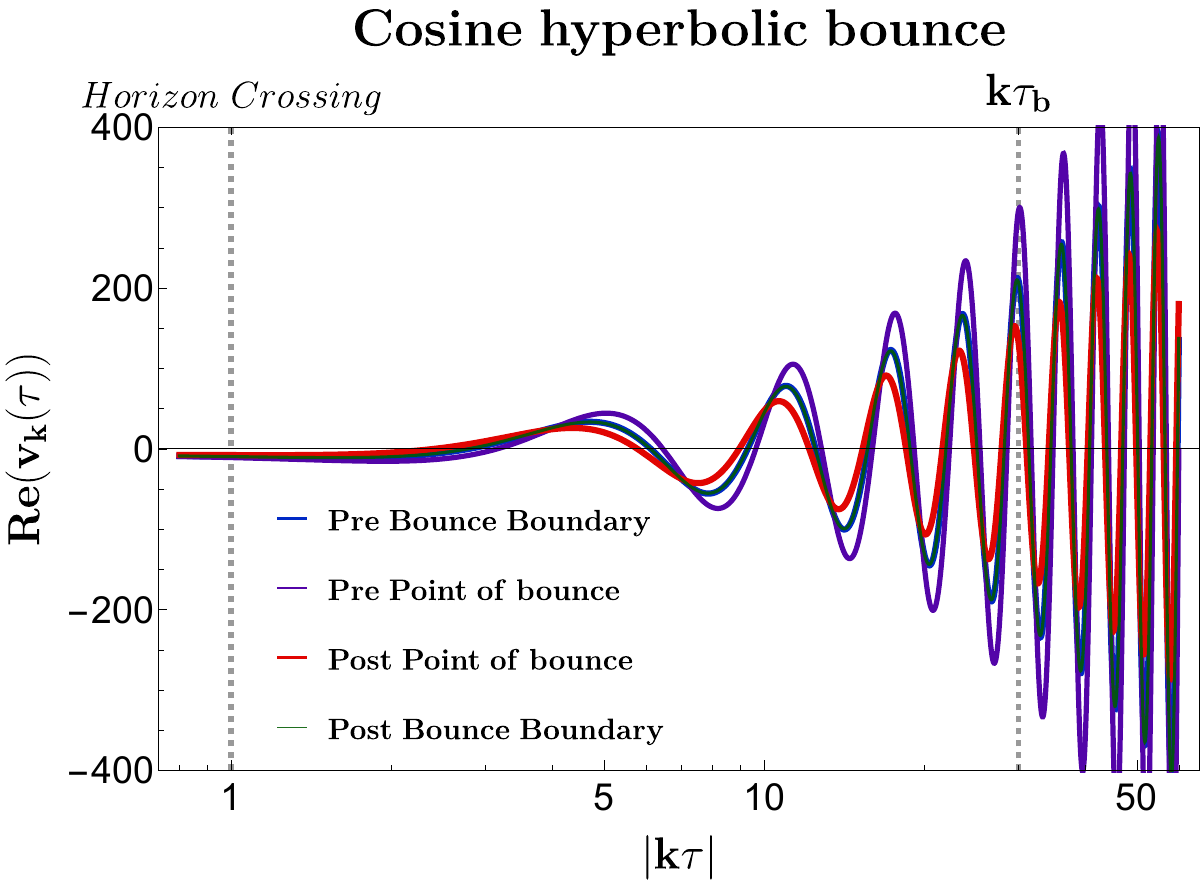}\label{fig:realvcosh}}
	\hfill
	\subfigure[][Model II]{\includegraphics[height=7cm,width=15cm]{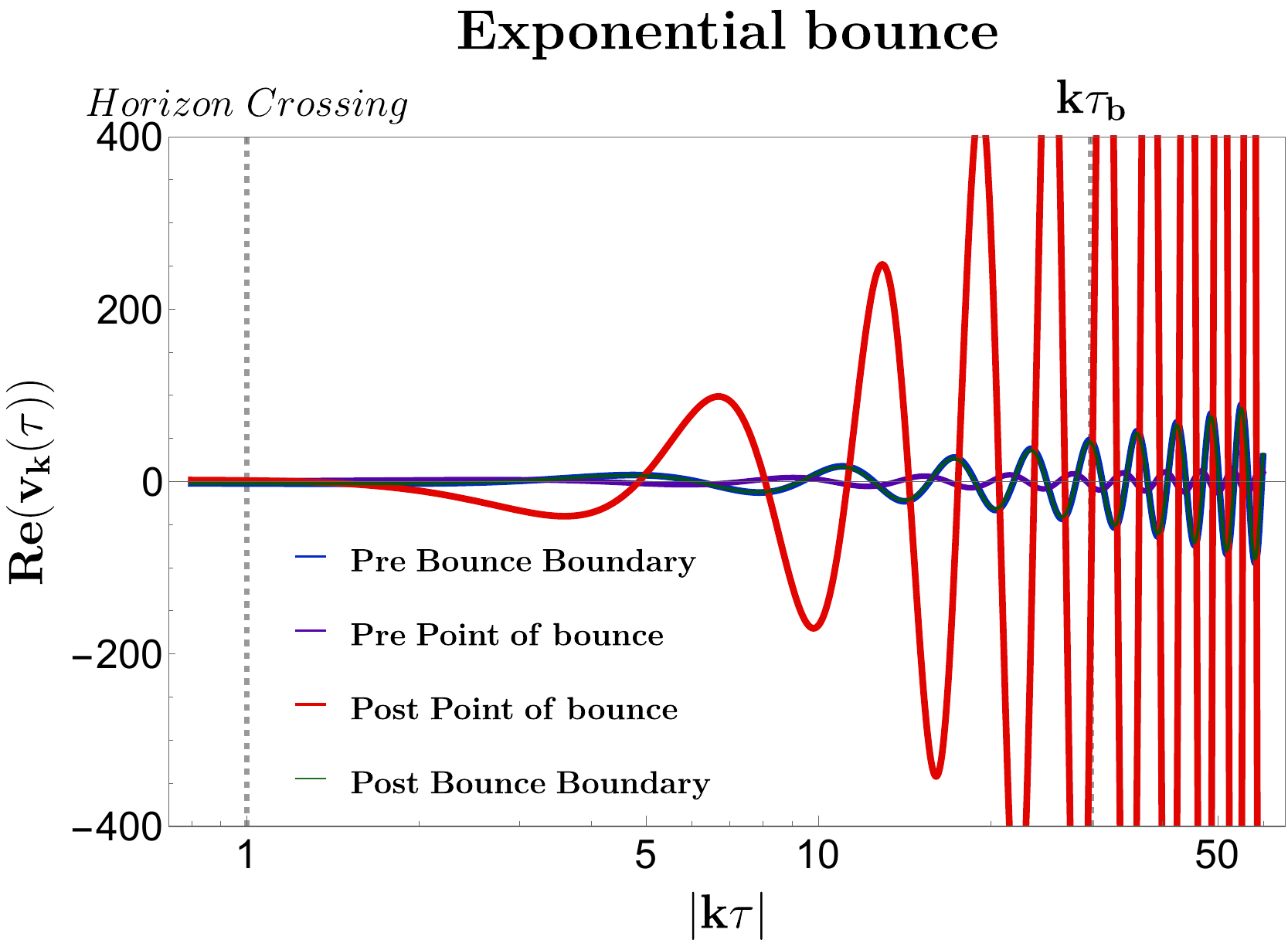}\label{fig:realvexp}}
	\caption{The real part of $v_k$ for the Cosine hyperbolic and the exponential bounce has been plotted. We can see the behaviour of $v_k$ in the sub-horizon region $|k\tau|\gg1$. Around the Horizon crossing, the $Re(v_k)$ is slightly negative for the Cosine hyperbolic bounce whereas it is almost zero for the exponential bounce. Far from the horizon crossing, the behaviour of both the models is almost identical. It slowly starts rising before becoming highly oscillatory with increasing amplitude and frequency as they approach the point of bounce $k\tau_b \approx 30$. The field variables in all regions behave almost similarly. However, a noticeable difference between the two models lies in the fact that the amplitude of the field variable for a particular region is in contrast for both the models. If the amplitude is maximum for a particular region for model-I then it is minimum for model-II.}
	\label{fig:revvstau}
\end{figure}

\begin{figure}[!htb]
	\centering
	\subfigure[][Model I]{\includegraphics[height=7cm,width=15cm]{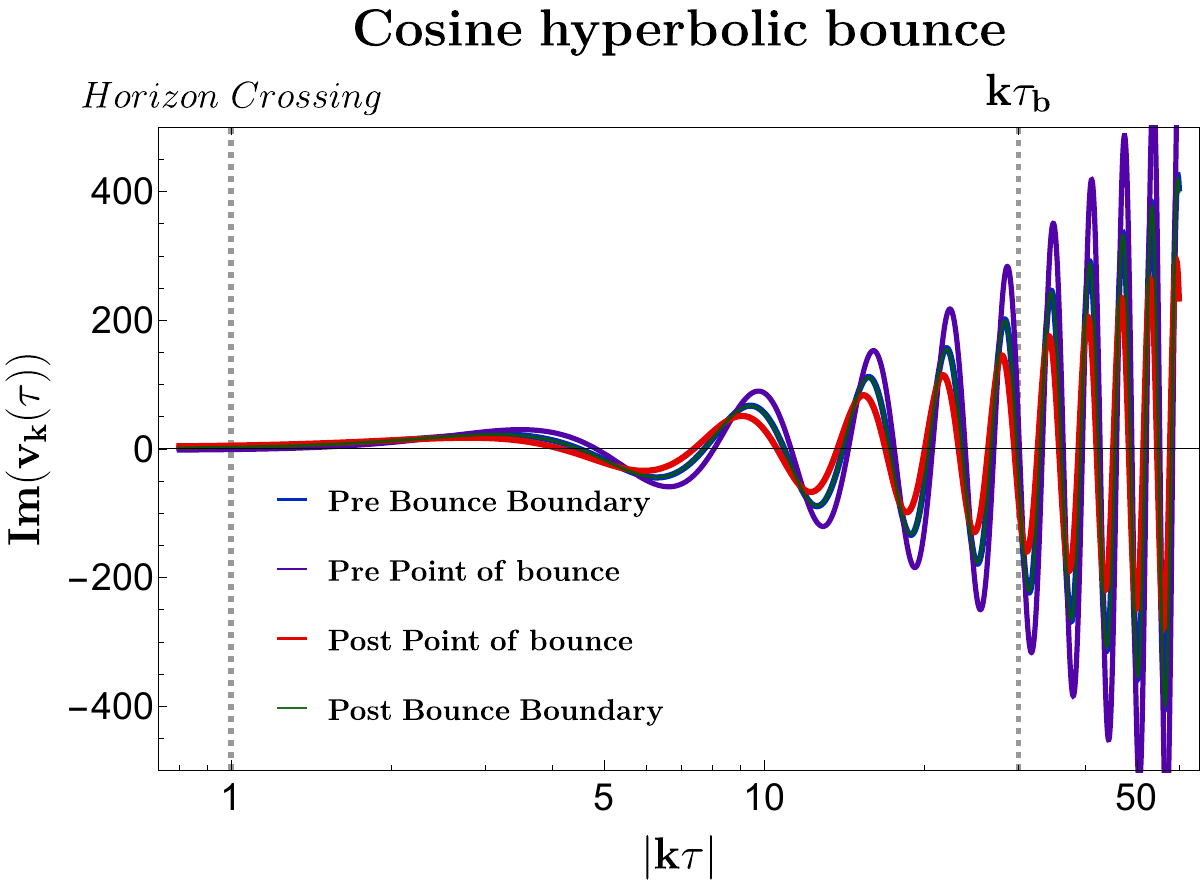}\label{fig:imvcosh}}
	\hfill
	\subfigure[][Model II]{\includegraphics[height=7cm,width=15cm]{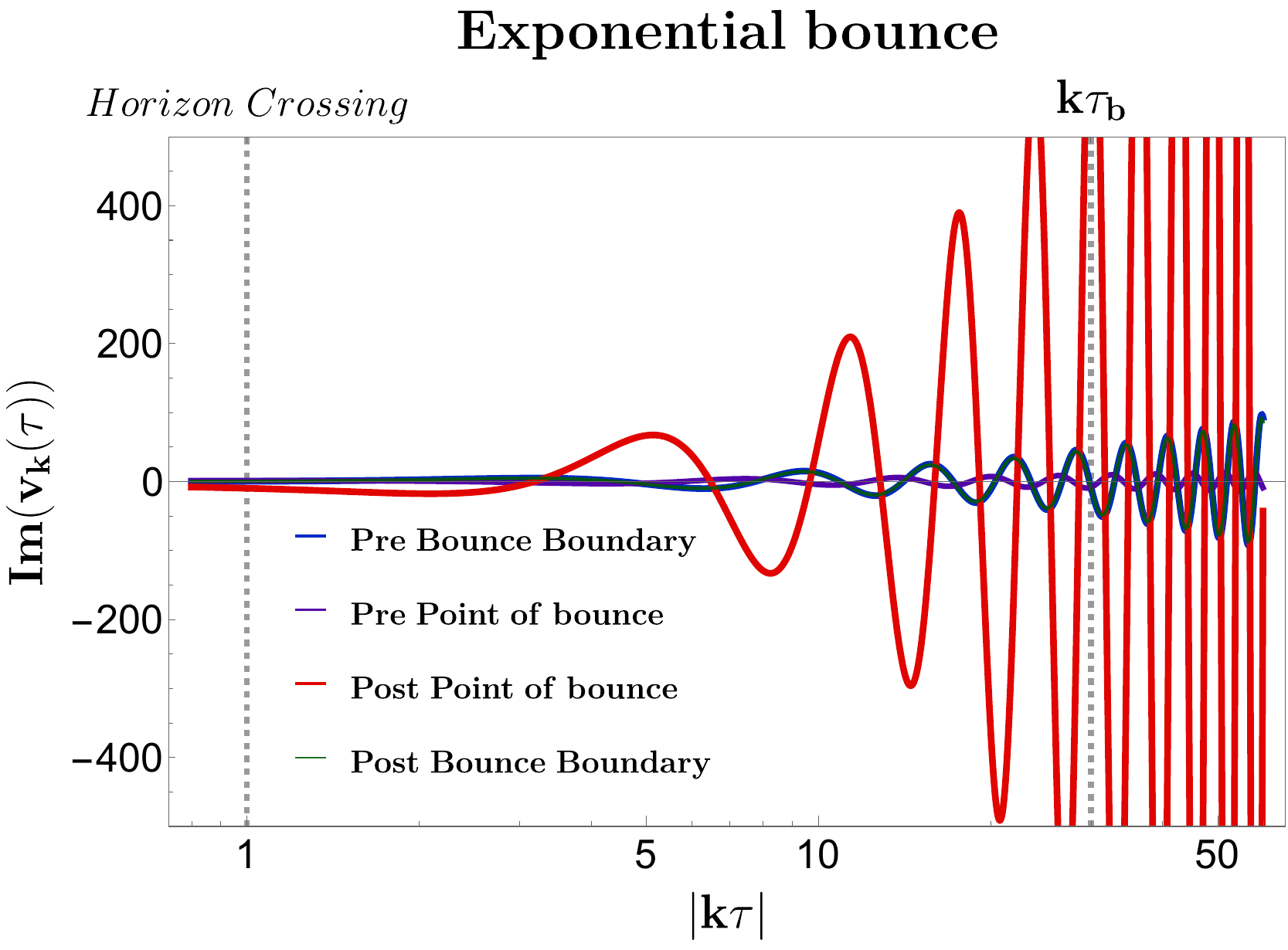}\label{fig:imvexp}}
	\caption{The imaginary part of $v_k$ for the Cosine hyperbolic and the exponential bounce has been plotted. We can see the behaviour of $v_k$ in the sub-horizon region $|k\tau|\gg1$. Around the Horizon crossing, the $Im(v_k)$ starts off positive for the Cosine hyperbolic bounce, whereas it starts off negative for the exponential bounce. Far from the horizon crossing, the behaviour of both the models are almost identical. It slowly starts rising before becoming highly oscillatory with increasing amplitude and frequency as they approach the point of bounce $k\tau_b \approx 30$. The field variables in all regions behave almost similarly. However a noticeable difference between the two models lies in the fact that the amplitude of the field variable for a particular region is in contrast for both the models. If the amplitude is maximum for a particular region for model-I, then it is minimum for model-II.}
	\label{fig:Imvvstau}
\end{figure}
Now, our next objective is to construct the classical Hamiltonian function studied for the present parametric oscillator problem. For this purpose, we need to find out the expression for the canonically conjugate momentum for the classical cosmologically perturbed scalar field variable appearing previously in the second order action perturbed action of the system that we have mentioned earlier in this section, and it is given by the following expression:
\bea \pi_{\bf k}(\tau):=\frac{\partial {\cal L}^{(2)}(v_{\bf k}(\tau),v'_{\bf k}(\tau),\tau)}{\partial v'_{\bf k}(\tau)}=v^{'*}_{{\bf k}}(\tau)-\Biggl(\frac{z'(\tau)}{z(\tau)}\Biggr)v_{{\bf k}}(\tau)\eea
Further, using the above mentioned results one can construct the expression for the classical Hamiltonian function from the present problem set up, which is given by:
\bea H(\tau)=\int d^3{\bf k}~\Biggl[\frac{1}{2}\left|\pi_{\bf k}(\tau)+\frac{z'(\tau)}{z(\tau)}v_{\bf k}(\tau)\right|^2+\frac{1}{2}\mu^2(k,\tau)|v_{\bf k}(\tau)|^2\Biggr],\eea
where the time dependent mass $\mu^2(k,\tau)$ of the parametric oscillator is given by the following expression:
\bea \mu^2(k,\tau):=\Biggl[k^2-\Biggl(\frac{z'(\tau)}{z(\tau)}\Biggr)^2\Biggr].\eea 
Next, using the previously mentioned solution of classical mode function we can further construct the quantum mechanical operators in the Heisenberg picture:
\bea \hat{v}({\bf x},\tau)&=&{\cal U}^{\dagger}(\tau,\tau_0)\hat{v}({\bf x},\tau_0){\cal U}(\tau,\tau_0)\nonumber\\
&=&\int \frac{d^3{\bf k}}{(2\pi)^3}~\left[v^{*}_{-{\bf k}}(\tau)~\hat{a}_{\bf k}+v_{{\bf k}}(\tau)~\hat{a}^{\dagger}_{-{\bf k}}\right]~\exp(i{\bf k}.{\bf x}),\\
\hat{\pi}({\bf x},\tau)&=&{\cal U}^{\dagger}(\tau,\tau_0)\hat{\pi}({\bf x},\tau_0){\cal U}(\tau,\tau_0)\nonumber\\
&=&\int \frac{d^3{\bf k}}{(2\pi)^3}~\left[\pi^{*}_{-{\bf k}}(\tau)~\hat{a}_{\bf k}+\pi_{{\bf k}}(\tau)~\hat{a}^{\dagger}_{-{\bf k}}\right]~\exp(i{\bf k}.{\bf x}).\eea
Now using the above mentioned quantum operator one can finally express the canonical Hamiltonian for the parametric oscillator in the following quantized form(See \Cref{sec:appendixAA}):
\bea \widehat{H}(\tau)&=&\int d^3{\bf k}~\Biggl[\frac{1}{2}\left|\left[v^{*'}_{-{\bf k}}(\tau)~\hat{a}_{\bf k}+v^{'}_{{\bf k}}(\tau)~\hat{a}^{\dagger}_{-{\bf k}}\right]+\frac{z'(\tau)}{z(\tau)}\left[v^{*}_{-{\bf k}}(\tau)~\hat{a}_{\bf k}+v_{{\bf k}}(\tau)~\hat{a}^{\dagger}_{-{\bf k}}\right]\right|^2 \nonumber\\
&&~~~~~~~~~~~~~~~~~~~~~~~~~~~~~~~~~~~~~~~~~~~~+\frac{1}{2}\mu^2(k,\tau)|\left[v^{*}_{-{\bf k}}(\tau)~\hat{a}_{\bf k}+v_{{\bf k}}(\tau)~\hat{a}^{\dagger}_{-{\bf k}}\right]|^2\Biggr]\nonumber\\
&=&\frac{1}{2}\int d^3{\bf k}\Biggl[~\underbrace{\Omega_{\bf k}(\tau)\left(\hat{a}^{\dagger}_{\bf k}\hat{a}_{\bf k}+\hat{a}^{\dagger}_{-{\bf k}}\hat{a}_{-{\bf k}}+1\right)}_{\textcolor{red}{\bf Contribution~from~the~free~term}}\nonumber\\
&&~~~~~~~~~~~~~~~~~~~~~~~~+i\underbrace{\lambda_{\bf k}(\tau)\Biggl(\exp(-2i\phi_{\bf k}(\tau))\hat{a}_{\bf k}\hat{a}_{-{\bf k}}-\exp(2i\phi_{\bf k}(\tau))\hat{a}^{\dagger}_{\bf k}\hat{a}^{\dagger}_{-{\bf k}}\Biggr)}_{\textcolor{red}{\bf Contribution~ from~ the~ Interaction~ term}}~\Biggr],~~~~~~~~~~~~~~\eea
where we define $\Omega_{\bf k}(\tau)$ and $\lambda_{\bf k}(\tau)$ by the following expressions:
\bea \Omega_{\bf k}(\tau):&=&\Biggl\{\left|v^{'}_{{\bf k}}(\tau)\right|^2+\mu^2(k,\tau)\left|v_{{\bf k}}(\tau)\right|^2\Biggr\},~~~~~
\lambda_{\bf k}(\tau):=\Biggl(\frac{z'(\tau)}{z(\tau)}\Biggr).\eea
Here $\Omega_{\bf k}(\tau)$ represents the conformal time dependent dispersion relation in the present bouncing cosmological set-up, and $\lambda_{\bf k}(\tau)$ basically captures the slowly conformal time varying function $\ln z(\tau)$, where $z(\tau)=a\sqrt{2\epsilon}$, is the {\it Mukhanov variable}, which appears during the computation of cosmological perturbation for scalar modes in the bouncing set-up. For the details of the computation, please refer to \Cref{sec:appendixA}

\textcolor{Sepia}{\subsection{\sffamily Time evolution of quantized scalar modes}}

\textcolor{Sepia}{\subsubsection{\sffamily Fixing the initial condition at horizon crossing}}
Here it is important to note that one can fix the initial condition in such a way that, at the time scale $\tau=\tau_0$, we get the following normalization:
\bea v_{\bf k}(\tau_0)&=&\frac{1}{\sqrt{2k}}~2^{\nu_{\rm B}-1}\left|\frac{\Gamma(\nu_{\rm B})}{\Gamma\left(\frac{3}{2}\right)}\right|~\exp\left(-i\left\{\frac{\pi}{2}\left(\nu_{\rm B}-2\right)-1\right\}\right),\\
\pi_{\bf k}(\tau_0)&=&i\sqrt{\frac{k}{2}}~2^{\nu_{\rm B}-\frac{3}{2}}\left|\frac{\Gamma(\nu_{\rm B})}{\Gamma\left(\frac{3}{2}\right)}\right|~\exp\left(-i\left\{\frac{\pi}{2}\left(\nu_{\rm B}-2\right)-1\right\}\right)~\nonumber\\
 &&~~~~~~~~~~~~~~\left[1-\sqrt{2}\frac{\displaystyle \left(\nu_{\rm B}-\frac{1}{2}\right)\left(\nu_{\rm B}+\frac{1}{2}+i\right)}{\displaystyle \left(\nu_{\rm B}+\frac{1}{2}\right)}\exp\left(-\frac{i\pi}{4}\right)\right],~\eea
provided we have imposed a constraint that, $k\tau_0=-1$, which basically represents the horizon crossing scale. Following this fact it is further expected that at any arbitrary later time scale $\tau$ in the Heisenberg picture one can write the associated quantum operators for the present problem as:
\bea \hat{v}_{\bf k}(\tau)&=&v_{\bf k}(\tau_0)\Biggl(a_{\bf k}(\tau)+a^{\dagger}_{-{\bf k}}(\tau)\Biggr),\\
\hat{\pi}_{\bf k}(\tau)&=&-\pi_{\bf k}(\tau_0)~\Biggl(a_{\bf k}(\tau)-a^{\dagger}_{-{\bf k}}(\tau)\Biggr), \eea
where both the creation and the annihilation operators at time $\tau$ can be expressed in terms of the results obtained from the initial time scale $\tau=\tau_0$ using the following unitary similarity transformation in the Heisenberg picture:
\bea && a_{\bf k}(\tau):={\cal U}^{\dagger}(\tau,\tau_0)a_{\bf k}{\cal U}(\tau,\tau_0),\\
&& a^{\dagger}_{-{\bf k}}(\tau):={\cal U}^{\dagger}(\tau,\tau_0)a^{\dagger}_{-{\bf k}}{\cal U}(\tau,\tau_0).\eea
Our next job is to determine the expression for the above mentioned unitary operator in the context of cosmological primordial perturbations of the scalar modes and to determine this expression, the well known {\it squeezed state formalism} used in the context of quantum mechanics will play a significant role.

\textcolor{Sepia}{\subsubsection{\sffamily Squeezed state formalism in Cosmology}}

The unitary evolution operator $\mathcal{U}$, produced by the previously mentioned  full quadratic quantized Hamiltonian function, can be factorized by following the proposal given in refs.~\cite{Albrecht:1992kf,Grishchuk:1990bj} and can be written as:
\begin{equation}
\label{eq:unitary}
\mathcal{U}(\tau,\tau_0) = \hat{\mathcal{S}}(r_{\bf k}(\tau,\tau_0) ,\phi_{\bf k}(\tau) )\hat{\mathcal{R}}(\theta_{\bf k}(\tau) ),
\end{equation}
where $\mathcal{R}$ is the two mode rotation operator, which is defined as:
\begin{equation}
\label{eq:rotationoperator}
\hat{\mathcal{R}}(\theta_{\bf k}(\tau) ) = \exp\Biggl( -i\theta_{k}(\tau)\big( \hat{a}_{\bf k}\hat{a}_{\bf k}^{\dagger} + \hat{a}_{-{\bf k}}^{\dagger}\hat{a}_{-{\bf k}} \big) \Biggr),
\end{equation}
and $\hat{\mathcal{S}}$ is the two-mode squeezing operator, defined as:
\begin{equation}
\label{eq:Squeezedoperator}
\hat{\mathcal{S}}(r_{\bf k}(\tau) ,\phi_{\bf k}(\tau) ) = \exp\Bigg(\frac{r_{\bf k}(\tau)}{2} \big[ \exp(-2i \phi_{\bf k}(\tau) )\hat{a}_{\bf k}\hat{a}_{-{\bf k}} - \exp(2i \phi_{\bf k}(\tau))\hat{a}_{-{\bf k}}^{\dagger}\hat{a}_{\bf k}^{\dagger} \big]\Bigg).
\end{equation}
Here the squeezing amplitude is represented by the time-dependent parameter, $r_{\bf k}(\tau)$ ,and the squeezing angle or the phase is represented by the time-dependent parameter $\phi_{\bf k}(\tau)$. Additionally, it is important to note that, the two-mode rotation operator, $\hat{\mathcal{R}}$ produces an irrelevant phase contribution $\exp(i\theta_{\bf k}(\tau))$ while acted upon the initial quantum vacuum state and can be ignored from our current analysis to avoid the appearance of unnecessary junks.
By recognizing that the interaction of the cosmological perturbation with the conformal time-dependent scale factor in the spatially flat FLRW background leads to
a conformal time-dependent frequency for the canonically normalized parametric oscillator, the appearance of a squeezed
quantum mechanical state for cosmological primordial perturbations is quite natural. The quantization of the conformal time dependent parametric oscillator is then described in terms of two-mode squeezed state formalism as introduced in ref.~ \cite{Albrecht:1992kf}.

For our further computation we choose the ground state of the free Hamiltonian as the initial quantum mechanical state:
\bea
\label{eq:initial vacuum}
\hat{a}_{\bf k}\ket{0}_{{\bf k}, -{\bf k}} = 0 ~~~~~ \forall~~ {\bf k},
\eea
which is basically a Poincare invariant vacuum state in the present context of discussion.

Now we are going to use the squeezed quantum operator $\hat{\mathcal{S}}$ which acts on the above mentioned initial vacuum state and produce a two-mode squeezed quantum vacuum state, as:
\bea
\ket{\Psi_{\bf sq}}_{{\bf k}, -{\bf k}} &=& \hat{\mathcal{S}}(r_{\bf k}(\tau),\phi_{\bf k}(\tau))\ket{0}_{{\bf k},-{\bf k}} \nonumber\\
&=& \frac{1}{\cosh r_{\bf k}(\tau)}\sum_{n=0}^{\infty}(-1)^{n}\exp(-2in~\phi_{\bf k}(\tau)\tanh^{n}r_{\bf k}(\tau)\ket{n_{\bf k}, n_{-{\bf k}}},~~~~
\eea
with the following two-mode excited or usually known as the occupation number state given by the following expression:
\begin{equation}
\label{eq:excitedmode}
\ket{n_{\bf k}, n_{-{\bf k}}} = \frac{1}{n!}\big( \hat{a}_{{\bf k}}^{\dagger} \big)^{n}\big( \hat{a}_{-{\bf k}}^{\dagger} \big)^{n}\ket{0}_{{\bf k},-{\bf k}} .
\end{equation}
Consequently, in the present context of discussion the full quantum wave function  can be expressed in terms of the product of the wave function for each two-mode pair as ${\bf k}, -{\bf k}$ given by the following expression:
\bea
\label{eq:fullwavefunction}
\ket{\Psi_{\bf sq}} &=& \bigotimes_{{\bf k}}\ket{\Psi_{\bf sq}}_{{\bf k}, -{\bf k}}\nonumber\\
&=& \bigotimes_{{\bf k}} \frac{1}{\cosh r_{\bf k}(\tau)}\Biggl(\sum_{n=0}^{\infty}\frac{(-1)^{n}}{n!}\exp(-2in~\phi_{\bf k}(\tau)\tanh^{n}r_{\bf k}(\tau)\big( \hat{a}_{{\bf k}}^{\dagger} \big)^{n}\big( \hat{a}_{-{\bf k}}^{\dagger} \big)^{n}\Biggr)\ket{0}_{{\bf k},-{\bf k}} ,\nonumber\\
&&
\eea

\textcolor{Sepia}{\subsubsection{\sffamily Time evolution in squeezed state formalism}}

Now we go back to the previous discussion where we have written the creation and the annihilation operators of the conformal time dependent parametric oscillator in the cosmological perturbation theory at any arbitrary time using the Heisenberg picture. This will help us to explicitly identify the time evolution of the perturbation field variable operator corresponding to the scalar modes and its associated canonically conjugate momentum operator. In terms of the above mentioned squeezed quantum state description one can further express the creation and annihilation operators in the present context as the unitary operator for the time evolution in the Heisenberg picture. The unitary operator can in turn be factorized in terms of the two-mode rotation operator and two-mode squeezed quantum state operator as we have discussed earlier. After performing the unitary similarity transformation in terms of the two-mode rotation and squeezed operator, one can write down the following expressions for the creation and the annihilation quantum operators at any arbitrary time scale $\tau$ as:
\bea \hat{a}_{\bf k}(\tau)&=&\hat{\cal U}^{\dagger}(\tau,\tau_0)~\hat{a}_{\bf k}~\hat{\cal U}(\tau,\tau_0)\nonumber\\
&=&\hat{\mathcal{R}}^{\dagger}(\theta_{\bf k}(\tau))\hat{\mathcal{S}}^{\dagger}(r_{\bf k}(\tau),\phi_{\bf k}(\tau))~\hat{a}_{\bf k}~\hat{\mathcal{R}}(\theta_{\bf k}(\tau))\hat{\mathcal{S}}(r_{\bf k}(\tau),\phi_{\bf k}(\tau))\nonumber\\
&=&\cosh r_{\bf k}(\tau)~\exp(-i\theta_{\bf k}(\tau))~\hat{a}_{\bf k}-\sinh r_{\bf k}(\tau)~\exp(i(\theta_{\bf k}(\tau)+2\phi_{\bf k}(\tau)))~\hat{a}^{\dagger}_{-{\bf k}},\\
\hat{a}^{\dagger}_{-{\bf k}}(\tau)&=&\hat{\cal U}^{\dagger}(\tau,\tau_0)~\hat{a}^{\dagger}_{-{\bf k}}~\hat{\cal U}(\tau,\tau_0)\nonumber\\
&=&\hat{\mathcal{R}}^{\dagger}(\theta_{\bf k}(\tau))\hat{\mathcal{S}}^{\dagger}(r_{\bf k}(\tau),\phi_{\bf k}(\tau))~\hat{a}^{\dagger}_{-{\bf k}}~\hat{\mathcal{R}}(\theta_{\bf k}(\tau))\hat{\mathcal{S}}(r_{\bf k}(\tau),\phi_{\bf k}(\tau))\nonumber\\
&=&\cosh r_{\bf k}(\tau)~\exp(i\theta_{\bf k}(\tau))~\hat{a}^{\dagger}_{-{\bf k}}-\sinh r_{\bf k}(\tau)~\exp(-i(\theta_{\bf k}(\tau)+2\phi_{\bf k}(\tau)))~\hat{a}_{{\bf k}}.~~~~\eea 
Consequently, the quantum operator associated with the cosmological perturbation field variable for the scalar fluctuation and the its canonically conjugate momenta can be expressed as:
\bea \hat{v}_{\bf k}(\tau)&=&v_{\bf k}(\tau_0)\Biggl(\hat{a}_{\bf k}(\tau)+\hat{a}^{\dagger}_{-{\bf k}}(\tau)\Biggr)\nonumber\\
&=&v_{\bf k}(\tau_0)\Biggl[\hat{a}_{\bf k}\Biggl(\cosh r_{\bf k}(\tau)~\exp(-i\theta_{\bf k}(\tau))-\sinh r_{\bf k}(\tau)~\exp(-i(\theta_{\bf k}(\tau)+2\phi_{\bf k}(\tau)))\Biggr)\nonumber\\
&&~~~~~~~+\hat{a}^{\dagger}_{-{\bf k}}\Biggl(\cosh r_{\bf k}(\tau)~\exp(i\theta_{\bf k}(\tau))-\sinh r_{\bf k}(\tau)~\exp(i(\theta_{\bf k}(\tau)+2\phi_{\bf k}(\tau)))\Biggr)\Biggr],\nonumber\\
&=&\left[v^{*}_{-{\bf k}}(\tau)~\hat{a}_{\bf k}+v_{{\bf k}}(\tau)~\hat{a}^{\dagger}_{-{\bf k}}\right],\\
\hat{\pi}_{\bf k}(\tau)&=&-\pi_{\bf k}(\tau_0)~\Biggl(a_{\bf k}(\tau)-a^{\dagger}_{-{\bf k}}(\tau)\Biggr)\nonumber\\
&=&-\pi_{\bf k}(\tau_0)\Biggl[\hat{a}_{\bf k}\Biggl(\cosh r_{\bf k}(\tau)~\exp(-i\theta_{\bf k}(\tau))+\sinh r_{\bf k}(\tau)~\exp(-i(\theta_{\bf k}(\tau)+2\phi_{\bf k}(\tau)))\Biggr)\nonumber\\
&&~~~~~~~-\hat{a}^{\dagger}_{-{\bf k}}\Biggl(\cosh r_{\bf k}(\tau)~\exp(i\theta_{\bf k}(\tau))+\sinh r_{\bf k}(\tau)~\exp(i(\theta_{\bf k}(\tau)+2\phi_{\bf k}(\tau)))\Biggr)\Biggr],\nonumber\\
&=&\left[\pi^{*}_{-{\bf k}}(\tau)~\hat{a}_{\bf k}+\pi_{{\bf k}}(\tau)~\hat{a}^{\dagger}_{-{\bf k}}\right]. \eea
Here we identify the classical mode function and the associated canonically conjugate momentum in terms of the squeezed parameters as:
\bea v_{{\bf k}}(\tau)&=&v_{\bf k}(\tau_0)\Biggl(\cosh r_{\bf k}(\tau)~\exp(i\theta_{\bf k}(\tau))-\sinh r_{\bf k}(\tau)~\exp(i(\theta_{\bf k}(\tau)+2\phi_{\bf k}(\tau)))\Biggr),~~~~~~~~\\
\pi_{{\bf k}}(\tau)&=&\pi_{\bf k}(\tau_0)\Biggl(\cosh r_{\bf k}(\tau)~\exp(i\theta_{\bf k}(\tau))+\sinh r_{\bf k}(\tau)~\exp(i(\theta_{\bf k}(\tau)+2\phi_{\bf k}(\tau)))\Biggr).\eea 
Further, the time evolution of the conformal time dependent quantum operators $\hat{\mathcal{R}}$ and $\hat{\mathcal{S}}$ are described by the Schr\"{o}dinger equation, which gives the following set of differential equations for the squeezing parameters in the present context:
\begin{align} 
\label{eq:diffneqns1}
&\frac{dr_{\bf k}(\tau)}{d\tau} = -\lambda_{\bf k}(\tau)~\cos(2\phi_{\bf k}(\tau)),\\
\label{eq:diffeqns2}
&\frac{d\phi_{\bf k}(\tau)}{d\tau} = \Omega_{\bf k}(\tau) + \lambda_{\bf k}(\tau)~\coth(2r_{\bf k}(\tau))\sin(2\phi_{\bf k}(\tau)),
\end{align}
where the time dependent factors, $\lambda_{\bf k}(\tau)$ and $\Omega_{\bf k}(\tau)$ in the squeezed state picture in the sub-Hubble region ($-k\tau\gg 1$) can be recast as:
\bea 
\lambda_{\bf k}(\tau):&=&\Biggl(\frac{z'(\tau)}{z(\tau)}\Biggr)={\cal H}\Biggl[\frac{1}{\epsilon(\tau)}-1+\epsilon(\tau)-\frac{1}{2}\frac{1}{\epsilon(\tau)}\frac{{\cal H}''}{{\cal H}^3}\Biggr],\\
 \Omega_{\bf k}(\tau):&=&\Biggl\{\left|\pi_{{\bf k}}(\tau)+\lambda_{\bf k}(\tau)v_{\bf k}(\tau)\right|^2+\left(k^2-\lambda^2_{\bf k}(\tau)\right)\left|v_{{\bf k}}(\tau)\right|^2\Biggr\}\nonumber\\
 &\approx & 3k~2^{2(\nu_{\rm B}-2)}~\left|\frac{\Gamma(\nu_{\rm B})}{\Gamma\left(\frac{3}{2}\right)}\right|^2. \eea
 Here it is important to note that in the sub-Hubble region the factor $\Omega_{\bf k}(\tau)$ is mainly controlled by the momentum scale of the scalar mode of the perturbation, $k$, and the slowly varying time dependence is taken care of by the conformal time dependent mass parameter $\nu_{\rm B}$, which can be approximately written by considering the contribution upto the next-to-leading order as:
 \bea \nu_{\rm B}\approx\left(\frac{1}{2}+\frac{{\cal H}''}{{\cal H}^2}\right),\eea
 where we have neglected the contributions of all higher order small correction terms for the computational simplicity. Now after substituting the above mentioned expression for the mass parameter $\nu_{\rm B}$ one can further write the following simplified form of the factor, $\Omega_{\bf k}(\tau)$ in the sub-Hubble region, as:
\bea \label{eq:Sigma_k} \Omega_{\bf k}(\tau)&\approx& \frac{3}{2}~k~2^{\displaystyle \left(\frac{2{\cal H}''}{{\cal H}^2}\right)}~\left|\frac{\displaystyle \Gamma\left(\frac{1}{2}+\frac{{\cal H}''}{{\cal H}^2}\right)}{\displaystyle \Gamma\left(\frac{1}{2}\right)}\right|^2\nonumber\\
&=&\frac{3}{2\pi}~k~2^{\displaystyle \left(\frac{2{\cal H}''}{{\cal H}^2}\right)}~\left|\displaystyle \Gamma\left(\frac{1}{2}+\frac{{\cal H}''}{{\cal H}^2}\right)\right|^2\nonumber\\
&\approx &\frac{3}{\pi}~k~\left[\left(1-\frac{1}{2}\gamma_{\rm E}\right)+2\left\{\left(1-\frac{1}{2}\gamma_{\rm E}\right)\ln2-1\right\} \left(\frac{{\cal H}''}{{\cal H}^2}\right)-4\ln 2\left(\frac{{\cal H}''}{{\cal H}^2}\right)^2+\cdots\right],~~~~~~~~~~~\eea
where $\gamma_{\rm E}$ is the {\it Euler-Mascheroni constant}, which is $\gamma_{\rm E}=0.577$. For a more detailed discussion on dispersion relation please refer to \Cref{sec:appendixB}.\\
\textcolor{Sepia}{\section{\sffamily Quantum complexity from squeezed quantum states in Bouncing cosmology} 
\label{sec:coscompwsqueezedQS}}
In this section, we compute the complexity from the squeezed cosmological perturbations studied in the previous section for the bouncing framework. We use the wave function formalism of computing circuit complexity developed by \cite{Jefferson:2017sdb,Guo:2018kzl} and used extensively in \cite{Bhattacharyya:2020kgu,Bhattacharyya:2020rpy,Bhattacharyya:2019kvj}. Computing the circuit complexity involves choosing a certain reference state and a target state. In the case of cosmological perturbations, a commonly chosen reference state is the two-mode quantum initial vacuum state $\ket{0}_{{\bf k},-{\bf k}}$, as mentioned in the previous section. The target quantum state is the squeezed two-mode vacuum state $\ket{\Psi_{\bf sq}}_{{\bf k},-{\bf k}}$. In ref.~\cite{Jefferson:2017sdb,Guo:2018kzl} the authors expressed the reference and the target states as Gaussian wave-functions. We follow an identical approach in this paper for further computation. We use the following field operator and its associated canonically conjugate momentum operator as:
\bea
	&&\hat{v}_{\bf k}(\tau) =v_{\bf k}(\tau_0)\left[\hat{a}_{\bf k}^{\dagger}(\tau)+\hat{a}_{\bf k}(\tau)\right] ,\\
	&& \hat{\pi}_{\vec{k}}(\tau) =\pi_{\bf k}(\tau_0)\left[\hat{a}_{\bf k}^{\dagger}(\tau)-\hat{a}_{\bf k}(\tau)\right]
\eea
where $v_{\bf k}(\tau_0)$ and $\pi_{\bf k}(\tau_0)$ fix the initial condition on the classical scalar mode and its associated canonically conjugate momentum at the horizon crossing scale, $-k\tau_0=1$. We have computed their explicit expressions in the previous section. Additionally, we have also computed the expressions for the associated quantum operators at any arbitrary time scale $\tau$ in terms of the squeezed conformal time dependent parameters $r_{\bf k}(\tau)$ and $\theta_{\bf k}(\tau)$ in the Heisenberg picture of quantum mechanics. At any arbitrary time scale $\tau$, these cosmological quantum operators satisfy the following well known equal time commutation relation (ETCR), given by:
\bea \left[\hat{v}_{\bf k}(\tau),\hat{\pi}_{\bf k'}(\tau)\right]=i\delta^3\left({\bf k}-{\bf k}'\right).\eea

The two-mode vacuum state wave function, which we choose as our reference state is defined as:
 \bea \hat{a}_{\bf k}\ket{0}_{{\bf k},-{\bf k}}=0~~~\forall~{\bf k}\eea
 which has the following usual Gaussian structure:
\bea
\label{eq:vacuumstate}
	\Psi_{\rm Ref}(v_{\bf k},v_{-{\bf k}}) :=\left( \frac{\Omega_{\bf k}}{\pi}\right)^{1/4} \exp\left(-\frac{\Omega_{\bf k}}{2}(v^2_{\bf k} + v^2_{-{\bf k}})\right)
\eea
where we have used the expression for $\Omega_{\bf k}$ in the sub-Hubble region, the approximated analytical expression of which we have already derived explicitly in the previous section.

The wave function of the target or the squeezed quantum state for the cosmological perturbation can be calculated by noting that a particular combination of the squeezing parameters along with the creation and annihilation operator annihilates the two mode squeezed vacuum state, constructed in the previous section. That particular combination is written as:
\bea
	\left(\cosh r_{\bf k}(\tau)~\hat{a}_{\bf k} +\exp(-2i\phi_{\bf k}(\tau)) \sinh r_{\bf k}(\tau)~ \hat{a}^{\dagger}_{-{\bf k}} \right) \ket{\Psi_{\bf sq}}_{{\bf k},-{\bf k}} =0.
\eea
The cosmological perturbed field space representation of the wave function is given by the following expression:
\bea
\label{eq:squeezedstate}
	\Psi_{\bf sq}(v_{\bf k}, v_{-{\bf k}})&=&\langle v_{\bf k},v_{-{\bf k}}|\Psi_{\bf sq}\rangle_{{\bf k},-{\bf k}}\nonumber\\
	& =& \frac{\exp\left(\mathcal{A}(\tau)~(v^2_{\bf k} + v^2_{-{\bf k}})-\mathcal{B}(\tau)~ v_{\bf k}~v_{-{\bf k}}\right)}{\cosh~r_{\bf k}(\tau) \sqrt{\pi (1-\exp(-4i\phi_{\bf k}(\tau))~\tanh^2 r_{\bf k}(\tau)-1)}},
\eea
where the coefficients $\mathcal{A}(\tau)$ and $\mathcal{B}(\tau)$ are the functions of the squeezing parameter $r_{\bf k}(\tau)$ and the squeezing angle $\phi_{\bf k}(\tau)$, and are explicitly given by the following expression:
\bea
\label{eq:AandB}
	&& {\cal A}(\tau):= \frac{\Omega_{\bf k}}{2}\biggl(\frac{\exp(-4i\phi_{\bf k}(\tau))~\tanh^2 r_{\bf k}(\tau)+1}{\exp(-4i\phi_{\bf k}(\tau))~\tanh^2 r_{\bf k}(\tau)-1}\biggr),\\
	&& {\cal B}(\tau):= 2\Omega_{\bf k}\biggl(\frac{\exp(-2i\phi_{\bf k}(\tau))~\tanh^2 r_{\bf k}(\tau)}{\exp(-4i\phi_{\bf k}(\tau))~\tanh^2 r_{\bf k}(\tau)-1}\biggr).
\eea
The vacuum reference and the target squeezed state written in \ref{eq:vacuumstate} and \ref{eq:squeezedstate} is eventually used to calculate the complexity from two types of cost functions namely the "linear weighting" ($\mathcal{C}_1$) and the "geodesic weighting" ($\mathcal{C}_2$) respectively within the framework of Cosmology and represented by the following expressions:
\bea
\label{eq:compform1} 
	\mathcal{C}_1(k) &=&\frac{1}{2}\biggl(\text{ln}\biggl|\frac{\Sigma_{\k}}{\omega_{\k}}\biggr|+ \text{ln}\biggl|\frac{\Sigma_{-\k}}{\omega_{-\k}}\biggr|+ \tan^{-1} \frac{\text{Im}~\Sigma_{\k}}{\text{Re}~\omega_{\k}} + \tan^{-1} \frac{\text{Im}~\Sigma_{-\k}}{\text{Re}~\omega_{-\k}}\biggr)\nonumber\\
	&=&\frac{1}{2}\Biggl(\text{ln}\biggl|\frac{\Sigma_{\k}}{\omega_{\k}}\biggr|+ \text{ln}\biggl|\frac{\Sigma_{-\k}}{\omega_{-\k}}\biggr|+ \tan^{-1}\frac{\displaystyle \frac{\text{Im}~\Sigma_{\k}}{\text{Re}~\omega_{\k}}+\frac{\text{Im}~\Sigma_{-\k}}{\text{Re}~\omega_{-\k}}}{\displaystyle 1-\frac{\text{Im}~\Sigma_{-\k}}{\text{Re}~\omega_{-\k}}\frac{\text{Im}~\Sigma_{-\k}}{\text{Re}~\omega_{-\k}}}\Biggr) \\
	\label{eq:compform2}
	\mathcal{C}_2(k) &=& \frac{1}{2}\sqrt{\biggl(\ln\biggl|\frac{\Sigma_{\k}(\tau)}{\omega_{\k}(\tau)}\biggr|\biggr)^2 + \biggl(\ln\biggl|\frac{\Sigma_{-\k}(\tau)}{\omega_{-\k}(\tau)}\biggr|\biggr)^2 + \biggl(\tan^{-1} \frac{\text{Im}~\Sigma_{\k}(\tau)}{\text{Re}~\omega_{\k}(\tau)} +\biggr)^2 + \biggl(\tan^{-1} \frac{\text{Im}~\Sigma_{-\k}(\tau)}{\text{Re}~\omega_{-\k}(\tau)}\biggr)^2 }.~\nonumber\\
	&&
\eea
A trivial generalisation of the complexity measure of the homogeneous and inhomogeneous family of cost functionals can be also be done in the present context. The expression of the complexity for the homogeneous family is given by:
\begin{align}
\label{eq:compkfamily}
	\mathcal{C}_{\kappa}(k) = \frac{1}{2}\biggl[\biggl(\ln\biggl|\frac{\Sigma_{\k}(\tau)}{\omega_{\k}(\tau)}\biggr|\biggr)^{\kappa} +\biggl(\ln\biggl|\frac{\Sigma_{-\k}(\tau)}{\omega_{-\k}(\tau)}\biggr|\biggr)^{\kappa}+\biggl(\tan^{-1} \frac{\text{Im}~\Sigma_{\k}(\tau)}{\text{Re}~\omega_{\k}(\tau)} +\biggr)^{\kappa}+\biggl(\tan^{-1} \frac{\text{Im}~\Sigma_{-\k}(\tau)}{\text{Re}~\omega_{-\k}(\tau)}\biggr)^{\kappa}\biggr]
\end{align} 
Similarly the expression of the complexity for the inhomogeneous family can be written as:
\begin{align}
\label{eq:compibykfamily}
\mathcal{C}_{\frac{1}{\kappa}}(k) = \biggl[\frac{1}{2}\biggl[\biggl(\ln\biggl|\frac{\Sigma_{\k}(\tau)}{\omega_{\k}(\tau)}\biggr|\biggr)^{\kappa} +\biggl(\ln\biggl|\frac{\Sigma_{-\k}(\tau)}{\omega_{-\k}(\tau)}\biggr|\biggr)^{\kappa}+\biggl(\tan^{-1} \frac{\text{Im}~\Sigma_{\k}(\tau)}{\text{Re}~\omega_{\k}(\tau)} +\biggr)^{\kappa}+\biggl(\tan^{-1} \frac{\text{Im}~\Sigma_{-\k}(\tau)}{\text{Re}~\omega_{-\k}(\tau)}\biggr)^{\kappa}\biggr]\biggr]^{\frac{1}{\kappa}}
\end{align} 
where we define the following functions:
\bea &&\Sigma_{\bf k}(\tau)={\cal B}(\tau)-2{\cal A}(\tau),\\
&&\Sigma_{-{\bf k}}(\tau)=-{\cal B}(\tau)-2{\cal A}(\tau),\\
&& \omega_{\bf k}(\tau)=\frac{1}{2}\Omega_{\bf k}=\omega_{-{\bf k}}(\tau).
\eea
It might happen that in some particular context, the measures $\mathcal{C}_1$ and $\mathcal{C}_2$ are not good enough to probe the underlying chaos and randomness of the system. Complexity calculated from the homogeneous and the non homogeneous family might come in handy in that scenario and may bring out some essential features which remains unidentified by $\mathcal{C}_1$ and $\mathcal{C}_2$. In this paper, we have mainly focused on the complexity measure calculated from the linear and the geodesic weighted cost functionals to comment on the chaoticity of the universe from the bouncing cosmological framework. We can express the complexity measures in terms of the squeezed state parameters.  
Substituting \ref{eq:AandB} in \ref{eq:compform1}, \ref{eq:compform2}, \ref{eq:compkfamily} and \ref{eq:compibykfamily} the complexity measures for the bouncing set up for two mode squeezed vacuum state can be written as:
\begin{align}
\label{eq:complexities}
	\mathcal{C}_1(k,\tau) &= \biggl|\text{ln}\biggl|\frac{1+\exp(-2i\phi_\k(\tau))~\tanh r_\k (\tau)}{1-\exp(-2i\phi_\k(\tau))~\tanh r_\k (\tau)}\biggr|\biggr|+ |\tanh^{-1}(\sin(2\phi_\k (\tau))\sinh (2r_\k (\tau)))| \\
	\mathcal{C}_2(k,\tau) &= \frac{1}{\sqrt{2}}\sqrt{\biggl(\text{ln}\biggl|\frac{1+\exp(-2i\phi_\k(\tau))~\tanh r_\k (\tau)}{1-\exp(-2i\phi_\k(\tau))~\tanh r_\k (\tau)}\biggr|\biggr)^2 + (\tanh^{-1}(\sin(2\phi_\k (\tau))\sinh (2r_\k (\tau))))^2}.\\
	\mathcal{C}_{\kappa}(k,\tau) &= \biggl|\text{ln}\biggl|\frac{1+\exp(-2i\phi_\k(\tau))~\tanh r_\k (\tau)}{1-\exp(-2i\phi_\k(\tau))~\tanh r_\k (\tau)}\biggr|\biggr|^{\kappa}+ |\tanh^{-1}(\sin(2\phi_\k (\tau))\sinh (2r_\k (\tau)))|^{\kappa}\\
	\mathcal{C}_{\frac{1}{\kappa}}(k,\tau) &= \biggl[\biggl|\text{ln}\biggl|\frac{1+\exp(-2i\phi_\k(\tau))~\tanh r_\k (\tau)}{1-\exp(-2i\phi_\k(\tau))~\tanh r_\k (\tau)}\biggr|\biggr|^{\kappa}+ |\tanh^{-1}(\sin(2\phi_\k (\tau))\sinh (2r_\k (\tau)))|^{\kappa}\biggr]^{\frac{1}{\kappa}}
\end{align} 
One can further derive some approximate analytical expressions for the cosmological complexity in different limiting situations, which are discussed below:\\
\begin{enumerate}
\item \underline{\textcolor{red}{\bf Small ~$r_{\bf k}(\tau)$~\&~ Small ~$\phi_{\bf k}(\tau)$}:}\\
For small $r_{\bf k}(\tau)$ and $\phi_{\bf k}(\tau)$ one can write:
\bea \exp(-2i\phi_{\bf k}(\tau))\approx 1,~~\sin(2\phi_{\bf k}(\tau))\approx 2\phi_{\bf k}(\tau),~~\tanh r_{\bf k}(\tau)\approx r_{\bf k}(\tau),~~\sinh(2r_{\bf k}(\tau))\approx 2r_{\bf k}(\tau).~~~~~~\eea
In this limit, we have the following simplified formulae of cosmological complexity for the bouncing set up for two mode squeezed vacuum state:
\begin{align}
	\mathcal{C}_1(k,\tau) &\approx 2|r_\k (\tau)|\left(1+2|\phi_\k (\tau)|\right) \\
	\mathcal{C}_2(k,\tau) &\approx \sqrt{2}|r_\k (\tau)|\sqrt{1 + 4(\phi_\k (\tau))^2}. \\
	\mathcal{C}_{\kappa}(k,\tau) &\approx (\mathcal{C}_1(k,\tau))^{\kappa} \\
	\mathcal{C}_{\frac{1}{\kappa}}(k,\tau) &\approx \mathcal{C}_1(k,\tau)\approx \left(\mathcal{C}_{\kappa}(k,\tau)\right)^{\frac{1}{\kappa}}
\end{align} 
\item \underline{\textcolor{red}{\bf Large ~$r_{\bf k}(\tau)$~\&~ Large ~$\phi_{\bf k}(\tau)$}:}\\
For large $r_{\bf k}(\tau)$ and  $\phi_{\bf k}(\tau)$ one can write:
\bea \exp(-2i\phi_{\bf k}(\tau))\approx 0.\eea
Consequently, the cosmological complexity for the bouncing set up for two mode squeezed vacuum state reduces to the following simplified expressions:
\begin{align}
	\mathcal{C}_1(k,\tau) &\approx |\tanh^{-1}(\sin(2\phi_\k (\tau))\sinh (2r_\k (\tau)))| \\
	\mathcal{C}_2(k,\tau) &\approx \frac{1}{\sqrt{2}}\tanh^{-1}(\sin(2\phi_\k (\tau))\sinh (2r_\k (\tau)))\\
	\mathcal{C}_{\kappa}(k,\tau) &\approx (\mathcal{C}_1(k,\tau))^{\kappa} \\
	\mathcal{C}_{\frac{1}{\kappa}}(k,\tau) &\approx \mathcal{C}_1(k,\tau)\approx \left(\mathcal{C}_{\kappa}(k,\tau)\right)^{\frac{1}{\kappa}}
\end{align} 
which will finally lead to the following approximated connecting relationship between the two cosmological complexities computed from different cost functions:
\bea |\mathcal{C}_2(k,\tau)|&\approx &\frac{1}{\sqrt{2}}~\mathcal{C}_1(k,\tau)\approx \frac{1}{\sqrt{2}}~\mathcal{C}_{\frac{1}{\kappa}}(k,\tau)\approx  \frac{1}{\sqrt{2}}~\left(\mathcal{C}_{\kappa}(k,\tau)\right)^{\frac{1}{\kappa}}.\eea
\item \underline{\textcolor{red}{\bf Small ~$r_{\bf k}(\tau)$~\&~ Large ~$\phi_{\bf k}(\tau)$}:}\\
For small $r_{\bf k}(\tau)$ and large $\phi_{\bf k}(\tau)$ one can write:
\bea \exp(-2i\phi_{\bf k}(\tau))\approx 0,~~\tanh r_{\bf k}(\tau)\approx r_{\bf k}(\tau),~~\sinh(2r_{\bf k}(\tau))\approx 2r_{\bf k}(\tau).~~~~~~\eea
Consequently, we have the following simplified formulae of cosmological complexity for the bouncing set up for two mode squeezed vacuum state:
\begin{align}
	\mathcal{C}_1(k,\tau) &\approx 2|r_\k (\tau)\sin(2\phi_\k (\tau))| \\
	\mathcal{C}_2(k,\tau) &= \sqrt{2}r_\k (\tau)\sin(2\phi_\k (\tau))\\
	\mathcal{C}_{\kappa}(k,\tau) &\approx (\mathcal{C}_1(k,\tau))^{\kappa} \\
	\mathcal{C}_{\frac{1}{\kappa}}(k,\tau) &\approx \mathcal{C}_1(k,\tau)\approx \left(\mathcal{C}_{\kappa}(k,\tau)\right)^{\frac{1}{\kappa}},
\end{align}  
which will finally lead to the following approximated relationship between the two cosmological complexities computed from different cost functions:
\bea |\mathcal{C}_2(k,\tau)|&\approx &\frac{1}{\sqrt{2}}~\mathcal{C}_1(k,\tau)\approx \frac{1}{\sqrt{2}}~\mathcal{C}_{\frac{1}{\kappa}}(k,\tau)\approx  \frac{1}{\sqrt{2}}~\left(\mathcal{C}_{\kappa}(k,\tau)\right)^{\frac{1}{\kappa}}.\eea
\item \underline{\textcolor{red}{\bf Large ~$r_{\bf k}(\tau)$~\&~ Small ~$\phi_{\bf k}(\tau)$}:}\\
For large $r_{\bf k}(\tau)$ and  small $\phi_{\bf k}(\tau)$ one can write:
\bea \exp(-2i\phi_{\bf k}(\tau))\approx 1,~~~~\sin(2\phi_\k (\tau))\approx 2\phi_\k (\tau).\eea
Consequently, we have the following simplified formulae of cosmological complexity for the bouncing set up for two mode squeezed vacuum state:
\begin{align}
	\mathcal{C}_1(k,\tau) &= \biggl|\text{ln}\biggl|\frac{1+\tanh r_\k (\tau)}{1-\tanh r_\k (\tau)}\biggr|\biggr|+ |\tanh^{-1}(2\phi_\k (\tau)\sinh (2r_\k (\tau)))| \\
	\mathcal{C}_2(k,\tau) &= \frac{1}{\sqrt{2}}\sqrt{\biggl(\text{ln}\biggl|\frac{1+\tanh r_\k (\tau)}{1-\tanh r_\k (\tau)}\biggr|\biggr)^2 + (\tanh^{-1}(2\phi_\k (\tau)\sinh (2r_\k (\tau))))^2}.\\
	\mathcal{C}_{\kappa}(k,\tau) &= \biggl|\text{ln}\biggl|\frac{1+\tanh r_\k (\tau)}{1-\tanh r_\k (\tau)}\biggr|\biggr|^{\kappa}+ |\tanh^{-1}(2\phi_\k (\tau)\sinh (2r_\k (\tau)))|^{\kappa} \\
	\mathcal{C}_{\frac{1}{\kappa}}(k,\tau) &= \biggl[\biggl|\text{ln}\biggl|\frac{1+\tanh r_\k (\tau)}{1-\tanh r_\k (\tau)}\biggr|\biggr|^{\kappa}+ |\tanh^{-1}(2\phi_\k (\tau)\sinh (2r_\k (\tau)))|^{\kappa}\biggr]^{\frac{1}{\kappa}}=\left[\mathcal{C}_{\kappa}(k,\tau)\right]^{\frac{1}{\kappa}}.
\end{align} 
\end{enumerate}

In the next section, we have done a detailed numerical analysis with the already introduced models of bounce to study their physical impacts on cosmological complexity from two types of cost functions and interpret the physical outcomes from those models.

\textcolor{Sepia}{\section{\sffamily Numerical results and interpretation: Connection with quantum chaos}\label{sec:tybouncesol}}

In this section our prime objective is to numerically solve the time evolution equations of the conformal time dependent squeezed state parameter $r_{\bf k}(\tau)$ and squeezed angle $\theta_{\bf k}(\tau)$, given in \Cref{eq:diffneqns1} and \Cref{eq:diffeqns2}. However, instead of using the conformal time $\tau$ as the dynamical variable, we have chosen the scale factor $a(\tau)$ to make the computation simpler and physically justifiable. To perform the change in variable from $\tau$ to $a(\tau)$ 
we have to replace the following differential operator in the above mentioned evolution equations using the chain rule, as:
\bea \tau\longrightarrow a(\tau):~~~~\frac{d}{d\tau}=\frac{d}{da(\tau)}\frac{da(\tau)}{d\tau}=a'(\tau)\frac{d}{da(\tau)}\eea
In general quantum field theory literature we usually identify such type of variable transformation as \textbf{field redefinition}. One can treat the scale factor $a(\tau)$ as a classical field and the same interpretation is valid in this context.
Consequently, the evolution of the squeezed state parameter $r_{\bf k}(a)$ and squeezed angle $\theta_{\bf k}(a)$, can be recast in terms of the newly defined dynamical variable $a(\tau)$ as:
\begin{align} \label{eq:diffeqnswa1}
&\frac{dr_{\bf k}(a)}{da} = -\frac{\lambda_{\bf k}(a)}{a'} \cos 2\phi_{\bf k}(a),\\
\label{eq:diffeqnswa2}
&\frac{d\phi_{\bf k}(a)}{da} = \frac{\Omega_{\bf k}}{a'} -\frac{\lambda_{\bf k}(a)}{a'} \coth2 r_{\bf k}(a)\sin 2\phi_{\bf k}(a) 
\end{align}
In the above set of evolution equations, since we do not need to care about the explicit conformal time dependence we have written the scale factor $a(\tau)$ as $a$, where $a$ itself is treated as a new dynamical variable. Once we numerically solve the evolution of the squeezed state parameter $r_{\bf k}(a)$ and squeezed angle $\theta_{\bf k}(a)$ in terms of the scale factor $a$, we can construct the target quantum state out of a Gaussian initial state. This will further help us to numerically compute and understand the quantum complexities in Eq (\ref{eq:compform1}) and Eq (\ref{eq:compform2}) within the framework of primordial cosmological perturbation theory, where the effects of the quantum fluctuations is treated in terms of the squeezed state parameter $r_{\bf k}(a)$ and squeezed angle $\theta_{\bf k}(a)$ in the {\it squeezed state formalism}. For the explicit computational details, we suggest the readers to look into the previous two sections very carefully where we have explicitly shown why and how these interesting connections can be established. Now since we have a good understanding of both the complexities, ${\cal C}_1(a)$ and ${\cal C}_2(a)$, we will compute them from the two previously mentioned cost functions and analyze the behaviour, from ${\cal C}_1(a)$ and ${\cal C}_2(a)$, vs scale factor $a$ plots, specifically in the exponentially rising region. Now, by studying the exponential rise in the complexities, ${\cal C}_1(a)$ and ${\cal C}_2(a)$, one can write the following approximated expression for the complexities:
\bea \label{aprox}&& {\cal C}_i(a)\approx c_i~\exp(\lambda_{i}a)~~~\forall~~~i=1,2,\eea
which are valid only in the domain of exponential rising with respect to the scale factor $a$. Additionally, it is important to note that, though the exponential growth feature is common in both the complexities, we have written the expressions for the two complexities separately because the overall amplitudes, which are represented by $c_1$ and $c_2$, and the slope of the previously mentioned plots, quantified by two factors, $\lambda_{1}$ and $\lambda_{2}$, are different which can be confirmed by comparing the features of both the plots. This can be demonstrated as:
\bea \label{eq:lyapexp}  \lambda_i=\left(\frac{d\ln{\cal C}_i(a)}{da}\right)_{a=a_{\rm grow}}~~~~~~\forall~~~~~i=1,2, \eea
where $a_{\rm grow}$ is the specified value of the scale factor from the region where exponential growth feature can be explicitly visible from the complexities vs scale factor plots.

Most importantly, Eq~(\ref{aprox}) is a conjectured relationship which we have proposed by seeing and comparing the numerical behaviour of the obtained plots from this analysis. For this reason, we have written $\approx$ symbol instead of using $=$. To know the complete evolution one needs to solve the system numerically which will give us an exact result, valid in all evolutionary regions of the scale factor $a$, and not only in the exponentially rising region. On the other hand, by doing the explicit computation of out-of time-ordered correlation (OTOC) functions obtained from the classical field $a$ and its canonically conjugate momenta $\pi_{a}$ one can find the following relationship:
\bea \label{eq:otoc} {\rm OTOC}\approx \exp(-c~\exp(\lambda a))~~~~\Longrightarrow~~~~c~\exp(\lambda a)=-\ln\left({\rm OTOC}\right),\eea
which is again valid in the specific region of interest. Here $\lambda$ is identified to be the {\it Quantum Lyapunov Exponent} which captures the effect of chaos in the quantum regime, and in ref.~\cite{Maldacena:2015waa}, the authors, {\it Juan Maldacena} (\textcolor{red}{\bf M}), {\it Stephen Shenker} (\textcolor{red}{\bf S}) and {\it Douglus Stanford} (\textcolor{red}{\bf S}) have found that for a generic quantum chaotic system {\it Quantum Lyapunov Exponent} has to be bounded by the following saturation upper bound, as given by:
\bea \textcolor{red}{\bf MSS~bound:}~~~~\lambda \leq \frac{2\pi}{\beta},~~~~{\rm where}~~\beta=\frac{1}{T},\eea
where $\beta$ is the inverse equilibrium temperature of the chaotic system during saturation of the OTOC at large evolutionary scale. The equality symbol in the {\it MSS} bound represents the maximal saturation of chaotic OTOC. 

Now further using Eq~(\ref{aprox}) and Eq~(\ref{eq:otoc}) together we get the following simplified results: 
\bea &&\Delta_1(a):=\frac{c_1}{c}\exp\left((\lambda_1-\lambda)a\right)= -\frac{{\cal C}_1(a)}{\ln\left({\rm OTOC}\right)},\\
&&\Delta_2(a):=\frac{c_2}{c}\exp\left((\lambda_2-\lambda)a\right)= -\frac{{\cal C}_2(a)}{\ln\left({\rm OTOC}\right)}.\eea
Now after studying  the above mentioned equations we can arrive at the following conclusion:
\bea {\cal C}_i(a)= -\ln\left({\rm OTOC}\right)^{\Delta_i (a)}~~~{\rm where}~~\Delta_{i}(a):=\frac{c_i}{c}\exp\left((\lambda_i-\lambda)a\right)~~~\forall~~i=1,2,~~\eea
which implies that the connection between OTOC and the two different measure of complexities are not strictly same.

Additionally, in the present context we have the following restriction:
\bea \Delta_{1}(a)\neq \Delta_{2}(a)~~~~~{\rm because}~~~{\cal C}_1 (a)\neq {\cal C}_2 (a).\eea

However, to have an universal feature it is expected that the following fact is also true in the present context:
\bea \Delta_{1}(a)\sim {\cal O}(1)~~{\rm and}~~\Delta_{2}(a)\sim {\cal O}(1)~~~{\rm even~when}~~\Delta_{1}(a)\neq \Delta_{2}(a),\eea
which is only true in the limit, $\lambda_i \rightarrow \lambda,~c_i\rightarrow c~~~\forall~~i=1,2$. In this limit, precisely we have:
\bea \Delta_{i}(a)\sim \underbrace{1}_{\textcolor{red}{\bf Leading~order~effect}}+\underbrace{\frac{c_i}{c}(\lambda_i-\lambda)a+\cdots}_{\textcolor{red}{\bf Negligibly~small~sub-leading~effects}}.\eea
This further implies that if we neglect all the extremely small sub-leading contributions, and restrict our attention to only the leading order term then it is possible to write down the following universal highlighting relationship between all possible measures of complexities and the OTOC, as:
\beq
\label{eq:otocuniv}
\textcolor{red}{\bf Universal~relation:}~~~\mathcal{C} = -\ln{({\rm OTOC})} \approx {\cal C}_i~~\forall ~~i=1,2.
\eeq
Here it is important to note that, the above mentioned {\it universal relation} is perfectly consistent with the ref.~\cite{Bhattacharyya:2020kgu}. The only difference is that, here we have achieved the universality using the dynamical variable, scale factor $a$ and in ref.~\cite{Bhattacharyya:2020kgu}, the authors have pointed such universality using the physical time variable $t$. Though, both the discussions hold good in their preferred choice of dynamical variables, ultimately both of them support the same chaotic behaviour during the exponential rise.  

Also it is observed that, when the universality is achieved we expect to get a saturation in the behaviour of complexities as well as in the OTOC with respect to the dynamical scale $a$. Now to have a precise agreement with consistency condition, which is described by the well known {\it MSS bound}, one needs to satisfy the following constraint, which will provide a cost function model dependent lower bound on the {\it Lyapunov exponent} appearing from the definition of the complexities:
\bea \lambda_i\precsim \lambda \leq \frac{2\pi}{\beta}~~\forall ~~i=1,2.\eea 
If the maximal saturation is achieved, then from this relation one can further get a lower bound on the equilibrium temperature of the quantum system of our universe under study during bouncing scenario, and this is given by:
\bea T\succsim \frac{\lambda_i}{2\pi}~~\forall ~~i=1,2~~~~\Longrightarrow~~~~T\succsim\frac{1}{2\pi}\left(\frac{d\ln{\cal C}_i(a)}{da}\right)_{a=a_{\rm grow}}~~\forall ~~i=1,2.\eea
Finally, when the universality as well the maximal saturation both have been achieved simultaneously in the above mentioned expression, the equality gives the exact estimation of the equilibrium temperature of the quantum system of the universe studied during bounce, which is valid at very large values of the evolutionary scale represented by $a$. Now from the above bound since $\lambda_i \sim \lambda$ and that $\lambda_i \neq \lambda$, it is also expected that the lower bound on the equilibrium temperature can have two predictions in terms of the two possibilities of the complexities originated from two possible cost functions in the present context. However, the numerical order of both of the predictions computed from the plots will be same and somewhat in a broader sense support the universality criteria, which tells us that both of the predicted temperature will not be much different. From the above obtained lower bound on the equilibrium temperature one more important aspect we want to point here is that, this result does not depend on any particular particle content or a specific model available during bounce and gives us a generic estimation of the equilibrium temperature.

Now if we are thinking about the more realistic cosmological observation then it is not very good to study the evolution with respect to the scale factor, because in the context of realistic cosmology the scale factor is not the direct physical observable which one can probe in the observation for various cosmological missions running (or supposed to run in the near future) to test the signatures of the primordial cosmological paradigm. In that case  instead of using the scale factor $a$ one can consider a more physically realistic variable, which is the rescaled number of e-foldings, ${\cal N}$, which one can use as a direct probe in various cosmological observations. In this specific situation one needs to use the following transformation for which the linear differential operator appearing in the evolutionary equations of the squeezed parameter and the squeezed angle will be modified as:
\bea a\rightarrow {\cal N}:~~~~~\frac{d}{d\ln a(\tau)}=\left(1-\epsilon(\tau)\right)~\frac{d}{d\ln |aH|}=\left(1-\epsilon(\tau)\right)~\frac{d}{d{\cal N}}=\frac{d}{dN},\eea
where we have used the following couple of facts for the above mentioned transformation:
\bea && dN=d\ln a(\tau),\\
&& d{\cal N}=d\ln|a H|=d\ln|{\cal H}|,\\
&& \frac{d{\cal N}}{dN}=\left(1-\epsilon(\tau)\right),\\
&& \epsilon(\tau)=-\frac{\dot{H}}{H^2}=1-\frac{{\cal H}'}{{\cal H}^2}.\eea
Here, $N$ is the actual number of e-foldings, ${\cal N}$ is the number of e-foldings in terms of the re-defined variables, and $\epsilon(\tau)$ is the slowly varying conformal time dependent parameter. 

Consequently, the evolution of the squeezed state parameter $r_{\bf k}({\cal N})$ and squeezed angle $\theta_{\bf k}({\cal N})$, can be recast in terms of the newly defined dynamical preferred choice of suitable variable ${\cal N}$ as:
\begin{align} \label{eq:diffeqnsnumwa1}
&\frac{dr_{\bf k}({\cal N})}{d{\cal N}} = -\frac{\lambda_{\bf k}({\cal N})}{\left(1-\epsilon(\tau)\right){\cal H}} \cos 2\phi_{\bf k}({\cal N}),\\
\label{eq:diffeqnsnumwa2}
&\frac{d\phi_{\bf k}({\cal N})}{d{\cal N}} = \frac{1}{\left(1-\epsilon(\tau)\right){\cal H}} \left[\Omega_{\bf k}-\lambda_{\bf k}({\cal N}) \coth2 r_{\bf k}({\cal N})\sin 2\phi_{\bf k}({\cal N}) \right].
\end{align}
In this context, $r_{\rm co}=(aH)^{-1}$ or $r_{\rm co}={\cal H}^{-1}$ represents the co-moving Hubble radius, which is extremely important quantity in terms of which the newly re-defined number of e-foldings have been expressed in terms of the good old definition of the number of e-foldings. So instead of solving these sets of first order coupled differential equations in terms of the dynamical variable $a$ here our further objective is to study the evolution numerically with respect to the re-defined dynamical variable, ${\cal N}$. Here, we additionally want to point out that by replacing the dynamical variable $a$ in terms of the re-defined expression for the number of e-foldings ${\cal N}$, we can write down similar type of conclusion which we have written earlier to interpret the exponential growth and then the saturation in the large scale. Here one can write:
\bea &&{\cal C}_i({\cal N})\approx c_{{\cal N},i}\exp(\lambda_{{\cal N},i}{\cal N})\Longrightarrow\lambda_{{\cal N},i}=\left(\frac{d\ln{\cal C}_{i}({\cal N})}{d{\cal N}}\right)_{{\cal N}={\cal N}_{\rm grow}}\Longrightarrow T\succsim\frac{1}{2\pi}\left(\frac{d\ln{\cal C}_{i}({\cal N})}{d{\cal N}}\right)_{{\cal N}={\cal N}_{\rm grow}}\nonumber\\
&&~~~~~~~~~~~~~~~~~~~~~~~~~~~~~~~~~~~~~~~~~~~~~~~~~~~~~~~~~~~~~~~~~~~~~~~~~~~~~~~~~~~~~~~~~~~~~~~~~~~~\forall~i=1,2.\eea 
Similarly one can derive the {\it universality relation} which will be same as the previous one. In the next two subsections, we will explicitly numerically solve the previously mentioned dynamical equations of the squeezed parameter and squeezing angle with respect both the dynamical variables, scale factor $a$ and the re-defined number of e-foldings ${\cal N}$ for cosine hyperbolic and exponential bouncing models that we have introduced in the first section of the paper. The explicit details of the analysis and the corresponding physical interpretation of the numerical results and the plots are discussed in the following two subsections. Discussion of the differential equations with respect to different dynamical variables is given in \Cref{sec:appendixC}.

Another important aspect that one can estimate numerically from our present set up, is the well known \textcolor{blue}{\it scrambling time scale}. Within the framework of quantum chaos this time scale plays very significant role to understand the underlying behaviour of the physical systems. There are several definitions associated with this quantity, in the theoretical physics community in different contexts for physical interpretation of various unknown phenomena. We will now quote the most frequently used definitions, and follow one of them in the present context to numerically estimate the order of \textcolor{blue}{\it scrambling time scale} from the bouncing cosmological scenario:
\begin{enumerate}
\item \underline{\textcolor{red}{\bf Definition I:}}\\
According to this definition this is the time which the OTOC takes to equilibriate. This is a very modern definition and directly associated with the phenomena of quantum mechanical chaos~\footnote{In ref.~\cite{Yoshida:2017non}, the authors have explicitly shown that this definition is sufficient enough for the {\it Heyden Preskill protocol}.}

\item \underline{\textcolor{red}{\bf Definition II:}}\\
According to this definition this is the time which a system takes starting in an arbitrary tensor product state to become nearly maximally entangled.

\end{enumerate}

Now according to Leonard Susskind \cite{Sekino:2008he} and later pointed in many other refs.~\cite{Hayden:2007cs} for the first scrambler the \textcolor{blue}{\it scrambling time scale} can be computed as:
\bea t_{\rm sc}\sim \frac{\beta}{2\pi}~\ln N,\eea
where $\beta$ is the inverse of the equilibrium temperature of the physical system which corresponds to the saturation of quantum chaos and $N$ represents the very large number of configurations. Now making use of the MSS bound one can further simplify the above mentioned expression and obtain a lower bound on the \textcolor{blue}{\it scrambling time scale} in terms of the quantum Lyapunov exponent:
\bea t_{\rm sc}\geq \frac{1}{\lambda}~\ln N.\eea
Here the equality holds good for the maximal saturation of chaos.

Now, within the present framework we have used the conformal time dependent scale factor $a$ and/or the number of e-foldings ${\cal N}$ as dynamical variable using which we have studied all the evolution of cosmological complexity and the OTOC in this paper (for the details see the next two subsections.). One can then ask a very justifiable question in this case that how we then define the \textcolor{blue}{\it scrambling time scale} within the framework of cosmology? Following the previous logical discussions and interpretations of the universality relation between the cosmological complexity and cosmological OTOC  by replacing the time with the scale factor one can define the  \textcolor{blue}{\it scale factor at scrambling time scale} or \textcolor{blue}{\it scarmbling scale factor}, which is given by:
\bea a_i(\tau_{\rm sc})\succsim  \frac{1}{\lambda_i}~\ln N~~~~~~\forall~~~i=1,2.\eea
Here the index $i=1,2$ is used to differentiate between the value of the scale factors obtained from the two definitions of complexities used in this paper. To hold the universality between the cosmological complexities and the OTOC we have previously shown the deviation from the results obtained from both of the definition has to lie within a very small numerical error range. It is expected that the same argument also holds here perfectly and in the next two subsections we are going to investigate this very carefully from the numerical plots to ensure the justifiability of this statement. Now, we have already computed  the expression for the scale factor in terms of the conformal time for both the models and also most of the quantum chaotic predictions are appearing ( for the details see the next two subsections.) from the bouncing solutions. For this reason using those definitions one can extract the information of the associated \textcolor{blue}{\it scrambling time scale} in the conformal coordinates within the framework of bouncing cosmological paradigm. Additionally, since we also know the connecting relationship between the physical time scale and the conformal time scale, then using this it is further possible to determine the \textcolor{blue}{\it scrambling time scale} in terms of the physical time coordinate in cosmology. In the next two subsections, for two different bouncing models we are going to estimate this time scale from the numerical plots. Finally, there is a confusion regarding the fact that in the present cosmological set up how can one give a numerical estimation of the factor $N$ which represents number of physical configurations. We are now going to give an estimate of this factor in the present context in terms of the known parameters. To obtain this estimate we start with the following relationship:
\bea {\cal C}_i \approx -\ln({\rm OTOC})=c_i~\exp(\lambda_i a)~~~~\forall~~~i=1,2.\eea
Using this relation and truncating the expression for OTOC in the second term we get:
\bea {\rm OTOC}\approx 1-c_i~\exp(\lambda_i a)+\cdots\eea
where in the usual quantum chaos literature one can identify:
\bea c_i \sim \frac{1}{N^2}~~~~\Longrightarrow~~~~~N\sim\frac{1}{\sqrt{c_i}}.\eea
The one can further write the expression for the \textcolor{blue}{\it scrambling scale factor} in terms of the known parameters as:
\bea \label{scramblingscale} a_i(\tau_{\rm sc})\succsim  \frac{1}{2\lambda_i}~\ln \left(\frac{1}{c_i}\right)~~~~~~\forall~~~i=1,2.\eea
From the numerical plots given in the next two subsections one can estimate both $\lambda_i$ and $c_i$ (for $i=1,2$) from both the bouncing models and from this relation it is possible to give a numerical estimation of the \textcolor{blue}{\it scrambling time scale} from the models of bouncing cosmology discussed in this paper. Additionally it is important to note that, in this connection the equivalent result can be obtained by considering the number of e-foldings as the dynamical variable instead of the scale factor within the framework of cosmology.
\textcolor{Sepia}{\subsection{\sffamily Cosine Hyperbolic bounce}}
We have numerically plotted the squeezing parameters and the derived complexity measures for cosine hyperbolic in four different regions - pre bounce boundary, pre point of bounce, post point of bounce and post bounce boundary against the scale factor~\footnote{{\bf Reading graphs vs scale factor:}~~Proper way to read the graph is going from right to left starting from much early times for \textcolor{blue}{\bf pre-bounce boundary line graph}, and crossing the pre-bounce boundary and again reading right to left for the \textcolor{violet}{\bf pre-point of bounce line graph} till the point of bounce. Now one goes from left to right with the \textcolor{Red}{\bf Post bounce region line} till the boundary, followed by a \textcolor{OliveGreen}{\bf post-bounce boundary line} till the present time to the right}. From \Cref{{fig:avstaucosh}} we can see that at present time and at a time much before the boundary ($\tau \rightarrow -\infty$) the value of scale factor $a = 1$. We have taken the value of pre-boundary and post-boundary parameters $r_{\bf k}(a=1) = 1, \phi_{\bf k}(a=1) = 1$ to set our initial conditions, and ensured continuity at $a_{boundary}$ as initial conditions for the bouncing region parameters for numerically solving differential equations with respect to scale factor (Eqs. \ref{eq:diffeqnswa1} and \ref{eq:diffeqnswa2}). For the analysis of Cosine hyperbolic bounce we have taken $-k\tau_b = 30$ and the range of $-k\tau$ goes from 0 to 60. The parameter $r_1$, appearing in the expression of the scale factor is related to the cosmological constant by the relation $r_1=2\Lambda/9$. We have fixed the value of $\Lambda$ to 10$^{-4}$ for our numerical analysis.

\begin{figure}[!htb]
	\centering
\includegraphics[width=15cm,height=8cm]{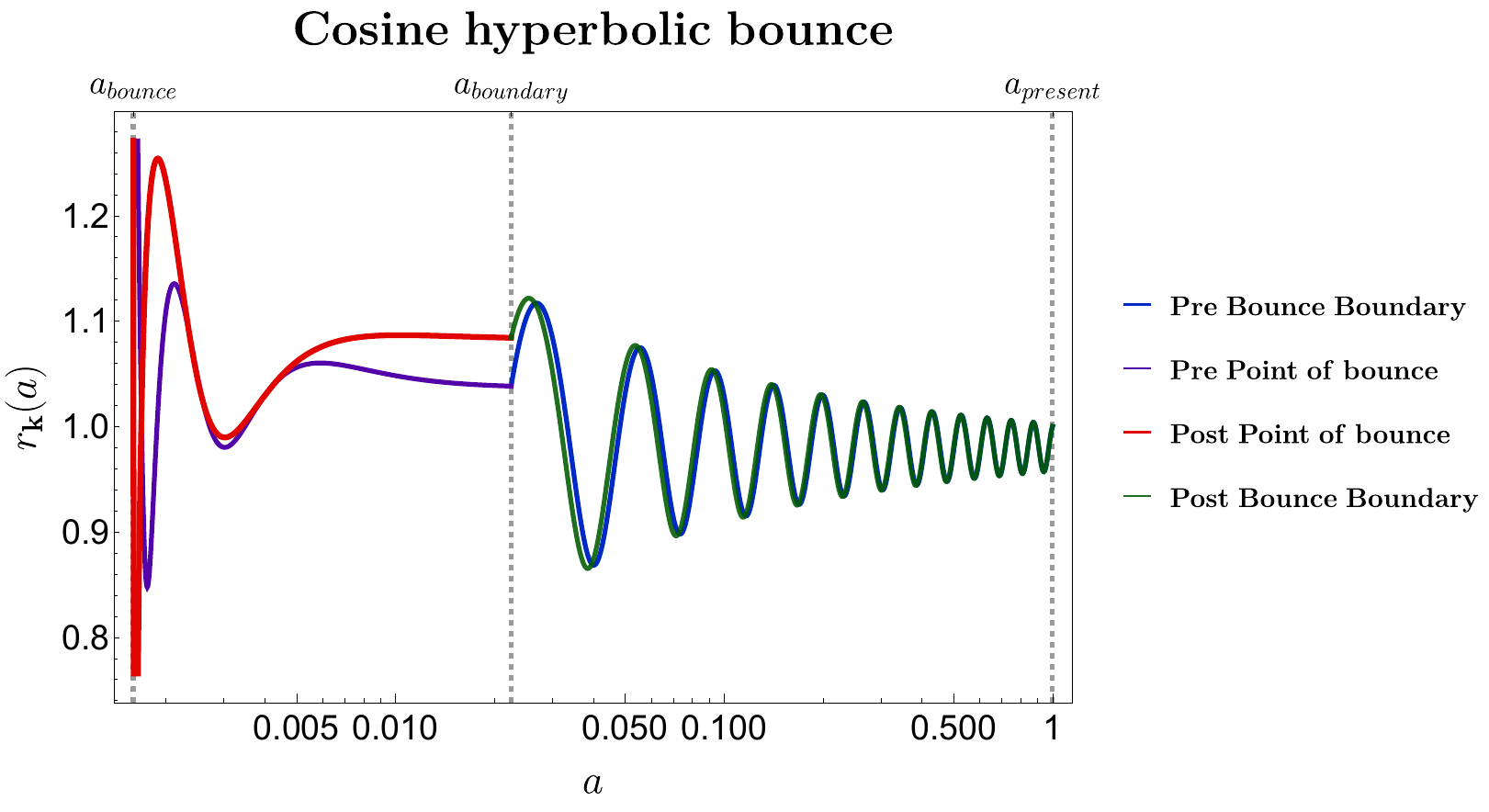}

	\caption{Squeezing parameter at different regions plotted against scale factor}\label{fig:rkvacosh}
\end{figure}

\begin{figure}[!htb]
	\centering
\includegraphics[width=15cm,height=8cm]{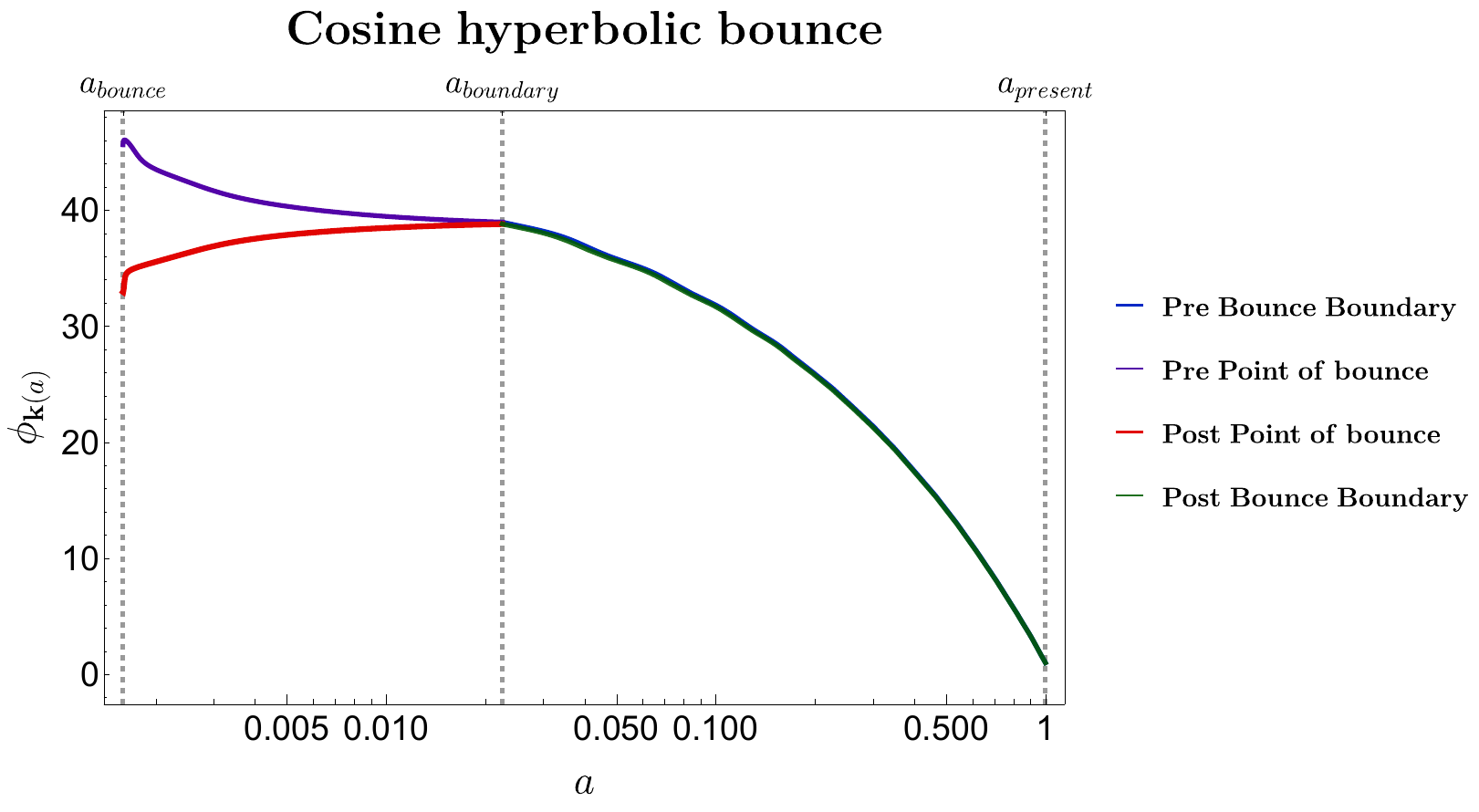}

	\caption{Squeezing angle plotted at different regions against scale factor}\label{fig:phivacosh}
\end{figure}

\begin{figure}[!htb]
	\centering
\includegraphics[width=15cm,height=8cm]{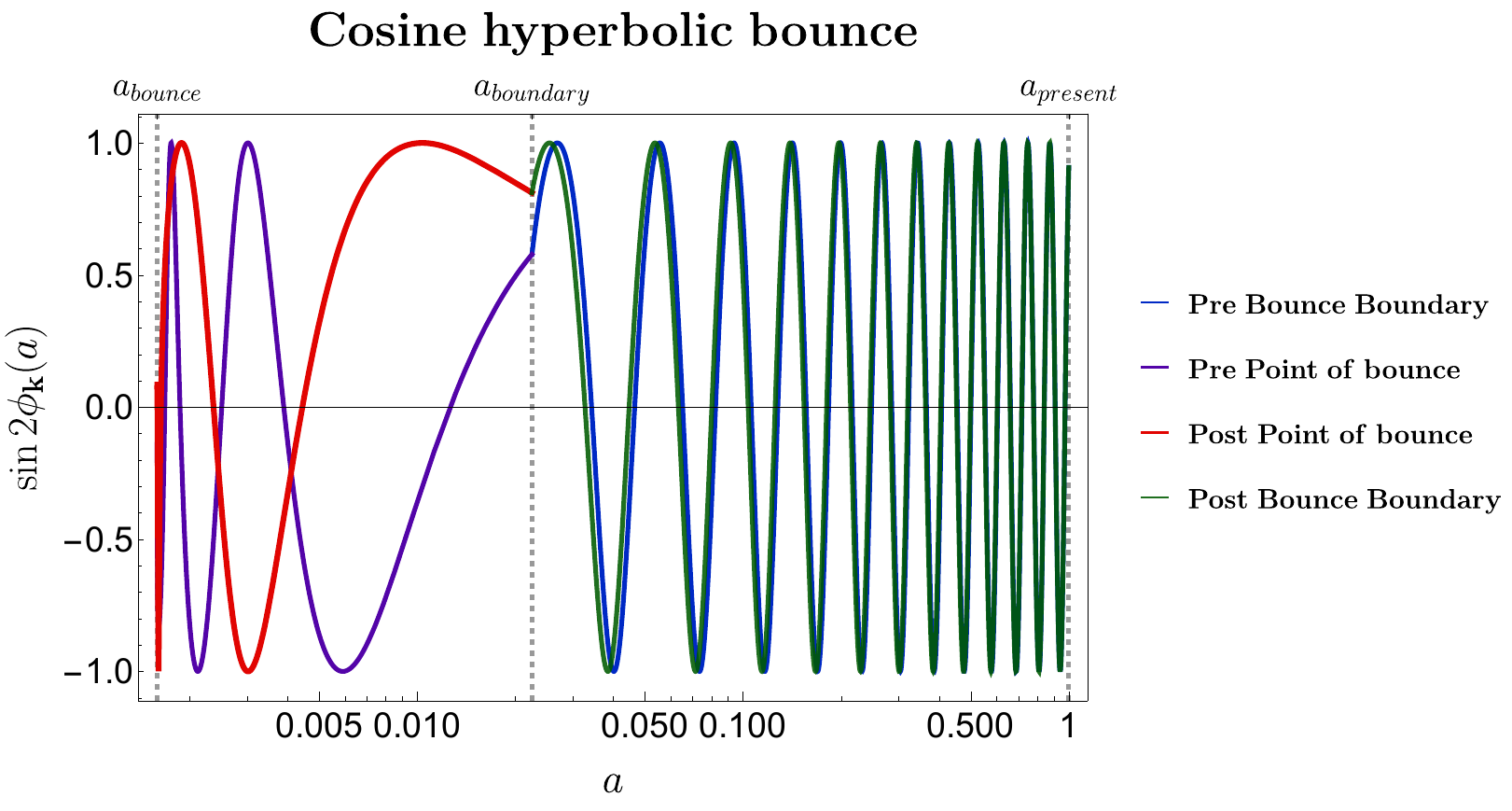}

	\caption{Sine of twice of squeezing angle at different regions plotted against scale factor }\label{fig:sin2pvacosh}

\end{figure}

For the \textbf{squeezing paramater} plotted in \Cref{fig:rkvacosh}
\begin{itemize}
	\item the pre-boundary and the post-boundary behaviour is oscillatory with decreasing amplitude as it approaches $a=1$(very early times in case of pre bounce boundary and present time in case of post bounce boundary),
	\item while inside the bouncing region we see highly oscillatory behaviour near the point of bounce(with very high amplitudes) that saturates into  a given value as it nears the boundary. This saturation behaviour of squeezing parameter near the boundary might point to saturation behaviour of the Complexities as we will see in further analysis.
\end{itemize} 
The \textbf{squeezing angle} and the \textbf{sine of twice its value} are also important to understand the Squeezing operator. See \Cref{fig:phivacosh} and \Cref{fig:sin2pvacosh}.
\begin{itemize}
	\item The has an exponential increase even against a logarithmic scale, with the rate of increase falling down while approaching the bounce boundary from earlier times in the pre-boundary region. The frequency of the $\sin 2\phi_\k$ corresponds to the rapid rate at which it increases, initially oscillating really fast to slow spaced oscillations at the boundary.
	\item Upon entering the bouncing region the angle just has a sturdy exponential rise till the point of bounce after which it exponentially increases with a slow rate till the boundary after crossing the point of bounce. The sine again behaves similarly with slowed down oscillation at the boundary, where we see saturated rate of change in the angle. 
	\item Outside the boundary the angle exponentially decreases at a rapid rate and the sine value correspondingly  increases in oscillations as we approach present time.
\end{itemize}

\begin{figure}[!htb]
	\centering
	\includegraphics[width=15cm,height=8cm]{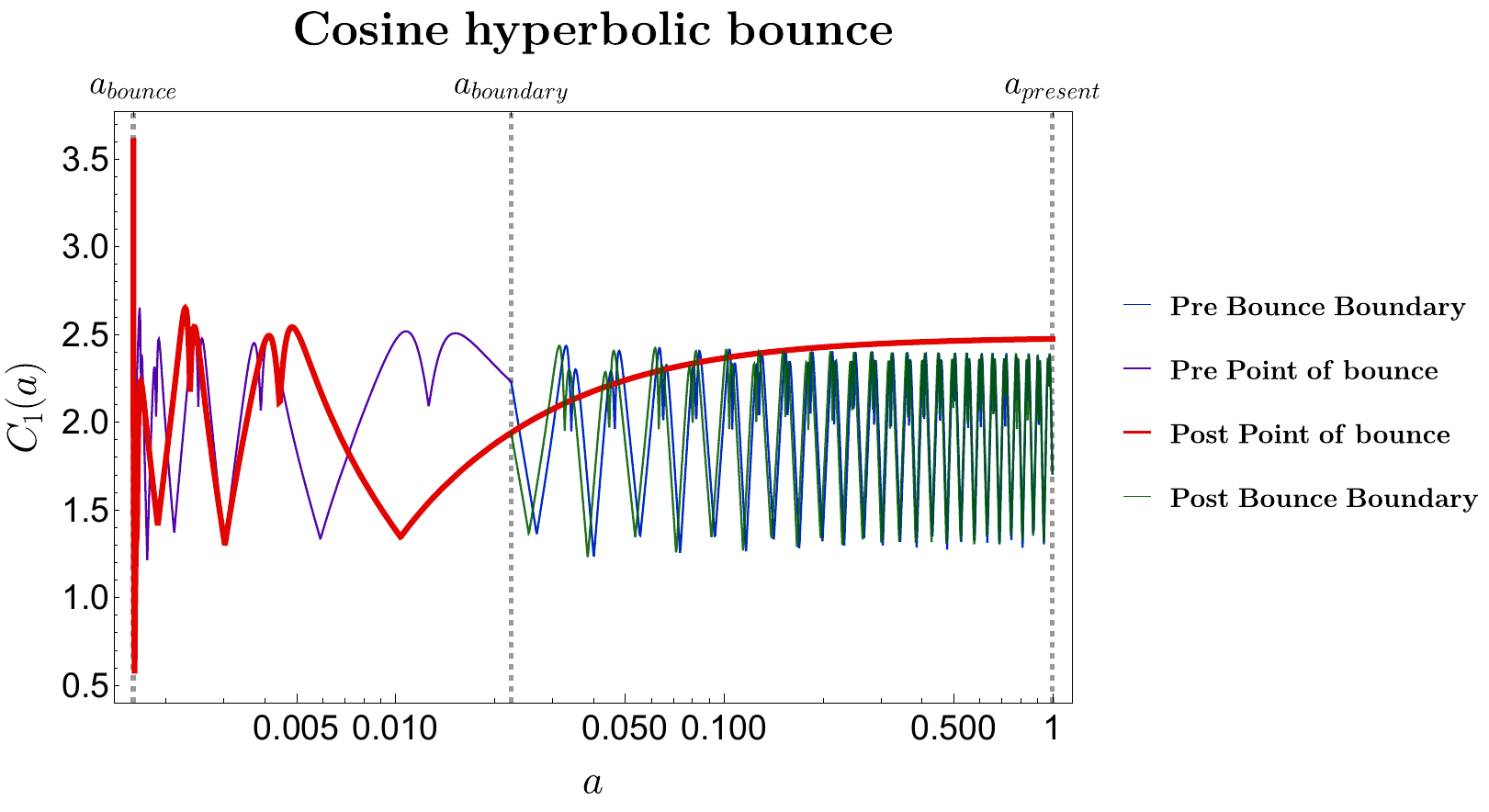}
	
	\caption{Linearly weighted Complexity value at different regions plotted against scale factor}\label{fig:c1vacosh}
\end{figure}

\begin{figure}[!htb]
	\centering
	\includegraphics[width=15cm,height=8cm]{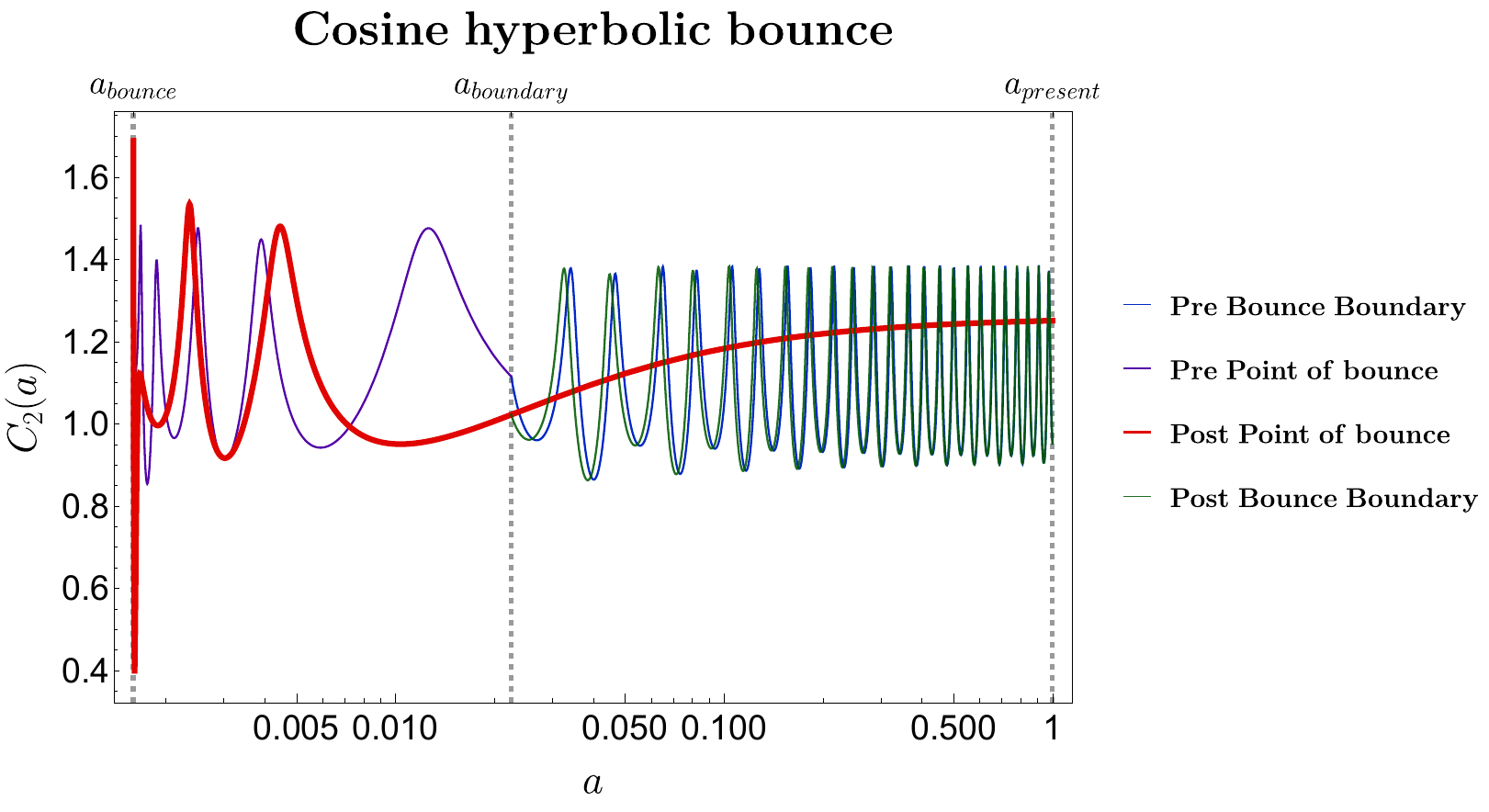}
	
	\caption{Geodesically weighted Complexity value at different regions plotted against scale factor}\label{fig:c2vacosh}
\end{figure}

\begin{figure}[!htb]
	\centering
	\includegraphics[width=15cm,height=8cm]{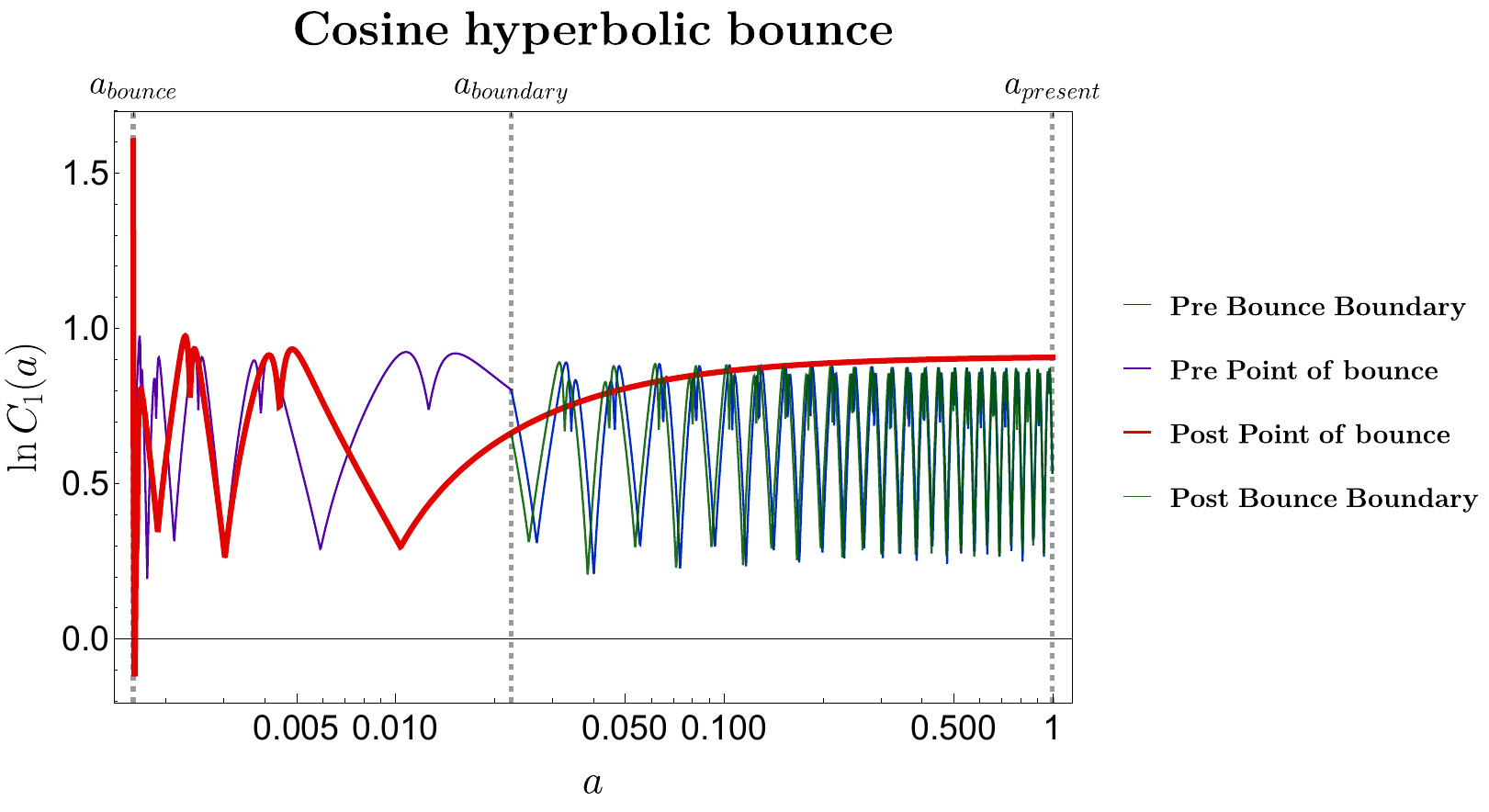}
	
	\caption{Logarithm of Linearly weighted Complexity value at different regions plotted against scale factor}\label{fig:lnc1vacosh}
\end{figure}

\begin{figure}[!htb]
	\centering
	\includegraphics[width=15cm,height=8cm]{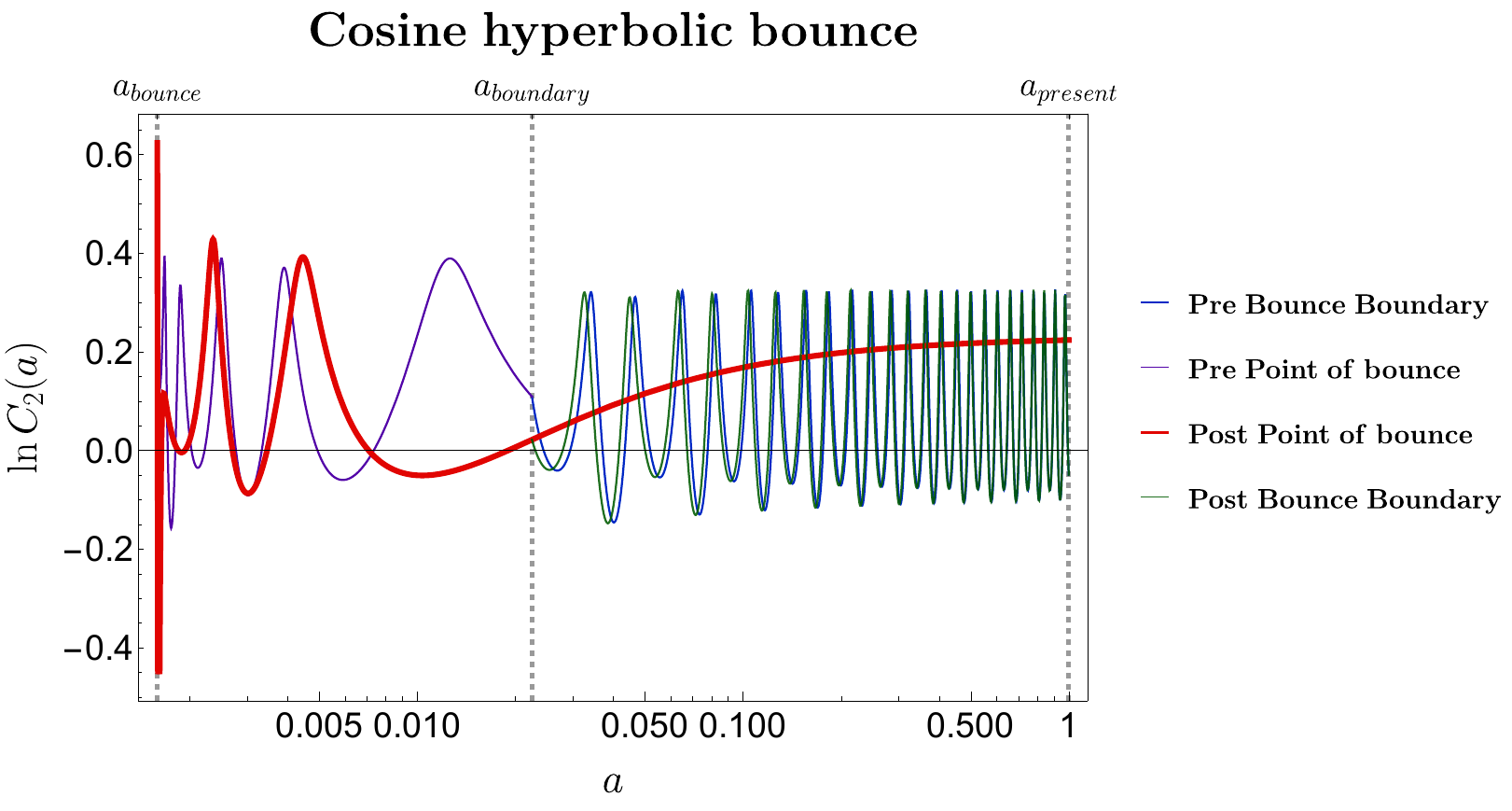}
	
	\caption{Logarithm of Geodesically weighted Complexity value at different regions plotted against scale factor}\label{fig:lnc2vacosh}
\end{figure}

\begin{table}[!htb]
\centering
\begin{tabular}{|P{1cm}||P{2cm}|P{2cm}|P{2cm}|P{2cm}|P{2cm}|}
\hline 
& Very early times & Entering bouncing region & Around point of bounce & Exiting bouncing region & Late or Present time \\ 
\hline 
$\C_1$ & 1.704 & 2.229 & 14.187 & 1.938 & 2.47 \\ 
\hline 
$\C_2$ & 0.951 & 1.115 & 8.99 & 1.021 & 1.25 \\ 
\hline 
\end{tabular} 
\caption{Complexity values at different points of interest with respect to scale factor}
\label{tab:compcosh}
\end{table}

The \textbf{complexity} of the two mode vacuum state from \Cref{eq:complexities} is used to analyze and plot $\C_1,\C_2$ along with their \textbf{log values}, and \textbf{predicted OTOC}. Though both $\C_1$ and $\C_2$ are extremely good measures of the circuit complexity, the linearly weighted complexity $\C_1$ shows similarity to the calculations from holographic side.

Both the complexity measures have very similar behaviour \Cref{fig:c1vacosh} and \Cref{fig:c2vacosh},
\begin{itemize}
	\item The value of complexity outside the bouncing boundary based on the respective squeezing parameters defined there for very early times and nearing present times is oscillatory with smaller frequency at the boundary. 
	\item Inside the bouncing region prior to the bounce both the complexities cross the boundary with a sturdy rise and then go on to become highly oscillatory and spike up. The post point of bounce values are of greater interest as they show a sturdy rise and a nice saturation when extrapolated to present time.
	\item Even though the post bounce boundary behaviour looks oscillatory it is important to note that the growing behaviour of complexity at post point of bounce. We see a sudden exponential rise near the boundary. The analysis of growing complexity observed by extrapolating the post point of bounce values at late times shows saturation after an initial rise across the boundary. We have written down the values in \Cref{tab:compcosh}.
\end{itemize}

We can see extremely high complexity values at the point of bounce. This points to highly complex transformations taking place between the reference and target quantum state during the bounce.

The slope of logarithm of complexity at the point of rise directly corresponds to the value of the quantum Lyapunov exponent as mentioned in Eq (\ref{eq:lyapexp}). 
To predict the slope of the logarithmic value of complexities we consider the change of y-axis value over the range of the x-axis value i.e. between point of rise and point of saturation. Mathematically it is represented by
\begin{equation}
	\lambda_i= \frac{ln~\C_i~(\text{point of saturation})- ln~\C_i~(\text{point of rise})}{a~(\text{point of saturation})-a~(\text{point of rise})}
\end{equation} For this we have plotted the logarithm of complexity values in \Cref{fig:lnc1vacosh} and \Cref{fig:lnc2vacosh}. We observe the qualitative features to be same as that of the complexity graphs, showing corresponding oscillatory, rising and saturation at the respective regions.
We calculate the Lyapunov exponent from the post point of bounce case as it shows exponential and saturation at late times  and this gives an estimation on the lower bound of temperature. 
\begin{table}[!htb]
	\centering
	\begin{tabular}{|P{3cm}||P{3cm}|P{3cm}|}
		\hline  
	ln $\mathcal{C}_i$	& point of rise  & point of saturation \\ 
		\hline 
		ln $\mathcal{C}_1$ & 0.29579  & 0.90557  \\ 
		\hline 
	ln $\mathcal{C}_2$ &  -0.05055 & 0.217005 \\ \hline
	\end{tabular} 
	\caption{Log of complexity values at point of rise and point of saturation. The point of saturation is considered to be that initial point from which the value upto second decimal place is constant. Point of rise $\C_1$ is $a=0.01037$ and for $\C_2$ it is $a=0.01002$ and the point of saturation for $\C_1$ is $a=1$ and for $\C_2$ it is $a=0.495$}
	\label{tab:lyacosh}
\end{table}

The Lyapunov exponent can be calculated from these values given in \Cref{tab:lyacosh}:
\begin{align*}
	\lambda_1= 0.616166 && \lambda_2= 0.551685
\end{align*}

The estimated lower bound on the temperature from the calculated values of the Lyapunov exponents are
\begin{align*}
T_{\mathcal{C}_1} \succsim 0.09806 && T_{\mathcal{C}_2} \succsim 0.08780
\end{align*}

Using Eq(\ref{scramblingscale}), we have numerically calculated the lower bound of scrambling time scale in terms of scale factor and conformal time. We have considered the region of saturation and taken the values of complexity at what we have perceived as the starting point and the ending point of the region of saturation. We have then calculated $\Delta a_i(\tau_{sc})$, which will give us the lower bound of the \textbf{scrambling interval in terms of the scale factor}. We have converted this in terms of conformal time for easy physical interpretation. In our numerical analysis we have extensively used the conformal time and we have normalized all other numerical measures with respect to conformal time in both models whereas the physical time is not normalized with respect to our numerical analysis. Hence calculating scrambling time scale in terms of physical time will not make much sense quantitatively in our case without appropriate normalization (and redoing complete analysis). In our cosine hyperbolic case we have normalized conformal time in such a way bounce occurs at $\tau_b = -3000$, and present day time is $0$, and hence we can interpret the values given in \Cref{tab:scramcosh}, qualitatively in terms of physical time too. We get conformal scrambling time scales around one-tenth of the time period since bounce till present. This roughly points to the time taken for OTOC to attain equilibrium as can be seen from the graph. A quicker scrambling time scale points to \textbf{smoother saturation of complexity}. A sense of time period in terms of physical time can then be qualitatively understood using this argument.

\begin{table}[!htb]
\centering
\begin{tabular}{|P{2cm}||P{4cm}|P{4cm}|P{2cm}|P{2cm}|}
\hline 
& $c_i$ at start of saturation & $c_i$ at end of saturation & $\Delta a_i(\tau_{sc})$ & $|\Delta \tau_{sc}|$  \\ 
\hline 
From $C_1$& 2.466 & 2.4746 & 0.002825 & 291.642 \\ 
\hline 
From $C_2$& 1.2404 & 1.2471 & 0.00517 & 377.35\\ 
\hline 
\end{tabular} 
\caption{Estimated lower bound values of scrambling time scales for the region of saturation in cosine hyperbolic case from the two different complexities}
\label{tab:scramcosh}
\end{table}

\begin{figure}[!htb]
	\centering
	\includegraphics[width=15cm,height=8cm]{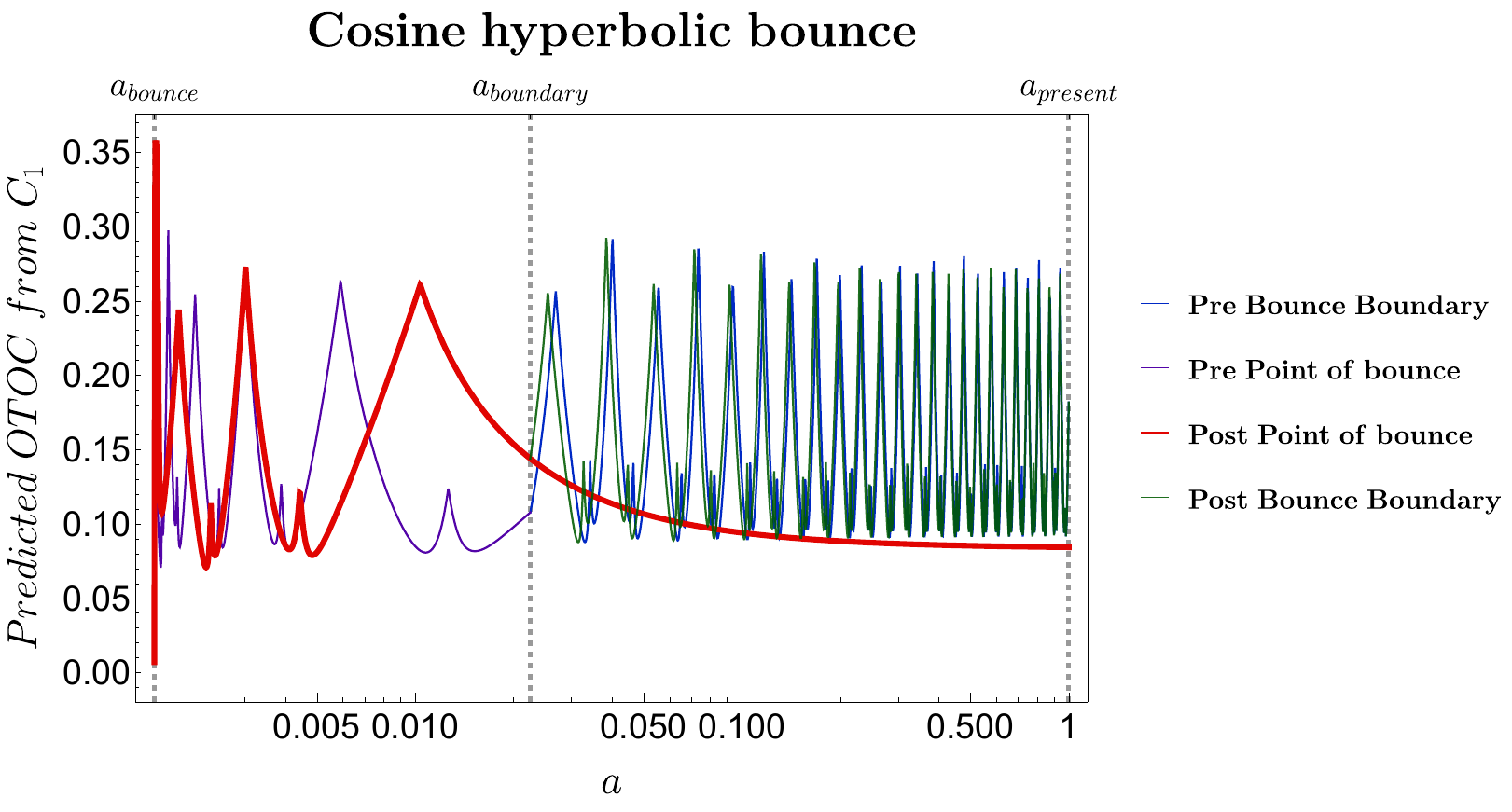}
	
	\caption{Predicted OTOC values from geodesically weighted Complexity at different regions against scale factor}\label{fig:otoc1vacosh}
\end{figure}

\begin{figure}[!htb]
	\centering
	\includegraphics[width=15cm,height=8cm]{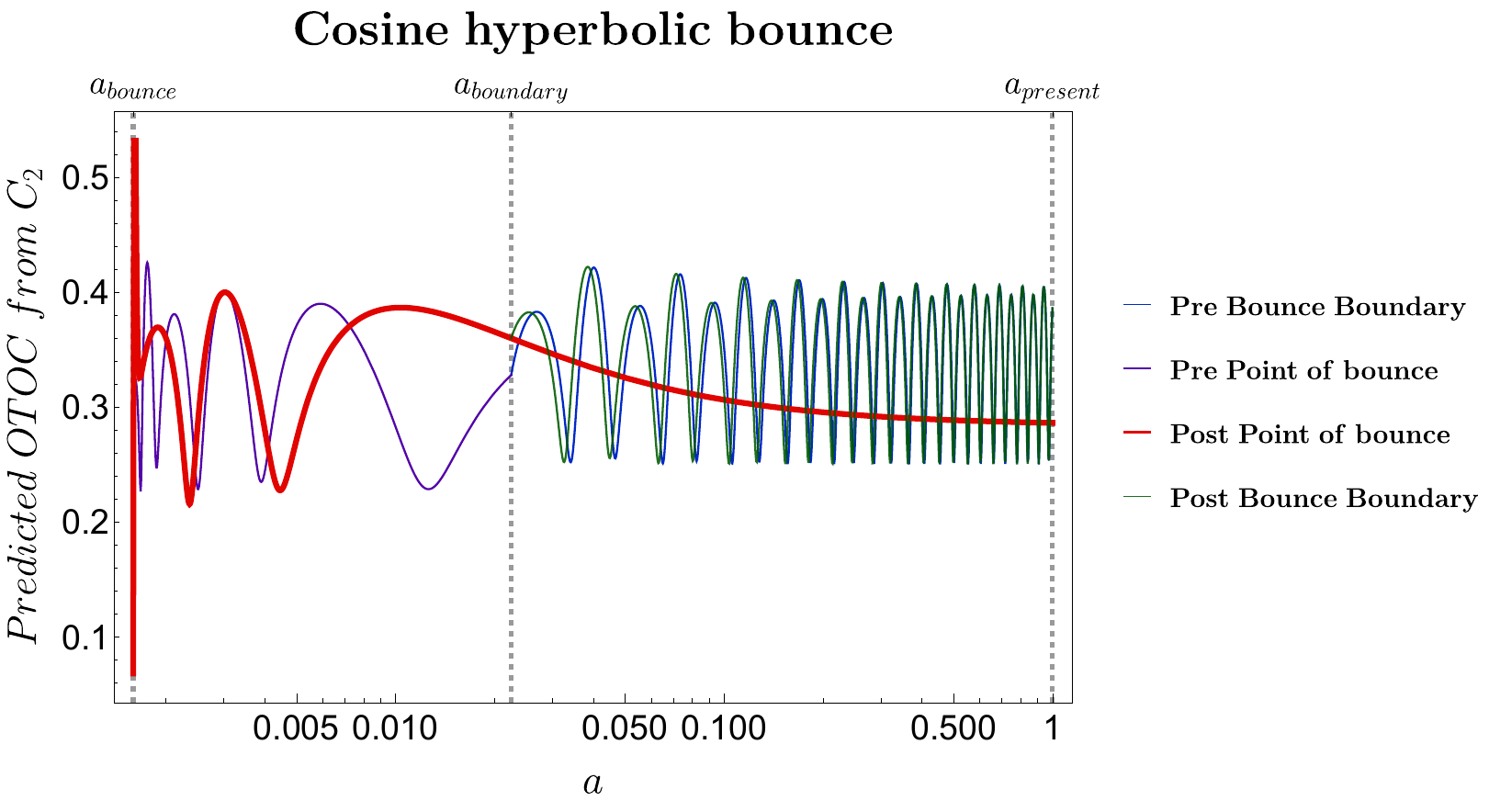}
	
	\caption{Predicted OTOC values from geodesically weighted Complexity at different regions against scale factor}\label{fig:otoc2vacosh}

\end{figure}

The OTOC values plots calculated from the universality relation mentioned in Eq (\ref{eq:otocuniv}). The behaviour observed is very similar to the complexity behaviour at different regions - from being random oscillations outside the bouncing region to settling at the boundary to again oscillating and spiking at the point of bounce. The OTOC values at different points have been written in \Cref{tab:otoccosh}. One noticeable observation is the really small value of the OTOC at the point of bounce from both the complexity measures.

\begin{table}[!htb]
\centering
\begin{tabular}{|P{2cm}||P{2cm}|P{2cm}|P{2cm}|P{2cm}|P{2cm}|}
\hline 
& Very early times & Entering bouncing region & Around point of bounce & Exiting bouncing region & Late or Present time \\ 
\hline 
$OTOC_1$& 0.182 & 0.107 & $6.9 \times 10^{-7}$ & 0.144 & 0.084 \\ 
\hline 
$OTOC_2$& 0.386 & 0.327 & $1.2 \times 10^{-4}$ & 0.356 & 0.286 \\ 
\hline 
\end{tabular} 
\caption{Predicted OTOC values at different points of interest with respect to scale factor}
\label{tab:otoccosh}
\end{table}

\begin{figure}[!htb]
	\centering
	\includegraphics[width=15cm,height=8cm]{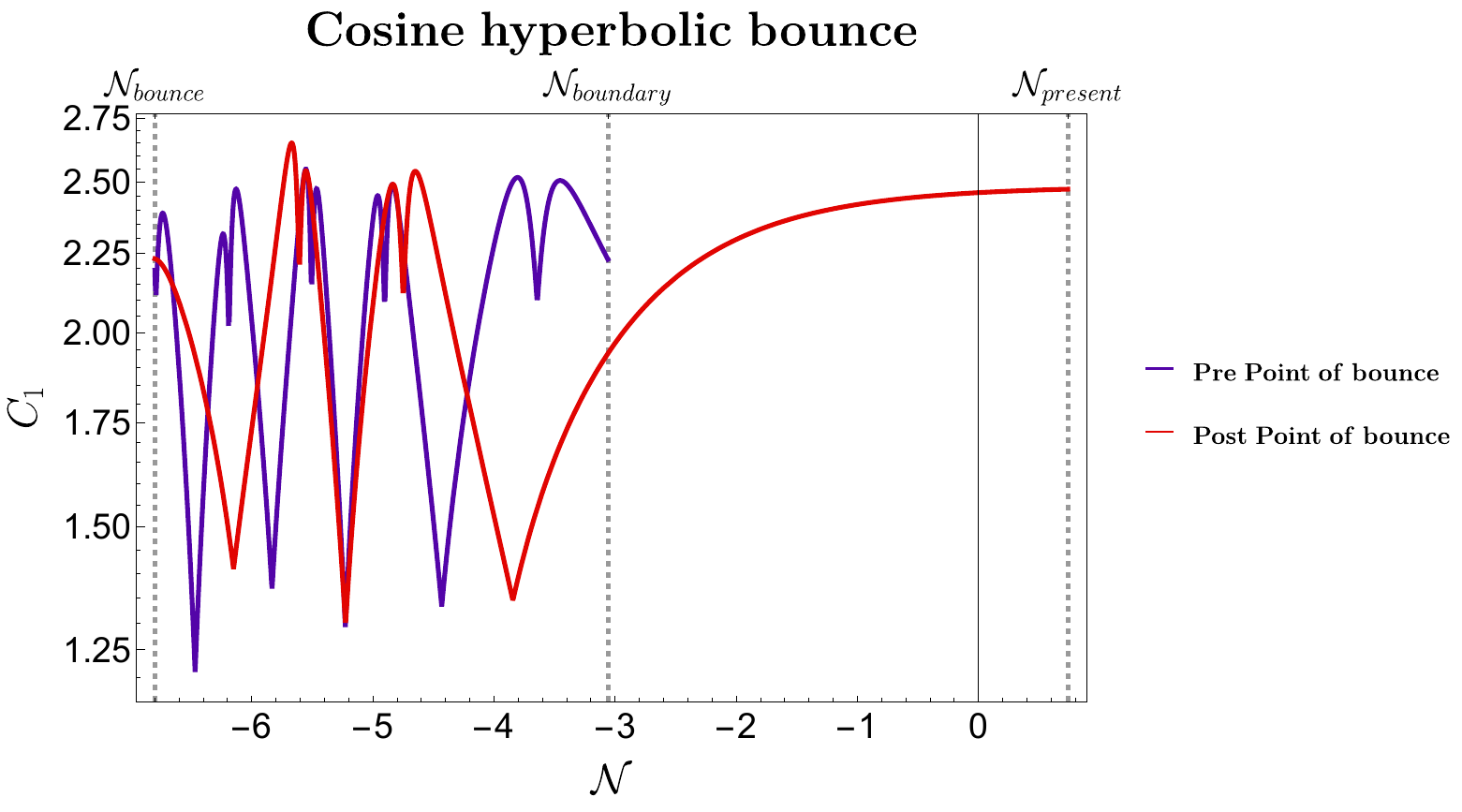}
	
	\caption{Variation of linearly weighted complexity inside bouncing region with respect to number of e-foldings}\label{fig:c1invzcosh}
\end{figure}

\begin{figure}[!htb]
	\centering
	\includegraphics[width=15cm,height=8cm]{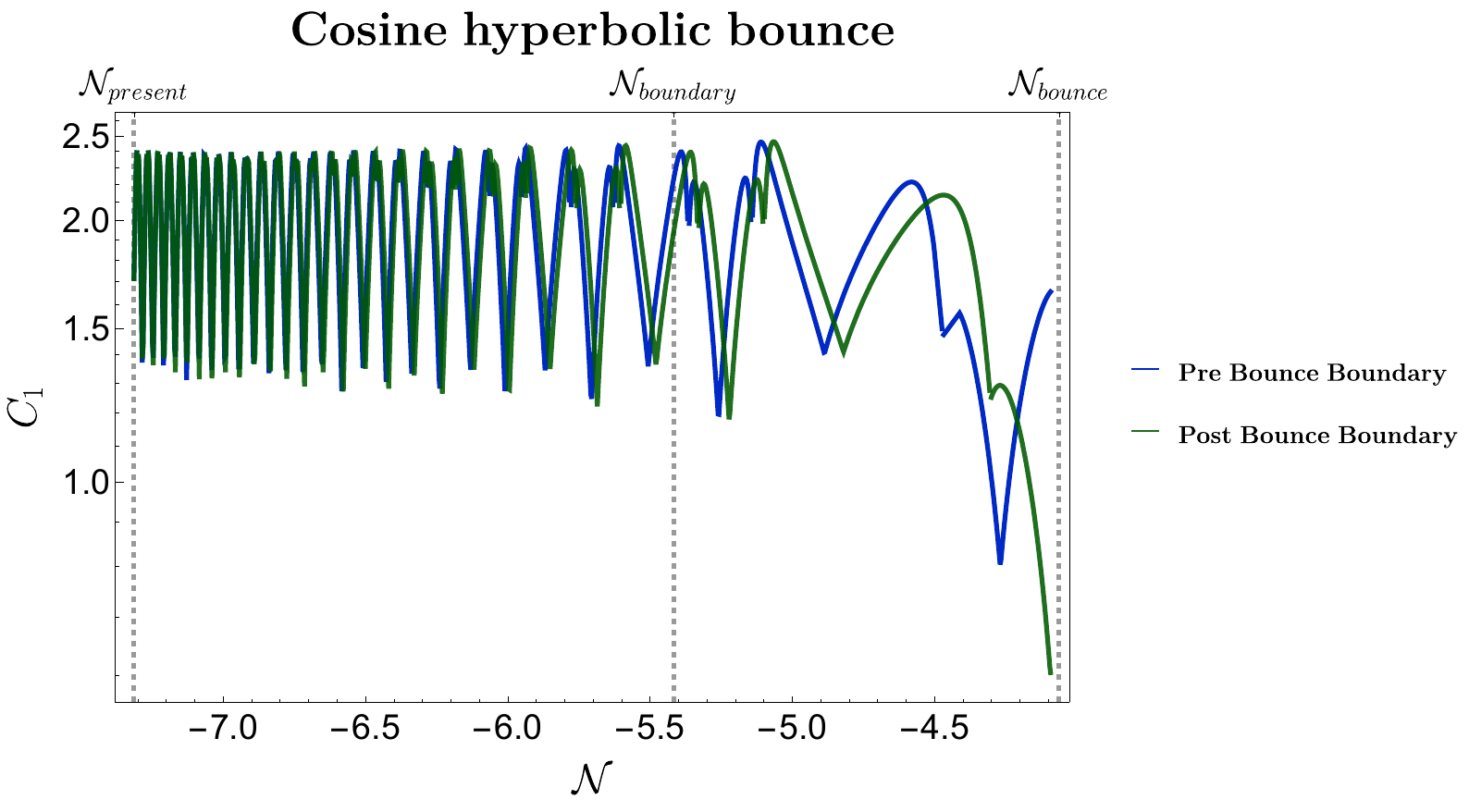}
	
		\caption{Variation of linearly weighted complexity outside bouncing region with respect to number of e-foldings}\label{fig:c1outvzcosh}
\end{figure}

We know that more than the scale factor the Number of e-foldings($\N = \log a\H$) is a measurable and interesting value. Using the simple relation we have also plotted Complexity and the OTOC against $\N$. One can see from the behaviour of $a,\H$ from \Cref{fig:avstaucosh} and \Cref{fig:Hvstaucosh}, that the direction of $\log a\H$ will be different inside the bouncing region and outside the bouncing region. For this reason the plots have been made separately to ensure readability~\footnote{{\bf Reading graphs vs $\N$:}~~Inside the bouncing region the value of $\N$ at bounce and boundary can be obtained and it is seen that $\N_{bounce} < \N_{boundary}$ and extrapolating the same one can get value at present time which is greater than both. Hence the should be read left to right(bounce to present) for \textcolor{Red}{Post point of bounce graph} and boundary to bounce for \textcolor{violet}{pre point of bounce line}. For outside the bouncing region  the values evolve in reverse and through extrapolation value at bounce is found. The graph should be read from left to right(very early times to entering boundary) for \textcolor{blue}{pre boundary line} whereas it should be read boundary to present times for \textcolor{OliveGreen}{post boundary line}}. 

\begin{figure}[!htb]
	\centering
	\includegraphics[width=15cm,height=8cm]{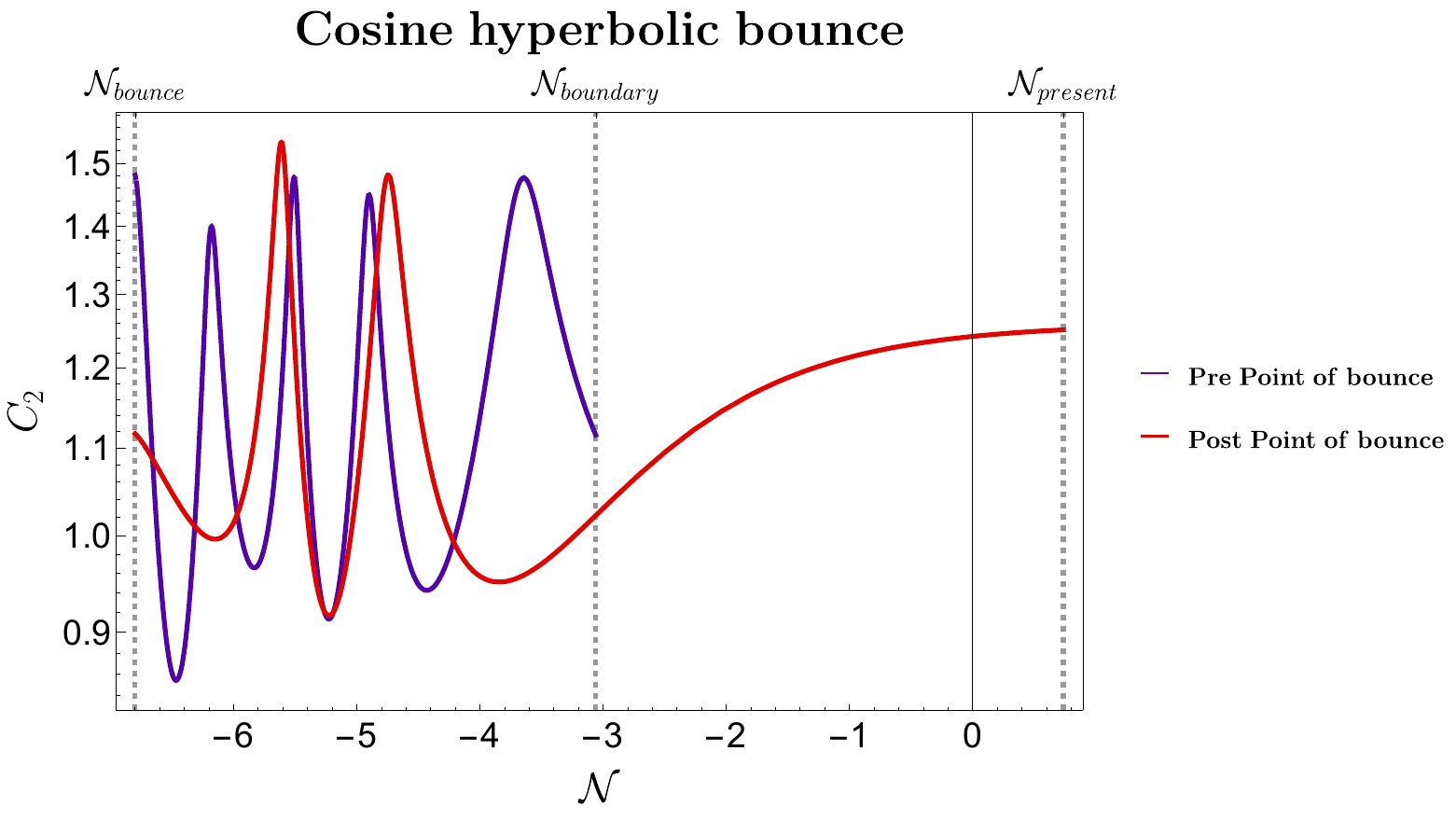}
	
	\caption{Variation of geodesically weighted complexity inside bouncing region with respect to number of e-foldings}\label{fig:c2invzcosh}

\end{figure}

\begin{figure}[!htb]
	\centering
	\includegraphics[width=15cm,height=8cm]{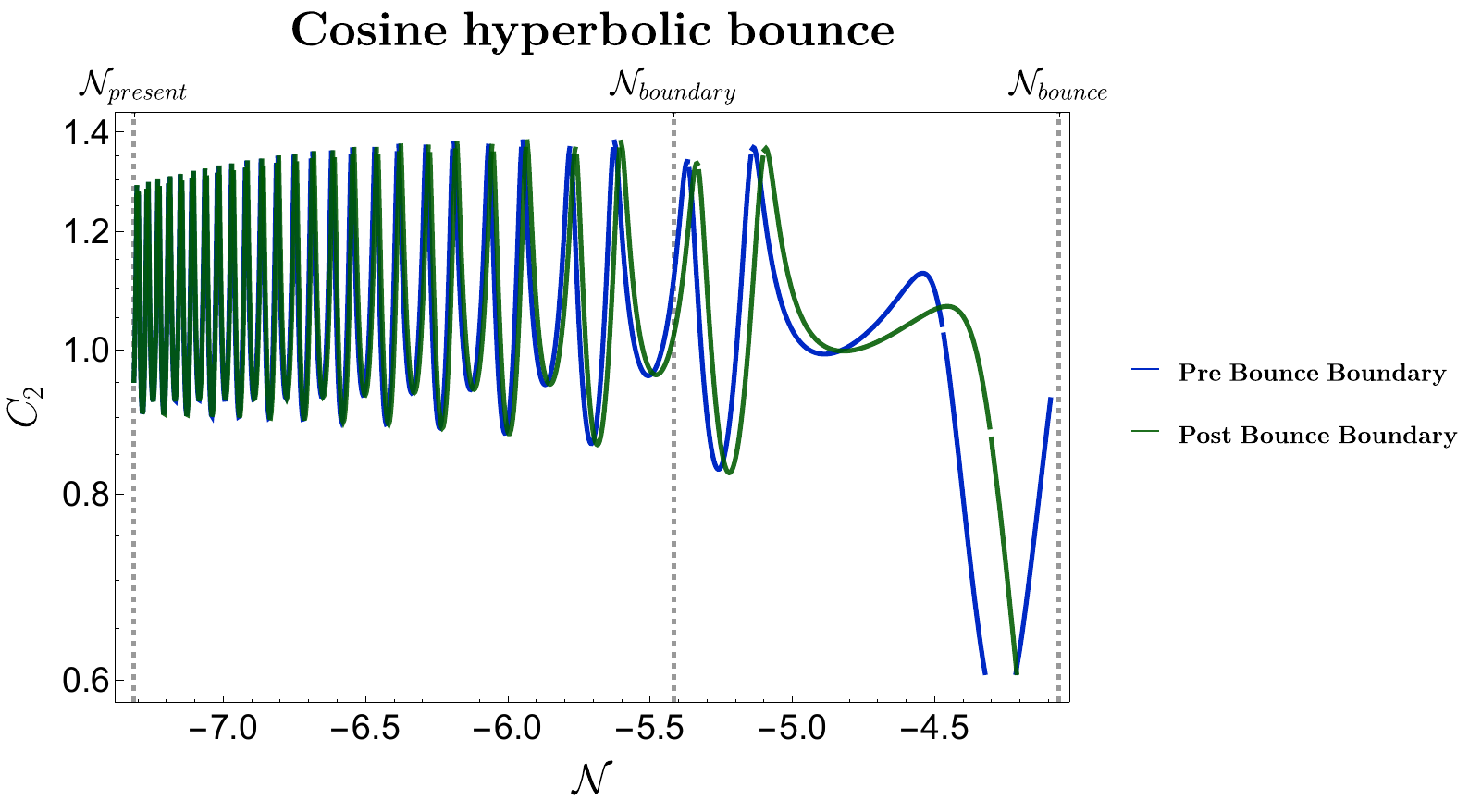}
	
	\caption{Variation of geodesically weighted complexity outside bouncing region with respect to number of e-foldings}\label{fig:c2outvzcosh}

\end{figure}

For both linearly weighted complexity $\C_1$ and geodesically weighted complexity $\C_2$: 
\begin{itemize}
	\item the outside bouncing region behaviour is oscillatory and random. The oscillations decrease near the boundary. We have extrapolated their graphs inside the boundary to show that their oscillatory behaviour drops to a certain/very low complexity. But this is not what is actually expected from complexity measures previously done inside the bouncing region. See \Cref{fig:c1outvzcosh} and \Cref{fig:c2outvzcosh}. 
	\item Inside the bouncing region we see that the pre point of bounce graph starts at a high value at boundary and oscillates randomly till bounce. From \Cref{fig:c1invzcosh} we observe continuity at point of bounce(at a higher value than whatever the extrapolated outside region lines pointed at) and see the post point of bounce graph follow random oscillations till nearing the boundary from which it starts rising and slowly goes on to saturates at late times when extrapolated. 
	\item From \Cref{fig:c2invzcosh} the $\C_2$ behaviour is seen to be similar although the oscillations have a single defined smoother trough and peak. We do not see continuity at point of bounce, but we see the similar rising behaviour of post point of bounce line as it approaches boundary and saturates upon extrapolation. 
\end{itemize}


\begin{figure}[!htb]
	\centering
	\includegraphics[width=15cm,height=8cm]{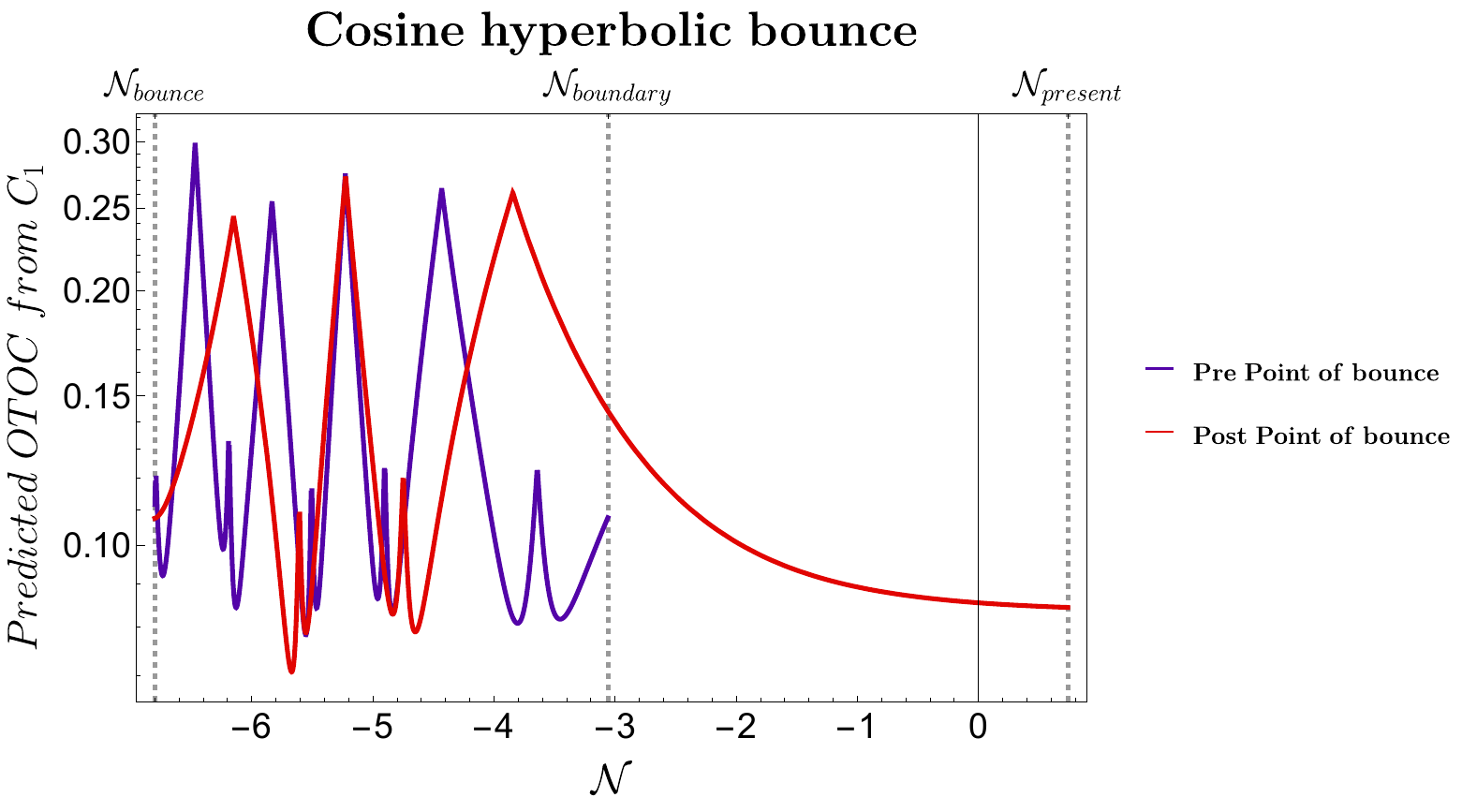}
	\caption{Predicted OTOC from linearly weighted complexity outside bouncing region with respect to number of e-foldings}\label{fig:otocc1invzcosh}

\end{figure}

\begin{figure}[!htb]
	\centering
	\includegraphics[width=15cm,height=8cm]{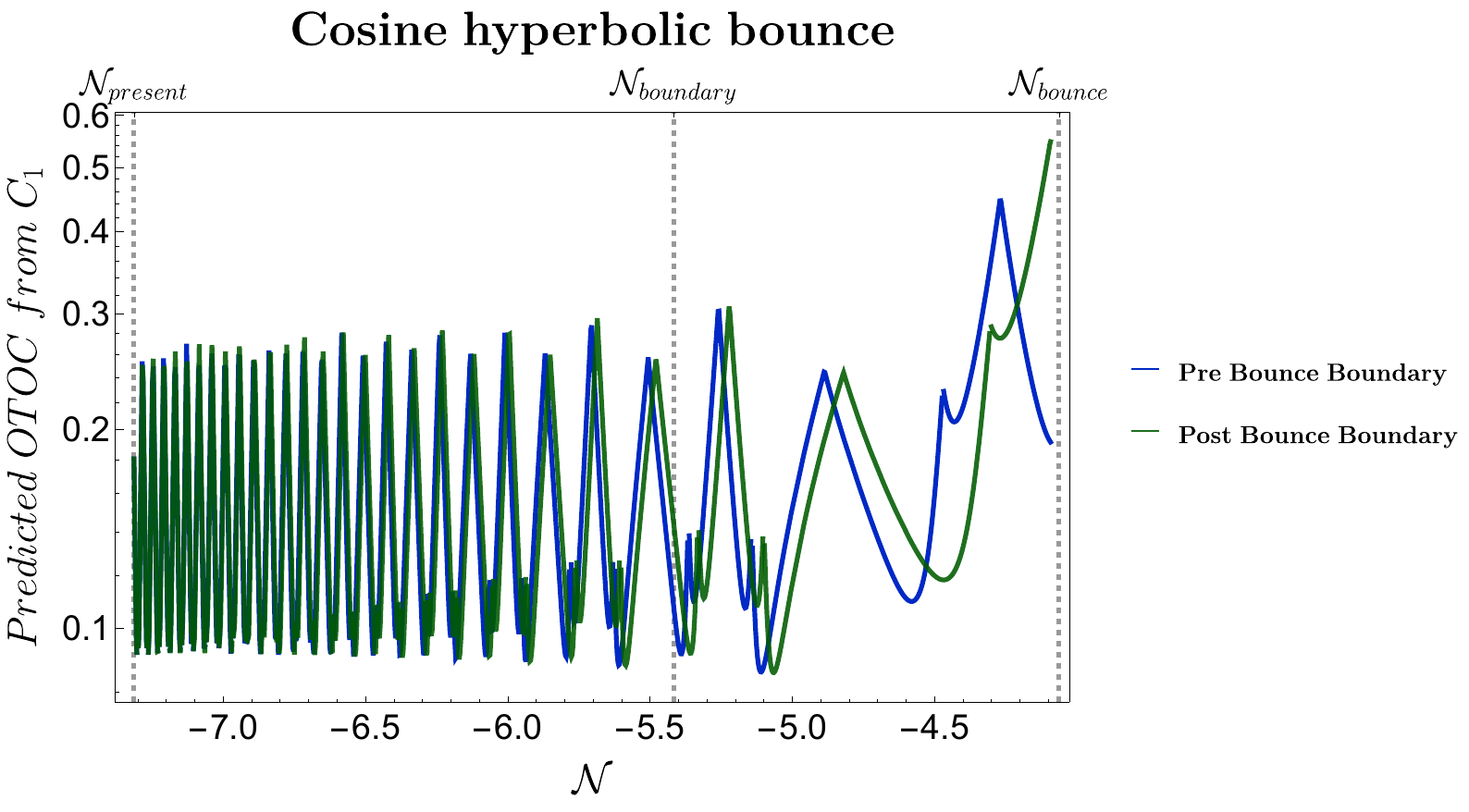}
	\caption{Predicted OTOC from linearly weighted complexity outside bouncing region with respect to number of e-foldings}\label{fig:otocc1outvzcosh}

\end{figure}

\begin{figure}[!htb]
	\centering
	\includegraphics[width=15cm,height=8cm]{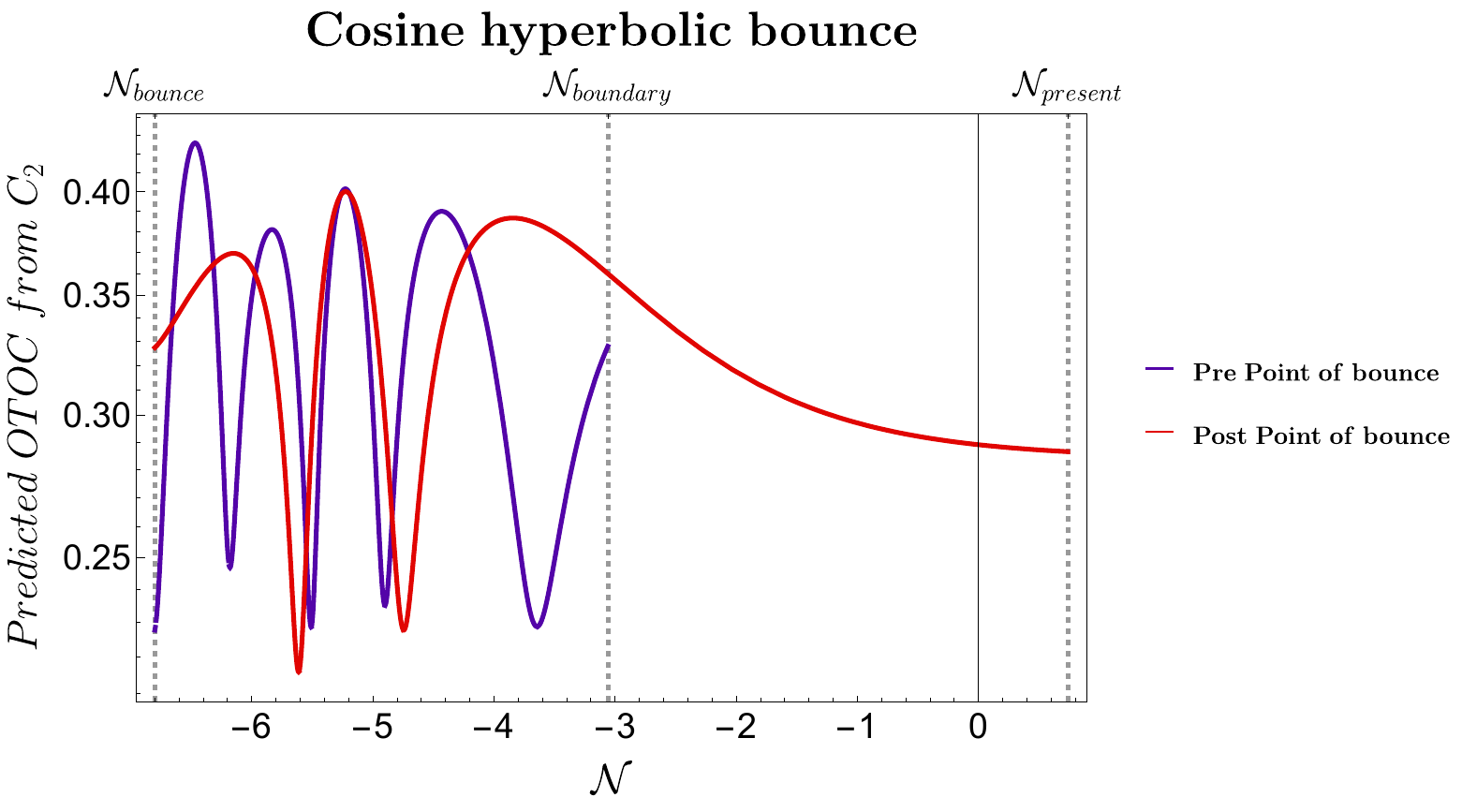}
	\caption{Predicted OTOC from geodesically weighted complexity inside bouncing region with respect to number of e-foldings}\label{fig:otocc2invzcosh}

\end{figure}

\begin{figure}[!htb]
	\centering
	\includegraphics[width=15cm,height=8cm]{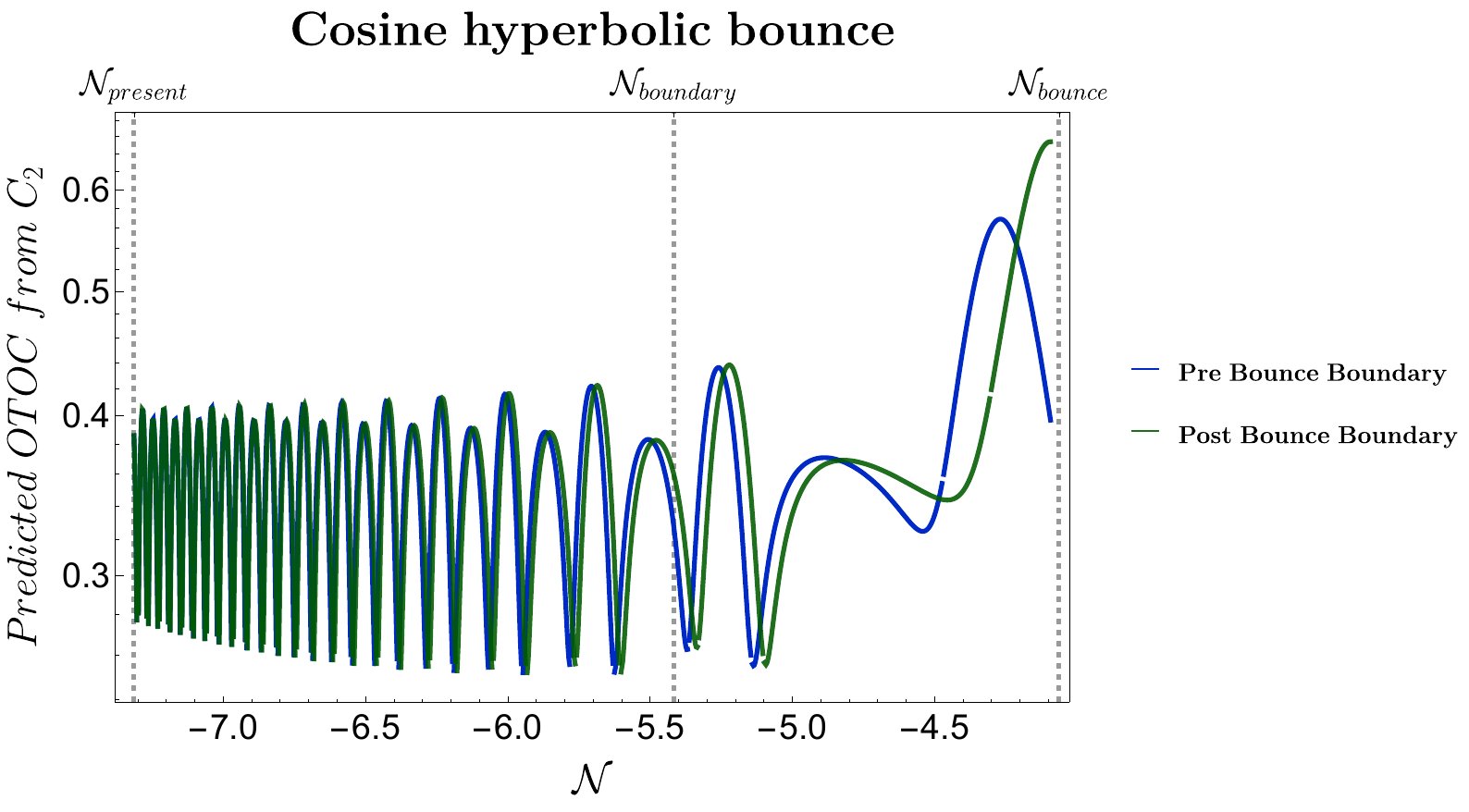}
	\caption{Predicted OTOC from geodesically weighted complexity outside bouncing region with respect to number of e-foldings}\label{fig:otocc2outvzcosh}
\end{figure}

We can also observe the behaviour of the predicted OTOC values from the complexities. The behaviour is very similar to that of the one we saw with scale factor. 
\begin{itemize}
	\item outside the bouncing region the values are highly oscillating and produce random fluctuations upon extrapolation. The oscillations are smoother(smooth peaks and troughs) in the case of OTOC from $\C_2$. See \Cref{fig:otocc1outvzcosh} and \Cref{fig:otocc2outvzcosh}. 
	\item Inside the bouncing region as before we have extrapolated the post point of bounce line that decreases steadily from near the boundary to the extrapolated value of $\N_{present}$ where it saturates as seen in \Cref{fig:otocc1invzcosh} and \Cref{fig:otocc2invzcosh} This is the signature of a chaotic system. The pre point of bounce line shows random fluctuations and an increasing value of OTOC if extrapolated.
	\end{itemize}

\newpage
\textcolor{Sepia}{\subsection{\sffamily Exponential bounce}}

We have numerically plotted the squeezing parameters and the derived Complexity measures for  Exponential bounce model in four different regions - pre bounce boundary, pre point of bounce, post point of bounce and post bounce boundary against the scale factor. In \Cref{{fig:avstauexp}}, the scale factor of the exponential model has been plotted with respect to the conformal time. It can be seen that at present time and at a time much before the boundary ($\tau \rightarrow -\infty$) the value of scale factor $a = 1$. We have taken the value of pre-boundary and post-boundary parameters $r_{\bf k}(a=1) = 1, \phi_{\bf k}(a=1) = 1$ to set our initial conditions, and ensured continuity at $a_{boundary}$ as initial conditions for the bouncing region parameters for numerically solving differential equations with respect to scale factor (Eqs. \ref{eq:diffeqnswa1} and \ref{eq:diffeqnswa2}). For the analysis of exponential bounce we have taken $-k\tau_b = 30$ and the range of $-k\tau$ goes from 0 to 60. We have fixed the value of $\Lambda$ to 10$^{-4}$ for our numerical analysis in this case as well.

\begin{figure}[!htb]
	\centering
\includegraphics[width=15cm,height=8cm]{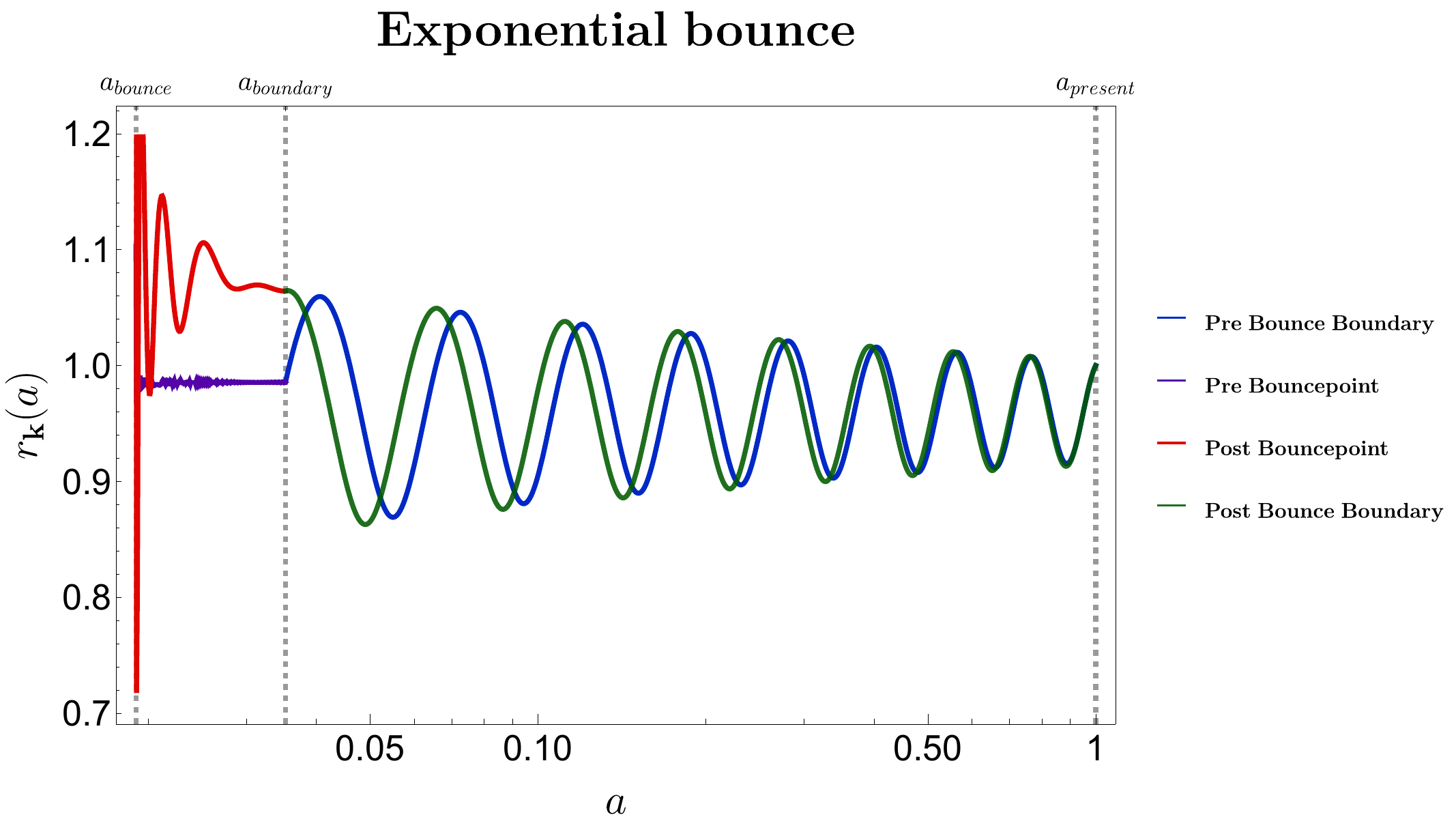}
	\caption{Squeezing parameter at different regions plotted against scale factor}
	\label{fig:rkvaexp}
\end{figure}

\begin{figure}[!htb]
	\centering
\includegraphics[width=15cm,height=8cm]{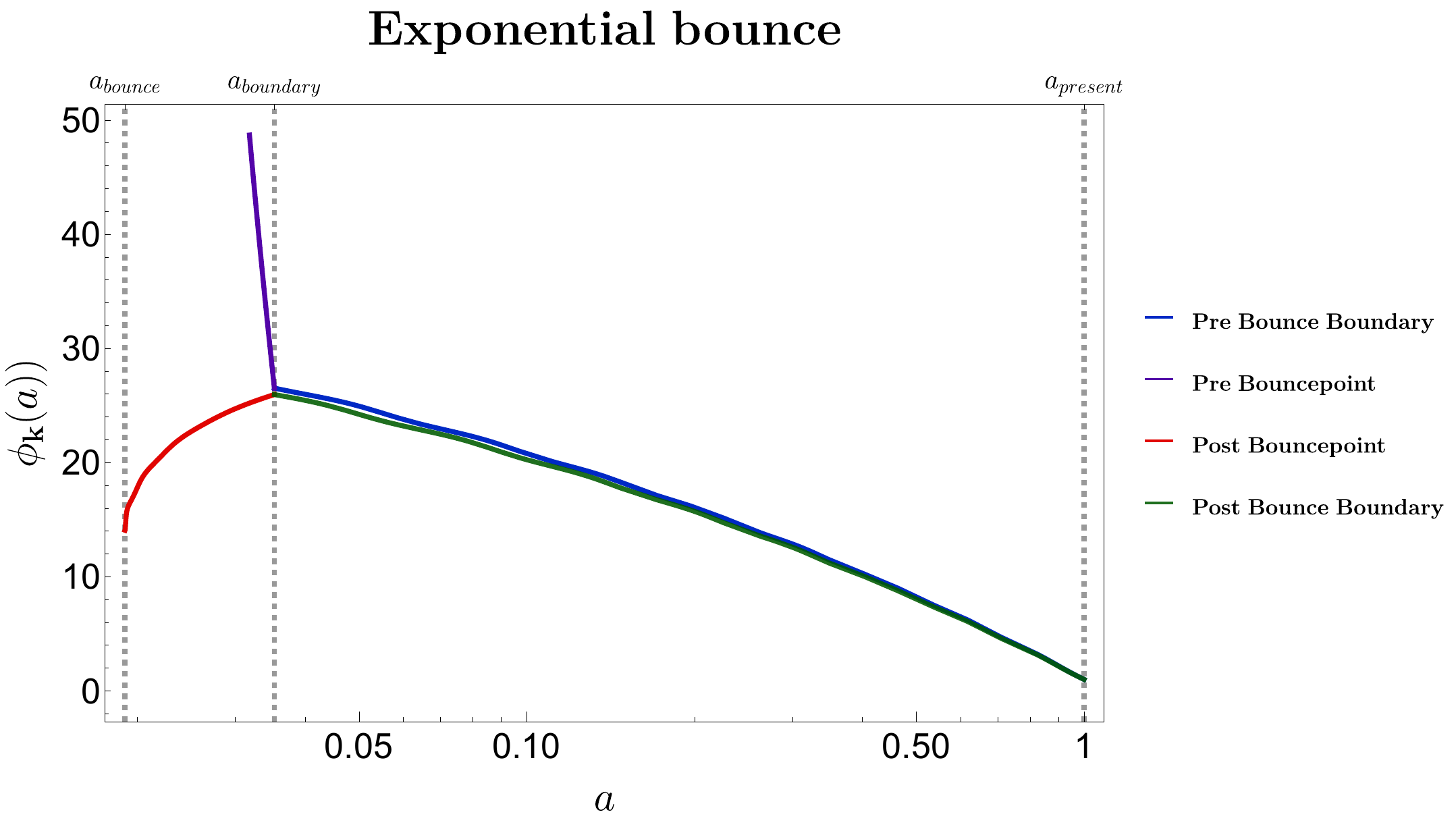}
	\caption{Squeezing angle at different regions plotted against scale factor}
	\label{fig:phivaexp}
\end{figure}

\begin{figure}[!htb]
	\centering
\includegraphics[width=15cm,height=8cm]{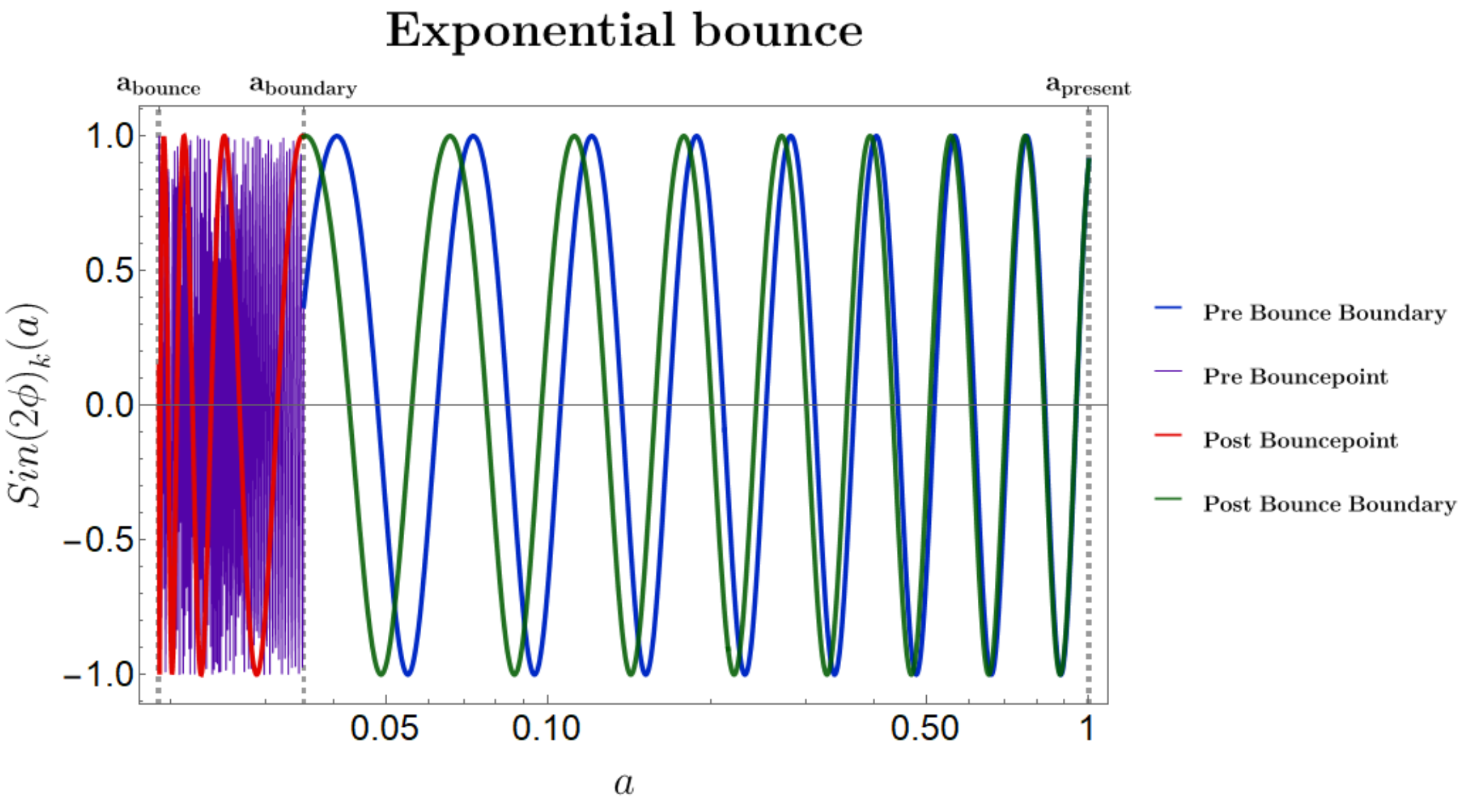}
	\caption{Sine of twice of squeezing angle at different regions plotted against scale factor}
	\label{fig:sin2pvaexp}
\end{figure}
\begin{itemize}
	\item In \Cref{fig:rkvaexp} the behaviour of the squeezing parameter $r_{k}$ has been plotted with respect to the scale factor of the model for four different regions of interest. We have taken the value of pre-boundary and post-boundary parameters $r_{\bf k}(a=1) = 1$ to set our initial conditions, and ensured continuity at $a_{boundary}$ as initial conditions for the bouncing region parameters for numerically solving Eq(\ref{eq:diffeqnswa1}), and Eq(\ref{eq:diffeqnswa2}). 
	\item The pre-boundary and the post-boundary behaviour of $r_k$ is oscillatory with decreasing amplitude as it approaches $a=1$. 
	\item The behaviour of $r_k$ for the post bounce region can be seen to be highly oscillatory near the point of bounce and the amplitude of oscillations reduces significantly near the boundary. However the behaviour of the squeezed parameter $r_k$ is almost constant with minor fluctuations in the region between pre bounce boundary and the point of bounce.
\end{itemize}
 In \Cref{fig:phivaexp} and \Cref{fig:sin2pvaexp}, the \textbf{squeezing angle} and the \textbf{sine of twice its value} has been plotted with respect to the scale factor.
\begin{itemize}
	\item The squeezed angle parameter $\phi_\k$ slows an exponential increase starting from almost zero, for the pre-bounce boundary region, with the rate of rise decreasing as the parameter approaches the boundary of the bouncing region from very early times. The sine of twice the angle of the squeezed parameter in this region is a periodic function with the frequency of oscillation decreasing as it approaches the boundary of the bouncing region.
	\item  $\phi_\k$ shows a asymptotic rise for the pre point of bounce region and the sine of twice the angle of $\phi_\k$ shows wild oscillation in this region.
	\item The squeezed angle parameter $\phi_\k$ slows an exponential increase for the post point of bounce region and the sine of twice the angle of $\phi_\k$ shows oscillatory behaviour with the frequency of oscillation larger near the point of bounce than near the boundary. 
	\item An exponential decay of the squeezed angle parameter can be seen for the post bounce boundary region with the value approaching zero for the present day. The sine of twice the angle shows regular oscillatory behaviour with the frequency of oscillations increasing as one approaches the present day.
\end{itemize}

\begin{figure}[!htb]
	\centering
	\includegraphics[width=15cm,height=8cm]{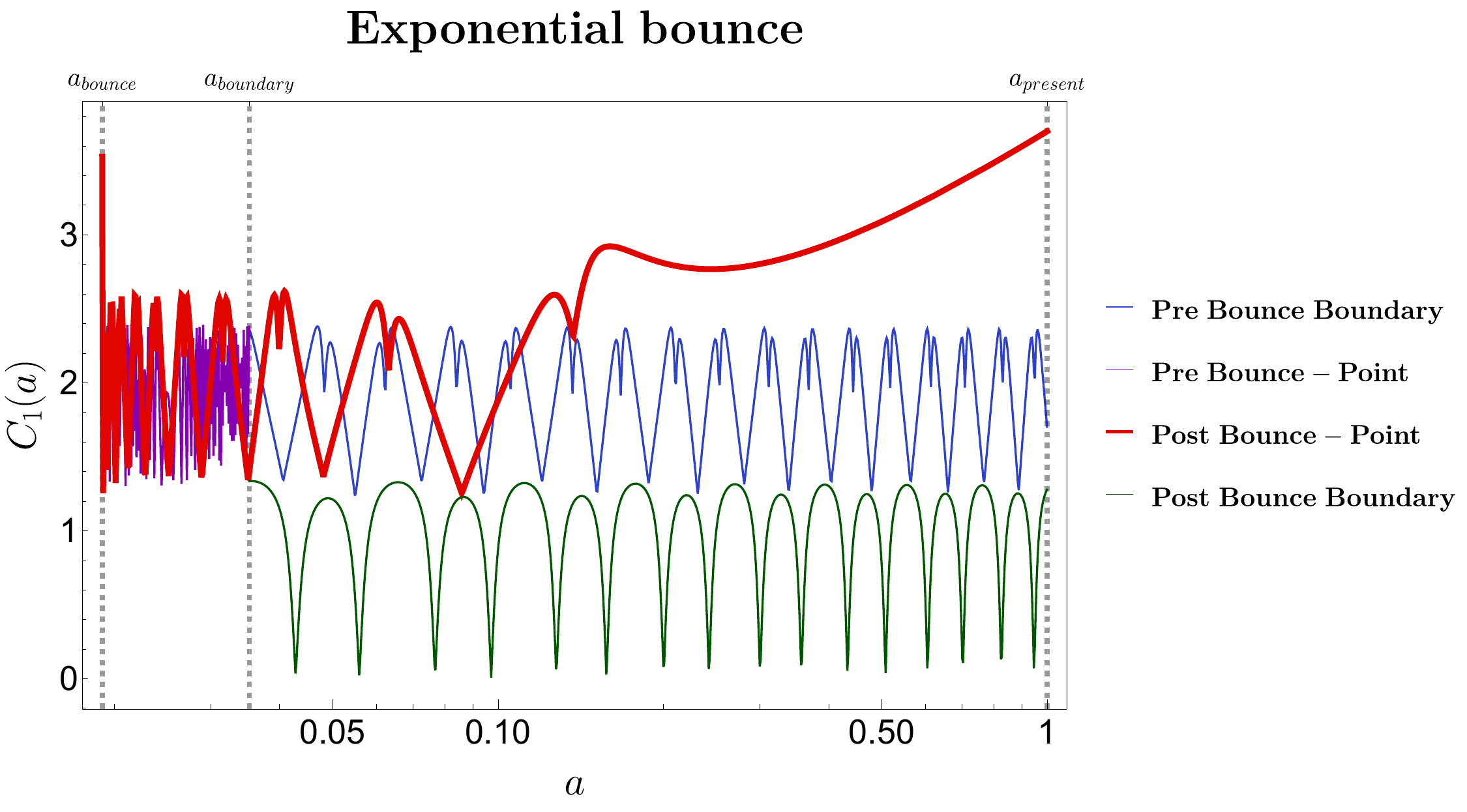}
	\caption{Linearly weighted complexity in different regions against scale factor}
	\label{fig:c1vaexp}
\end{figure}

\begin{figure}[!htb]
\centering
\includegraphics[width=15cm,height=8cm]{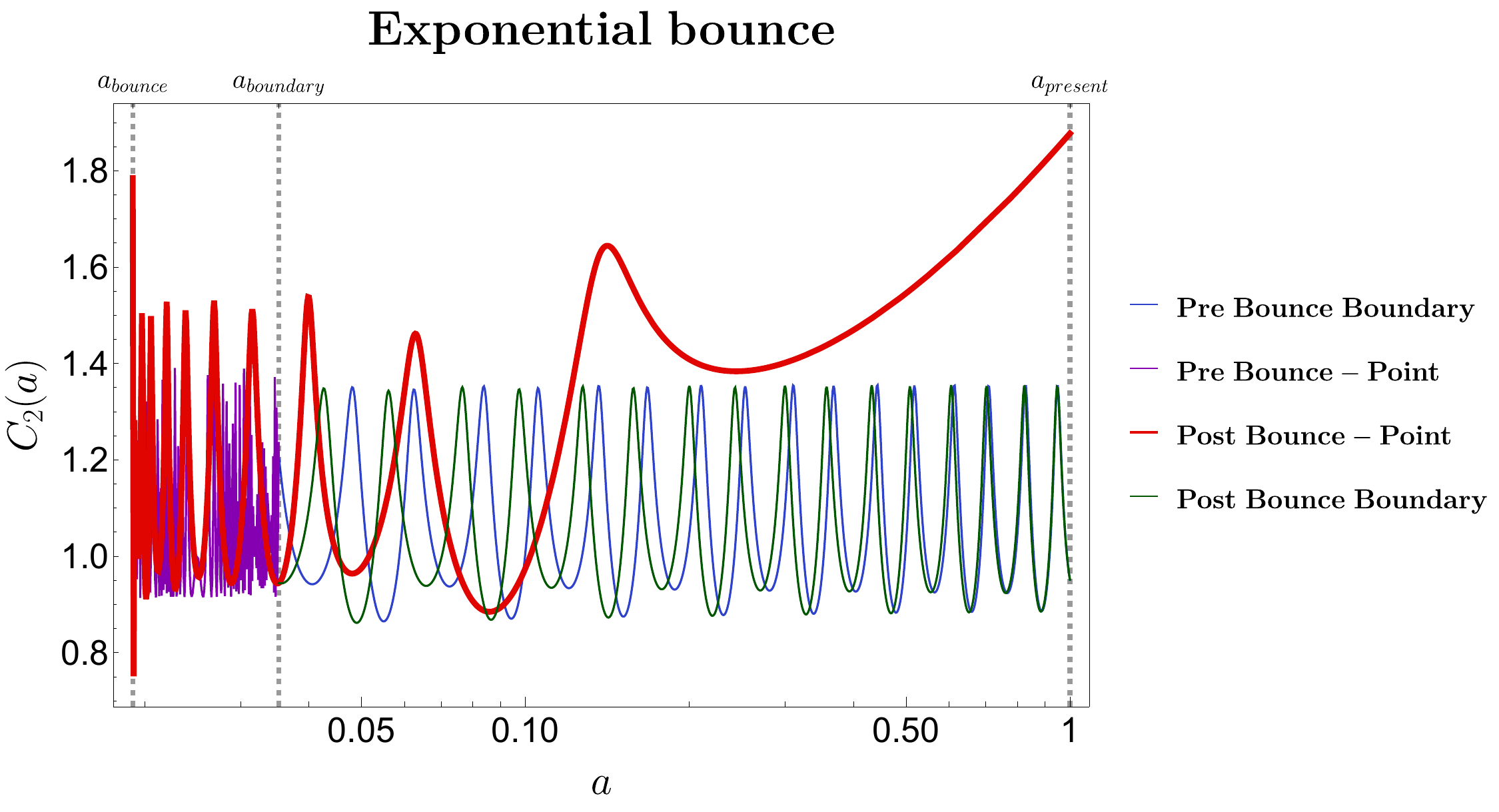}
	\caption{Geodesically weighted complexity in different regions against scale factor}
	\label{fig:c2vaexp}
\end{figure}

\begin{figure}[!htb]
	\centering
	\includegraphics[width=15cm,height=8cm]{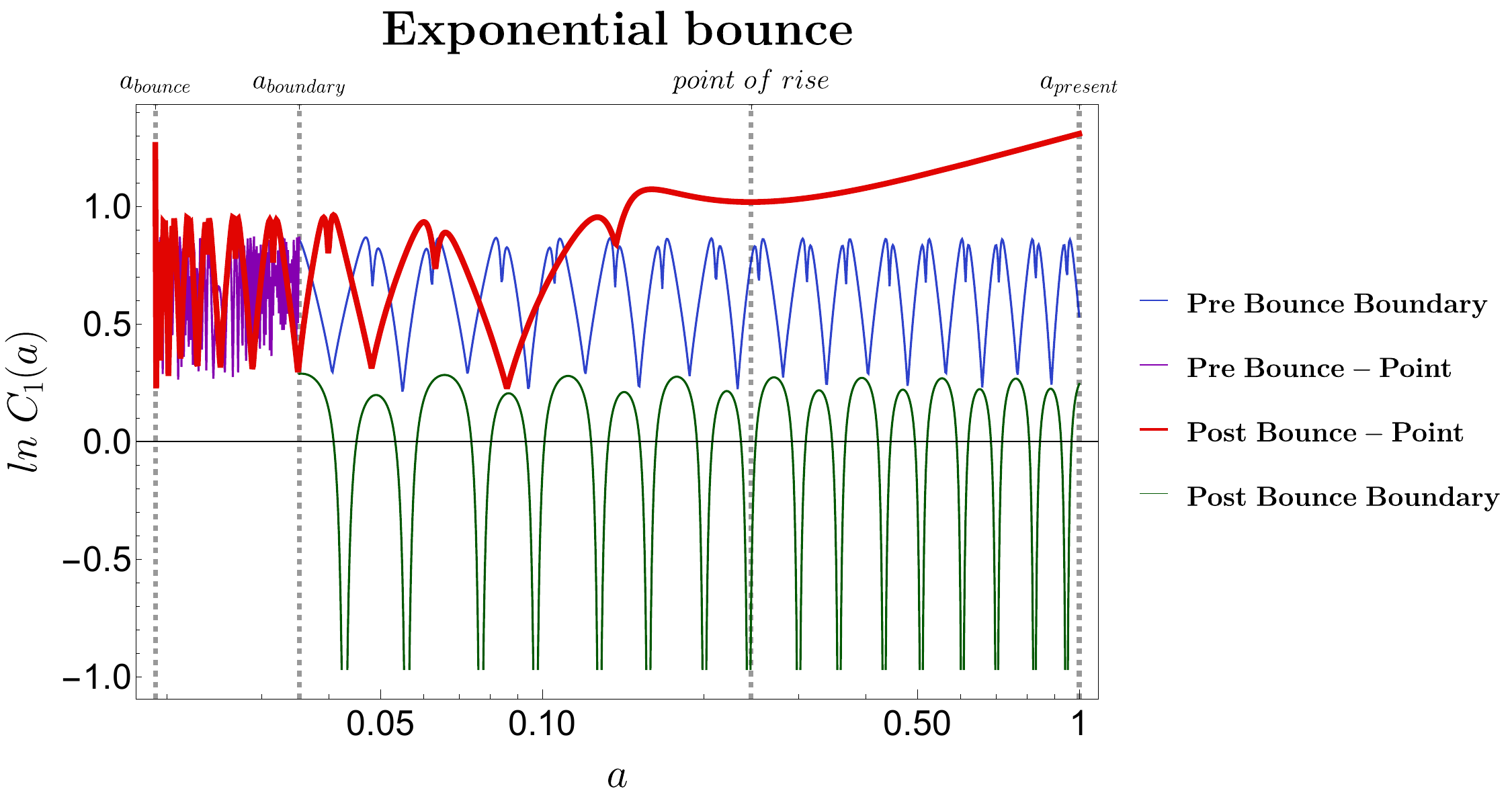}
	\caption{Logarithm of linearly weighted complexity in different regions against scale factor}
	\label{fig:lnc1vaexp}
\end{figure}

\begin{figure}[!htb]
	\centering
	\includegraphics[width=15cm,height=8cm]{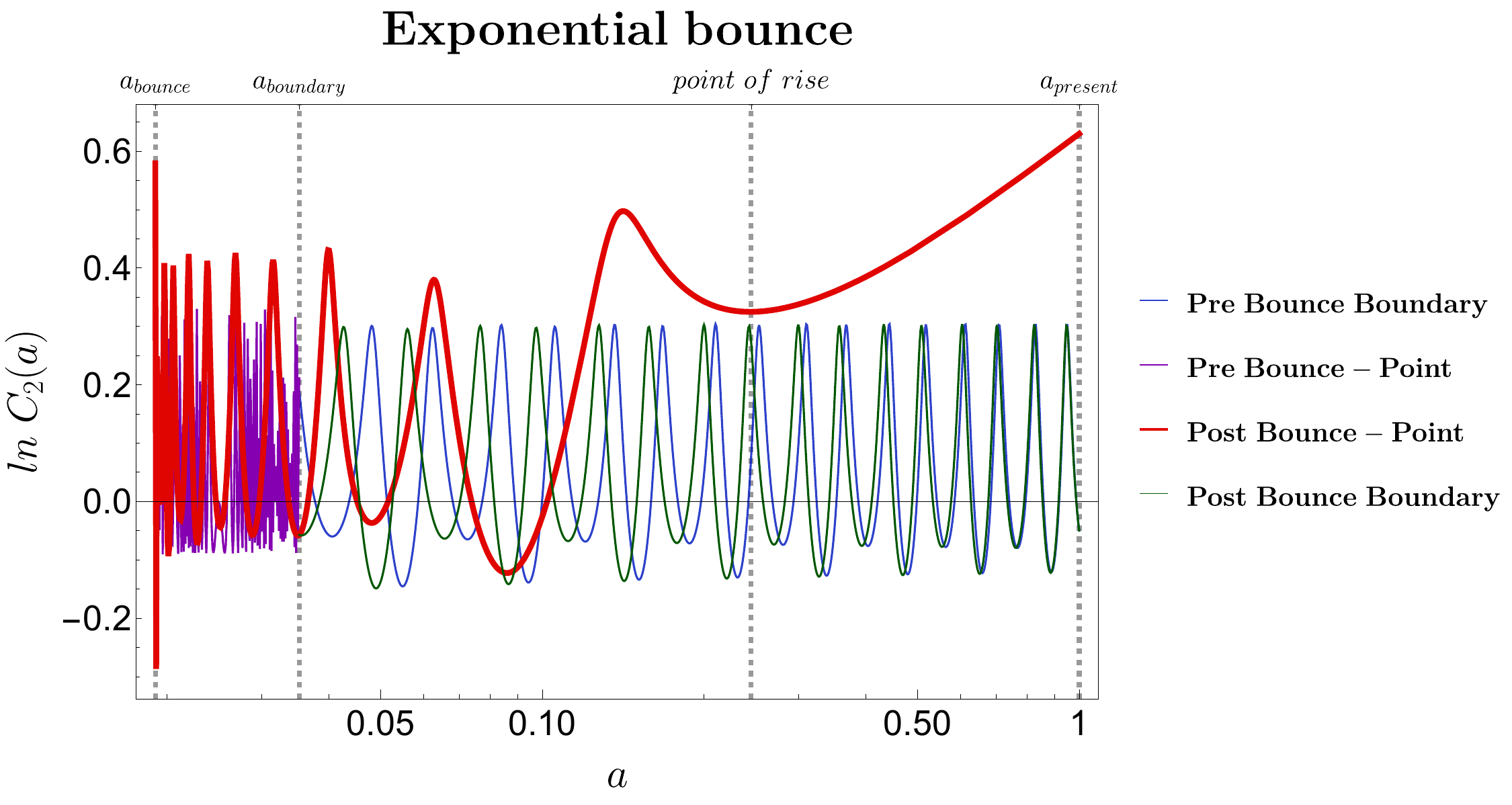}
	\caption{Logarithm of geodesically weighted complexity in different regions against scale factor}
	\label{fig:lnc2vaexp}
\end{figure}

In \Cref{fig:c1vaexp} and \Cref{fig:c2vaexp} the complexity measures $\mathcal{C}_1$ and $\mathcal{C}_2$ have been plotted with respect to the scale factor. The behaviour of both the complexity measures are almost identical. 
\begin{itemize}
	\item  The value of complexity outside the bouncing boundary based on the respective squeezing parameters defined there for very early times and nearing present times is oscillatory with smaller frequency at the boundary. The amplitude of oscillation however remains almost identical for the early and nearing present times to that at the boundary. 
	\item Inside the bouncing region the complexity before the pre-bounce shows wild oscillations.
	\item The behaviour of the post point of bounce shows interesting features on extrapolation to the present day. Although within its domain i.e inside the bouncing boundary region the behaviour is oscillatory as the pre-point of bounce behaviour but extrapolation to the present day shows exponential increase in the complexity after a certain time. This is different to what we observed in the cosine hyperbolic model where the rising behaviour of complexities was seen inside the bouncing region before the post bounce boundary itself.
\end{itemize}

\begin{table}[!htb]
	\centering
	\begin{tabular}{|P{1cm}||P{2cm}|P{2cm}|P{2cm}|P{2cm}|P{2.2cm}|}
		\hline 
		& Very early times & Entering bouncing region & Around point of bounce & Exiting bouncing region & Extrapolated Present time \\ 
		\hline 
		$\C_1$ & 1.701 & 2.357 & 12.696 & 1.402 & 3.698 \\ 
		\hline 
		$\C_2$ & 0.0951 & 1.208 & 7.944 & 0.944 & 1.877 \\ 
		\hline 
	\end{tabular} 
	\caption{Complexity values at different points of interest with respect to scale factor}
	\label{tab:compexp}
\end{table}
In \Cref{tab:compexp} we have presented the values of complexity at various time scales as observed in this exponential bouncing cosmology model.

From the complexity values shown in \Cref{tab:compexp}, one can interpret that the system tends towards a highly chaotic behaviour near the point of bounce, which is understood from the maximum complexity value at that point. In the language of squeezed quantum states, the rapid oscillation of the squeezing parameters near the point of bounce may be an indirect way of signifying chaos.

Since we observe most interesting features from the post point of bounce plots(on extrapolating to the present time), the prediction of Lyapunov exponent from that case is extremely useful as it gives an estimation on the lower bound of temperature. 
The slope of logarithm of complexity at the point of rise directly corresponds to the value of the quantum Lyapunov exponent as mentioned in Eq (\ref{eq:lyapexp}). 
To predict the slope of the logarithmic value of complexities we consider the change of y-axis value over the range of the x-axis value i.e. between point of rise and point of saturation. Mathematically it is represented by
\begin{equation}
\lambda_i= \frac{ln~\C_i~(\text{present time})- ln~\C_i~(\text{point of rise})}{a~(\text{present time})-a~(\text{point of rise})}
\end{equation} For this we have plotted the logarithm of complexity values in \Cref{fig:lnc1vaexp} and \Cref{fig:lnc2vaexp}. We observe the qualitative features to be same as that of the complexity graphs, showing corresponding oscillatory, rising and saturation at the respective regions.
We calculate the Lyapunov exponent from the post point of bounce case as it shows exponential feature and this gives an estimation on the lower bound of temperature.

\begin{table}[!htb]
	\centering
	\begin{tabular}{|P{3cm}||P{3cm}|P{3cm}|}
		\hline  
	ln $\mathcal{C}_i$	& point of rise  & present time  \\ 
		\hline 
		ln $\mathcal{C}_1$ & 1.0180  & 1.3079  \\ 
		\hline 
	ln $\mathcal{C}_2$ &  0.32469 & 0.62969 \\ \hline
	\end{tabular} 
	\caption{Log of complexity values at point of rise and present time. In this case we do not observe any saturation as such, hence we compute the slope between the point of rise and the present time. Point of rise for both $\C_1$ and $\C_2$ is $a=0.245$ and scale factor corresponding to the present time is $a=1$}
	\label{tab:lyaexp}
\end{table}
 
In \Cref{tab:lyaexp} we have written the numerical values of the logarithm of the complexity values at the region of a where complexity shows an exponential rise. The Lyapunov exponent calculated from these values are:
\begin{align*}
	\lambda_1= 0.3839 && \lambda_2= 0.4039
\end{align*}

The estimated lower bound on the temperature from the calculated values of the Lyapunov exponents are
\begin{align*}
T_{\mathcal{C}_1} \succsim 0.06109 && T_{\mathcal{C}_2} \succsim 0.06428
\end{align*}

Using Eq(\ref{scramblingscale}), similar to the cosine hyperbolic case we have also computed lower bound of scrambling time intervals for the exponential case. The main difference is that as we have seen for the exponential case the complexity values do not actually saturate even at much later times. Hence we have calculated the scrambling time in the region of rise (the same region that we have numerically considered for calculating the Lyapunov exponent in exponential case). It is unclear whether the physical interpretation of the scrambling time will remain same as there is no given region of saturation in the exponential case as we had for the cosine hyperbolic case. Nevertheless an \textbf{estimated value for the same in the region of rise} is given in \Cref{tab:scramexp}. Since our normalization for conformal time at the exponential case is different with bounce at $\tau_b = -150$ and present time at $0$, we can see that the scrambling period is more than one-half of time from bounce to present day. Such a high scrambling time (more time for OTOC to attain equilibrium) can be due to the lack of saturation and late and perpetually rising complexity values, and a never hence a never saturating OTOC. Hence the interpretation of scrambling time in rising region might point to fact that it takes really long (almost never) for OTOC to attain equilibrium hence hinting at the \textbf{lack of saturation in the chaotic behaviour} that we have seen in complexity in the case of cosine hyperbolic model. 

\begin{table}[!htb]
\centering
\begin{tabular}{|P{2cm}||P{3.5cm}|P{3.5cm}|P{2cm}|P{2cm}|}
\hline 
& $c_i$ at $a=0.75$ & $c_i$ at $a=1$ & $\Delta a_i(\tau_{sc})$ & $|\Delta \tau_{sc}|$  \\ 
\hline 
From $C_1$ & 3.427 & 3.698 & 0.09937 & 82.5063 \\ 
\hline 
From $C_2$ & 1.724 & 1.87704 & 0.1052 & 82.9205\\ 
\hline 
\end{tabular} 
\caption{Estimated lower bound scrambling time scales for the region of rise due to lack of saturation region for the exponential model from the two different complexities}
\label{tab:scramexp}
\end{table}

\begin{figure}[!htb]
	\centering
	\includegraphics[width=15cm,height=8cm]{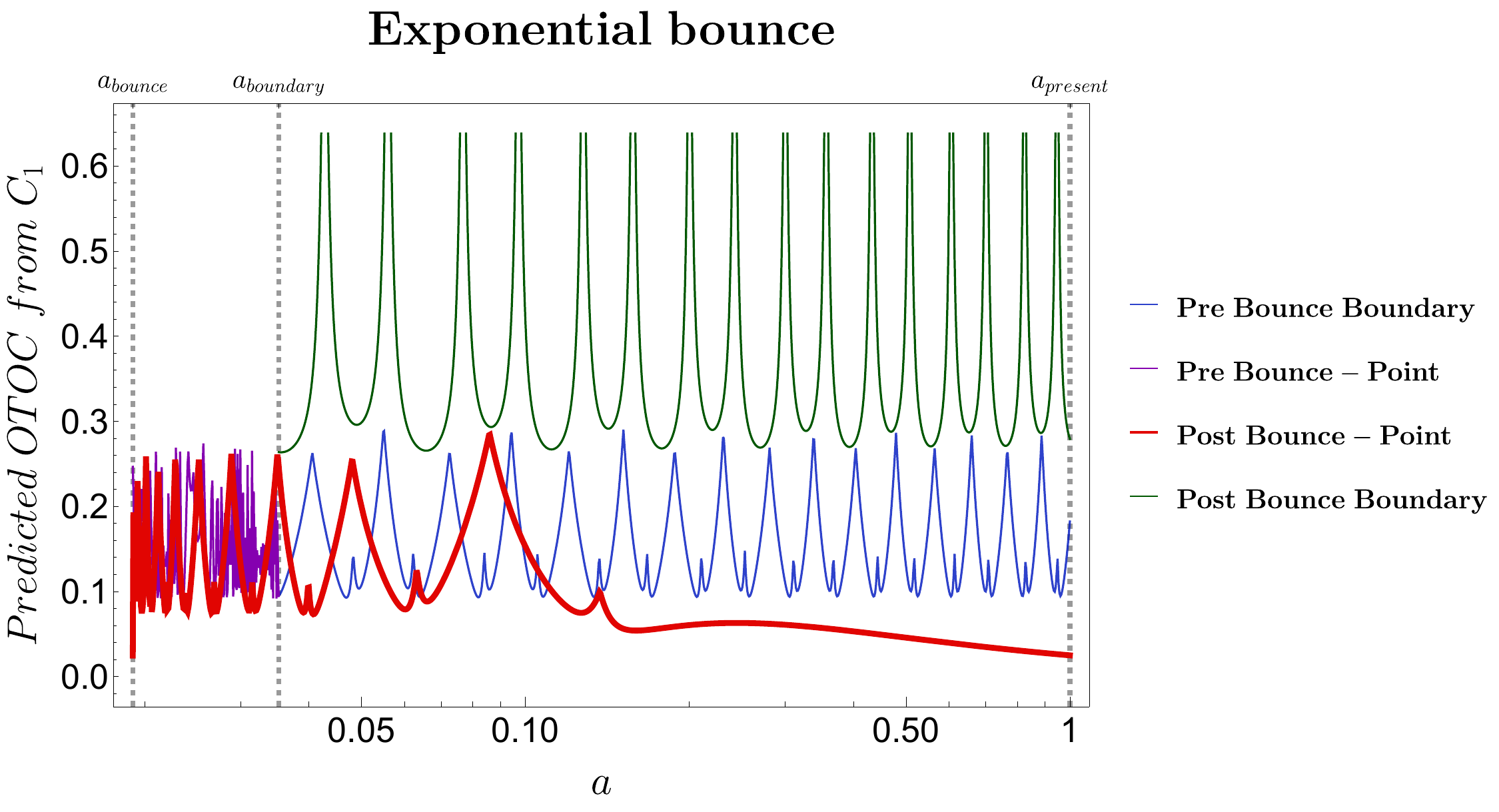}
	\caption{Predicted OTOC from linearly weighted complexity in different regions against scale factor}
	\label{fig:otoc1vaexp}
\end{figure}

\begin{figure}[!htb]
	\centering
	\includegraphics[width=15cm,height=8cm]{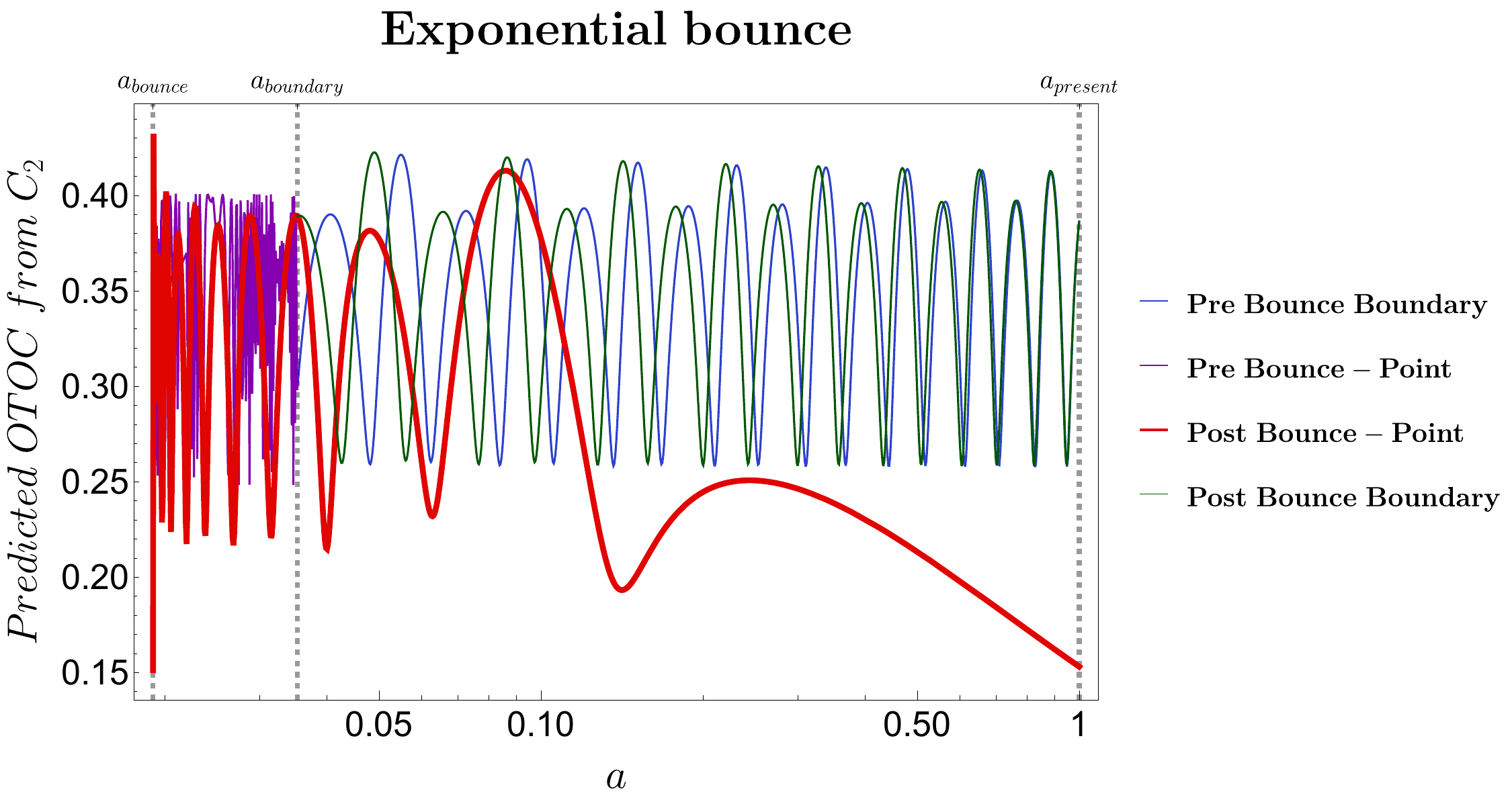}
	\caption{Predicted OTOC from geodesically weighted complexity in different regions against scale factor}
    \label{fig:otoc2vaexp}
\end{figure}

\begin{itemize}
	\item In \Cref{fig:otoc1vaexp} the predicted OTOC from complexity measure $\mathcal{C}_1$ has been plotted with respect to the scale factor. We observe that the outer bouncing boundary curves shows similar features and is oscillatory with the amplitude of post bounce boundary region always lying above the pre-bounce boundary region. Inside the boundary region OTOC shows rapid oscillations for both the pre and the post point of bounce. However interesting features is observed from the post point of bounce curve. The frequency of oscillations starts decreasing near the boundary of the bouncing region. On extrapolation to the present time the OTOC predicted from $\mathcal{C}_1$ actually shows an exponential decay. 
	\item In \Cref{fig:otoc2vaexp} the predicted OTOC from complexity measure $\mathcal{C}_2$ has been plotted with respect to the scale factor. We observe that the outer bouncing boundary curves shows similar features and is oscillatory but unlike the OTOC predicted from $\mathcal{C}_1$. The amplitude of post bounce boundary region is almost identical to the pre-bounce boundary region. Inside the boundary region OTOC shows rapid oscillations for both the pre and the post point of bounce. The frequency of oscillations for the post point of bounce curve starts decreasing near the boundary of the bouncing region. On extrapolation it to the present time the OTOC predicted from $\mathcal{C}_2$ actually shows a similar exponential decay as OTOC predicted from $\mathcal{C}_1$ . 
\end{itemize}

In \Cref{tab:otocexp} the numerical values of the predicted OTOC's from both the complexity measures from our present analysis is written. Again at the point of bounce the OTOC shows a drastic reduction in the values.
\begin{table}[!htb]
	\centering
	\begin{tabular}{|P{2cm}||P{2cm}|P{2cm}|P{2cm}|P{2cm}|P{2.2cm}|}
		\hline 
		& Very early times & Entering bouncing region & Around point of bounce & Exiting bouncing region & Extrapolated present  \\ 
		\hline 
		$OTOC_1$& 0.018 & 0.094 & $3.06 \times 10^{-6}$ & 0.246 & 0.024 \\ 
		\hline 
		$OTOC_2$& 0.386 & 0.298 & $3.54 \times 10^{-4}$ & 0.389 & 0.153 \\ 
		\hline 
	\end{tabular} 
	\caption{Predicted OTOC values at different points of interest with respect to scale factor}
	\label{tab:otocexp}
\end{table} 

As discussed in the previous section it is better to relate the complexity with some quantity which is observable. The number of e-foldings is one such observable quantity. Again we plot the complexity corresponding to within the bouncing region and the outside boundary region separately as done for Model-I.

\begin{figure}[!htb]
	\centering
	\includegraphics[width=15cm,height=8cm]{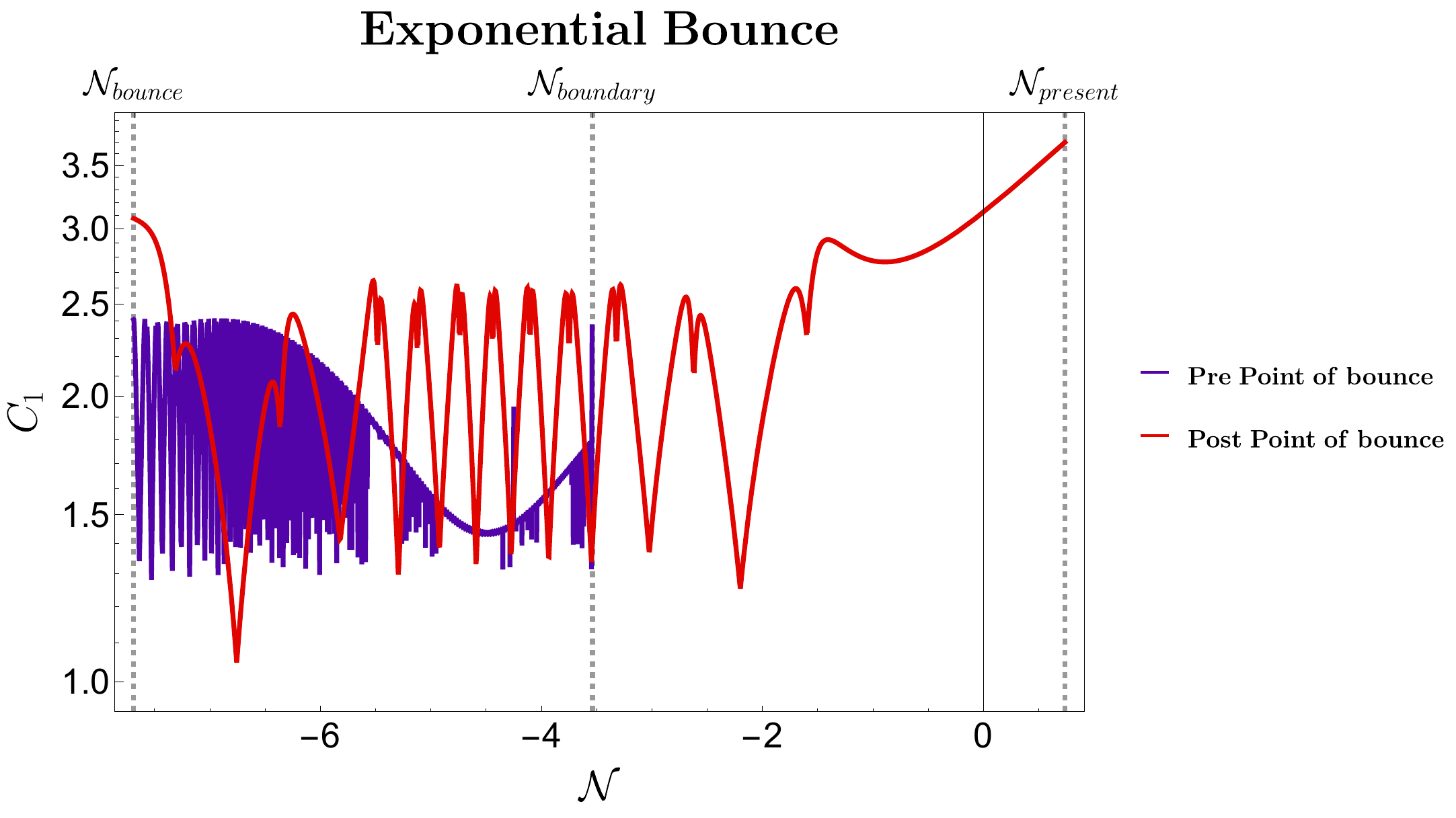}
	
	\caption{Variation of linearly weighted complexity inside bouncing region with respect to number of e-foldings}\label{fig:c1invzexp}
\end{figure}

\begin{figure}[!htb]
	\centering
	\includegraphics[width=15cm,height=8cm]{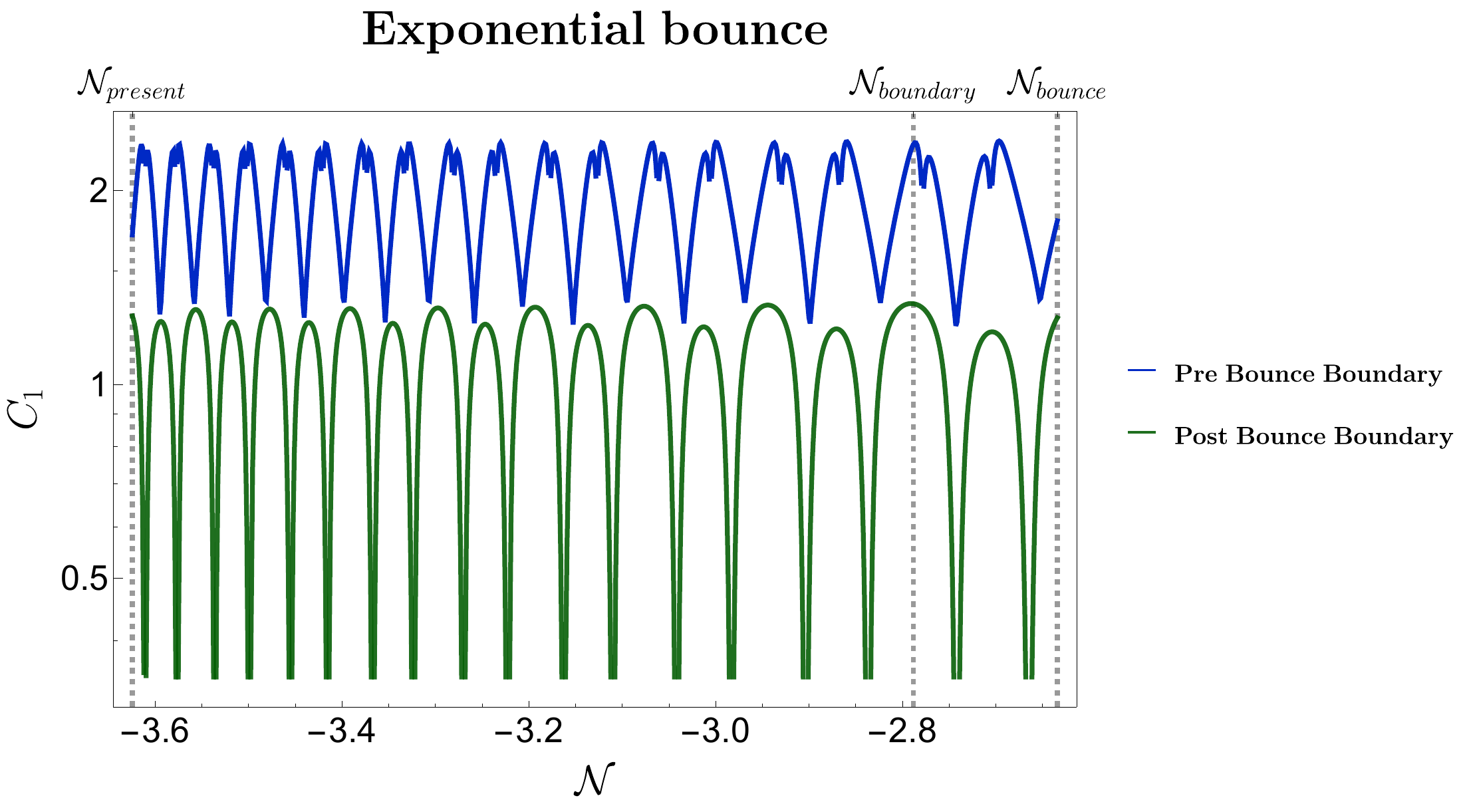}
	
	\caption{Variation of linearly weighted complexity outside bouncing region with respect to number of e-foldings}\label{fig:c1outvzexp}
\end{figure}

\begin{itemize}
	\item In \Cref{fig:c1invzexp} the linearly weighted complexity measure $\mathcal{C}_1$ has been plotted with respect to the number of e-foldings inside the bouncing region. We observe that the behaviour near the boundary is some random fluctuations of negligible amplitudes, however as it approaches the point of bounce it fluctuates wildly. Similarly the behaviour post point of bounce is arbitrary and random near the point of bounce whereas it takes a regular periodic shape on approaching the boundary. However the interesting part can be realised on extrapolating the post bounce behaviour to the present times. We see a sudden exponential rise as it approaches the present times. 
	\item In \Cref{fig:c1outvzexp} we have plotted the complexity measure $\mathcal{C}_1$ as a function of the number of e-foldings for outside the bouncing region. For both the pre and the post boundary region we see a smooth, regular and periodic behaviour of the complexity even on extrapolating it inside the boundary region. However an important feature to notice is that the frequency of oscillation for the pre boundary region decreases when it approaches the boundary whereas the post boundary behaviour shows a contrasting behaviour as the complexity approaches the present time. Also the value of the complexity for the prebounce boundary region is always greater than the post bounce boundary region.
\end{itemize}

\begin{figure}[!htb]
	\centering
	\includegraphics[width=15cm,height=8cm]{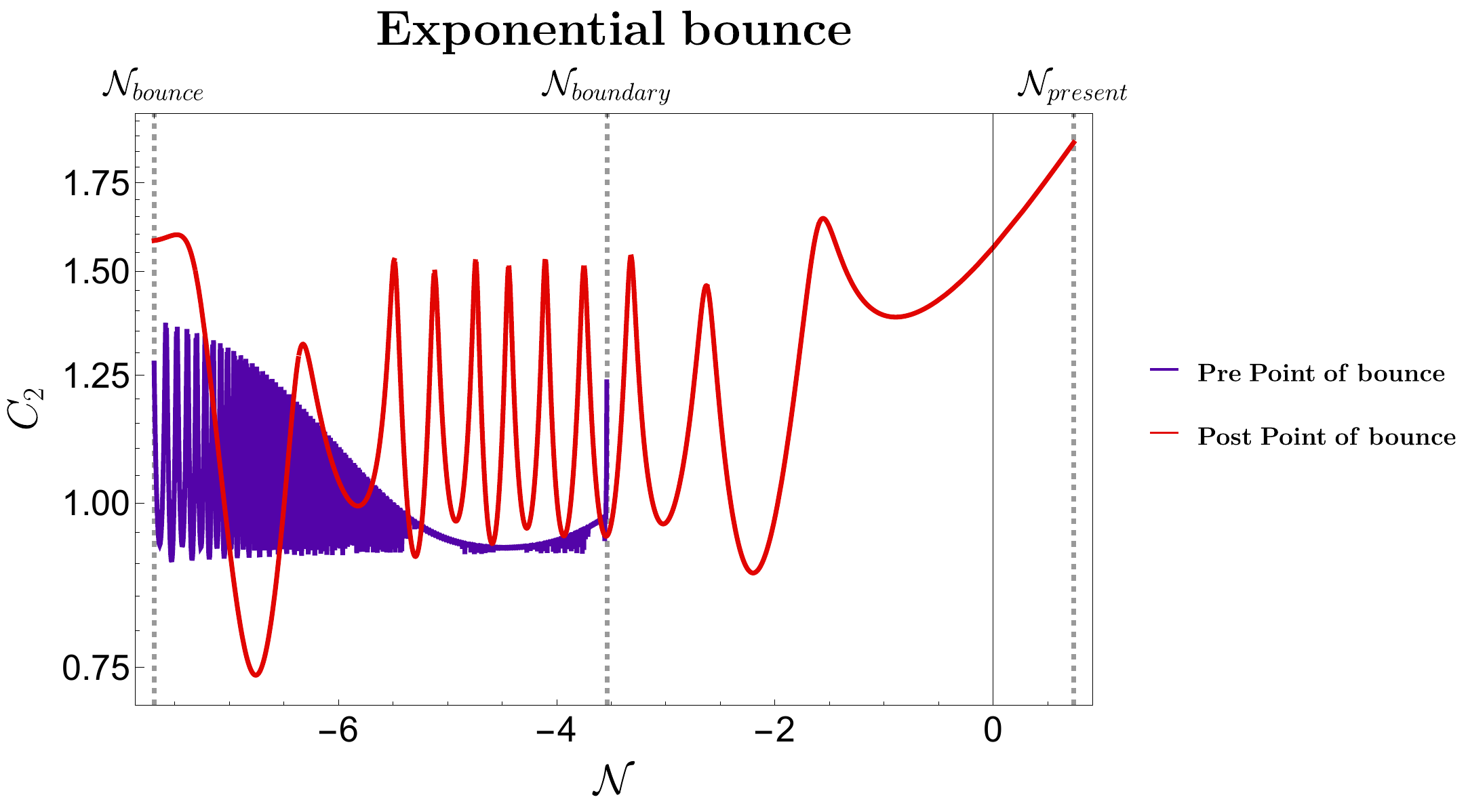}
	
	\caption{Variation of geodesically weighted complexity inside bouncing region with respect to number of e-foldings}\label{fig:c2invzexp}
\end{figure}

\begin{figure}[!htb]
	\centering
	\includegraphics[width=15cm,height=8cm]{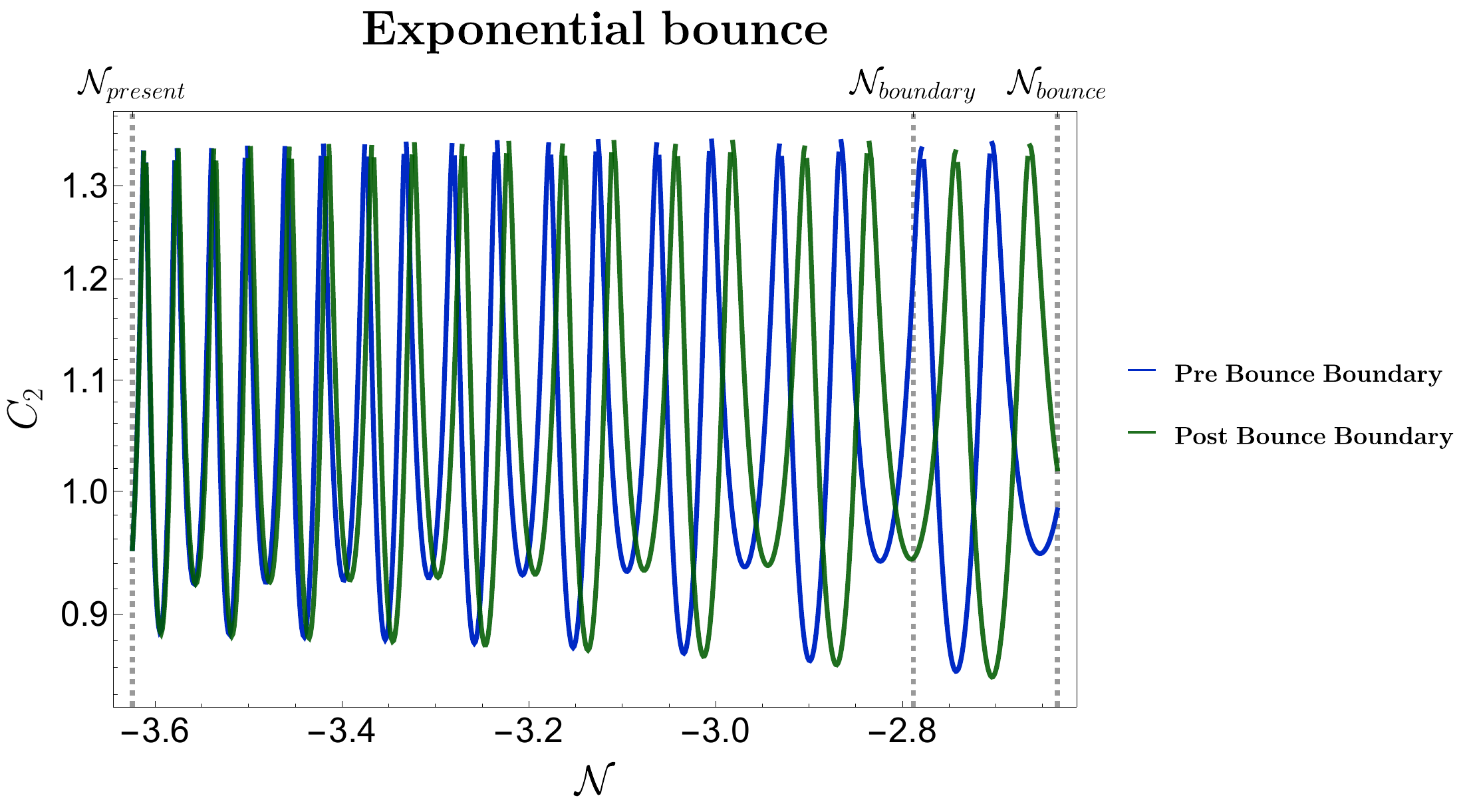}
	
	\caption{Variation of geodesically weighted complexity outside bouncing region with respect to number of e-foldings}\label{fig:c2outvzexp}
\end{figure}
\begin{figure}[!htb]
	\centering
	\includegraphics[width=15cm,height=8cm]{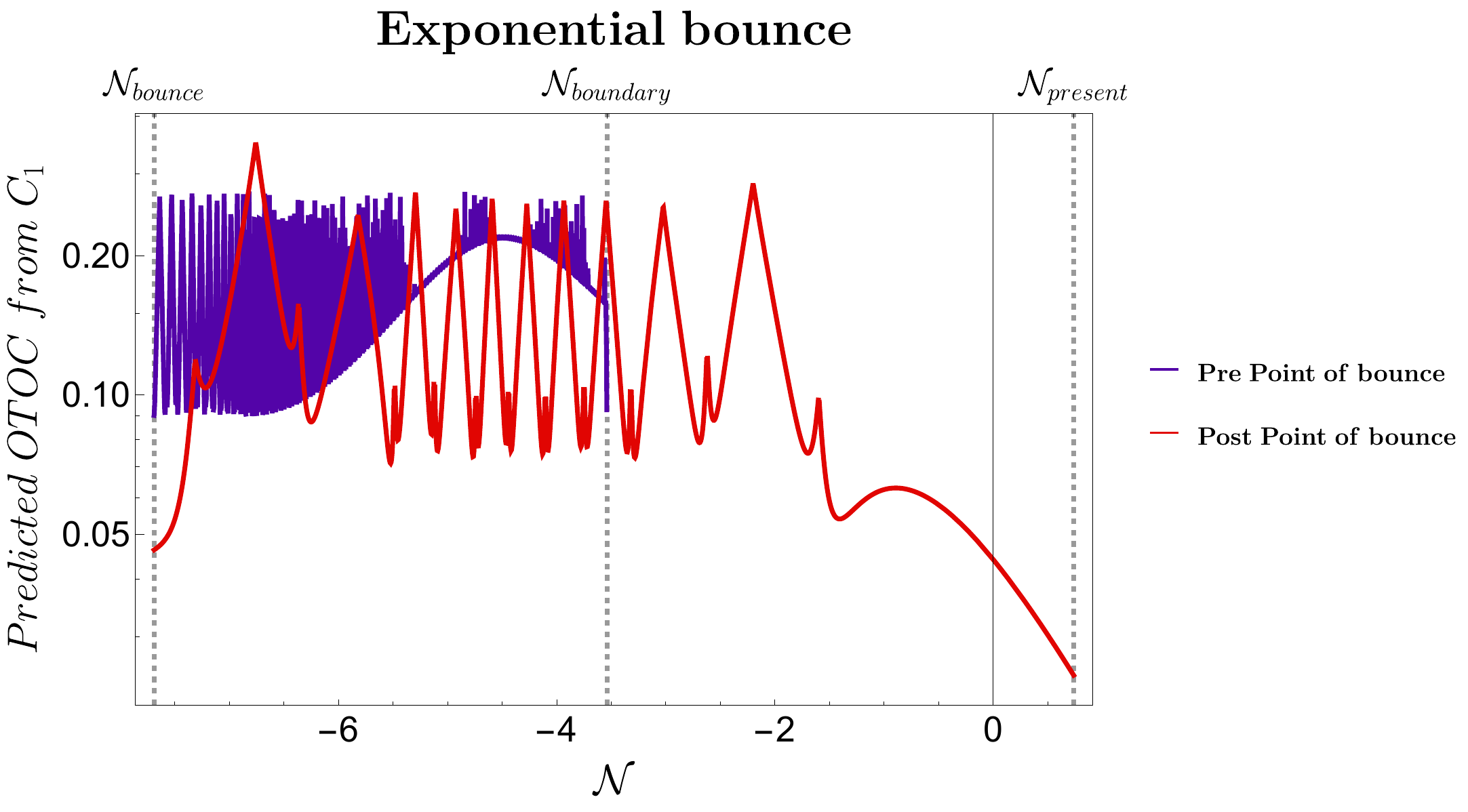}
	\caption{Predicted OTOC from linearly weighted complexity inside bouncing region with respect to number of e-foldings}\label{fig:otocc1invzexp}
	
\end{figure}

\begin{figure}[!htb]
	\centering
	\includegraphics[width=15cm,height=8cm]{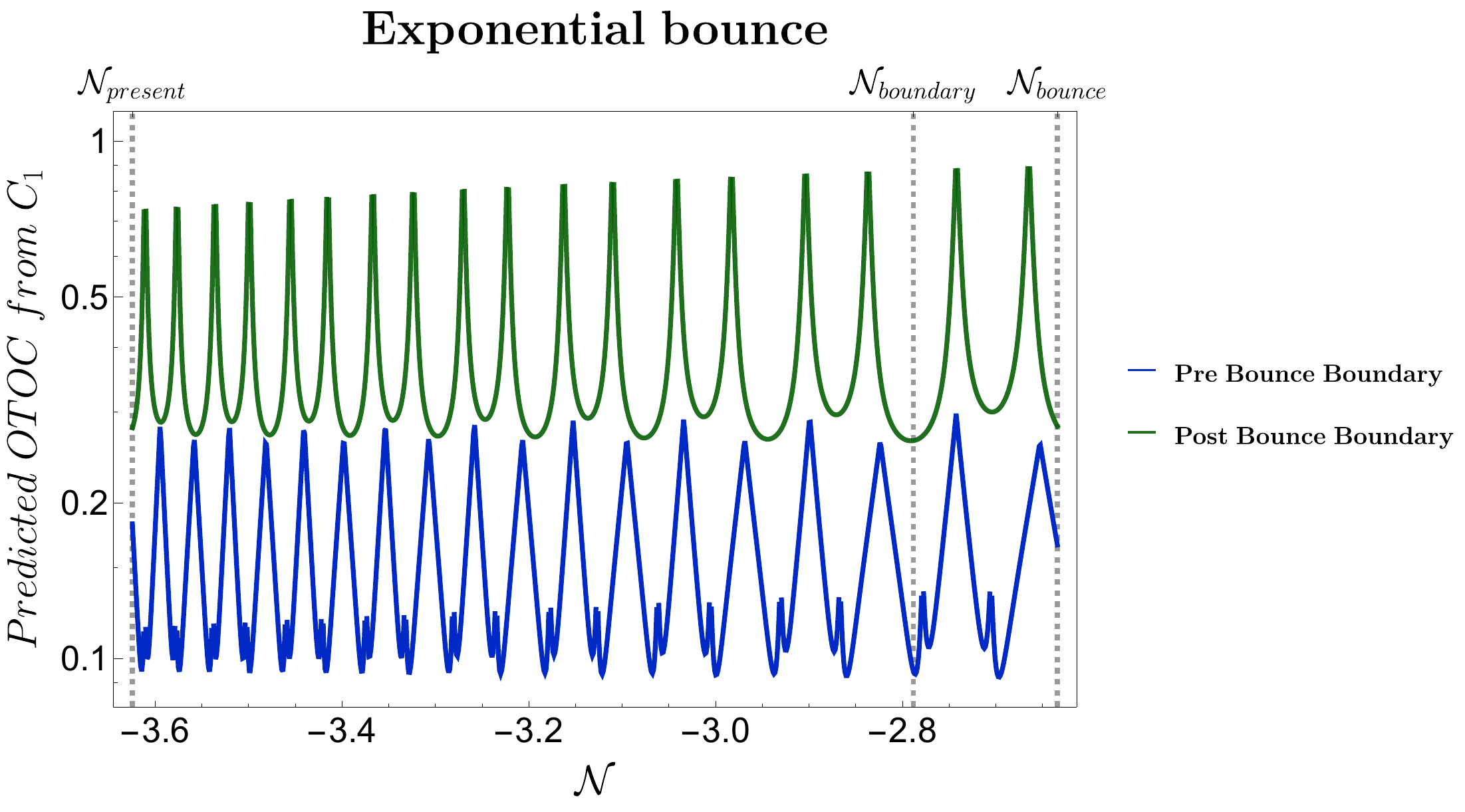}
	\caption{Predicted OTOC from linearly weighted complexity outside bouncing region with respect to number of e-foldings}\label{fig:otocc1outvzexp}
	
\end{figure}

\begin{figure}[!htb]
	\centering
	\includegraphics[width=15cm,height=8cm]{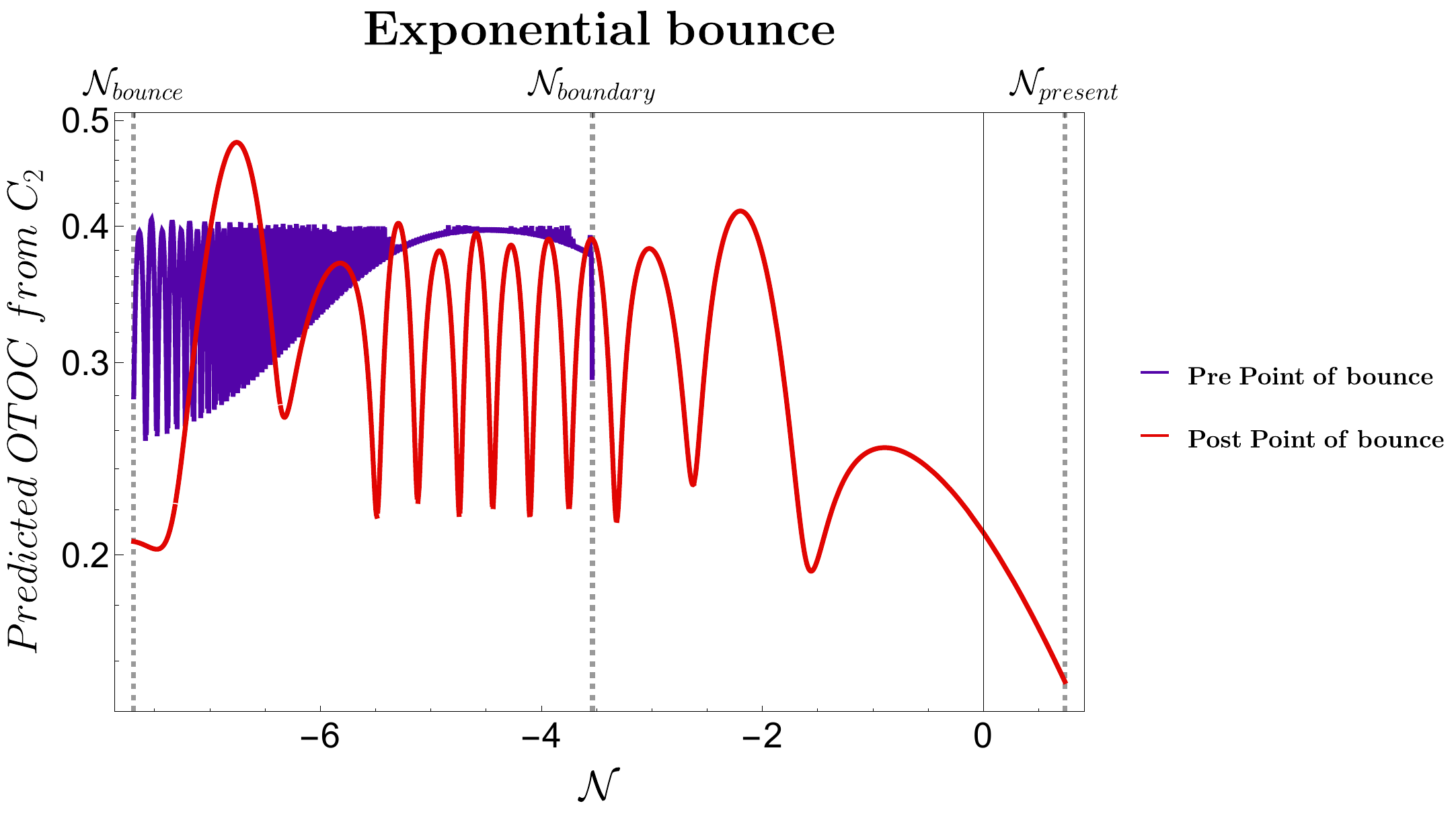}
	\caption{Predicted OTOC from geodesically weighted complexity inside bouncing region with respect to number of e-foldings}\label{fig:otocc2invzexp}
	
\end{figure}

\begin{figure}[!htb]
	\centering
	\includegraphics[width=15cm,height=9cm]{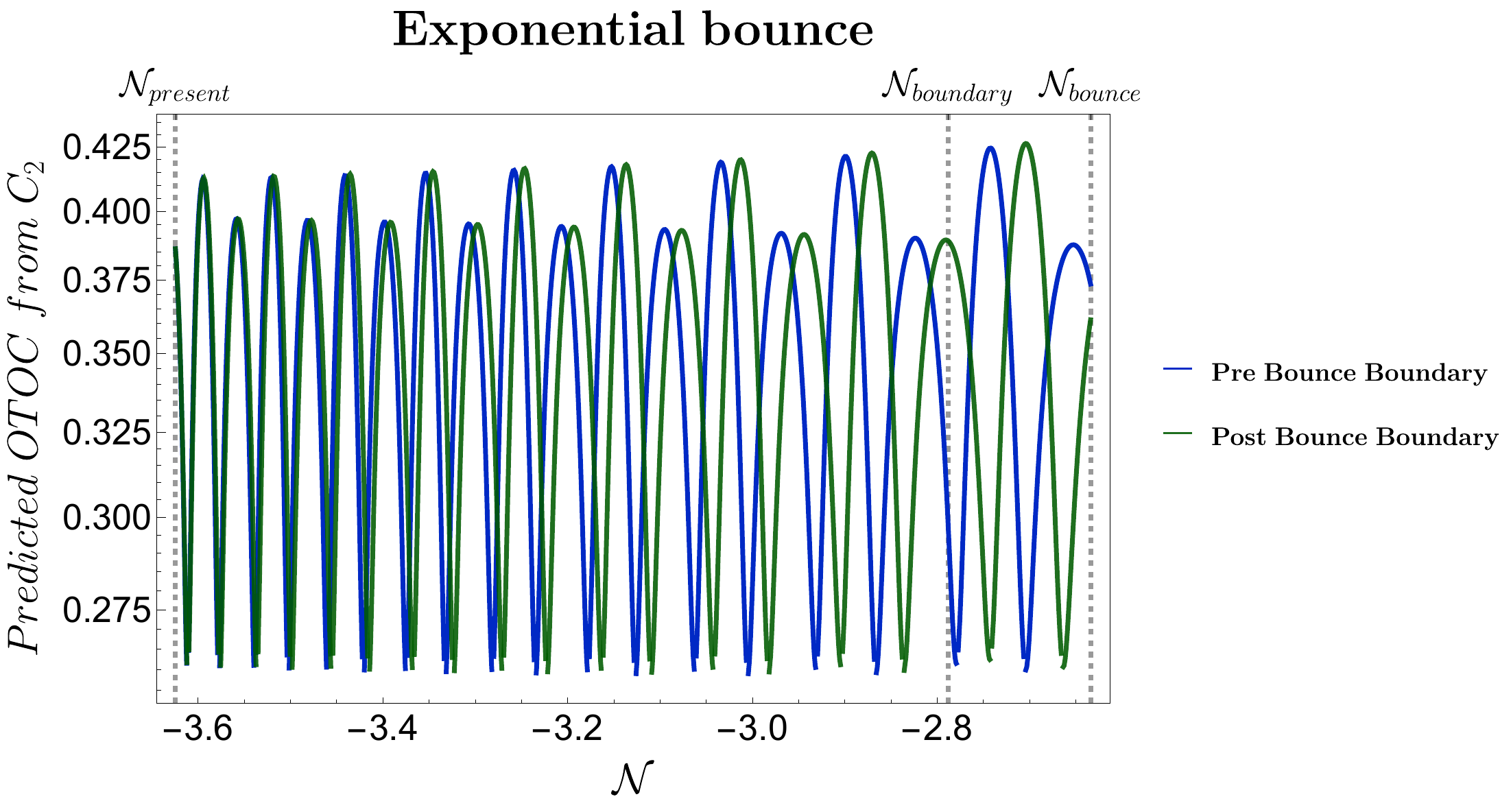}
	\caption{Predicted OTOC from geodesically weighted complexity outside bouncing region with respect to number of e-foldings}\label{fig:otocc2outvzexp} 
\end{figure}

\begin{itemize}
	\item In \Cref{fig:c2invzexp} we have plotted the geodesically weighted measure of complexity with respect to the number of e-folds for inside the bouncing region. We observe that the behaviour of $\mathcal{C}_2$ in this region is almost identical to the behaviour of the linearly weighted measure of complexity. Even the extrapolated behaviour of the post point of bounce is identical to that of $\mathcal{C}_1$.
	\item  We observe similar periodic behaviour in the complexity measure $\mathcal{C}_2$ as $\mathcal{C}_1$. However an important observation is that unlike $\mathcal{C}_1$, $\mathcal{C}_2$ has equal values for all regions even when extrapolated inside the bouncing region.
\end{itemize}

\begin{table}[h!]
	\centering
	\footnotesize
	\begin{tabular}{|||||m{1.5cm}||m{1.8cm}||m{2.2cm}||m{2.2cm}||m{3cm}||m{2.2cm}|||||}
		\hline \hline \hline \hline \hline
		& \textcolor{blue}{\bf Chaos and Complexity measure} & \textcolor{red}{\bf Pre-bounce boundary} & \textcolor{red}{\bf Pre-point of bounce} & \textcolor{red}{\bf Point point of bounce} & \textcolor{red}{\bf Post bounce boundary} \\  \hline \hline \hline \hline
		\bf{Model-I} & Complexity measures $\mathcal{C}_1$ and $\mathcal{C}_2$ & Periodic oscillations with decreased frequency near the boundary & Random oscillations near the point of bounce & Random oscillations near the point of bounce,\textbf{ rises near boundary, saturates at extrapolated present times}  &  Periodic oscillations with decreased frequency near the boundary \\ \hline \hline\hline
		& OTOC from $\mathcal{C}_1$ and $\mathcal{C}_2$ &  Periodic oscillations with decreased frequency near the boundary & Random oscillations near the point of bounce & Random oscillations near the point of bounce \textbf{starts to fall near boundary} till extrapolated present times & Periodic oscillations with decreased frequency near the boundary \\ \hline \hline \hline\hline \hline
		\bf{Model-II} & Complexity measures $\mathcal{C}_1$ and $\mathcal{C}_2$ & Well behaved, periodic oscillations with slight decrease in frequency near the boundary & Wild and random fluctuations near the point of bounce & Random fluctuations near the point of bounce,\textbf{ rises outside the boundary}, till the extrapolated present times, \textbf{no saturation observed}  &  Periodic oscillations with decreased frequency near the boundary \\ \hline \hline\hline
		& OTOC from $\mathcal{C}_1$ and $\mathcal{C}_2$ &  Periodic oscillations with slightly decrease in frequency near the boundary & Wild and random fluctuations near the point of bounce & Random fluctuations near the point of bounce \bf{starts to falls after the boundary} till extrapolated present times & Periodic oscillations with decreased frequency near the boundary \\ \hline \hline \hline\hline \hline
	\end{tabular} 
	\caption{Behaviour of the cosmological complexity measures and cosmological OTOC in different region of interest for the two models of bouncing paradigm.}
	\label{tab:observation}
\end{table} 

\Cref{tab:observation} contains all the key features of the \textbf{complexity measures} $\mathcal{C}_1$ and $\mathcal{C}_2$ and the \textbf{Out of time ordered correlation functions} predicted from them for both of the Bouncing cosmological models.
\textcolor{Sepia}{\section{\sffamily Conclusions}\label{sec:conclusion}}
From our study of the complexity measures computed from the linearly weighted and geodesically weighted cost functionals within the framework of bouncing cosmology we have the following final remarks:
\begin{itemize}
	\item \underline{\textcolor{red}{\bf Remark~I:}}\\
	The complexity measure calculated from two different types of cost functionals has an overall identical behaviour for both the models of bouncing cosmology under consideration with some noticeable differences. Though their behaviour is identical, it is evident from the plots that for a particular value of the scale factor, the linearly weighted complexity measure ($\mathcal{C}_1$), is always greater than the geodesically weighted one ($\mathcal{C}_2$). We see this feature in both the models.\\
	\item  \underline{\textcolor{red}{\bf Remark~II:}}\\
	We observe that the complexity measure calculated for \DB{\textbf{the post-point of bounce}} for both the models is the most interesting one. Though we observe random quantum fluctuations inside the bouncing region, once extrapolated to the present day we observe an exponential rise in the complexity measures followed by its saturation. An important point worth noting is that for the \textit{Cosine Hyperbolic bounce} model, the starting point of the exponential rise in the two complexity measures is observed inside the bouncing region itself and the saturation is well observed on extrapolation. However, the rise in the two complexities for the \textit{Exponential bounce} model is observed only after extrapolating it outside the bouncing region. We do not observe saturation as such in the complexity measures even on extrapolating it to the present day for the Exponential bounce model. We observe similar behaviour of the complexities when the analysis is done concerning the observationally important quantity known as the number of e-folds.
	\item  \underline{\textcolor{red}{\bf Remark~III:}}\\
	The behaviour of the complexity measure \HS{\textbf{outside the bouncing region}} is not of prime significance as we observe smooth, well behaved and periodic nature in those regions. Though the periodicity may not be equal near the boundary of the bouncing region and the present time, it does not give us any random or chaotic behaviour in those regions. However, the behaviour of complexity measure in the \HS{\textbf{pre-point of bounce region}} is of some significance because we observe random oscillations of the complexity in this region. For the exponential case, these oscillations are extremely wild and it can be attributed to quantum mechanical fluctuations.
	\item  \underline{\textcolor{red}{\bf Remark~IV:}}\\
	We observe an \HS{\textbf{exponential decay}} in the predicted OTOC's computed from the complexity measures $\mathcal{C}_1$ and $\mathcal{C}_2$, which is in accordance with the recently established predictions of OTOC's in the context of Cosmology\cite{Choudhury:2020yaa}. This behaviour of the OTOC's is observed not only with respect to the theoretical measure of scale factor but also with the respect to the observational measure of the number of e-folds.
	\item  \underline{\textcolor{red}{\bf Remark~V:}}\\
	The \DB{\textbf{Lyapunov exponent}} calculated from both complexities at the region of rising only differ in the second decimal place for both models. The fractional variation between the Lyapunov exponents is observed to be less than ten percent. We expect the variation to be much lesser if we consider higher-order terms in the dispersion relation for the numerical analysis.
	\item  \underline{\textcolor{red}{\bf Remark~VI:}}\\
	Since we have solved the same dynamical equations with respect to the Number of e-foldings, another calculation of Lyapunov exponent with respect to the number of e-foldings, is expected to have the same order and hence a separate calculation is redundant. 
	\item  \underline{\textcolor{red}{\bf Remark~VII:}}\\
	We get a theoretical prediction for the \DB{\textbf{lower bound of temperature}} for both models in the region of rising which falls from before the boundary till late times for the hyperbolic cosine case and completely outside the boundary and at much late times for the exponential case. 
	\item   \underline{\textcolor{red}{\bf Remark~VIII:}}\\
	The \HS{\textbf{choice of initial conditions}} at the horizon crossing that we have chosen for evaluating the perturbed action can have significant changes to the obtained results in the form of complexity, OTOC, and the Lyapunov exponent.
	\item   \underline{\textcolor{red}{\bf Remark~IX:}}\\
	The \DB{\textbf{scrambling time scale}} for both cases has been estimated. For the cosine hyperbolic case (estimated in the region of saturation) it turns out to be one-tenth of the conformal time period from bounce to present day, whereas for the exponential case the estimated scrambling time (in the region of rise), turns out to be more than one-half of the conformal bounce-to-present day time period, which might signify the lesser time required for OTOC to attain equilibrium in the cosine hyperbolic case (also pointing towards saturation of complexity), whereas a long time required for exponential OTOC to attain equilibrium hence hinting at a never saturating complexity value at late times.
	
\end{itemize}

The future prospects of the present work are appended below:
\begin{itemize}
	\item  \underline{\textcolor{red}{\bf Prospect~I:}}\\
	The framework for bouncing cosmology that we have mentioned along with the generalized perturbed action, dispersion relations, and the Hamiltonian can be used in a more general way without any truncation of higher-order terms for investigating cosmological complexity in any models. Though we have focused on the sub-Hubble region due to its quantum fluctuations, numerically one can use the full solution, rather approximating it in the sub-Hubble region. One can also apply the same analysis for the generic inflationary paradigm, which has not been considered in any work yet in an appropriate way.
	
	\item \underline{\textcolor{red}{\bf Prospect~II:}}\\
	We are working currently on applying the same framework to certain models in Island Cosmology as discussed in \cite{Hartman:2020khs,Chen:2020tes}. We will be planning to work out on this framework with different quantum initial conditions which will appear very soon in an upcoming paper. 
	
	\item \underline{\textcolor{red}{\bf Prospect~III:}}\\
	An interesting study would be to see how this complexity can be used to study the non equilibrium phenomenon and chaoticity in various entangled systems \cite{Choudhury:2017qyl,Choudhury:2017bou,Cao:2016mst,Cao:2017hrv,Choudhury:2016cso,Choudhury:2016pfr,Choudhury:2020ivj}. This measure might be used to see if the long-range correlation between systems induces chaoticity and quantum randomness or not.
	
	\item \underline{\textcolor{red}{\bf Prospect~IV:}}\\
	It is naturally expected that chaos and randomness might be an inherent property of Open quantum systems \cite{Akhtar:2019qdn,Bhattacherjee:2019eml,Banerjee:2020ljo} depending on the properties of the quantum dissipation and its impact. Complexity finds another use in this direction of research where one might be inclined to study chaos in nature from a more realistic point of view. 
	
	\item \underline{\textcolor{red}{\bf Prospect~V:}}\\
	It is also possible to find a general representation of squeezed state formalism for the multi-field interacting scenario in the context of cosmological perturbation theory written in a spatially flat FLRW background. But till now there are no concrete results available on this in the cosmology literature because the result is completely dependent on the type and strength of the interaction and extremely model dependent. For this reason, it is extremely difficult to deal with such types of computations within the framework of the quantum field theory of cosmological perturbation theory. But apart from having these mentioned difficulties, if one can write a general structure of the squeezed state formalism at least for two interacting fields in the spatially flat FLRW cosmological background by considering all possible general renormalizable coupling and interactions in the perturbative regime of the quantum field theory then it is possible to compute many physical observables out of those results. Using the interacting two field squeezed state formalism it is also good to understand the role of quantum mechanical chaos and complexity within the framework of cosmological perturbation theory. The future aim should be to carry forward such computations and explore some of the important unknown important underlying physical features of cosmological perturbation theory in presence of interacting quantum mechanical fluctuations.
	
	\item \underline{\textcolor{red}{\bf Prospect~VI:}}\\
	In this paper, we have restricted our analysis only for scalar mode quantum fluctuations generated from a cosmological perturbation in the spatially flat FLRW cosmological background. However, a similar analysis can be extended for the primordial gravitational waves appearing from the tensor mode fluctuations in the same cosmological background set up. It would be really nice and also important to check how the primordial gravitation waves and related tensor mode fluctuations get affected by two modes squeezed state formalism and also important to study how that will further put a stringent constraint on the phenomena of quantum mechanical chaos and complexity. 
	
	\item \underline{\textcolor{red}{\bf Prospect~VII:}}\\
	Generally, people try to explain this concept of chaos and complexity through various models. However a completely model-independent notion of complexity can be given from the perspective of effective field theory\cite{Choudhury:2017glj,Naskar:2017ekm}. It is in general possible to start from the single EFT action and derive all the models under various constraints satisfied by the parameters of the action. Squeezed state formalism for such a universal action can be developed to generalize an give a model-independent prescription of complexity.
\end{itemize}

\subsection*{Note:}
The background Gaussian image in the table of contents is inspired from \cite{Green:2020whw}.
	\subsection*{Acknowledgements}
	SC is thankful to  Robert Myers, Daniel Green, Sumit Ranjan Das, Igor R. Klebanov, Eva Silverstein, Leonardo Senatore, Subhashish Banerjee, Anupam Mazumdar, Savan Kharel for enormous helpful discussions, suggestions and support. SC would like to specially thank his very good friend Vincent Vennin for providing constant motivation, unconditional support and various constructive suggestions in many contexts. The Post-Doctoral research fellowship of SC is supported by the ERC Consolidator grant 772295 ``Qosmology" of Professor Jean-Luc Lehners. Also SC take this opportunity to thank sincerely to
Jean-Luc Lehners for his constant support through the funding in the post-doctoral fellowship and providing huge inspiration throughout his stay at Max Planck Institute for Gravitational Physics (Albert Einstein Institute), Potsdam, Germany. SC thank 
 Latham Boyle for inviting at Perimeter Institute for Theoretical Physics (PITP), Zohar Komargodski for inviting at Simons Center for Geometry and Physics (SCGP), Stony Brook University, Leonardo Senatore  for inviting at  Institute for Theoretical Physics, Stanford University, Juan Martín Maldacena for inviting at  Workshop on Qubits and Spacetime, Institute for Advanced Studies (IAS), Princeton, Paul Joseph Steinhardt  for inviting at  Department of Physics, Princeton University, Martin Bojowald  for inviting at 
The Institute for Gravitation and the Cosmos (IGC),  Department of Physics, Eberly College of Science, Pennsylvania State University (University Park campus), Sudhakar Panda  for inviting at School of Physical Sciences, National Institute of Science Education and Research (NISER), Bhubaneswar, Abhishek Chowdhury  for inviting at Department of Physics, Indian Institute of Technology (IIT), Bhubaneswar, Anjan Sarkar  for inviting at Department of Astrophysics, Raman Research Institute, Bengaluru, Aninda Sinha and Banibrata Mukhopadhyay for inviting at Center for High Energy Physics (CHEP) and Department of Astronomy and Astrophysics, Indian Institute of Science, Bengaluru, Uma Shankar for inviting at Department of Physics, Indian Institute of Technology (IIT), Bombay, Shiraz Minwalla for inviting at Department of Theoretical Physics, Tata Institute of Fundamental Research, Mumbai, Abhishek Mahapatra for inviting at National Institute of Technology (NIT), Rourkela, for official academic visit. Part of this work was presented as a talk, titled "Cosmology from Condensed Matter Physics: A study of out-of-equilibrium physics" and " Cosmology Meets Condensed Matter Physics" at Perimeter Institute for Theoretical Physics (PITP) (See the link: \textcolor{red}{http://pirsa.org/19110117/}), Simons Center for Geometry and Physics (SCGP), Stony Brook University (See the link: \textcolor{red}{http://scgp.stonybrook.edu/video portal/video.php?id=4358}), Department of Physics, Princeton University, The Institute for Gravitation and the Cosmos (IGC),  Department of Physics, Eberly College of Science, Pennsylvania State University (University Park campus), workshop on "Advances in Astroparticle Physics and Cosmology, AAPCOS-2020" at Saha Institute of Nuclear Physics, Kolkata on the occasion of the 100 years of Saha Ionisation Equation by Prof. Meghnad Saha, Department of Physics, Scottish Church College, Kolkata, School of Physical Sciences, National Insitute of Science Education and Research (NISER), Bhubaneswar, Department of Physics, Indian Institute of Technology (IIT), Bhubaneswar, Department of Astrophysics, Raman Research Institute, Bengaluru, Center for High Energy Physics (CHEP), Indian Institute of Science, Bengaluru, Department of Physics, Indian Institute of Technology (IIT), Bombay, National Institute of Technology (NIT), Rourkela. SC would like to thank Quantum Gravity and Unified Theory and Theoretical Cosmology
Group, Max Planck Institute for Gravitational Physics, Albert Einstein Institute (AEI), Perimeter Institute for Theoretical Physics (PITP), Simons Center for Geometry and Physics (SCGP), Stony Brook University, Institute for Theoretical Physics, Stanford University, 
The Institute for Gravitation and the Cosmos (IGC),  Department of Physics, Eberly College of Science, Pennsylvania State University (University Park campus), School of Physical Sciences, National Insitute of Science Education and Research (NISER), Bhubaneswar, Department of Astrophysics, Raman Research Institute, Bengaluru, Department of Physics, Indian Institute of Technology (IIT), Bombay, Quantum Space-time Group (Earlier known as String Theory and Mathematical Physics Group), Department of Theoretical Physics, Tata Institute of Fundamental Research, Mumbai and National Institute of Technology (NIT), Rourkela for providing financial support for the academic visits at Canada, U.S.A. and India. SC would like to thank the natural beauty of Prague, Dresden, Hamburg, Leipzig, Potsdam, Berlin, Mumbai, Bangalore, Kolkata, Bhubaneswar which inspires to do work very hard during the weekend trips and various academic visits. Particularly SC want to give a separate credit to all the members of the EINSTEIN KAFFEE Berlin Alexanderplatz for providing work friendly environment, good espresso shots, delicious chocolate and caramel cakes and cookies, which helped to write the most of the part of the paper in that coffee shop in the last few months. SC also thank all the members of our newly formed virtual international non-profit consortium ``Quantum Structures of the Space-Time \& Matter" (QASTM) for elaborative discussions. SC also would like to thank all the speakers of QASTM zoominar series from different parts of the world  (For the uploaded YouTube link look at: \textcolor{red}{https://www.youtube.com/playlist?list=PLzW8AJcryManrTsG-4U4z9ip1J1dWoNgd}) for supporting my research forum by giving outstanding lectures and their valuable time during this COVID pandemic time. Parth Bhargava, Satyaki Chowdhury, Anurag Mishra, Sachin Panner Selvam, Gabriel D. Pasquino would like to thank  RWTH Aachen University, NISER Bhubaneswar, NIT Rourkela, BITS Hyderabad, University of Waterloo respectively for providing support. SP acknowledges
the J. C. Bose National Fellowship for support of his research. Last but not the least, we would like to acknowledge our debt to 
the people belonging to the various part of the world for their generous and steady support for research in natural sciences. 
	
\clearpage
	\appendix
	\textcolor{Sepia}{\section{\sffamily Quantization of Hamiltonian for scalar modes in terms of squeezed parameters in cosmological perturbation theory }\label{sec:appendixAA}}	
	
	In this appendix, we present the details of the quantization of the Hamiltonian for scalar modes obtained from the cosmological perturbation theory within the framework of bouncing cosmology. To serve this purpose, let us start with the expression for the classical Hamiltonian function already derived earlier in the paper and it is represented by:
\bea H(\tau)=\int d^3{\bf k}~\Biggl[\frac{1}{2}\left|v'_{\bf k}(\tau)\right|^2+\frac{1}{2}\mu^2(k,\tau)|v_{\bf k}(\tau)|^2\Biggr],\eea
where the conformal time dependent mass $\mu^2(k,\tau)$ of the parametric oscillator is given by the following expression:
\bea \mu^2(k,\tau):=\Biggl[k^2-\lambda^2_{\bf k}(\tau)\Biggr]~~~{\rm where~we~define}~~~~\lambda_{\bf k}(\tau):=\Biggl(\frac{z'(\tau)}{z(\tau)}\Biggr).\eea 
Also one can express the field velocity with respect to the canonically conjugate momentum density in the Fourier space as:
\bea v'_{\bf k}(\tau):=\pi_{\bf k}(\tau)+\lambda_{\bf k}(\tau)v_{\bf k}(\tau).\eea
Next, using the classical mode function we can further construct the quantum mechanical operators:
\bea \hat{v}_{\bf k}(\tau)&=&\left[v^{*}_{-{\bf k}}(\tau)~\hat{a}_{\bf k}+v_{{\bf k}}(\tau)~\hat{a}^{\dagger}_{-{\bf k}}\right],\\
\hat{\pi}_{\bf k}(\tau)&=&
\left[\pi^{*}_{-{\bf k}}(\tau)~\hat{a}_{\bf k}+\pi_{{\bf k}}(\tau)~\hat{a}^{\dagger}_{-{\bf k}}\right].\eea
Now we evaluate the following quantities:
\bea \left|v'_{\bf k}(\tau)\right|^2&=&\left|\left[\pi^{*}_{-{\bf k}}(\tau)~\hat{a}_{\bf k}+\pi^{}_{{\bf k}}(\tau)~\hat{a}^{\dagger}_{-{\bf k}}\right]+\lambda_{\bf k}(\tau)\left[v^{*}_{-{\bf k}}(\tau)~\hat{a}_{\bf k}+v_{{\bf k}}(\tau)~\hat{a}^{\dagger}_{-{\bf k}}\right]\right|^2\nonumber\\
&=&\left|\left[\pi^{*}_{-{\bf k}}(\tau)~\hat{a}_{\bf k}+\pi^{}_{{\bf k}}(\tau)~\hat{a}^{\dagger}_{-{\bf k}}\right]\right|^2+\lambda^2_{\bf k}(\tau)\left|\left[v^{*}_{-{\bf k}}(\tau)~\hat{a}_{\bf k}+v_{{\bf k}}(\tau)~\hat{a}^{\dagger}_{-{\bf k}}\right]\right|^2\nonumber\\
&&~~~~~~+\lambda_{\bf k}(\tau)\left[\pi^{*}_{-{\bf k}}(\tau)~\hat{a}_{\bf k}+\pi_{{\bf k}}(\tau)~\hat{a}^{\dagger}_{-{\bf k}}\right]^{\dagger}\left[v^{*}_{-{\bf k}}(\tau)~\hat{a}_{\bf k}+v_{{\bf k}}(\tau)~\hat{a}^{\dagger}_{-{\bf k}}\right]\nonumber\\
&&~~~~~~+\lambda_{\bf k}(\tau)\left[v^{*}_{-{\bf k}}(\tau)~\hat{a}_{\bf k}+v_{{\bf k}}(\tau)~\hat{a}^{\dagger}_{-{\bf k}}\right]^{\dagger}\left[\pi^{*}_{-{\bf k}}(\tau)~\hat{a}_{\bf k}+\pi_{{\bf k}}(\tau)~\hat{a}^{\dagger}_{-{\bf k}}\right]\nonumber\\
&=&\Biggl\{\left|\pi_{{\bf k}}(\tau)\right|^2+\lambda^2_{\bf k}(\tau)\left|v_{{\bf k}}(\tau)\right|^2+\lambda_{\bf k}(\tau)\left(\pi^{*}_{\bf k}(\tau)v_{\bf k}(\tau)+v^{*}_{\bf k}(\tau)\pi_{\bf k}(\tau)\right)\Biggr\}\left(\hat{a}^{\dagger}_{\bf k}\hat{a}_{\bf k}+\hat{a}^{\dagger}_{-{\bf k}}\hat{a}_{-{\bf k}}+1\right)\nonumber\\
&&+\lambda_{\bf k}(\tau)\left(\pi^{*}_{\bf k}(\tau)v^{*}_{-{\bf k}}(\tau)~\hat{a}_{\bf k}\hat{a}_{-{\bf k}}+\pi_{-{\bf k}}(\tau)v_{\bf k}(\tau)~\hat{a}^{\dagger}_{\bf k}\hat{a}^{\dagger}_{-{\bf k}}\right)\nonumber\\.\eea  
and 
\bea \mu^2(k,\tau)|v_{\bf k}(\tau)|^2&=&\mu^2(k,\tau)\left|\left[v^{*}_{-{\bf k}}(\tau)~\hat{a}_{\bf k}+v_{{\bf k}}(\tau)~\hat{a}^{\dagger}_{-{\bf k}}\right]\right|^2\nonumber\\
&=&\left(k^2-\lambda^2_{\bf k}(\tau)\right)\left|v_{{\bf k}}(\tau)\right|^2\left(\hat{a}^{\dagger}_{\bf k}a_{\bf k}+\hat{a}^{\dagger}_{-{\bf k}}\hat{a}_{-{\bf k}}+1\right)\eea 
Now using the above mentioned quantum operator one can finally express the canonical Hamiltonian for the parametric oscillator in the following quantized form:
\bea \widehat{H}(\tau)&=&\frac{1}{2}\int d^3{\bf k}\Biggl[~\underbrace{\Omega_{\bf k}(\tau)\left(\hat{a}^{\dagger}_{\bf k}\hat{a}_{\bf k}+\hat{a}^{\dagger}_{-{\bf k}}\hat{a}_{-{\bf k}}+1\right)}_{\textcolor{red}{\bf Contribution~from~the~free~term}}\nonumber\\
&&~~~~~~~~~~~~~~~~~~~~~~~~+i\underbrace{\lambda_{\bf k}(\tau)\Biggl(\exp(-2i\phi_{\bf k}(\tau))\hat{a}_{\bf k}\hat{a}_{-{\bf k}}-\exp(2i\phi_{\bf k}(\tau))\hat{a}^{\dagger}_{\bf k}\hat{a}^{\dagger}_{-{\bf k}}\Biggr)}_{\textcolor{red}{\bf Contribution~ from~ the~ Interaction~ term}}~\Biggr],~~~~~~~~~~~~~~\eea
where we define $\Omega_{\bf k}(\tau)$ and $\phi_{\bf k}(\tau)$ by the following expressions:
\bea \Omega_{\bf k}(\tau):&=&\Biggl\{\left|v^{'}_{{\bf k}}(\tau)\right|^2+\mu^2(k,\tau)\left|v_{{\bf k}}(\tau)\right|^2\Biggr\},\\
i\exp(-2i\phi_{\bf k}(\tau)):&=&\pi^{*}_{\bf k}(\tau)v^{*}_{-{\bf k}}(\tau).\eea
Here $\Omega_{\bf k}(\tau)$ represents the conformal time dependent dispersion relation in the present bouncing cosmological set-up and $\phi_{\bf k}(\tau)$ is squeezing angle appearing in the squeezed state formalism as discussed earlier in the text.

	\textcolor{Sepia}{\section{\sffamily Hamilton's equations in the Heisenberg picture in cosmological perturbation for scalar modes}\label{sec:appendixA}}	
	Next, using the previously mentioned solution of classical mode function we can further construct the quantum mechanical operators in the Heisenberg picture:
\bea \hat{v}({\bf x},\tau)&=&{\cal U}^{\dagger}(\tau,\tau_0)\hat{v}({\bf x},\tau_0){\cal U}(\tau,\tau_0)\nonumber\\
&=&\int \frac{d^3{\bf k}}{(2\pi)^3}~\left[v^{*}_{-{\bf k}}(\tau)~\hat{a}_{\bf k}+v_{{\bf k}}(\tau)~\hat{a}^{\dagger}_{-{\bf k}}\right]~\exp(i{\bf k}.{\bf x}),\\
\hat{\pi}({\bf x},\tau)&=&{\cal U}^{\dagger}(\tau,\tau_0)\hat{\pi}({\bf x},\tau_0){\cal U}(\tau,\tau_0)\nonumber\\
&=&\int \frac{d^3{\bf k}}{(2\pi)^3}~\left[\pi^{*}_{-{\bf k}}(\tau)~\hat{a}_{\bf k}+\pi_{{\bf k}}(\tau)~\hat{a}^{\dagger}_{-{\bf k}}\right]~\exp(i{\bf k}.{\bf x}).\eea 
Now, our objective is to find out the fact that whether the mode functions for both the field variables and their associated momenta in Fourier space satisfy the well known, {\it Hamilton equations} or not. In the Heisenberg picture one can write down the following equations~\footnote{Here we have explicitly used the following operator identity which is valid in the Heisenberg quantum mechanical picture:
\begin{eqnarray}
\frac{\partial \hat{A}(\tau)}{\partial \tau}=\hat{A}^{'}(\tau)=-i\left[\hat{A}(\tau),\hat{H}(\tau)\right].
\end{eqnarray}}
\begin{eqnarray}
\label{H1}\hat{v}^{'}_{{\bf k}}(\tau)&=&-i\left[\hat{v}_{{\bf k}}(\tau),\hat{H}_{\bf k}(\tau)\right],\\
\label{H2}\hat{\pi}^{'}_{{\bf k}}(\tau)&=&-i\left[\hat{\pi}_{{\bf k}}(\tau),\hat{H}_{\bf k}(\tau)\right],
\end{eqnarray}
where the field operator and the corresponding canonically conjugate momentum operator can be expressed in terms of the creation and annihilation operators of conformal time dependent parametric oscillator as:
\bea \hat{v}_{{\bf k}}(\tau)&=&\left[v^{*}_{-{\bf k}}(\tau)~\hat{a}_{\bf k}+v_{{\bf k}}(\tau)~\hat{a}^{\dagger}_{-{\bf k}}\right],\\
\hat{\pi}_{{\bf k}}(\tau)&=&\left[\pi^{*}_{-{\bf k}}(\tau)~\hat{a}_{\bf k}+\pi_{{\bf k}}(\tau)~\hat{a}^{\dagger}_{-{\bf k}}\right].\eea

Also the Hamiltonian operator in Fourier space can be expressed explicitly in terms of creation and annihilation operators as:
\bea \hat{H}_{\bf k}(\tau)&=&\Biggl[\frac{1}{2}\left|\left[v^{*'}_{-{\bf k}}(\tau)~\hat{a}_{\bf k}+v^{'}_{{\bf k}}(\tau)~\hat{a}^{\dagger}_{-{\bf k}}\right]+\frac{z'(\tau)}{z(\tau)}\left[v^{*}_{-{\bf k}}(\tau)~\hat{a}_{\bf k}+v_{{\bf k}}(\tau)~\hat{a}^{\dagger}_{-{\bf k}}\right]\right|^2 \nonumber\\
&&~~~~~~~~~~~~~~~~~~~~~~~~~~~~~~~~~~~~~~~~~~~~+\frac{1}{2}\mu^2(k,\tau)|\left[v^{*}_{-{\bf k}}(\tau)~\hat{a}_{\bf k}+v_{{\bf k}}(\tau)~\hat{a}^{\dagger}_{-{\bf k}}\right]|^2\Biggr]\nonumber\\
&=&\frac{1}{2}\Biggl[~\underbrace{\Omega_{\bf k}(\tau)\left(\hat{a}^{\dagger}_{\bf k}\hat{a}_{\bf k}+\hat{a}^{\dagger}_{-{\bf k}}\hat{a}_{-{\bf k}}+1\right)}_{\textcolor{red}{\bf Contribution~from~the~free~term}}\nonumber\\
&&~~~~~~~~~~~~~~~~~~~~~~~~+i\underbrace{\lambda_{\bf k}(\tau)\Biggl(\exp(-2i\phi_{\bf k}(\tau))\hat{a}_{\bf k}\hat{a}_{-{\bf k}}-\exp(2i\phi_{\bf k}(\tau))\hat{a}^{\dagger}_{\bf k}\hat{a}^{\dagger}_{-{\bf k}}\Biggr)}_{\textcolor{red}{\bf Contribution~ from~ the~ Interaction~ term}}~\Biggr],~~~~~~~~~~~~~~\eea
where we define the dispersion relation $\Omega_{\bf k}(\tau)$ and $\lambda_{\bf k}(\tau)$ by the following expressions:
\bea \Omega_{\bf k}(\tau):&=&\Biggl\{\left|v^{'}_{{\bf k}}(\tau)\right|^2+\mu^2(k,\tau)\left|v_{{\bf k}}(\tau)\right|^2\Biggr\},\\
\lambda_{\bf k}(\tau):&=&\Biggl(\frac{z'(\tau)}{z(\tau)}\Biggr),~~~{\rm where} ~~z(\tau)=a\sqrt{2\epsilon(\tau)}~~~{\rm with}~~\epsilon(\tau)=\left(1-\frac{{\cal H}'}{{\cal H}^2}\right).\eea

Further substituting all of the above mentioned  expressions in Eq~(\ref{H1}) and Eq~(\ref{H2}), we get the following result:
\begin{eqnarray}
\label{H1}v^{*'}_{-{\bf k}}(\tau)~\hat{a}_{\bf k}+v^{'}_{{\bf k}}(\tau)~\hat{a}^{\dagger}_{-{\bf k}}&=&-i\left[v^{*}_{-{\bf k}}(\tau)~\hat{a}_{\bf k}+v_{{\bf k}}(\tau)~\hat{a}^{\dagger}_{-{\bf k}},\hat{H}_{\bf k}(\tau)\right],\\
\label{H2}\pi^{*'}_{-{\bf k}}(\tau)~\hat{a}_{\bf k}+\pi^{'}_{{\bf k}}(\tau)~\hat{a}^{\dagger}_{-{\bf k}}&=&-i\left[\pi^{*}_{-{\bf k}}(\tau)~\hat{a}_{\bf k}+\pi_{{\bf k}}(\tau)~\hat{a}^{\dagger}_{-{\bf k}},\hat{H}_{\bf k}(\tau)\right],
\end{eqnarray}
After doing considerable amount of algebraic manipulations we finally get the following simplified form of the {\it Hamilton equations} associated with the cosmological perturbation theory of scalar mode fluctuation:
\bea \label{g1}\left(\frac{d}{d\tau}-\lambda_{\bf k}(\tau)\right)v_{\bf k}(\tau)&=&\pi_{\bf k}(\tau),\\
\label{g2} \left(\frac{d}{d\tau}+\lambda_{\bf k}(\tau)\right)\pi_{\bf k}(\tau)&=&-\Omega^2_{\bf k}(\tau)v_{\bf k}(\tau).
\eea
Further using Eq~(\ref{g1}) in Eq~(\ref{g2}), we get:
\bea  \left(\frac{d}{d\tau}+\lambda_{\bf k}(\tau)\right)\left(\frac{d}{d\tau}-\lambda_{\bf k}(\tau)\right)v_{\bf k}(\tau)&=&-\Omega^2_{\bf k}(\tau)v_{\bf k}(\tau)\nonumber\\
\Longrightarrow\frac{d^2v_{\bf k}(\tau)}{d\tau^2}+\left(\Omega^2_{\bf k}(\tau)-\lambda^2_{\bf k}(\tau)\right)v_{\bf k}(\tau)=0&&.\eea
Now we can write:
\bea \lambda^2_{\bf k}(\tau)=\Biggl(\frac{z''(\tau)}{z(\tau)}\Biggr)-\lambda'_{\bf k}(\tau)\approx \Biggl(\frac{z''(\tau)}{z(\tau)}\Biggr),\eea
and finally we define:
\bea \mu^2(k,\tau)=\left(\Omega^2_{\bf k}(\tau)-\lambda^2_{\bf k}(\tau)\right)=\Biggl[\Omega^2_{\bf k}(\tau)- \Biggl(\frac{z''(\tau)}{z(\tau)}\Biggr)\Biggr].\eea
As a result we get the following simplified form of the equation of motion:
\bea \frac{d^2v_{\bf k}(\tau)}{d\tau^2}+\mu^2(k,\tau)v_{\bf k}(\tau)=0,\eea
which is the generalized version of the well known {\it Mukhanov Sasaki equation}. In the sub-Hubble region ($-k\tau\gg 1$) one can simplify the expression for the dispersion relation, $\Omega_{\bf k}(\tau)$, which is explicitly discussed in the next section. Now considering the leading order contribution we get the following expression for the conformal time dependent frequency parameter:
\bea \mu^2(k,\tau)\approx \Biggl[k^2- \Biggl(\frac{z''(\tau)}{z(\tau)}\Biggr)\Biggr],\eea
which is exactly appearing in the {\it Mukhanov Sasaki equation}.

\textcolor{Sepia}{\section{\sffamily Dispersion relation in terms of squeezed parameters}\label{sec:appendixB}}	
In this appendix, our prime objective is to derive the expression for the dispersion relation in terms of the squeezed parameter $r_{\bf k}(\tau)$ and the squeezed angle $\phi_{\bf k}(\tau)$, where the dispersion relation appears in the Hamiltonian after quantization that we studied in the paper explicitly.

Let us first write down the expression for the conformal time dependent dispersion relation $\Omega_{\bf k}$ in terms of the canonical field variable and its associated canonically conjugate momentum that appears after performing the cosmological perturbation theory for a single scalar field: 

\bea \label{eq:Sigma_k} \Omega_{\bf k}(\tau):&=&\Biggl\{\left|v^{'}_{{\bf k}}(\tau)\right|^2+\mu^2(k,\tau)\left|v_{{\bf k}}(\tau)\right|^2\Biggr\}\nonumber\\
&=&\Biggl\{\left|\pi_{{\bf k}}(\tau) + \lambda_{\bf k}(\tau)v_{\bf k}(\tau)\right|^2+\left(k^2  - \lambda^2_{\bf k}(\tau)\right)\left|v_{{\bf k}}(\tau)\right|^2\Biggr\} \\
&=&\Biggl\{\left|\pi_{{\bf k}}(\tau)\right|^2+ k^2 \left|v_{{\bf k}}(\tau)\right|^2 +\lambda_{\bf k}(\tau)\Biggl(\pi^*_{{\bf k}}(\tau)v_{{\bf k}}(\tau) + v^*_{{\bf k}}(\tau)\pi_{{\bf k}}(\tau)\Biggr)\Biggr\} , \nonumber\eea

Now, we plug in the expressions for $\pi_{{\bf k}}(\tau)$ and $v_{{\bf k}}(\tau)$, which are reproduced here for convenience : 

\bea 
v_{{\bf k}}(\tau)&=&v_{\bf k}(\tau_0)\Biggl(\cosh r_{\bf k}(\tau)~\exp(i\theta_{\bf k}(\tau))-\sinh r_{\bf k}(\tau)~\exp(i(\theta_{\bf k}(\tau)+2\phi_{\bf k}(\tau)))\Biggr),~~~~~~~~\\
\pi_{{\bf k}}(\tau)&=&\pi_{\bf k}(\tau_0)\Biggl(\cosh r_{\bf k}(\tau)~\exp(i\theta_{\bf k}(\tau))+\sinh r_{\bf k}(\tau)~\exp(i(\theta_{\bf k}(\tau)+2\phi_{\bf k}(\tau)))\Biggr),\eea 
and after doing a bit of algebraic manipulation we finally get:
 
 \bea
 \begin{split}\label{eq:2}
    \Omega_{\bf k}(\tau) ={}& \Biggl(\left|\pi_{{\bf k}}(\tau_0)\right|^2+ k^2 \left|v_{{\bf k}}(\tau_0)\right|^2\Biggr) \Biggl( \cosh^2 r_{\bf k}(\tau) + \sinh^2 r_{\bf k}(\tau)\Biggr)\\
         & + \sinh r_{\bf k}(\tau)\cdot \cos2\phi_{\bf k}(\tau)\Biggl(\left|\pi_{{\bf k}}(\tau_0)\right|^2 - k^2 \left|v_{{\bf k}}(\tau_0)\right|^2\Biggr)\\
         & + \lambda_{\bf k}(\tau)\Biggl\{\Biggl(\pi^*_{{\bf k}}(\tau_0)v_{{\bf k}}(\tau_0) + v^*_{{\bf k}}(\tau_0)\pi_{{\bf k}}(\tau_0)\Biggr) \\
         & ~~~~~~~~~~~~~~~~+ i \sinh 2r_{\bf k}(\tau)\sin2\phi_{\bf k}(\tau)\Biggl( \pi^*_{{\bf k}}(\tau_0)v_{{\bf k}}(\tau_0) - v^*_{{\bf k}}(\tau_0)\pi_{{\bf k}}(\tau_0) \Biggr) \Biggr\}.
\end{split} \eea

Here we have chosen the initial condition at the time scale $\tau=\tau_0$ by considering the horizon crossing scale, $-k\tau_0=1$. We impose this condition on the perturbation field variable and on the canonically conjugate momentum obtained for scalar fluctuation. We finally get:
\bea v_{\bf k}(\tau_0)&=&\frac{1}{\sqrt{2k}}~2^{\nu_{\rm B}-1}\left|\frac{\Gamma(\nu_{\rm B})}{\Gamma\left(\frac{3}{2}\right)}\right|~\exp\left(-i\left\{\frac{\pi}{2}\left(\nu_{\rm B}-2\right)-1\right\}\right),\\
\pi_{\bf k}(\tau_0)&=&i\sqrt{\frac{k}{2}}~2^{\nu_{\rm B}-\frac{3}{2}}\left|\frac{\Gamma(\nu_{\rm B})}{\Gamma\left(\frac{3}{2}\right)}\right|~\exp\left(-i\left\{\frac{\pi}{2}\left(\nu_{\rm B}-2\right)-1\right\}\right)~\nonumber\\
 &&~~~~~~~~~~~~~~\left[1-\sqrt{2}\frac{\displaystyle \left(\nu_{\rm B}-\frac{1}{2}\right)\left(\nu_{\rm B}+\frac{1}{2}+i\right)}{\displaystyle \left(\nu_{\rm B}+\frac{1}{2}\right)}\exp\left(-\frac{i\pi}{4}\right)\right].~\eea 
 
 Neglecting the phase factors in the above equation and also noting that $\nu_B \approx \frac{1}{2}+\cdots$, we get a pretty simplified expression for $\Omega_{\bf k}(\tau)$, i.e.
 \bea
 \begin{split}\label{eq:2}
    \Omega_{\bf k}(\tau) ={}& 2^{2\nu_{\rm B}-2}\left|\frac{\Gamma(\nu_{\rm B})}{\Gamma\left(\frac{3}{2}\right)}\right|^2 \Biggl[\frac{3k}{4} \Biggl( \cosh^2 r_{\bf k}(\tau) + \sinh^2 r_{\bf k}(\tau) \Biggr) - \frac{k}{4} \sinh r_{\bf k}(\tau)\cos2\phi_{\bf k}(\tau)\\
    & ~~~~~~~~~~~~~~~~~~~~~~~~~~~~~~~~~~~~~~~~~~~~~~~-\frac{1}{\sqrt 2} \lambda_{\bf k}(\tau)~\sinh 2r_{\bf k}(\tau)\sin2\phi_{\bf k}(\tau)\Biggr].
\end{split} \eea
 
 \textcolor{Sepia}{\subsection{\sffamily Sub-Hubble limiting result}\label{sec:subappendix}}
 In the sub-Hubble limit, $-k\tau\gg 1$, it is expected to have very small contribution from the squeezed parameter, $r_{\bf k}(\tau)$ for which one can use the following approximations:
 \bea \cosh r_{\bf k}(\tau)\approx 1,~~~~~\sinh r_{\bf k}(\tau)\approx r_{\bf k}(\tau).\eea
Consequently, in the limit $r_{\bf k}(\tau)\rightarrow 0$, we get the following result for the dispersion relation in the sub-Hubble region:
 \bea
 \begin{split}\label{eq:2sh}
    \Omega^{\rm Sub}_{\bf k}(\tau) ={}&\frac{3k}{4} 2^{2\nu_{\rm B}-2}\left|\frac{\Gamma(\nu_{\rm B})}{\Gamma\left(\frac{3}{2}\right)}\right|^2 =3k~2^{2(\nu_{\rm B}-2)}~\left|\frac{\Gamma(\nu_{\rm B})}{\Gamma\left(\frac{3}{2}\right)}\right|^2,
\end{split} \eea
which is basically dependent on the co-moving wave number and a very slowly varying time dependent quantity $\nu_{\rm B}$ at the sub-Hubble scale. In the previous ref.~\cite{Bhattacharyya:2020rpy,Bhattacharyya:2020kgu} the authors have not considered this additional slowly varying time dependence appearing from the parameter $\nu_{\rm B}$, which is not appropriate if we want to extract the information regarding quantum correlation in the out-of-equilibrium phase where random fluctuations play significant role. Now we will explicitly show how the slow time dependence is appearing in the parameter $\nu_{\rm B}$. In the sub-Hubble region the conformal time dependent mass parameter $\nu_{\rm B}$ can be approximately written by considering the contribution upto the next-to-leading order as:
 \bea \nu_{\rm B}\approx\left(\frac{1}{2}+\frac{{\cal H}''}{{\cal H}^2}+\cdots\right),\eea
 where we have neglected the contributions of all higher order small correction terms appearing as $\cdots$ for the computational simplicity. But out of all the terms in the correction part, ${\cal H}''/{\cal H}^2$ term gives the most significant contribution. Due to slowly varying feature with respect to the conformal time, neglecting this term is not physically justifiable.~\footnote{ In the previous works, where people did the analysis for inflation, this contribution was dropped as they have taken exact de Sitter solution, which is for inflation represented by the mass parameter, $\nu_{\rm  B}=3/2$. But as we know if we do that then one cannot stop inflation, to stop inflation at a specific point in the field space one need to include the contribution of slow-roll correction terms which serves the purpose. One can explicitly show that this contribution for inflation is given by, \bea \nu_{\rm B}=\frac{3}{2}+\epsilon+\cdots=\frac{3}{2}+\left(1-\frac{{\cal H}'}{{\cal H}^2}\right)+\cdots=\frac{5}{2}-\frac{{\cal H}'}{{\cal H}^2}+\cdots\eea
 where the first term represent the exact de Sitter solution and the second term represents the amount of deviation from that which is required to stop the inflation.} It needs to be incorporated to stop the bouncing phase and go either to the post-bounce or to the pre-bounce region in the field space. So for the bouncing cosmological paradigm the contribution ${\cal H}''/{\cal H}^2$ is explicitly needed to stop bounce and go to the next phase in the evolution. Now after substituting the above mentioned expression for the mass parameter $\nu_{\rm B}$ one can further write the following simplified form of the factor, $\Omega_{\bf k}(\tau)$ in the sub-Hubble region, which is given by:
\bea  \Omega^{\rm Sub}_{\bf k}(\tau)&\approx& \frac{3}{2}~k~2^{\displaystyle \left(\frac{2{\cal H}''}{{\cal H}^2}\right)}~\left|\frac{\displaystyle \Gamma\left(\frac{1}{2}+\frac{{\cal H}''}{{\cal H}^2}\right)}{\displaystyle \Gamma\left(\frac{1}{2}\right)}\right|^2\nonumber\\
&=&\frac{3}{2\pi}~k~2^{\displaystyle \left(\frac{2{\cal H}''}{{\cal H}^2}\right)}~\left|\displaystyle \Gamma\left(\frac{1}{2}+\frac{{\cal H}''}{{\cal H}^2}\right)\right|^2\nonumber\\
&\approx &\frac{3}{\pi}~k~\left(1+ 2\ln2~\left(\frac{{\cal H}''}{{\cal H}^2}\right)+\cdots\right)\left[\left(1-2\left(\frac{{\cal H}''}{{\cal H}^2}\right)\right)-\frac{1}{2}\gamma_{\rm E}+\cdots\right]\nonumber\\
&=&\frac{3}{\pi}~k~\left[\left(1-\frac{1}{2}\gamma_{\rm E}\right)+2\left\{\left(1-\frac{1}{2}\gamma_{\rm E}\right)\ln2-1\right\} \left(\frac{{\cal H}''}{{\cal H}^2}\right)-4\ln 2\left(\frac{{\cal H}''}{{\cal H}^2}\right)^2+\cdots\right],~~~~~~~~~~\eea
where $\gamma_{\rm E}$ is the {\it Euler-Mascheroni constant}, which is $\gamma_{\rm E}=0.577$.

Here for the above computation we have used the following important results for the series expansion:
\bea  2^{\displaystyle \left(\frac{2{\cal H}''}{{\cal H}^2}\right)}&=&\left(1+ 2\ln2~\left(\frac{{\cal H}''}{{\cal H}^2}\right)+\cdots\right),\\
\left|\displaystyle \Gamma\left(\frac{1}{2}+\frac{{\cal H}''}{{\cal H}^2}\right)\right|^2 &=&\left[\left(1-2\left(\frac{{\cal H}''}{{\cal H}^2}\right)\right)-\frac{1}{2}\gamma_{\rm E}+\cdots\right].\eea 
\textcolor{Sepia}{\subsection{\sffamily Super-Hubble limiting result}\label{sec:subappendix2}}
Though, we have not explicitly performed any numerical computation using the super-Hubble limiting solution, described by $-k\tau\ll 1$, but for completeness we provide the expression for the dispersion relation in this region. Previously we have only provided the solution of the mode function and its conformal time derivative in the super-Hubble region. 

 In the super-Hubble limit, $-k\tau\ll 1$, it is expected to have very small contribution from the squeezed angle $\phi_{\bf k}(\tau)$, and consequently we have the following conditions:
 \bea \cosh 2\phi_{\bf k}(\tau)\approx 1,~~~~~\sinh 2\phi_{\bf k}(\tau)\approx 2\phi_{\bf k}(\tau).\eea
Consequently, in the limit $\phi_{\bf k}(\tau)\rightarrow 0$, we get the following result for the dispersion relation in the super-Hubble region:
\bea
\label{eq:2xx}
    \Omega^{\rm Sup}_{\bf k}(\tau) &=&\frac{3k}{4} 2^{2\nu_{\rm B}-2}\left|\frac{\Gamma(\nu_{\rm B})}{\Gamma\left(\frac{3}{2}\right)}\right|^2 \Biggl[ \Biggl( \cosh^2 r_{\bf k}(\tau) + \sinh^2 r_{\bf k}(\tau) \Biggr) - \frac{1}{3} \sinh r_{\bf k}(\tau)\Biggr]\nonumber\\
  &=&3k~2^{2(\nu_{\rm B}-2)}~\left|\frac{\Gamma(\nu_{\rm B})}{\Gamma\left(\frac{3}{2}\right)}\right|^2 \Biggl[ \Biggl( \cosh^2 r_{\bf k}(\tau) + \sinh^2 r_{\bf k}(\tau) \Biggr) - \frac{1}{3} \sinh r_{\bf k}(\tau)\Biggr]\nonumber\\
 &=&\Omega^{\rm Sub}_{\bf k}(\tau)\Biggl[ 1+2 \sinh r_{\bf k}(\tau)\left(\sinh r_{\bf k}(\tau)    - \frac{1}{6} \right)\Biggr],
 \eea
 where $\Omega^{\rm Sub}_{\bf k}(\tau)$ is the dispersion relation derived in the previous sub section in the sub-Hubble region. In the super-Hubble region one needs to consider the contributions appearing in the bracketed terms in the above derived expression during the study of the evolution with respect to any dynamical parameters involved in the system. 
 
 One can further consider a more simpler situation in the super-Hubble region, which is described by very small value of the squeezed parameter, $r_{\bf k}(\tau)$ which is fixed by the following contribution:
 \bea \sinh r_{\bf k}(\tau)\approx r_{\bf k}(\tau).\eea
 Here we consider $r_{\bf k}(\tau)$ to be small but not approaching zero and also we neglect the quadratic contribution in $r_{\bf k}(\tau)$ due to smallness approximation. As a result, finally we get the following simplified answer for the dispersion relation in this specific situation:
 \bea
\label{eq:2xx}
    \Omega^{\rm Sup}_{\bf k}(\tau) 
 &\approx &\Omega^{\rm Sub}_{\bf k}(\tau)\left[ 1- \frac{1}{3}r_{\bf k}(\tau)  \right],
 \eea
where one need to consider the contribution from the second term in the evolution equations and this ensures the fact that, $ \Omega^{\rm Sup}_{\bf k}(\tau) \neq \Omega^{\rm Sub}_{\bf k}(\tau)$ in this limiting situation.
 
\textcolor{Sepia}{\subsection{\sffamily Matching condition at the horizon}\label{sec:subappendix3}}
Finally, in this section we have to present the matching condition at the horizon crossing, which is represented by $-k\tau_0=1$ at the time scale $\tau=\tau_0=-k^{-1}$ and this implies at this point the dispersion relation obtained in the sub-Hubble and super-Hubble region has to match. This is given by:
\bea   \Omega^{\rm Sup}_{\bf k}(\tau_0)=  \Omega^{\rm Sub}_{\bf k}(\tau_0),\eea
which further implies the following crucial fact:
\bea \sinh r_{\bf k}(\tau_0)\left(\sinh r_{\bf k}(\tau_0)    - \frac{1}{6} \right)=0.\eea
The above condition satisfy iff we have:
\bea  \sinh r_{\bf k}(\tau_0)=0~~~~\longrightarrow~~~~r_{\bf k}(\tau_0)=n\pi,~~~~~\forall ~n\in\mathbb{Z},\eea
or we have:
\bea \left(\sinh r_{\bf k}(\tau_0)    - \frac{1}{6} \right)=0~~~~\longrightarrow~~~~r_{\bf k}(\tau_0)={\rm sinh}^{-1}\left(\frac{1}{6}\right).\eea
In the above mentioned discussion it is clearly evident that to match the dispersion relation obtained from the sub-Hubble and super-Hubble region at the horizon crossing the squeezed parameter has to be either, $r_{\bf k}(\tau_0)=n\pi,~\forall n\in\mathbb{Z}$, or $r_{\bf k}(\tau_0)={\rm sinh}^{-1}\left(\frac{1}{6}\right)$.

\textcolor{Sepia}{\section{\sffamily Equivalent representations of the evolution equations in two-mode squeezed state formalism}\label{sec:appendixC}}	
In this section, we will discuss about three equivalent representations of the evolution equation of the squeezed parameter and squeezed angle using which one can study the impact of the two mode squeezed state formalism in the present bouncing cosmological set up which is described in the spatially flat cosmological FLRW background. The details of  each of the three representation has been discussed in the following three consecutive subsections respectively.

 \textcolor{Sepia}{\subsection{\sffamily Representation I: In terms of conformal time}\label{sec:subrep1}}

The time evolution equations of the conformal time dependent squeezed state parameter $r_{\bf k}(\tau)$ and squeezed angle $\theta_{\bf k}(\tau)$ are given by:
\begin{align} \label{eq:difftau1}
&\frac{dr_{\bf k}(\tau)}{d\tau} = -\lambda_{\bf k}(\tau) \cos 2\phi_{\bf k}(\tau),\\
\label{eq:difftau2}
&\frac{d\phi_{\bf k}(\tau)}{d\tau} = \Omega_{\bf k}(\tau) -\lambda_{\bf k}(\tau) \coth2 r_{\bf k}(\tau)\sin 2\phi_{\bf k}(\tau) 
\end{align}
The above set of evolution equations, are coupled differential equations of squeezed state parameter $r_{\bf k}(\tau)$ and squeezed angle $\theta_{\bf k}(\tau)$ where in both conformal time derivatives are involved. We choose the initial condition is at the horizon crossing scale, $-k\tau_0=1$ at $\tau=\tau_0$ and also consider the sub-Hubble ($-k\tau\gg 1$) region for the computational purpose, where the scalar modes for two momenta ${\bf k}$ and $-{\bf k}$ having all possible values becomes quantum in nature. Using these information one can numerically solve these equations to construct the target quantum state out of a Gaussian initial state. This will further help us to  numerically compute and understand the quantum complexities in Eq (\ref{eq:compform1}) and Eq (\ref{eq:compform2}) within the framework of primordial cosmological perturbation theory, where the effects of the quantum fluctuations is treated in terms of the squeezed state parameter $r_{\bf k}(\tau)$ and squeezed angle $\theta_{\bf k}(\tau)$ in the {\it squeezed state formalism}.  

Now we will discuss about the strong and the weak coupling region and the behaviour and the physical outcome of these evolution equation:

\begin{enumerate}
\item \underline{\textcolor{red}{\bf Strong coupling region and freeze-out phenomena :}}\\
In the strong coupling region the effect of squeezing phenomena become maximum because:
 \bea &&\lambda_{\bf k}(\tau)\gg \Omega_{\bf k}(\tau),\nonumber\\
 &&\phi_{\bf k}(\tau)\rightarrow {\rm Stable~fixed~point~(freeze~out)},\nonumber\\
 && r_{\bf k}(\tau)\rightarrow {\rm Monotonical~growth~in~time}.\eea
 As a result we get following simplified form of the evolution equations:
 \begin{align} \label{eq:difftau1}
&\frac{dr_{\bf k}(\tau)}{d\tau} = -\lambda_{\bf k}(\tau) \cos 2\phi_{\bf k}(\tau),\\
\label{eq:difftau2}
&\frac{d\phi_{\bf k}(\tau)}{d\tau} = 0.
\end{align}
Consequently, we get the following analytical solution:
\bea &&\phi_{\bf k}(\tau_*)={\rm Constant}\equiv D,\\
        &&r_{\bf k}(\tau)=-\cos D~\int~d\tau'~\lambda_{\bf k}(\tau')\nonumber\\
        &&~~~~~~~~\approx -\cos D~\int~d\tau'~\sqrt{\lambda^2_{\bf k}(\tau')-\Omega^2_{\bf k}(\tau)}.\eea

\item \underline{\textcolor{red}{\bf Weak coupling region and oscillation phenomena :}}\\        
In the weak coupling region the effect of oscillation phenomena become maximum because:        
    \bea &&\lambda_{\bf k}(\tau)\ll \Omega_{\bf k}(\tau),\nonumber\\
 && \lambda_{\bf k}(\tau),\Omega_{\bf k}(\tau),\phi_{\bf k}(\tau)\rightarrow {\rm Constant}.\eea     
 As a result we approximate:
 \bea \tan \phi_{\bf k}(\tau)\approx \cos \beta_{\bf k}\tan\left[-\Omega_{\bf k}\left(\tau-\tau_0\right)+\beta_{\bf k}\right]-\tan\beta_{\bf k}~~~~{\rm where}~~\beta_{\bf k}=\sin^{-1}\left(\frac{\lambda_{\bf k}}{\Omega_{\bf k}}\right)\ll 1.~~~~~~\eea
  As a result we get following simplified form of the evolution equations:
 \begin{align} \label{eq:difftau1}
&\frac{dr_{\bf k}(\tau)}{d\tau} = -\lambda_{\bf k}(\tau) \cos 2\phi_{\bf k}(\tau),\\
\label{eq:difftau2}
&\frac{d\phi_{\bf k}(\tau)}{d\tau} = \Omega_{\bf k}.
\end{align}
Consequently, we get the following analytical solution:
\bea &&\phi_{\bf k}(\tau)=\phi_{\bf k}(\tau_0)+\Omega_{\bf k}(\tau-\tau_0),\\
&& r_{\bf k}(\tau)=r_{\bf k}(\tau_0)-\frac{1}{2}\sin\beta_{\bf k}\sin 2\Omega_{\bf k}(\tau-\tau_0)\nonumber\\
&&~~~~~~~~=r_{\bf k}(\tau_0)-\frac{1}{2}\frac{\lambda_{\bf k}}{\Omega_{\bf k}}\sin 2\left(\phi_{\bf k}(\tau)-\phi_{\bf k}(\tau_0)\right).\eea
For the cosmological models when the modes appearing from the cosmological perturbation lies within the horizon, the above mentioned solutions works perfectly well. On average, the squeezing parameter $ r_{\bf k}(\tau)$ during this time is almost constant and the perturbation do not grow at all.
\end{enumerate}

 \textcolor{Sepia}{\subsection{\sffamily Representation II: In terms of scale factor}\label{sec:subrep2}}
 
 In this section instead of using the conformal time $\tau$ as the dynamical variable, we have chosen the scale factor $a(\tau)$ to make the computation simpler and physically justifiable. To perform the change in variable from $\tau$ to $a(\tau)$ 
we have to replace the following differential operator in the above mentioned evolution equations using the chain rule, as:
\bea \tau\longrightarrow a(\tau):~~~~\frac{d}{d\tau}=\frac{d}{da(\tau)}\frac{da(\tau)}{d\tau}=a'(\tau)\frac{d}{da(\tau)}\eea
Consequently, the evolution of the squeezed state parameter $r_{\bf k}(a)$ and squeezed angle $\theta_{\bf k}(a)$, can be recast in terms of the newly defined dynamical variable $a(\tau)$ as:
\begin{align} \label{eq:diffeqnswa1}
&\frac{dr_{\bf k}(a)}{da} = -\frac{\lambda_{\bf k}(a)}{a'} \cos 2\phi_{\bf k}(a),\\
\label{eq:diffeqnswa2}
&\frac{d\phi_{\bf k}(a)}{da} = \frac{\Omega_{\bf k}}{a'} -\frac{\lambda_{\bf k}(a)}{a'} \coth2 r_{\bf k}(a)\sin 2\phi_{\bf k}(a) 
\end{align}
Once we numerically solve the evolution of the squeezed state parameter $r_{\bf k}(a)$ and squeezed angle $\theta_{\bf k}(a)$ in terms of the scale factor $a$, we can construct the target quantum state out of a Gaussian initial state. This will further help us to  numerically compute and understand the quantum complexities in Eq (\ref{eq:compform1}) and Eq (\ref{eq:compform2}) within the framework of primordial cosmological perturbation theory.
 
  \textcolor{Sepia}{\subsection{\sffamily Representation III: In terms of co-moving Hubble radius/ number of e-foldings}
  \label{sec:subrep3}}

Now if we think about the more realistic cosmological observation then it is not very good to study the evolution with respect to the scale factor, because in the context of realistic cosmology the scale factor is not the direct physical observable that can be probed in the observation for various cosmological missions running (or supposed to run in the near future) to test the signatures of the primordial cosmological paradigm. In this specific situation one needs to use the following transformation for which the linear differential operator appearing in the evolutionary equations of the squeezed parameter and the squeezed angle will be modified as:
\bea a\rightarrow {\cal N}:~~~~~\frac{d}{d\ln a(\tau)}=\left(1-\epsilon(\tau)\right)~\frac{d}{d\ln |aH|}=\left(1-\epsilon(\tau)\right)~\frac{d}{d{\cal N}}=\frac{d}{dN},\eea
where we have used the following couple of facts for the above mentioned transformation:
\bea && dN=d\ln a(\tau),\\
&& d{\cal N}=d\ln|a H|=d\ln|{\cal H}|,\\
&& \frac{d{\cal N}}{dN}=\left(1-\epsilon(\tau)\right),\\
&& \epsilon(\tau)=-\frac{\dot{H}}{H^2}=1-\frac{{\cal H}'}{{\cal H}^2}.\eea
Here, $N$ is the actual number of e-foldings, ${\cal N}$ is the number of e-foldings in terms of the re-defined variables, and $\epsilon(\tau)$ is the slowly varying conformal time dependent parameter. 

Consequently, the evolution of the squeezed state parameter $r_{\bf k}({\cal N})$ and squeezed angle $\theta_{\bf k}({\cal N})$, can be recast in terms of the newly defined dynamical preferred choice of suitable variable ${\cal N}$ as:
\begin{align} \label{eq:diffeqnsnumwa1}
&\frac{dr_{\bf k}({\cal N})}{d{\cal N}} = -\frac{\lambda_{\bf k}({\cal N})}{\left(1-\epsilon(\tau)\right){\cal H}} \cos 2\phi_{\bf k}({\cal N}),\\
\label{eq:diffeqnsnumwa2}
&\frac{d\phi_{\bf k}({\cal N})}{d{\cal N}} = \frac{1}{\left(1-\epsilon(\tau)\right){\cal H}} \left[\Omega_{\bf k}-\lambda_{\bf k}({\cal N}) \coth2 r_{\bf k}({\cal N})\sin 2\phi_{\bf k}({\cal N}) \right].
\end{align}
In this context, $r_{\rm co}=(aH)^{-1}$ or $r_{\rm co}={\cal H}^{-1}$ represents the co-moving Hubble radius, which is extremely important quantity in terms of which the newly re-defined number of e-foldings have been expressed in terms of the good old definition of the number of e-foldings.

\newpage
\bibliography{references_coco}
\bibliographystyle{utphys}

\end{document}